\documentclass[a4paper,11pt]{article}
\usepackage{jcappub,graphicx,float,mathtools,caption,subcaption,cancel}
\usepackage[linktocpage=true]{hyperref}
\hypersetup{colorlinks=true, linkcolor=blue, filecolor=magenta,  urlcolor=blue, citecolor=red,}
\arxivnumber{2604.02240}

\usepackage[normalem]{ulem}
\usepackage{bigints}

 \usepackage{macros}
 \graphicspath{{./figs/}}

\subheader{CERN-TH-2026-081}

\title{Fluid perturbations from expanding bubbles in first-order phase transitions}

\author[a,b]{Chiara Caprini,}
\author[a,1]{Antonino S.~Midiri,\note{Corresponding author: \href{mailto:antonino.midiri@unige.ch}{antonino.midiri@unige.ch}}}
\author[a]{Simona Procacci,}
\author[a,2]{Alberto Roper Pol\,\note{Corresponding author: \href{mailto:alberto.roperpol@unige.ch}{alberto.roperpol@unige.ch}}}

\affiliation[a]{D\'epartement de Physique Th\'eorique, Universit\'e de Gen\`eve. \\ 24 Quai Ernest--Ansermet,
CH-1211 Gen\`eve, Switzerland}

\affiliation[b]{Theoretical Physics Department, CERN. \\ Esplanade des Particules 1, CH-1211 Gen\`eve, Switzerland}

\emailAdd{alberto.roperpol@unige.ch}
\emailAdd{antonino.midiri@unige.ch}
\emailAdd{simona.procacci@proton.me}
\emailAdd{chiara.caprini@cern.ch}

\abstract{
We study the power spectrum of the velocity field induced during a
first-order phase transition occurring in the radiation-dominated era.
We focus on the phase of bubble expansion,
assuming that it ends with the onset of the sound-wave regime.
The main result we present is a refined template for the velocity
spectrum at the beginning of the sound-wave phase,
which can be used for studying the resulting anisotropic stresses
and gravitational wave production.
In particular, we find that the breaks in the velocity spectrum are not
associated to the bubble size and the sound shell thickness, as previously proposed,
but to the position of the discontinuities.
This distinction is particularly relevant for supersonic deflagrations,
as it implies that the intermediate slope is more pronounced and 
the two breaks
are more separated
when the wall velocity approaches the Chapman-Jouget speed, instead of the sound speed.
We also show that the  asymptotic branches of the velocity power
spectrum are  determined by the integral over the single-bubble
profiles at large scales, and by the discontinuities of the velocity profiles at small scales.
Furthermore, we study the dependence of the two breaks and the intermediate slope
on the distribution function of the times of bubble nucleation (exponential
and simultaneous). 
All the results presented in this work have been included in the public Python package CosmoGW.
}

\begin{document}

\maketitle
\tableofcontents

\section{Introduction}
\label{intro}

Phase transitions in the early Universe can occur
when a scalar field undergoes
a process of spontaneous symmetry breaking, presenting a new vacuum state that becomes degenerate to the previous state as
the  critical temperature $T_c$ is reached.
At later times, the temperature continues to drop and the symmetry-broken state corresponds to the true vacuum of the
scalar-field potential.
When the phase transition is of first order,
the potential barrier separating the symmetric and broken phases is crossed
via tunneling, leading to the formation of broken-phase bubbles
\cite{Kirzhnits:1972ut,Coleman:1977py,Linde:1981zj,Steinhardt:1981ct},
which expand and collide with other bubbles, leading to the production of gravitational waves (GWs) 
\cite{Witten:1984rs,Hogan:1986qda,Kosowsky:1991ua,Kosowsky:1992vn,Caprini:2018mtu}.
According to the Standard Model (SM), both the electroweak \cite{Kajantie:1996qd}
and the QCD \cite{Stephanov:2006dn,Aoki:2006we} phase transitions occur as crossovers and,
therefore, they would not lead to significant GW production.
However, first-order phase transitions are found in extensions of the SM that have been proposed,
for example, to provide the required conditions for baryogenesis at the electroweak scale
or to produce particles that could act as dark matter candidates
(see reviews for the electroweak phase transition in 
refs.~\cite{Caprini:2019egz,vandeVis:2025efm,Athron:2023xlk} and references therein).
If this is the case, the resultant GW signal, produced
at the electroweak scale, could be probed by the Laser
Interferometer Space Antenna (LISA) \cite{Caprini:2015zlo,LISA:2017pwj,Caprini:2019egz,LISACosmologyWorkingGroup:2022jok,LISA:2024hlh,Caprini:2024hue}.
Furthermore, if the lepton asymmetry in the early Universe is large, the QCD
phase transition could become of first order
\cite{Schwarz:2009ii,Wygas:2018otj}
and could be probed by Pulsar Timing Array (PTA) experiments \cite{Sazhin:1978myk,Detweiler:1979wn,Deryagin:1986qq}.
The reported evidence for a GW background
detection in 2023 by the NANOGrav \cite{NANOGrav:2023gor,NANOGrav:2023hvm}, the EPTA and InPTA \cite{EPTA:2023fyk, EPTA:2023xxk}, the PPTA \cite{Reardon:2023gzh}, and the CPTA \cite{Xu:2023wog}
collaborations
has led to an increasing interest in looking
for beyond the SM (BSM) physics also around the QCD epoch \cite{Neronov:2020qrl,NANOGrav:2021flc,NANOGrav:2023hvm,Figueroa:2023zhu,Madge:2023dxc,Sesana:2025udx}.

A cosmological first-order phase transition in the radiation-dominated era
features a first stage when scalar field bubbles
in the broken phase nucleate and expand.
As the bubbles expand, the primordial plasma around
the bubbles is perturbed
in regions confined to concentric fluid shells
around the bubble walls \cite{Kosowsky:1992vn,Caprini:2009fx,Espinosa:2010hh},
and the friction with the primordial plasma can make the 
bubble walls reach a terminal velocity $\xi_w$ \cite{Ignatius:1993qn,Bodeker:2009qy,Bodeker:2017cim}.
If the friction is not strong enough, the bubbles might run away
and not reach a terminal
velocity.
Determining which terminal velocity the bubbles reach (if any)
is a complicated task that is under
study in the literature \cite{Bodeker:2009qy,Bodeker:2017cim,Laurent:2020gpg,Laurent:2022jrs,Ai:2023see,Ekstedt:2024fyq,vandeVis:2025plm} but, in general, we expect bubbles in
a first-order electroweak phase transition to reach a terminal velocity \cite{Bodeker:2017cim}.

GWs start to be sourced when bubbles (or the associated
fluid shells perturbed around the bubble walls) collide,
breaking the spherical symmetry of an ensemble of expanding
bubbles.
The GW sources are the anisotropic stresses associated with
the gradients of the scalar field \cite{Witten:1984rs,Hogan:1986qda,Kosowsky:1991ua,Kosowsky:1992vn,Caprini:2007xq,Huber:2008hg,Jinno:2016vai,Jinno:2017fby,Cutting:2018tjt} and with the
fluid perturbations induced in the primordial plasma by the broken-phase bubbles,
in the form of acoustic
\cite{Hindmarsh:2013xza,Hindmarsh:2015qta,Hindmarsh:2016lnk,Hindmarsh:2017gnf,Konstandin:2017sat,Niksa:2018ofa,RoperPol:2019wvy,Hindmarsh:2019phv,Cutting:2019zws,Jinno:2020eqg,Dahl:2021wyk,Jinno:2022mie,RoperPol:2023dzg,Sharma:2023mao,Cai:2023guc,Dahl:2024eup,Caprini:2024gyk,Correia:2025qif,Stomberg:2025kxf} 
and vortical motion
\cite{Kosowsky:2001xp,Nicolis:2003tg,Dolgov:2002ra,Kahniashvili:2005qi,Kahniashvili:2008er,Caprini:2009yp,RoperPol:2018sap,RoperPol:2019wvy,Kahniashvili:2020jgm,RoperPol:2021gjc,RoperPol:2021xnd,Brandenburg:2021bvg,RoperPol:2022iel,RoperPol:2022hxn,Auclair:2022jod,RoperPol:2023bqa}.
Eventually, the phase transition completes, and all the Universe is in the
symmetry-broken phase.
GW production by the bubble wall collisions stops, but
fluid perturbations continue to source GWs
even after the phase transition is completed.
Therefore, unless the amount of energy transferred from the false vacuum
to the fluid is negligible (e.g., when the
bubbles run away), the production of GWs is expected to be dominated by
the fluid perturbations \cite{Kosowsky:1992rz,Caprini:2019egz}.
In the limit of small fluid perturbations,
these can be described as a superposition of sound waves, as found in numerical simulations
of the fluid-scalar field
dynamics
\cite{Hindmarsh:2013xza,Hindmarsh:2015qta,Hindmarsh:2017gnf,Jinno:2020eqg,Jinno:2022mie,Caprini:2024gyk}.
This is the assumption at the base of the sound shell model
\cite{Hindmarsh:2016lnk,Hindmarsh:2019phv}, which has been extensively
used to evaluate the detectability of GWs from first-order phase transitions with different experiments \cite{Caprini:2015zlo,Caprini:2019egz,Giese:2020znk,Hindmarsh:2020hop,NANOGrav:2021flc,Gowling:2021gcy,Giese:2021dnw,LISACosmologyWorkingGroup:2022jok,Gowling:2022pzb,NANOGrav:2023hvm,RoperPol:2023bqa,Figueroa:2023zhu,Caprini:2024hue,Caprini:2024ofd,Ajith:2024mie,Guo:2024gmu,Tian:2024ysd,Tian:2025zlo,LIGOScientific:2025kry}.
In recent work, the GW spectrum from sound waves
was revisited \cite{RoperPol:2023dzg,Sharma:2023mao}, extending its applicability to all frequencies and
including the effect of the Universe expansion \cite{RoperPol:2023dzg,Caprini:2024gyk}.
The sound shell model has also been applied to deviations with respect
to an ultrarelativistic equation of state, i.e., for
$\cs^2 \neq \tfrac{1}{3}$ \cite{Giombi:2024kju,Giombi:2025tkv,Maki:2025ykv},
to include next-to-leading
order corrections due to curvature perturbations \cite{Giombi:2025tkv},
and to inverse phase transitions \cite{Barni:2024lkj,Barni:2025gnm}.
Since the GW spectrum in the sound-wave regime
increases with the sourcing time,
it is usually assumed to saturate at the time when nonlinearities develop
\cite{Hindmarsh:2013xza,Caprini:2019egz}.
Indeed, due to the expected formation of shocks
\cite{Pen:2015qta}, nonlinearities leading to
vorticity become relevant and eventually
lead to the development of turbulence \cite{Caprini:2006jb,Caprini:2009yp,Brandenburg:2016odr,Brandenburg:2017neh,RoperPol:2025lgc}.
Therefore, at the time of shock formation or
as the strength of the phase transition increases, nonlinearities are expected
to arise, leading to a decay of the total kinetic energy
that has an 
impact on the resulting GW amplitude \cite{Cutting:2019zws,Dahl:2021wyk,Caprini:2024gyk,Dahl:2024eup,Correia:2025qif}.
Furthermore, due to the interactions between the scalar field and the fluid,
numerical simulations show that the bubbles might slow
down before colliding in strong deflagrations,
leading to a suppression of the GW spectrum \cite{Cutting:2019zws,Correia:2025qif},
which has been included in
GW studies in refs.~\cite{Gowling:2021gcy,Gowling:2022pzb,Maki:2025ykv}.

In this work,
which is a follow up of ref.~\cite{RoperPol:2023dzg},
we investigate the properties of the velocity perturbations
and of the bubble nucleation history
that ultimately will determine the
GW spectrum from sound waves in the linear regime of fluid perturbations.
We investigate the origin of the
double broken-power-law
structure
reported in previous work
\cite{Hindmarsh:2013xza,Hindmarsh:2015qta,Hindmarsh:2016lnk,Hindmarsh:2017gnf,Hindmarsh:2019phv,Jinno:2020eqg,Jinno:2022mie,RoperPol:2023bqa,Sharma:2023mao,Cai:2023guc,Caprini:2024gyk,Stomberg:2025kxf}.
We focus on the characteristic scales of the
velocity spectrum at the initial time
of the sound-wave regime,
which will then enter in the GW spectrum via the
associated anisotropic stresses.
Here, we show how the velocity spectrum scales arise from the properties of the single-bubble
profiles \cite{Espinosa:2010hh}.
As sound waves develop and evolve in time,
the initial velocity
perturbations convert into both velocity and density
perturbations.
The resulting impact of the density perturbations, and the precise relation between the scales of the GW spectrum and those of the fluid velocity
spectrum will be presented
in a separate publication \cite{part2}.
A thorough study of the asymptotic
limits of the velocity spectrum provides us analytical expressions of its
different amplitudes, which are
based on the discontinuities of the velocity profiles,
and allows us to make predictions on the spectral breaks and the intermediate
slope.
We also investigate and reconcile different findings for the
tensor structure of the two-point correlation function of the velocity field
in previous work \cite{Caprini:2007xq, Jinno:2016vai}.
Our analytical and numerical templates can be used to reconstruct
the velocity spectrum arising from first-order phase transitions
during the bubble expansion phase and during the sound-wave regime, according
to the sound shell model.
We include our models and computations in the public Python package {\sc CosmoGW}
\cite{CosmoGW_GH,cosmogw_manual}, together with a tutorial to reproduce
the results presented in this work and in ref.~\cite{RoperPol:2023dzg}.

\paragraph{Outline.}
In \Sec{sec_1bubble}, we
study the hydrodynamics of the first stage of
uncollided expanding bubbles
and review the fluid self-similar single-bubble profiles in \App{1d_profiles}.
We then characterize
the function ${f'}^2$ that corresponds to the
velocity power spectral density before collisions.
This function is crucial since it also determines the
velocity spectrum after collisions and, hence, the GW spectrum
\cite{Hindmarsh:2019phv,RoperPol:2023dzg}.
In \Sec{subsec_RiemannL}, we investigate the asymptotic limits of ${f'}^2$ and relate them to
the discontinuities that arise in the fluid self-similar profiles.
We then provide single and double broken power law templates for
${f'}^2$ in \Sec{sec:f_template}.
We show in \Sec{sec:toymodel} that quadratic and linear-constant toy models
(see also \App{appendix_toy})
can accurately reproduce
the relevant spectral features of ${f'}^2$ for subsonic deflagrations
(quadratic), hybrids (linear-constant), and detonations (quadratic).
The analytical study of the correlation and spectral functions of the velocity field produced by the superposition of bubbles is then provided in
\Sec{kinetic_sp_bubbles}.
We review the general properties of the correlation and spectral functions of homogeneous and isotropic
fields in \App{appendix_correlation}.
In particular, we study the relations between the velocity field correlation function
in coordinate space and its power spectral density in momentum space,
and test results that apply for generic
irrotational fields to our case.
In \Secs{bij_realspace}{FL_momentum}, we take
the ensemble average over randomly distributed locations of bubble nucleations,
leading to a statistically homogeneous and isotropic velocity field.
In \Sec{ensemble_times}, we study the time
evolution of the power spectral density
of the velocity field, considering two common distributions for the
times of bubble nucleation (exponential and simultaneous).
We review the details and present computations for
these distributions in \App{time_dist}.
To conclude the treatment of the bubble expansion phase, we show in \Sec{anis_stress1} how spherical symmetry implies that the anisotropic stresses that source GWs before collisions are identically zero.
In \Sec{across_collisions},
we deal with energy conservation and the evolution
of the kinetic energy density in the limit of small perturbations,
for a single expanding bubble in \Sec{vrms_single}, and
for multiple bubbles in \Sec{velocity_rms_mult}.
Noting that the kinetic
energy density during the sound-wave regime in the sound shell model takes
the same value as for a single bubble, in \Sec{EffCollTime} we define
an effective starting time of the sound-wave regime,
$t_{\rm sw}$.
In \Sec{FL_template}, we propose an approach based on this deterministic time
to compute the velocity spectrum
and compare it to the one obtained in the sound shell model,
which models collisions via the introduction of a bubble lifetime
distribution function (see \Sec{distribution_lifetimes} for details).
We present the analytical asymptotic limits
in \Sec{asymptotic_FL}, and provide in \Sec{template_FL} the position of the spectral
breaks at the beginning of the sound-wave regime, as a function
of the breaks found in \Sec{subsec_RiemannL}, and the
considered nucleation time distributions.
We then apply the single and double broken power law fits, already used in \Sec{sec:f_template}
for ${f'}^2$, to the power spectral density of the velocity at the beginning
of the sound-wave regime.
The result is an analytical template that minimally depends on the properties
of the single-bubble fluid profiles and the nucleation time statistics.
The velocity spectrum eventually determines the anisotropic stresses sourcing GWs in the sound-wave
regime \cite{Hindmarsh:2016lnk,Hindmarsh:2019phv,RoperPol:2023dzg,Sharma:2023mao,
Giombi:2024kju,Giombi:2025tkv,Barni:2025gnm,part2}. 
We will explore how the results
presented in this work impact the resulting GW spectrum 
in a separate publication \cite{part2}.
Finally, we present our conclusions in \Sec{concls}.

\paragraph{Fourier convention.} In this paper, we use the following Fourier convention for a generic field $u(\xx)$,
\begin{equation}
    u (\kk) = \int u (\xx) \, e^{i \kk \cdot \xx} \dd^3 \xx\,,
    \quad \quad u (\xx) = \int \frac{u(\kk)}{(2 \pi)^3} \,
    e^{-i\kk \cdot \xx} \dd^3 \kk \,,
    \label{fourierconvention}
\end{equation}
distinguishing fields in coordinate and Fourier space only by their arguments.

\section{Fluid perturbations from expanding bubbles}\label{sec_1bubble}

In this section, we focus on characterizing the
initial perturbations induced in the fluid
around expanding uncollided broken-phase bubbles.
The analytical study of the velocity field correlations in this regime
(see \Sec{kinetic_sp_bubbles}) and across collisions
is a crucial step towards an accurate estimate of the velocity
spectrum in the sound-wave regime \cite{Hindmarsh:2016lnk,Hindmarsh:2019phv,RoperPol:2023dzg,part2}
and the resulting GW production after percolation \cite{Caprini:2019egz,Hindmarsh:2020hop,Athron:2023xlk,Caprini:2024hue,Caprini:2024ofd}.
In \Sec{vel_1d}, we describe the velocity and enthalpy fields induced by a superposition of expanding bubbles and investigate the
Fourier transform of the single-bubble profiles in \Sec{vel_1d_fourier}.
The radial profiles are computed following ref.~\cite{Espinosa:2010hh},
reviewed in \App{1d_profiles}, using {\sc CosmoGW} \cite{cosmogw_manual,CosmoGW_GH}.
In \Sec{subsec_RiemannL}, we study the asymptotic limits of the
function $f'$, which describes the Fourier transform of the velocity field,
and show that they are related to the presence of discontinuities in the velocity profiles,
and to the causality properties of the velocity field.
The function $f'$ is a key element since, as we demonstrate in \Sec{kinetic_sp_bubbles},
it determines the shape of the velocity spectral density.
We provide in \Sec{sec:f_template} templates for the envelope
of ${f'}^2$ in terms of single and double broken power laws.
The properties of $f'$ can be recovered using simplified toy models for the velocity profiles. 
We
present the analytical calculations related to these toy models in \Sec{sec:toymodel}:
they are expressed in terms of simple analytical functions that can be found by fixing
a few constants given the
single-bubble velocity profiles.
In \Sec{FL_template}, we evaluate how the scales and slopes that appear in the templates of ${f'}^2$
determine the power spectral density of the velocity field at the starting
time of the sound-wave regime, $\FL$. 

Denoting respectively as $p$, $e$, and $w = p + e$ the pressure, energy density,
and enthalpy of the fluid, the energy-momentum tensor of the cosmological fluid, $T_\munu$,
under the assumption of local thermal equilibrium, takes the perfect fluid (pf) form, 
\begin{equation}
    T_\munu^{\rm pf} = w \, U_\mu U_\nu + p \, g_\munu \,,
    \label{perf_fluid}
\end{equation}
where the metric tensor $g_{\mu \nu}$ reduces, in flat space-time, to the Minkowski one
$\eta_{\mu \nu} = {\rm diag} \{-1, 1, 1, 1\}$.
The 4-velocity is $U^\mu \equiv \gamma (1, \vv)$, 
where $\vv \equiv \dd \xx(t)/\dd t$ 
is the fluid velocity, with $\xx (t)$ representing the local position of a fluid parcel,
and
$\gamma \equiv (1 - v^2)^{-1/2}$ is the Lorentz factor.
The normalization condition is $U^\mu \, U_\mu = -1$.

Before bubble collisions, the total energy-momentum tensor of the
primordial plasma includes both the perfect fluid and the scalar
field contributions,
\begin{align}
    T_{\mu \nu} = w
    \, U_{\mu} U_{\nu}  + p \, g_{\mu \nu} \, +  \partial_{\mu} \varphi \,
    \partial_{\nu} \varphi  \, - \half g_{\mu \nu}\,
    \partial_{\alpha} \varphi \, \partial^{\alpha} \varphi\,,
\end{align}
where $\varphi$ describes the order parameter driving the phase transition,
$w=T \frac{\partial p}{\partial T} $ is the enthalpy, and $p=  - V(\varphi, T)$ is the pressure. 
The effective potential $V(\varphi, T)$ is often decomposed as the sum of the tree-level zero-temperature part $V_0(\varphi)$ and the thermal part $V_{T}(\varphi, T)$.
To model the fluid perturbations due to the 
expansion of bubbles, we make the following simplifying assumptions:
\begin{itemize}
    \item We assume that {\em the duration of the phase transition is short} compared to the time scale given by the Hubble rate $H_*$, i.e., $\beta/H_* \gg 1$.
    The characteristic duration of the phase transition is estimated as the inverse
    of the bubble nucleation rate $\beta$ \cite{Caprini:2019egz}
    (see \App{time_dist}
    for details on the nucleation time distribution functions).
    Therefore, we fix $a\approx a_*$ and $H\approx H_* \ll \beta$,
    and we neglect the expansion of the Universe
    in the bubble expansion phase.
    Hence, the conservation of the energy-momentum tensor is described in flat space-time $\partial_\mu T^{\mu\nu} = 0$.
    \item We restrict ourselves to {\em the bag equation of state}
    \cite{Chodos:1974je,Espinosa:2010hh} to describe the
    pressure and energy density of the perfect fluid.
    In the bag equation of state,
    the pressure is described with a thermal
    contribution from radiation and a vacuum
    contribution
    from the scalar field,
    \begin{equation}
        p_\pm = - V (\varphi_\pm, T_\pm)  = \tfrac{1}{3} a_\pm T_\pm^4 - \epsilon_\pm\,,
        \label{bag_eos_intro}
    \end{equation}
    where $T$ is the local temperature of the plasma and
    $\epsilon_{\pm}$ represents the vacuum energy in  the broken ($-$) and the symmetric ($+$) phases.
    In the bag equation of state, it is further assumed that $\epsilon_-=0$ and $\epsilon_+=\epsilon$,
    where the bag constant $\epsilon$ only depends on the background nucleation
    temperature $T_n$ and characterizes the difference in the effective potential between the broken
    and symmetric phases.
    The enthalpy and energy density are parameterized as
    \begin{equation}
    w_{\pm} = T_\pm \frac{\partial p_\pm}{\partial T} = \tfrac{4}{3} a_{\pm} T^{4}_{\pm}\,, \qquad 
    e_{\pm} = w_\pm - p_\pm = a_\pm T_\pm^4 + \epsilon_\pm\,.
    \end{equation}
    Under this assumption, we restrict the dynamics of the scalar field to the
    bubble wall thickness, assuming that, out of the bubble wall, the scalar field takes
    the value corresponding to either the broken or the symmetric phase.
    Hence, we will describe the fluid perturbations at both sides of the bubble walls considering
    the energy-momentum of the perfect fluid given in \Eq{perf_fluid},
    since the gradients of the scalar field vanish
    outside the bubble walls, and we will treat the bubble walls as a discontinuity in the
    fluid solutions (see \App{1d_profiles} for details).
\end{itemize}

\subsection{Superposition of single-bubble profiles}
\label{vel_1d}

In a first-order phase transition, once the broken-phase bubbles are nucleated, 
they expand due to the potential difference between
the true and the false vacua, described by the bag constant in the bag equation
of state.
The friction exerted by the plasma on the bubbles can balance this 
effect, resulting in a steady expansion at a terminal bubble wall velocity $\xi_w$.
In this case, the velocity and enthalpy perturbations of the plasma around the single-bubble
wall reach self-similar radial profiles, $v_\ip (\xi)$ and $w_\ip (\xi)$, where 
$\xi \equiv r^{(n)}/t^{(n)}$ is the self-similar variable,
with $r^{(n)}\equiv |\xx-\xx_0^{(n)}|$ being the radial distance to the nucleation center
of the $n$-th bubble $\xx_0^{(n)}$ and $t^{(n)} \equiv t - t_0^{(n)}$ being
the respective duration since its nucleation time
$t_0^{(n)}$ \cite{Espinosa:2010hh}.
The subindex ip indicates the ``individual profiles'' induced by a single expanding bubble.

We review the calculation of the radial profiles $v_{\rm ip} (\xi)$ and $w_\ip (\xi)$ in \App{1d_profiles},
following the description of refs.~\cite{Espinosa:2010hh,Hindmarsh:2019phv}.
As discussed in \Sec{intro}, since the duration of the phase transition
is assumed to be much shorter than the Hubble time, $\beta/H_* \gg 1$, we neglect
the expansion of the Universe in the phase of bubble expansion.
We provide, as part of the public Python package {\sc CosmoGW}\footnote{\url{https://cosmogw-manual.readthedocs.io/}} \cite{CosmoGW_GH,cosmogw_manual},
the {\tt hydro\_bubbles} library and a tutorial to compute
the self-similar profiles for different wall velocities and strengths of the phase transition,
together with the calculations required to compute the velocity spectrum presented in this work,
as well as the resulting GW spectrum, following
the description of ref.~\cite{RoperPol:2023dzg}.
We note that other numerical tools for the computation of the velocity profiles, as well as of related quantities including the GW
spectrum from sound waves in the sound shell model,
are available in the literature
(cf.~PTtools\footnote{\url{https://pttools.readthedocs.io/}} \cite{Maki:2025ykv}).

The resulting velocity and enthalpy fields can be expressed as the superposition of
the fluctuations induced by $N_b$ individual expanding
spherically symmetric
bubbles in a volume $V$ \cite{Hindmarsh:2016lnk},
\begin{align}
    \vv (t,\xx) = \sum_{n=1}^{N_b} \vv^{(n)}(t,\xx) = \sum_{n = 1}^{N_b}
    \hat \rr^{(n)} v_\ip (\xi)  \,, \qquad
    w (t,\xx) =
    \sum_{n = 1}^{N_b} w_\ip (\xi) \, ,
    \label{super_vel_w}
\end{align}
where we stress that $\xi \equiv r^{(n)}/t^{(n)}$ depends on the bubble $n$.

Since the bubble wall thickness is usually several orders of magnitude smaller
than the characteristic scale of the fluid \cite{Ignatius:1993qn,Espinosa:2010hh},
the fluid fields are approximated to be discontinuous across the phase boundary.
The boundary conditions at each side of the bubble wall depend on the specific value of
the wall velocity and the strength of the phase transition, leading to three types of solutions: 
subsonic deflagrations, supersonic deflagrations (hybrids), and detonations \cite{Espinosa:2010hh}.
The matching conditions across the bubble wall and the different types of
hydrodynamical solutions are respectively presented in \Secs{matching}{class_sols}.

\begin{figure}[t]
    \centering
    \includegraphics[width=.495\textwidth]{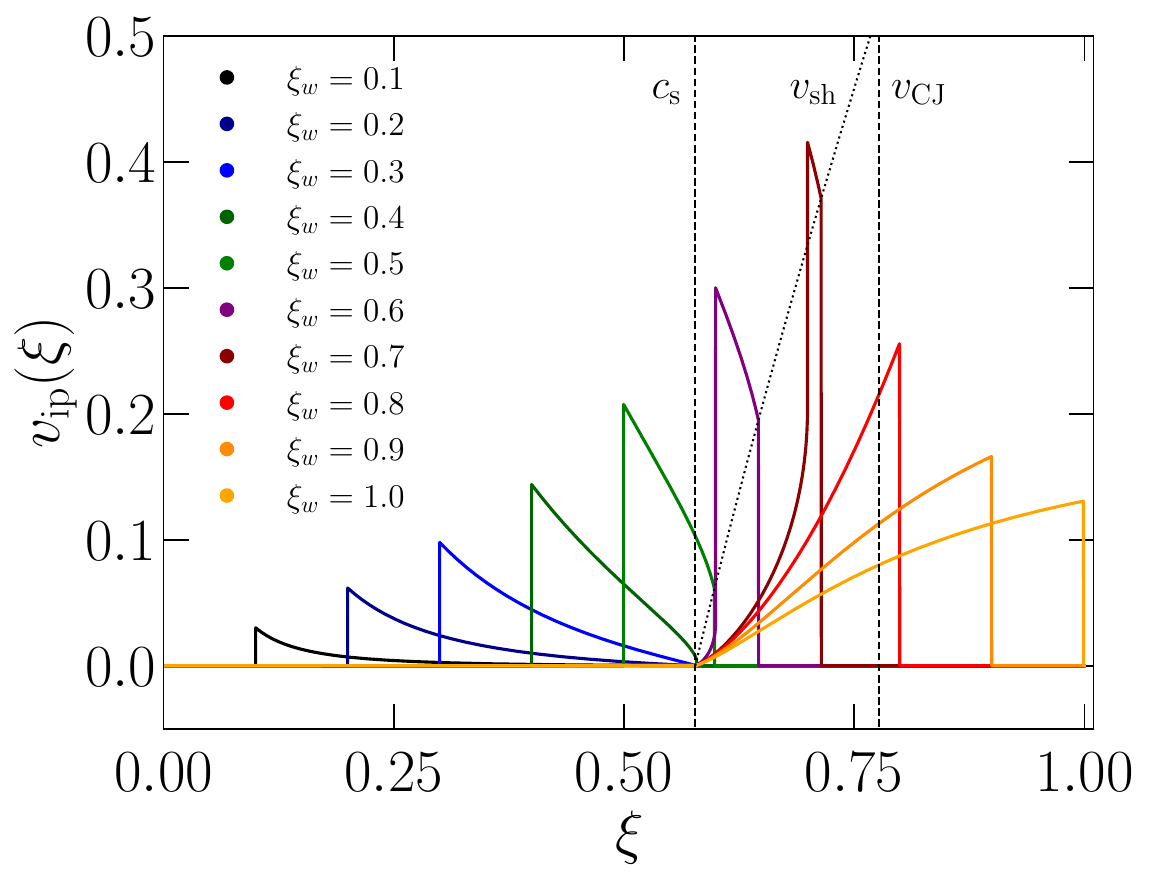}
    \includegraphics[width=.49\textwidth]{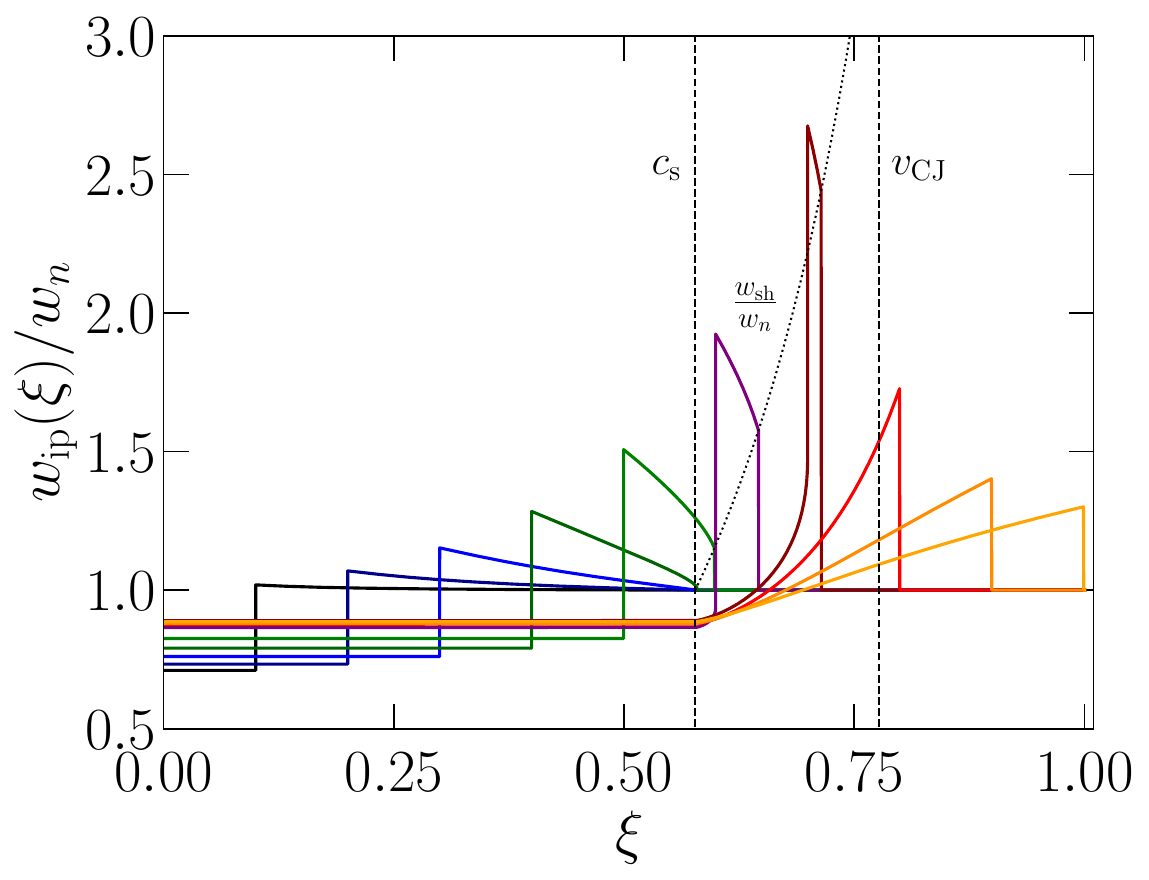}
    \caption{
    Solid lines illustrate the radial
    velocity $v_\ip (\xi)$ (left)
    and enthalpy $w_\ip (\xi)$ (right) profiles
    for different values of the bubble wall velocity $\xi_w$
    and the benchmark phase transition strength $\alpha = 0.1$.
    Vertical black, dashed lines are at the sound speed
    $\cs = \tfrac{1}{\sqrt{3}}$ and at the
    Chapman-Jouget speed $\vCJ$ [see \Eq{chapman2}],
    indicating the boundaries between
    subsonic deflagrations, hybrids, and detonations.
    Dotted lines correspond to the positions $\xi_\sh$ where a shock forms
    for deflagration solutions [see \Eqs{cond_shock}{cond_shock_w}].
    }
    \label{fig_ip}
\end{figure}

The radial profiles of the velocity and enthalpy fields, $v_{\rm ip} (\xi)$ and $w_\ip (\xi)$,
for the different types of solutions are shown in \Fig{fig_ip} for a benchmark value of $\alpha = 0.1$
and a range of wall velocities $\xi_w$.
The parameter $\alpha \equiv \tfrac{4}{3} \, \Delta \theta/w_n = \tfrac{4}{3} \, \epsilon/w_n$
characterizes the strength of the phase transition
in terms of the energy-momentum tensor trace anomaly,
$\Delta \theta\equiv \fourth \eta_\munu (T^\munu_+ - T^\munu_-) =
\fourth (\Delta e - 3 \Delta p) = \epsilon$, and $w_n$ is the
background enthalpy of the fluid at the nucleation temperature $T_n$ \cite{Caprini:2019egz}.
The profiles are computed using the bag equation of state as discussed above.

\begin{figure}[t]
    \centering
    \includegraphics[width=0.6\linewidth]{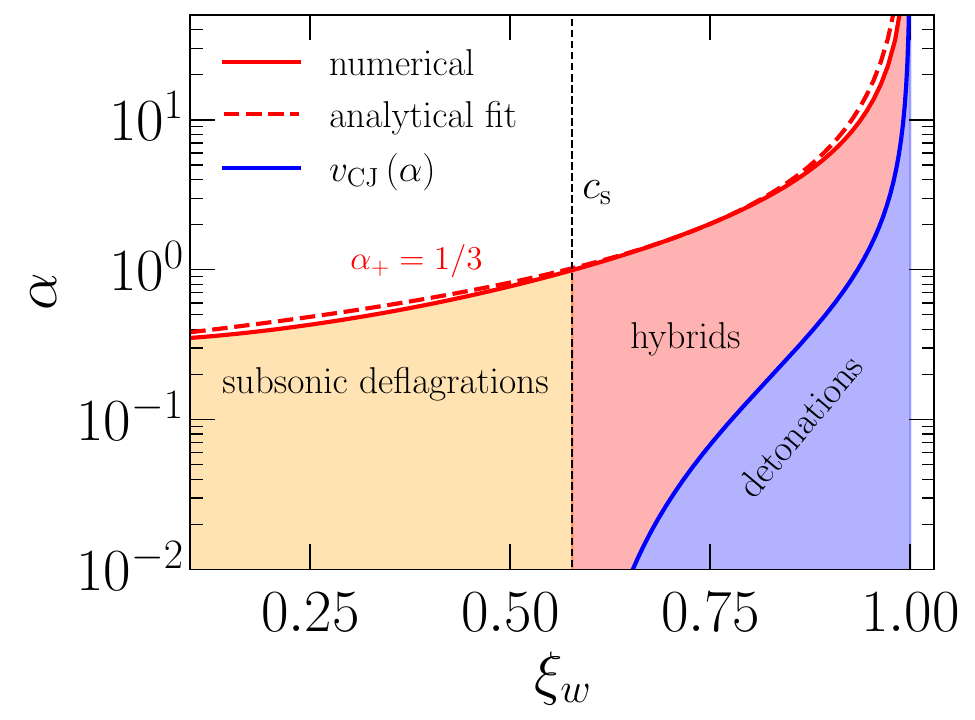}
    \caption{Regions of allowed solutions for the phase transition strength $\alpha$, evaluated at the nucleation temperature and the wall velocity $\xi_w$.
    The upper bound on $\alpha$ corresponds to $\alpha_+ = \tfrac{1}{3}$,
    being the largest value allowed for deflagrations (see discussion in \App{1d_profiles}).
    This bound is computed numerically and 
    compared to the analytical fit, $\alpha_{\rm max} \simeq \onethird (1 - \xi_w)^{-13/10}$ \cite{Espinosa:2010hh}. The boundary between subsonic and supersonic deflagrations is given by the dashed
    vertical line, indicating the speed of sound, $\cs = \tfrac{1}{\sqrt{3}}$.
    The boundary between supersonic deflagrations (hybrids) and detonations is indicated by the Chapman-Jouget speed, $\vCJ
    \, (\alpha)$ [see \Eq{chapman2}].}
    \label{limit_alphas}
\end{figure}

\FFig{limit_alphas} 
shows the different types of solutions that are
allowed as a function of $\alpha$ and $\xi_w$.
The boundary between subsonic and supersonic deflagrations is indicated
by the speed of sound, $\cs$, while the Chapman-Jouget speed, which
depends on $\alpha$ [see \Eq{Chapman}],
\begin{equation}
    \vCJ (\alpha) = \frac{1 + \sqrt{\alpha \, (2 + 3 \alpha)}}
    {\sqrt{3} \, (1 + \alpha)}\,,
    \label{chapman2}
\end{equation}
determines the boundary between supersonic deflagrations and
detonations.
As previously indicated in ref.~\cite{Espinosa:2010hh}, deflagrations (either subsonic
or supersonic)
are only allowed when $\alpha_+ \leq \tfrac{1}{3}$,
where $\alpha_+ \equiv \tfrac{4}{3} \, \Delta \theta/w_+ = \tfrac{4}{3} \, \epsilon/w_+$,
being $w_+$ the enthalpy evaluated
immediately in front of the bubble wall (see details in \App{1d_profiles}).
The energy released by the phase transition
can increase the enthalpy in the perturbed regions.
Therefore, in general $w_+ \geq w_n$, only having $w_+ = w_n$
for detonations as the region outside the
bubbles is not perturbed in this case.
This implies that $\alpha \geq \alpha_+$ and the upper bound on $\alpha_+ \leq \onethird$
corresponds to an upper bound on $\alpha < \alpha_{\rm max} \simeq \onethird (1 - \xi_w)^{-13/10}$, found in ref.~\cite{Espinosa:2010hh}
(see \Fig{limit_alphas} for a comparison between the numerically computed $\alpha_{\rm max}$ and this analytical fit).

\subsection{Self-similar fluid profiles in Fourier space: the $f' (z)$ function}
\label{vel_1d_fourier}

The Fourier transform
of the velocity field of a single expanding bubble,
$v_i^{(n)} (t, \xx) = \hat r_i^{(n)} v_\ip (\xi)$, is
\begin{align}
    v_i^{(n)}(t, \kk) =
    \int v_i^{(n)}(t,\xx) \, e^{i\kk \cdot\xx} \dd^3 \xx \label{vel_fz_fourier} =
    [t^{(n)}]^3 e^{i\kk\cdot\xx_0^{(n)}}
    \int \hat \xi_i \, v_\ip (\xi) \, e^{i {\pmb z} \cdot {\pmb \xi}} \dd^3 {\pmb \xi}   \,, 
\end{align}
where $z_i\equiv t^{(n)}\, k_i$, $z = |\zz|$, $\xi_i = r_i^{(n)}/t^{(n)}$,
$\rr = \xx - \xx_0^{(n)}$,
and $\hat \xi_i = \xi_i/\xi = \hat r_i^{(n)}$.
Using partial integration, we can write \cite{Hindmarsh:2016lnk,Hindmarsh:2019phv}
\begin{equation}
    v_i^{(n)}(t,\kk) = - 
    i\, [t^{(n)}]^3 \, e^{i\kk\cdot\xx_0^{(n)}} \, \hat{k}_i \, f'(z) \,.
    \label{velo_fz}
\end{equation}
The function $f(z)$
corresponds to the following one-dimensional integral
along the radial coordinate $\xi$,
\begin{align}
    f(z)  = 
    4 \pi \int_0^\infty j_0 (z\xi) \, \xi \, v_\ip (\xi) \dd \xi
    \quad \Rightarrow
    \quad
    f'(z) = -4 \pi \int_0^\infty j_1 (z\xi) \, \xi^2 \, v_\ip (\xi) \dd \xi \,,
    \label{def_fpz}
\end{align}
where $j_0(x) = \sin x/x$ and $j_1(x) = \sin x/x^2 - \cos x/x$ are spherical
Bessel functions of the first kind.
The velocity field has compact support in $\xi \in (\xib, \xif)$,
where $\xif$ and $\xib$ indicate the front and back positions
of the fluid shell, such that $v_\ip = 0$ at $\xi \in (0, \xib) \cup (\xif, 1)$ (see \Fig{fig_ip}).
Therefore, the integral in \Eq{def_fpz} always converges.

As discussed in \Sec{class_sols}
and summarized in \Tab{tab:scales_fpz},
$\xib = \min(\xi_w, \cs)$
in general.
For detonations, $\xif = \xi_w$,
whereas
for deflagrations,
$\xif = \xi_\sh$,
being $\xi_\sh$ the position
where
a shock forms
(see the velocity profiles in \Fig{fig_ip}).
While detonations only present one discontinuity
at $\xi_w$,
all deflagrations present two discontinuities: one discontinuity at $\xi_w$ and a second
discontinuity at $\xi_\sh$.
For subsonic deflagrations with $\xi_w \lesssim \tfrac{1}{2} \, \vCJ$, while
a shock still
forms at $\xi_\sh \simeq \cs$,
the jump in the velocity
becomes negligibly small.

\begin{table}[b!]
    \centering
    \caption{Discontinuities in the velocity profiles and their relation to the scales of $\fpsq$
    [see \Eqs{fit_z1}{fit_z2}].
    }
    \label{tab:scales_fpz}
    \resizebox{\columnwidth}{!}{
    \begin{tabular}{lcccccc}
        \hline
        & $\xif$ & $\xib$ & $\xi_w$ discontinuity? & $\xi_{\mathrm{sh}}$ discontinuity? &
        $\xi_1$ & $\xi_2$ \\
        \hline
        Subsonic deflagrations & $\xi_\sh \geq \cs$ & $\xi_w$ & yes & yes & $\tfrac{4}{3} \, (\xi_\sh+\xi_w)$
        & $2\, (\xi_\sh-\xi_w)$ \\
        Hybrids & $\xi_\sh > \cs$ & $\cs$ & yes & yes & $\tfrac{4}{3} \, (\xi_\sh+\xi_w)$
        & $2 \, (\xi_\sh-\xi_w)$ \\
        Detonations & $\xi_w$ & $\cs$ & yes  & no & $ \tfrac{5}{3}\, (\xi_w + \cs)$ & $\tfrac{4}{3}\, (\xi_w - \cs)$ \\
        \hline
    \end{tabular}
    }
\end{table}

The function $f'(z)$ is a crucial element in understanding the resulting
fluid perturbations.
As we will see in \Sec{kinetic_sp_bubbles}, ${f'}^2 (z)$ 
is involved in determining the velocity field
spectrum during the bubble expansion phase.
Furthermore, ${f'}^2(z)$ also enters in the velocity spectral density of the sound-wave phase,
according to the sound shell model \cite{Hindmarsh:2016lnk,Hindmarsh:2019phv,RoperPol:2023dzg,Giombi:2024kju,Giombi:2025tkv}.
As a consequence, the resulting GW spectrum is affected by the
characteristic scale dependence of ${f'}^2(z)$.
In particular,
it is usually argued that the GW spectrum produced by sound waves
is characterized by two length scales: the mean size of the bubbles at collisions,
$R_\ast$, and the fluid shell thickness $R_\ast \Delta \xi = R_\ast(\xif - \xib)$,
leading to a two-scale
structure of the spectral shape
\cite{Caprini:2007xq,Jinno:2016vai,Jinno:2017fby,Hindmarsh:2013xza,Hindmarsh:2015qta,Hindmarsh:2016lnk,Hindmarsh:2017gnf,Hindmarsh:2019phv,Hindmarsh:2020hop,Jinno:2020eqg,Cai:2023guc,RoperPol:2023bqa,RoperPol:2023dzg,Caprini:2024gyk,Caprini:2024hue,Caprini:2019egz,Caprini:2015zlo,Athron:2023xlk,Stomberg:2025kxf,Jinno:2022mie,Sharma:2023mao,Dahl:2021wyk,Dahl:2024eup}.
Motivated by understanding this structure,
we investigate the velocity field spectrum
and its characteristic scales,
finding that, in general,
the latter differ from
$R_\ast$ and $R_\ast \Delta \xi$.

\subsection{Asymptotic limits of $f' (z)$}
\label{subsec_RiemannL}

In this section, we analyze the behavior of $f'(z)$ and determine the scales at which its slope changes.
We first study the asymptotic large-scale ($z \to 0$) and small-scale ($z \to \infty$) limits of
$f'(z)$,
and investigate at which scales $z_1$ and $z_2$ do these limits 
respectively take over.
In \Sec{sec:f_template},
we then provide a template for the envelope
of the $\fpsq$ function.
We find that, in general,
$z_2$ is determined
by a new scale 
$\tilde \Delta \xi$.
We confirm that,
for detonations,
this is indeed the fluid shell thickness,
$\tilde \Delta \xi = \Delta \xi \equiv \xif - \xib$.
However,
for deflagrations, we find that $\tilde \Delta \xi$ is instead
the distance between discontinuities
in the velocity profile,
i.e., 
$\tilde \Delta \xi = \xi_\sh - \xi_w$.
Note that $\Delta \xi$ and $\tilde \Delta \xi$ are equal
for subsonic deflagrations, 
but differ for hybrids (see \Tab{tab:scales_fpz}).
Furthermore, we find that,
contrary to what has been
proposed previously in the literature,
also $z_1$
depends on the self-similar profiles via
the linear combination
$\xi_+ = \xif + \xib = \xi_w + \cs$ for detonations and
$\xi_+ = \xif+\xi_w = \xi_\sh + \xi_w$ for deflagrations.
In general, when two discontinuities are present, as in deflagrations, $z_1$ and $z_2$ are determined by the positions
of these discontinuities, $\xi_w$ and $\xi_\sh$. 
For subsonic deflagrations, these correspond to the positions of the front ($\cs$) and the back ($\xi_w$) of the fluid shell, while for supersonic deflagrations, the back of the shell ($\cs$) differs from the position of the discontinuity ($\xi_w$).
When only one discontinuity is present, as in detonations, $z_1$ and $z_2$ are determined by the positions of the front ($\xi_w$) and the back ($\cs$) of the fluid shell.
Then, $z_1$ is given by the sum of these scales and $z_2$ by their difference.
Based on these findings, we 
provide analytical templates of the function $f'(z)$ that depend on
$z_1$ and $z_2$.
In \Secss{kinetic_sp_bubbles}{FL_template} we will demonstrate precisely how $f'(z)$ determines the
velocity field power spectrum
at the initial time
of the sound-wave regime.
As we will see, an average over both  bubble nucleation 
locations and times is necessary. 
It is only after this averaging procedure, that the scales $z_1$ and $z_2$ can be related to their corresponding wavenumbers $k_1$ and $k_2$, and therefore can be compared to averaged quantities like, e.g., $R_*$. 

\subsubsection{Large-scale limit}
\label{fpz_large_scales_limit}

\begin{figure}[t]
    \centering
    \includegraphics[width=.32\textwidth]{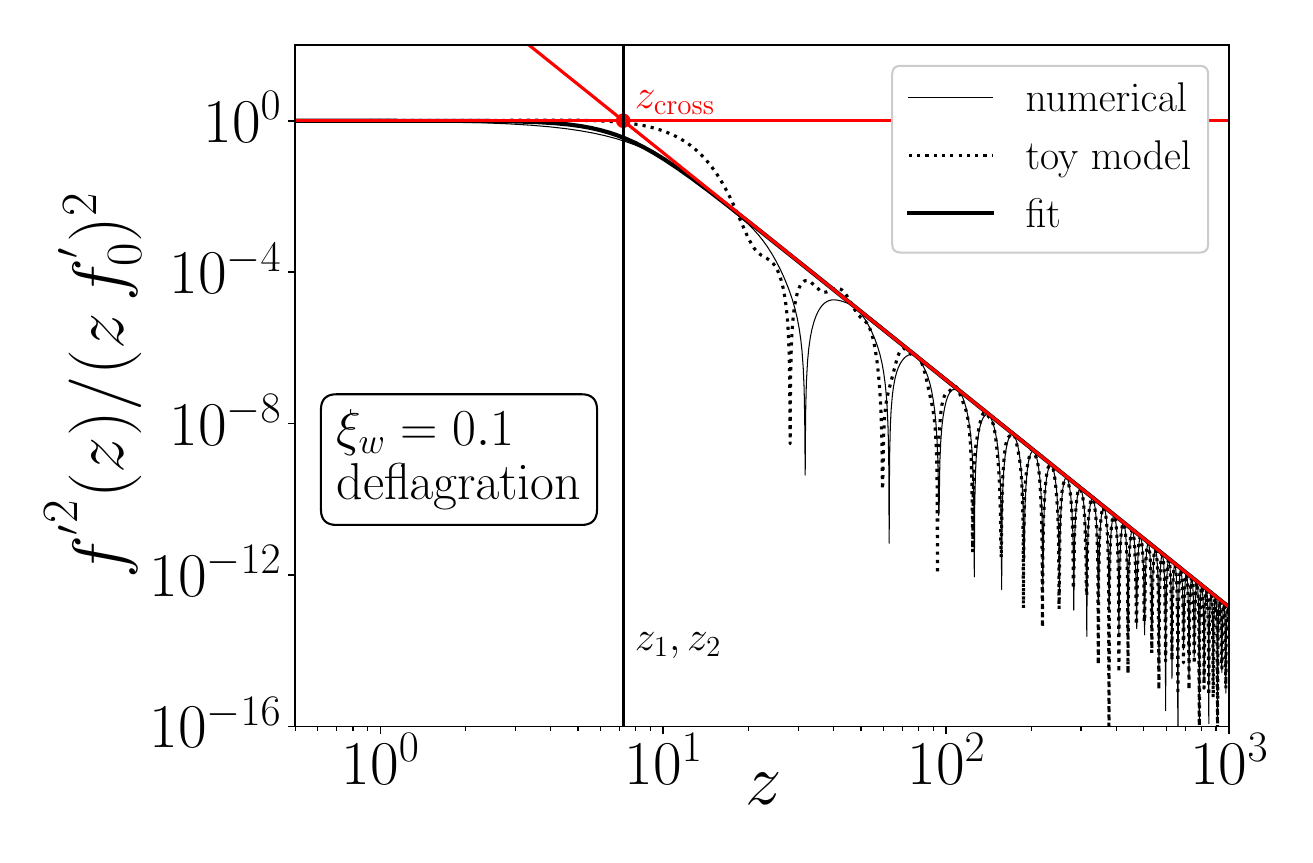}
    \includegraphics[width=.32\textwidth]{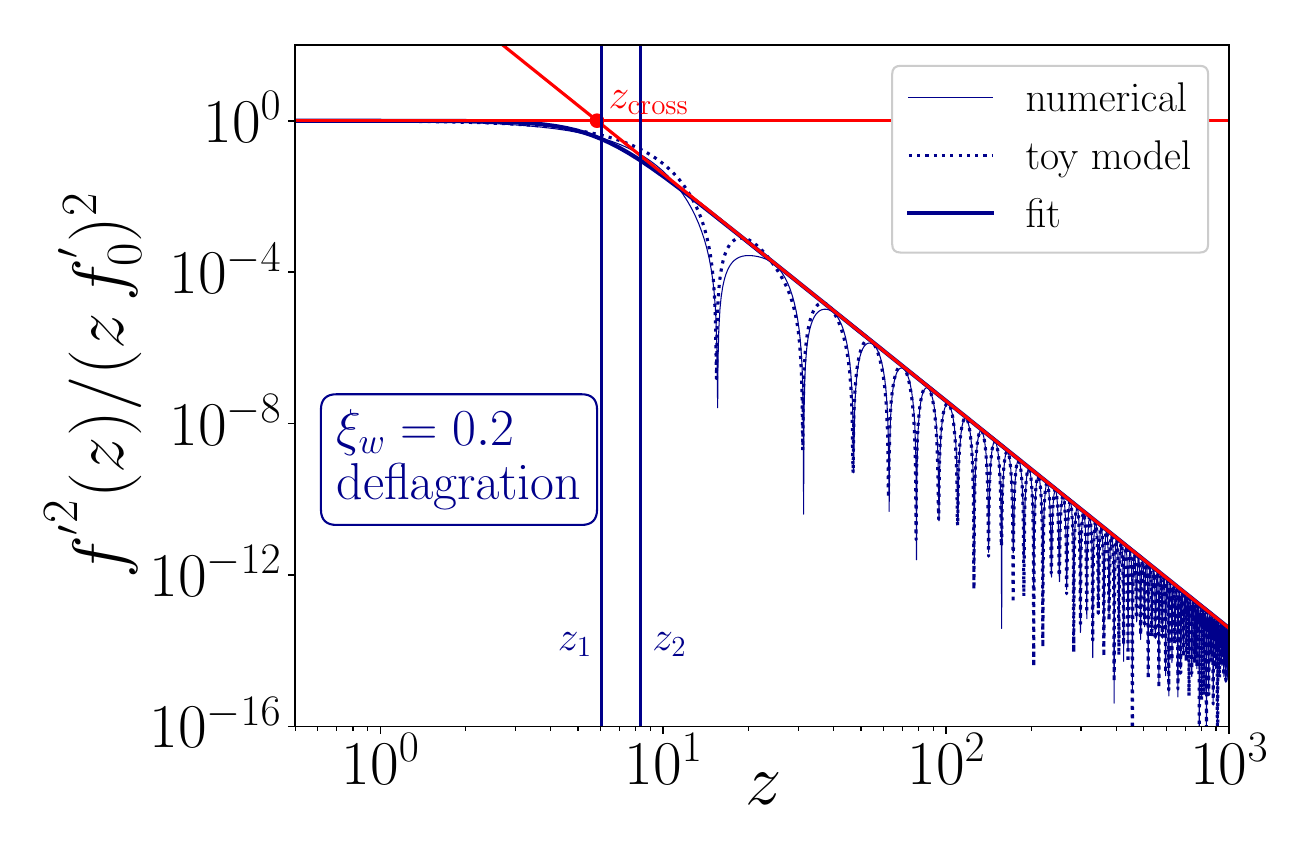}
    \includegraphics[width=.32\textwidth]{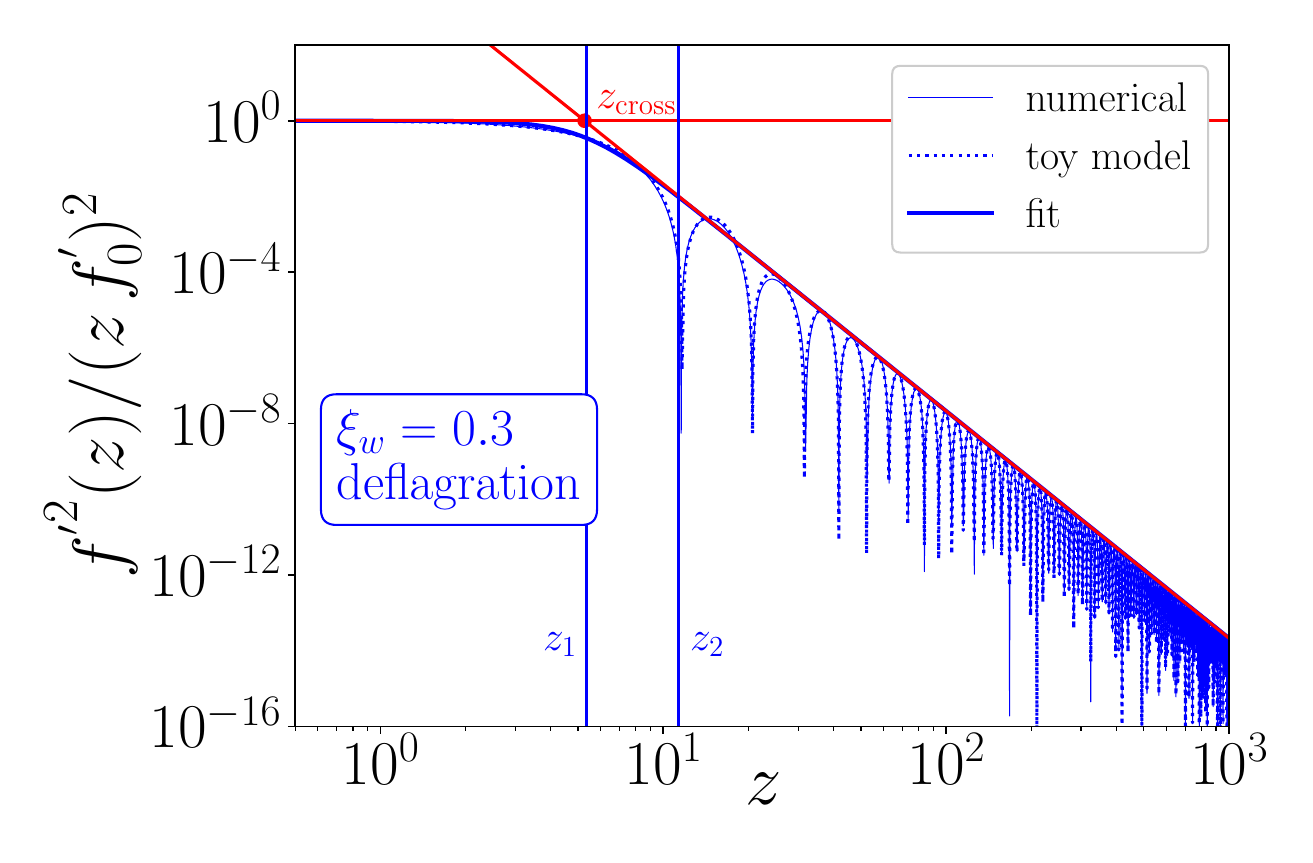}
    \includegraphics[width=.32\textwidth]{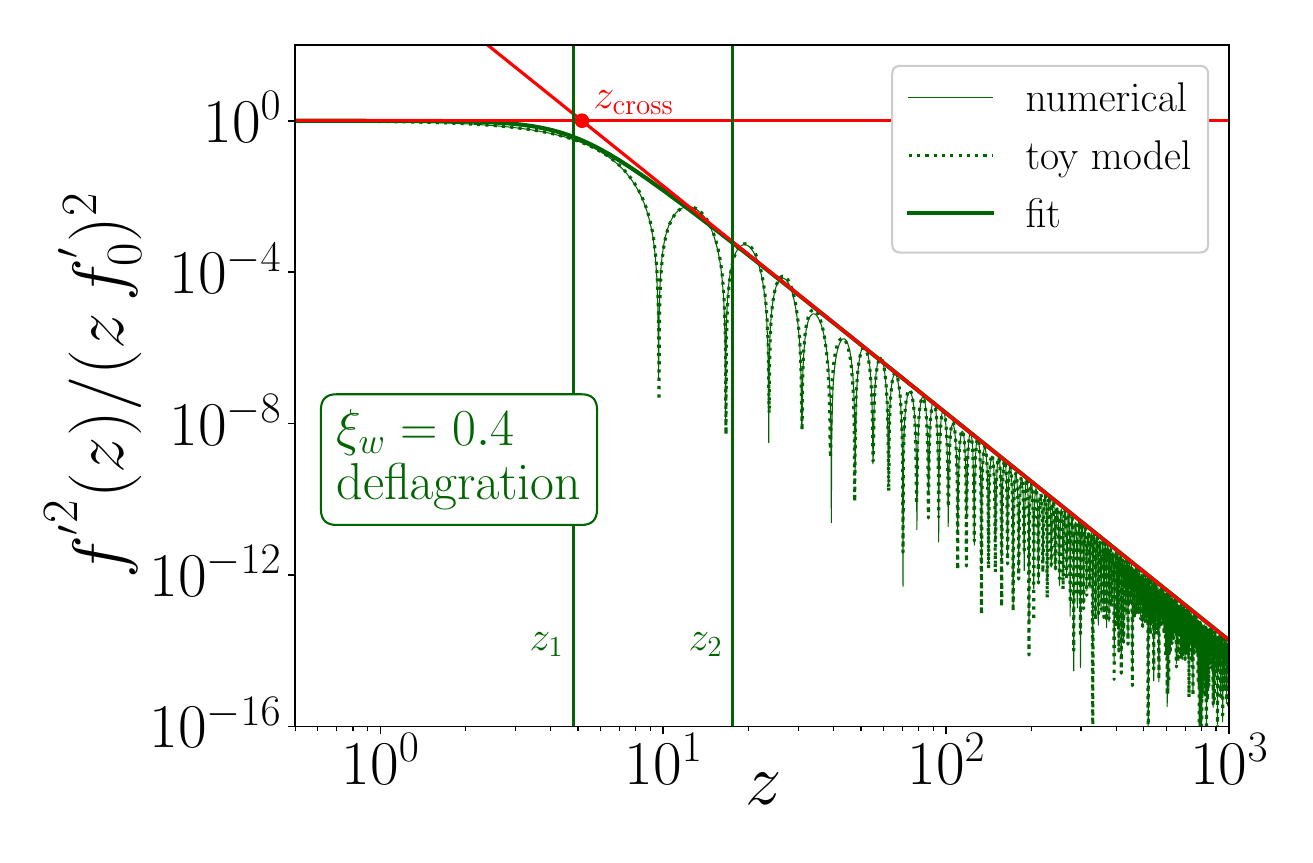}
    \includegraphics[width=.32\textwidth]{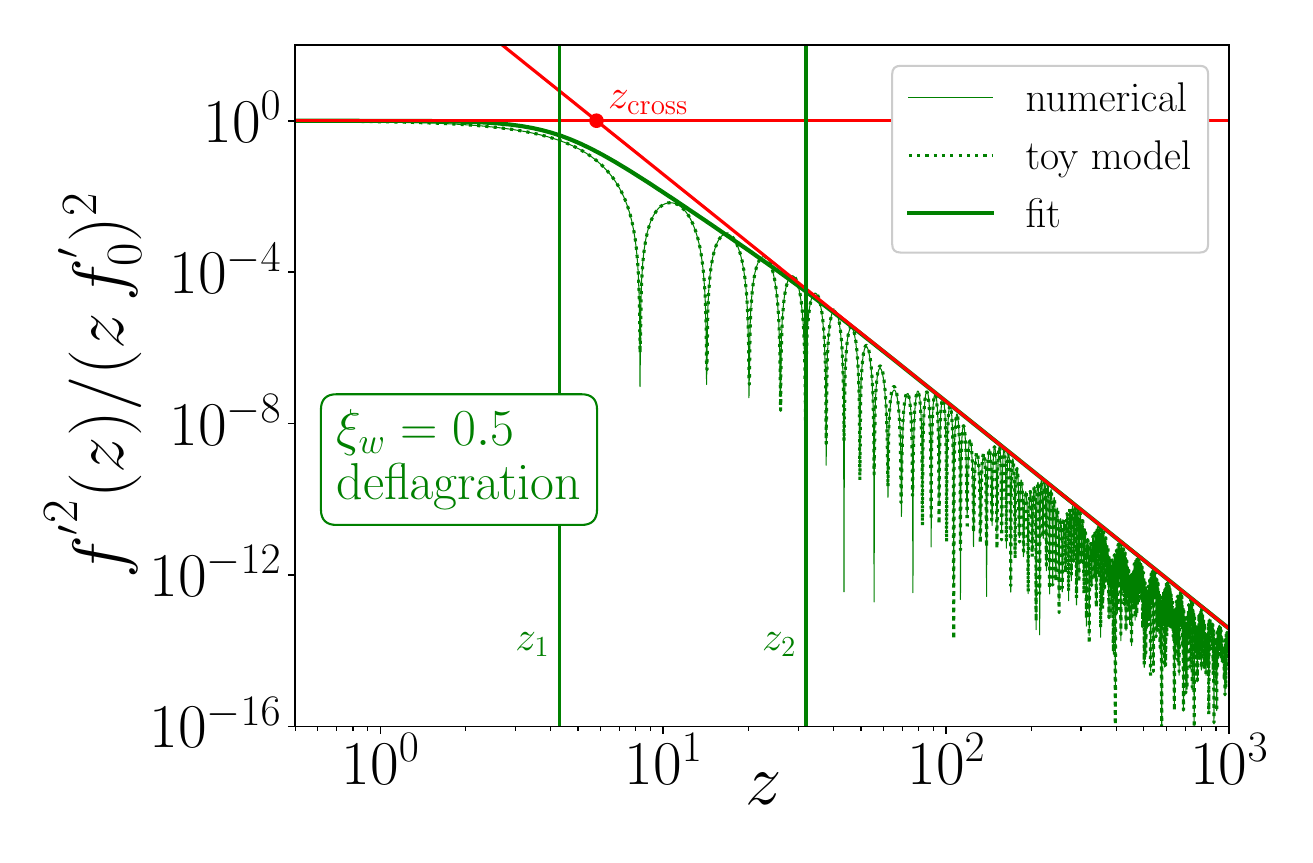}
    \includegraphics[width=.32\textwidth]{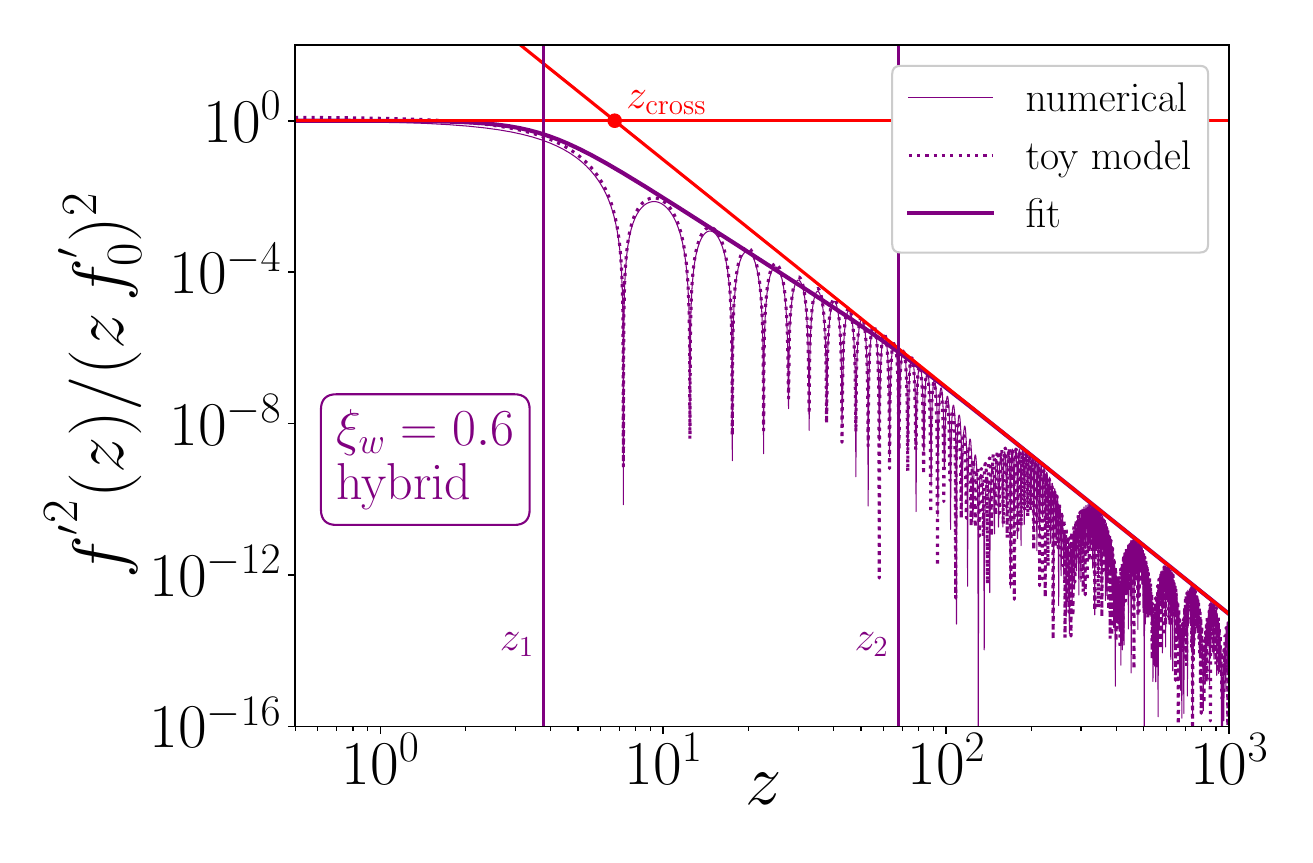}
    \includegraphics[width=.32\textwidth]{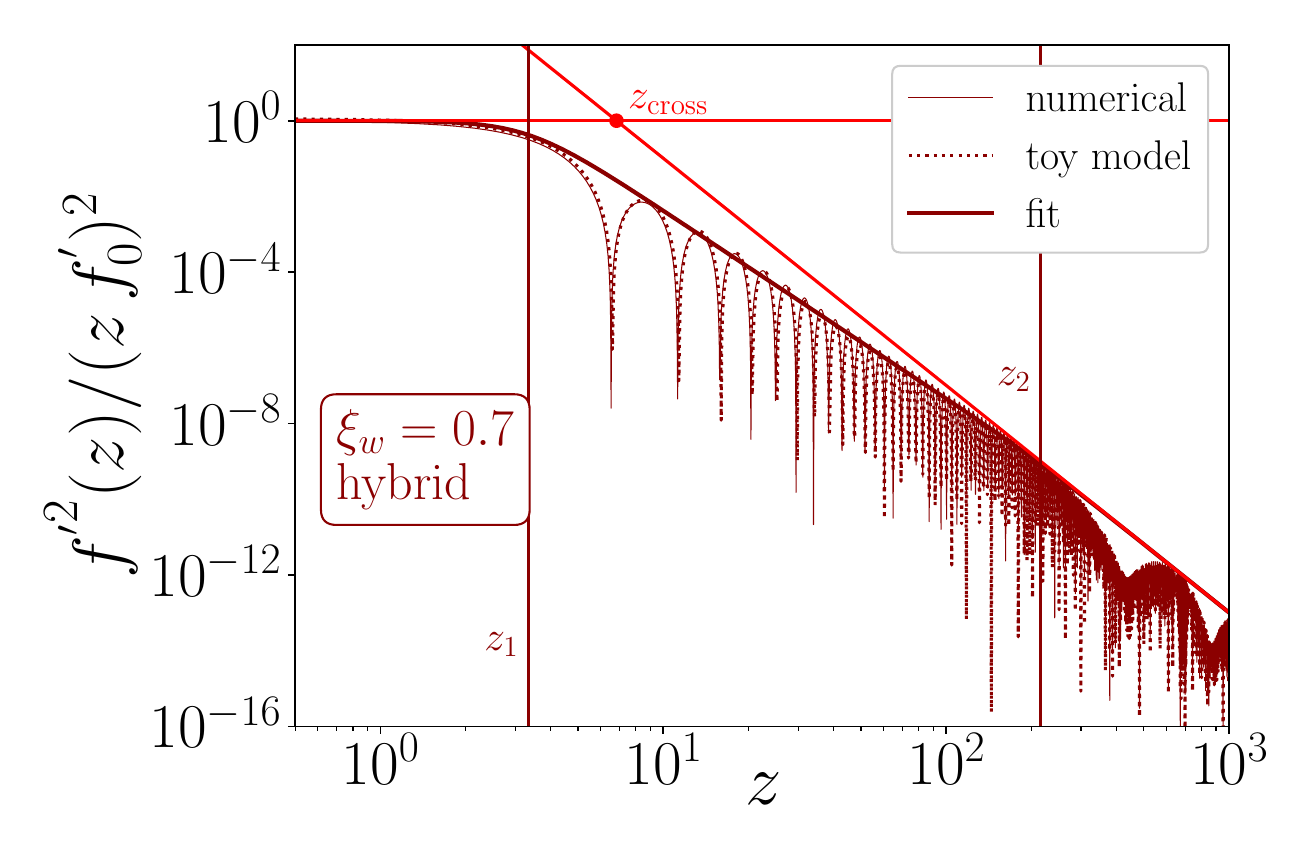}
    \includegraphics[width=.32\textwidth]{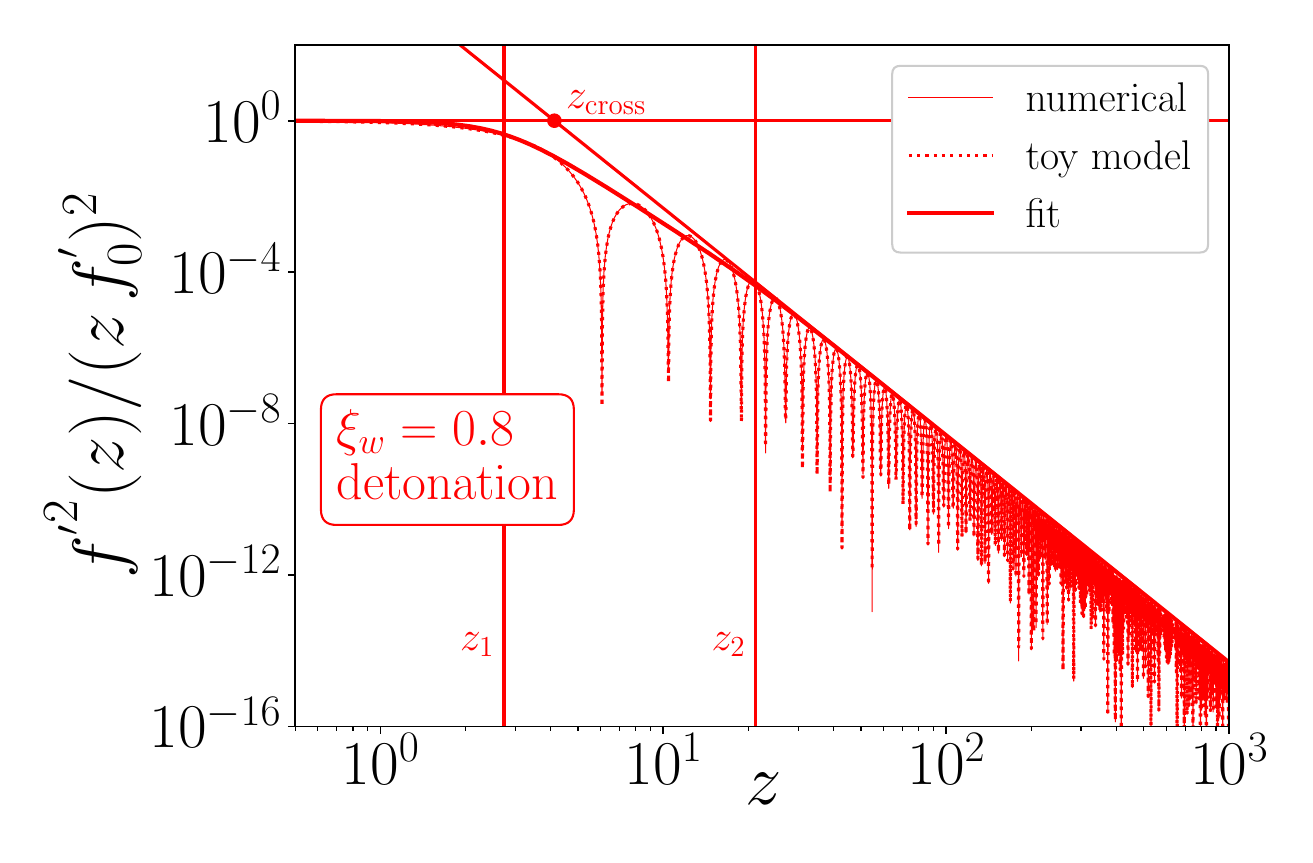}
    \includegraphics[width=.32\textwidth]{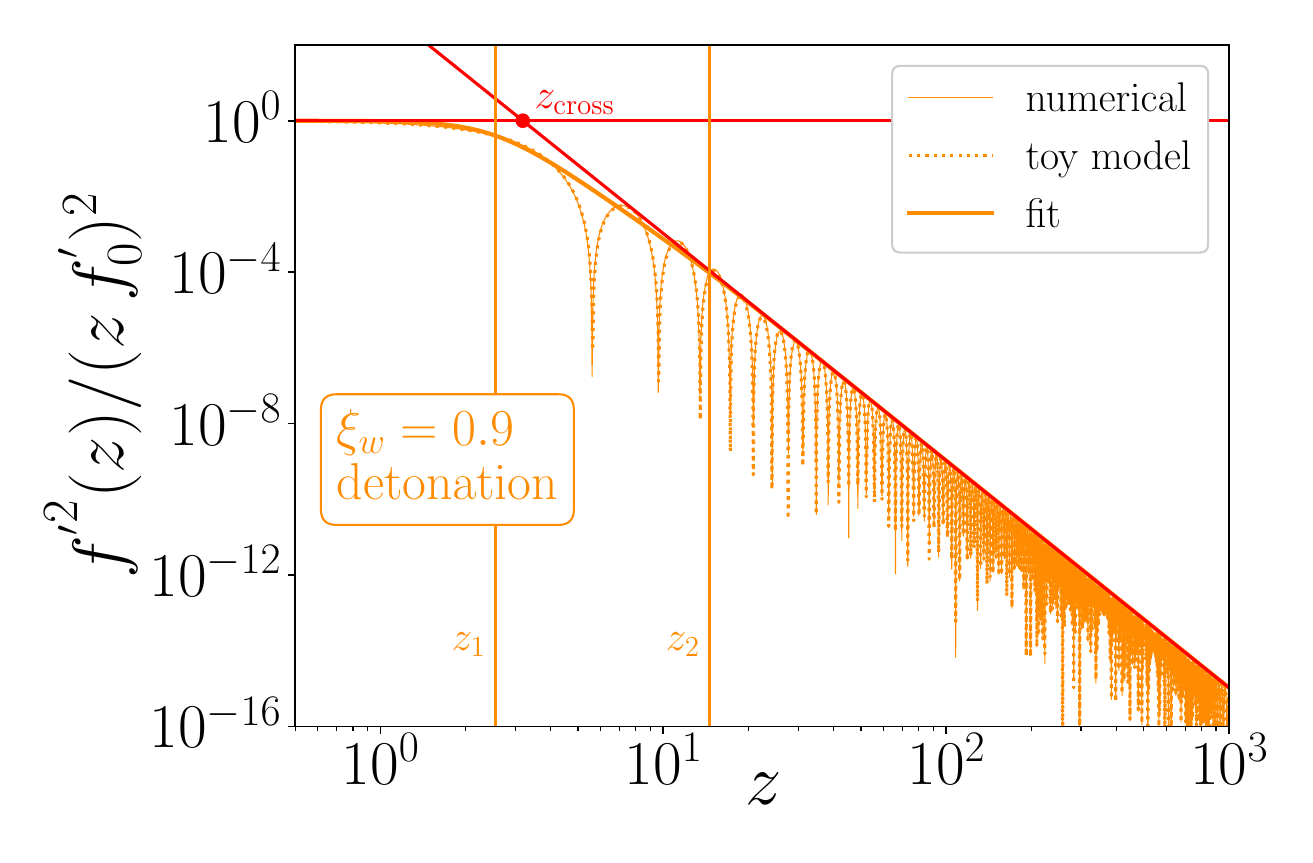}
    \caption{
    The function $\fpsq$ characterizes the spectral
    density of the velocity field. It is shown, normalized by its asymptotic form
    in the $z \to 0$
    limit, $\fpsqz z^2$ [see \Eq{asymptotic_fpz}], in
    colored solid lines, evaluated
    at different wall velocities $\xi_w$ for a benchmark
    phase transition strength $\alpha = 0.1$. 
    The thick solid lines represent the single or double broken power
    law fits given in \Sec{sec:f_template}.
    The dotted lines are computed using the
    corresponding toy models: quadratic profiles for subsonic deflagrations
    and detonations, and linear-constant profiles for hybrids 
    (see \Sec{sec:toymodel}).
    The positions of the scales $z_1$ and $z_2$, where the function $\fpsq$
    transitions to its asymptotic forms for $z \to 0$ and $z \to \infty$ respectively,
    are indicated with vertical lines and computed using the fits provided in \Eqs{fit_z1}{fit_z2}.
    The asymptotic limit $z \to 0$
    and the envelope $\fpenvsq$ in the $z \to \infty$ limit [see \Eq{fpenv_inf}]
    are shown in red solid lines.
    Their intersection corresponds to $z_{\rm cross}$, defined in \Eq{slope_z}.
    The type of solution is determined by $\xi_w$, where subsonic deflagrations
    occur for $\xi_w < \cs$, hybrids for $\cs \leq \xi_w \leq \vCJ$,
    and detonations for $\xi_w > \vCJ$, being $\vCJ \geq \cs$ the Chapman-Jouget speed,
    which takes a value $\vCJ \simeq 0.78$ for $\alpha = 0.1$ [see \Eq{chapman2}].
    }
    \label{fp2_fp20}
\end{figure}

To understand the large-scale limit,
we Taylor expand $f' (z)$ around $z \rightarrow 0$,
\begin{equation}
    f'(z) = f_0' \, z + {\cal O} (z^3)\,, \quad {\rm with \ \ } f_0' = -
    \frac{4 \pi}{3} \int_0^\infty
    \xi^3 \, v_{\rm ip} (\xi)  \dd \xi\,,
    \label{asymptotic_fpz}
\end{equation}
which converges due to the compact support of the velocity profiles.
Hence, in the large-scale limit, the function
$\fpsq$ has the following behavior, 
\begin{align}
    {f'}^2(z) = {f'_0}^2\, z^2 + {\cal O} (z^4)
    \quad \text{for \ } z \to 0 \,,
\end{align}
with the $z^2$ slope being a consequence of causality.
Indeed, we show in \Sec{kinetic_sp_bubbles} that $\fpsq$ provides the longitudinal component of
the velocity field power spectral density, and, since the velocity
field is irrotational, it should
scale proportional to $z^2$ in the $z \to 0$ limit, as required by causality (see \App{appendix_correlation}
for a review of the properties
of spectral functions of statistically homogeneous and isotropic fields).
\FFig{fp2_fp20}
shows the
function $\fpsq$ divided by $\fpsqz z^2$ for benchmark phase transitions with $\alpha = 0.1$
and a range of $\xi_w$. 
The numerical $\fpsq$ obtained using the velocity profiles
shown in \Fig{fig_ip} are compared to the analytical fits and toy models that
we study, respectively,
in \Secs{sec:f_template}{sec:toymodel}.

We find from the numerical evaluations of $\fpsq$ that
deviations from the branch growing as $z^2$ occur at
$z_1 = 2\pi/\xi_1$ (indicated by vertical solid lines in \Fig{fp2_fp20})
with
\begin{align}
      \xi_1 \equiv \frac{2 \pi}{z_1}
      \simeq \begin{cases}
         \, \tfrac{4}{3} \, (\xif + \xi_w) \ &\rm{for \ deflagrations}\,, \\
        \, \tfrac{5}{3} \, (\xi_w + \cs ) \ &\rm{
        for \ detonations}\,.
    \end{cases}
    \label{fit_z1}
\end{align}
As discussed above, we confirm that the position $z_1$
only depends on the sum of the
discontinuity locations
for deflagrations, $\xi_+ = \xif + \xi_w$, and
the sum of the front and back locations for detonations, $\xi_+ = \xif + \xib = \xi_w + \cs$.

\subsubsection{Small-scale limit}
\label{fpz_small_scales_limit}

In this section, we investigate
the small-scale limit
$z \to \infty$ of $f'(z)$.
This limit depends on the
discontinuities of the velocity self-similar profiles.
As we have seen,
they have compact support on $\xi \in (\xib,\xif)$, and present either one or two 
discontinuities: one at the bubble wall and another one due to the formation of a shock
at the front of the fluid shell in deflagrations (see discussion in \Sec{vel_1d_fourier} and
more details in \App{1d_profiles}).
Integrating by parts \Eq{def_fpz},
we can write
\begin{align}
    f'(z)
    = \frac{4 \pi}{z} \, \Biggl[
    \sum_{\xi_0}
    \xi_0^2\, j_2(z\xi_0)\, \Delta v_\ip(\xi_0)
    + \int_{\mathcal{C}}
    j_2(z\xi)\, \xi^3 \,\partial_\xi \left( \frac{ v_\ip(\xi) }{\xi} \right) \dd \xi \Biggr] \,,
    \label{eq:f1_firstPI}
\end{align}
where $\mathcal{C}\equiv
(\xi_b , \xi_w) \cup (\xi_w,\xi_f)$ is the manifold where the velocity profile $v_\ip (\xi)$ is smooth,
$\xi_0 = \{\xi_w, \, \xi_\sh\}$ is the potential set of discontinuities in the velocity,
$\Delta v_\ip (\xi_0) \equiv v_\ip (\xi_0^+) - v_\ip (\xi_0^-)$
is the set of
jumps at the discontinuities,
and $j_2(x) \equiv (3 - x^2) \sin x/x^3 - 3 \cos x/x^2$ is the
second-order spherical Bessel function of the first kind.
Integrating by parts again, the second term in \Eq{eq:f1_firstPI} yields higher powers of $z$ in the denominator,
such that the leading contribution in the small-scale limit is
\begin{subequations}
\label{eq:f1_zInfty_dependence}
\begin{align}
    \label{eq:f1_zInfty_dependence_a}
     f' (z)
    & =  
    \frac{f'_\infty (z)}{z^2} + \mathcal{O}(z^{-3}) \quad \text{for } z \to \infty \,,
\end{align}
where 
we have kept the leading-order term
of the spherical Bessel function $j_2(x) = -\sin x/x  + {\cal O}(x^{-2})$
and we have defined 
the oscillatory amplitude
\begin{align}
    f'_\infty (z) \equiv &\,
    -4 \pi \sum_{\xi_0} \xi_0 \sin (z\xi_0)\,
    \Delta v_\ip (\xi_0)
    \nonumber \\ = &\,  -4\pi
    \bigl[
    \xi_\sh \, \sin(z\xi_\sh) \,
    \Delta v_\ip(\xi_\sh)
    + \xi_w \, \sin(z\xi_w)\,  \Delta v_\ip(\xi_w)
    \bigr]
    \,. \label{eq:f1_zInfty}
\end{align}
\end{subequations}
The jump of the velocity at $\xi_\sh$
is $\Delta v_\ip (\xi_\sh) =
- v_\sh^- = - v_\sh$ [see \Eq{cond_shock}]
when a shock forms for deflagrations.
For subsonic deflagrations,
the jump in the velocity profile
at $\cs$ decays with $\xi_w$, so its contribution to the asymptotic limit
becomes negligible for small $\xi_w$.
The jump at $\xi_w$, $\Delta v_\ip (\xi_w) = v_+ - v_-$, is
positive for deflagrations and negative for detonations.
We summarize in \Tab{tab:scales_fpz} the discontinuities present for each type of solution.
In \Secs{matching}{class_sols} of \App{1d_profiles}, we describe how
to compute $v_+$ and $v_-$ for each type of solution
and how to determine the position ($\xi_\sh$) and amplitude of the shock, $v_\sh$ [see \Eq{cond_shock}].

\EEq{eq:f1_zInfty_dependence_a} justifies
the $z^{-4}$ asymptotic behavior $\fpsq$ at small scales, shown in \Fig{fp2_fp20}.
Furthermore, \Eq{eq:f1_zInfty} allows us to predict the amplitude of the asymptotic branch $z \to \infty$ as a function of the discontinuity jumps in the velocity profiles.
The envelope of this asymptotic branch is
\begin{equation}
    \fpenvsq (z) = \frac{\fpenvsq}{z^4}\,, \qquad \text{with \ }
    f'_\env \equiv
    4\pi \, (\xi_\sh \, v_\sh +
    \xi_w \, |v_+-v_-|) \,. 
    \label{fpenv_inf}
\end{equation}

We infer numerically that the onset of the $z^{-4}$ regime, indicated by vertical dotted lines in \Fig{fp2_fp20}, occurs at $z_2 = 2 \pi/\xi_2$ with
\begin{align}
     \xi_2 \equiv \frac{2 \pi}{z_2} \simeq \begin{cases}
         \, 2 \, (\xi_\sh - \xi_w) \ &\text{for deflagrations}\,, \\
        \, \tfrac{4}{3}\, (\xi_w - \cs) \ &\rm{for \ detonations}\,.
    \end{cases}
    \label{fit_z2}
\end{align}
For deflagrations,
$\xi_2 \propto \tilde \Delta \xi = \xi_\sh - \xi_w$
is determined by
the distance between the two velocity field discontinuities.
We note that for subsonic
deflagrations with $\xi_w \lesssim \half \vCJ$,
$z_2$ approaches $z_1$
and, hence, they are not clearly
distinguishable
in $\fpsq$, as it can
be seen in \Fig{fp2_fp20}.
For detonations,
there is only one discontinuity at $\xi_{\rm f} = \xi_w$, and we find
that the second scale $z_2$ depends on
the fluid shell thickness $\tilde \Delta \xi = \Delta \xi =
\xi_w - c_{\rm s}$. 

\begin{figure}[t]
    \centering
    \includegraphics[width=.32\textwidth]{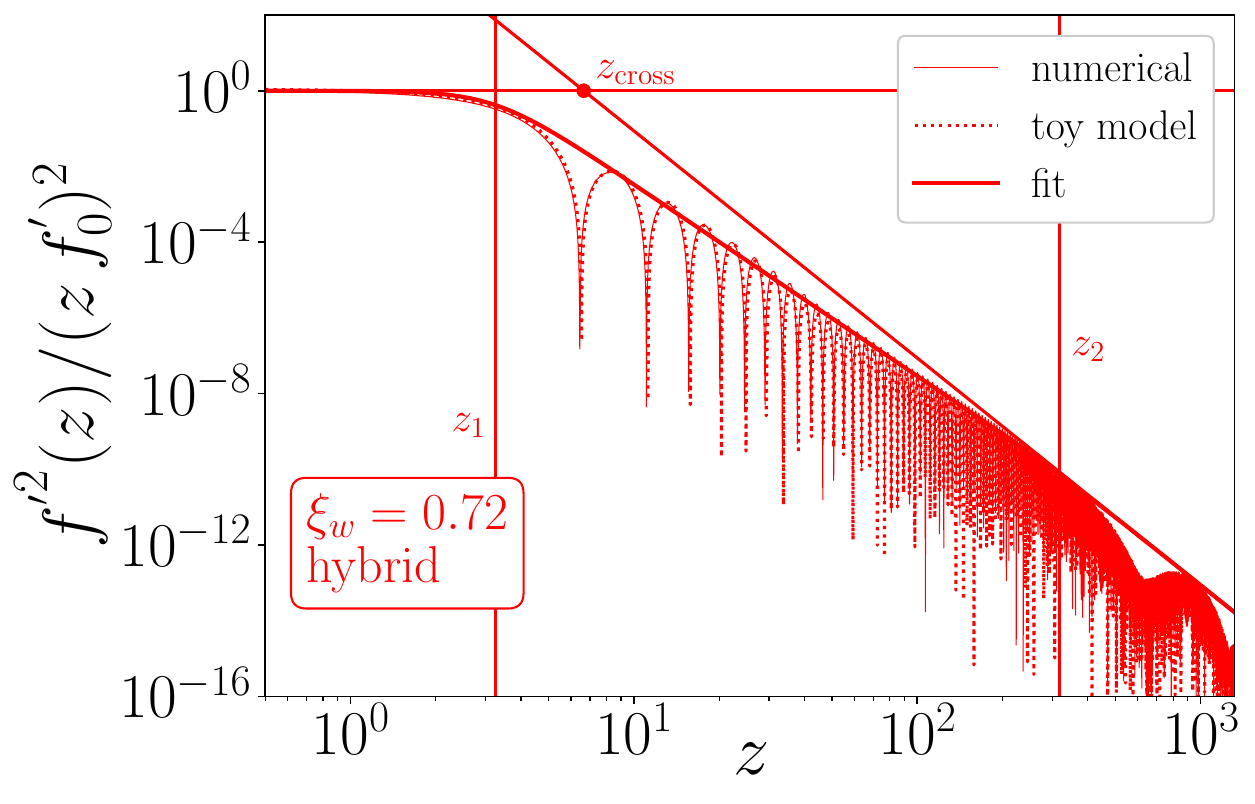}
    \includegraphics[width=.32\textwidth]{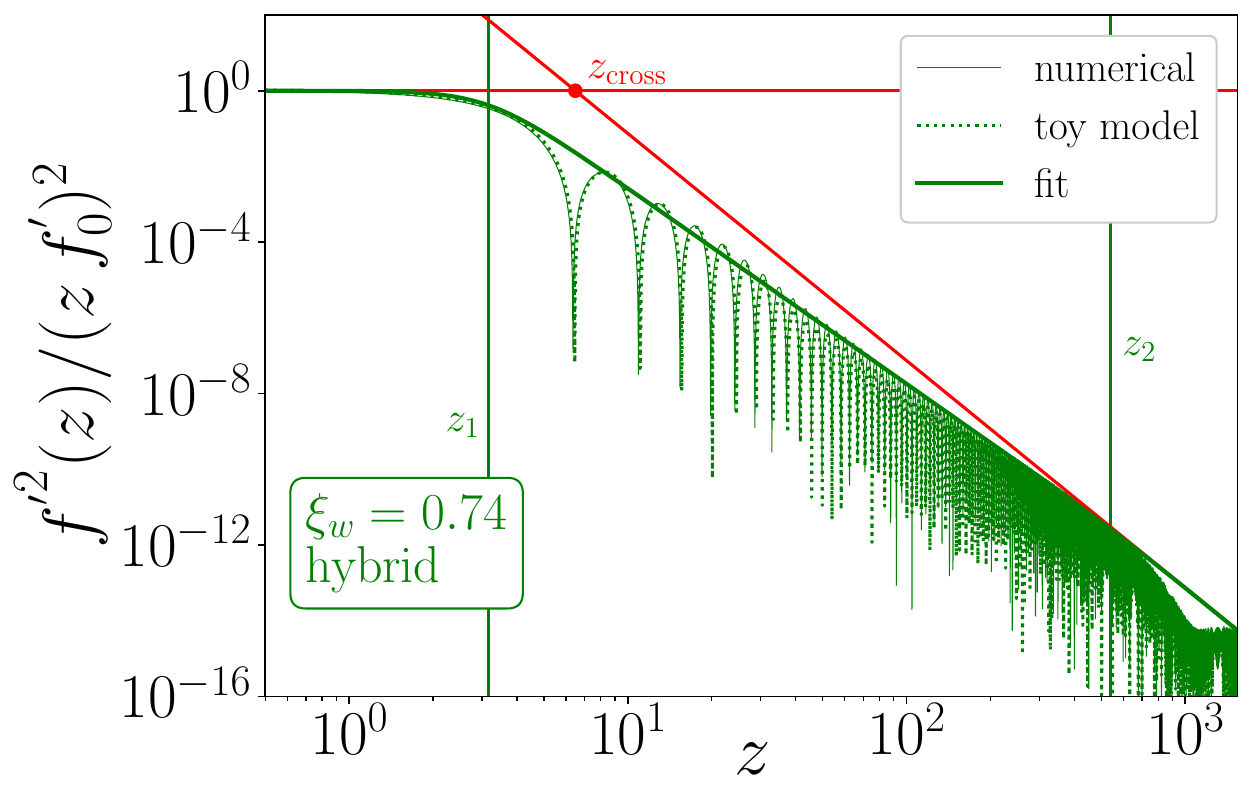}
    \includegraphics[width=.32\textwidth]{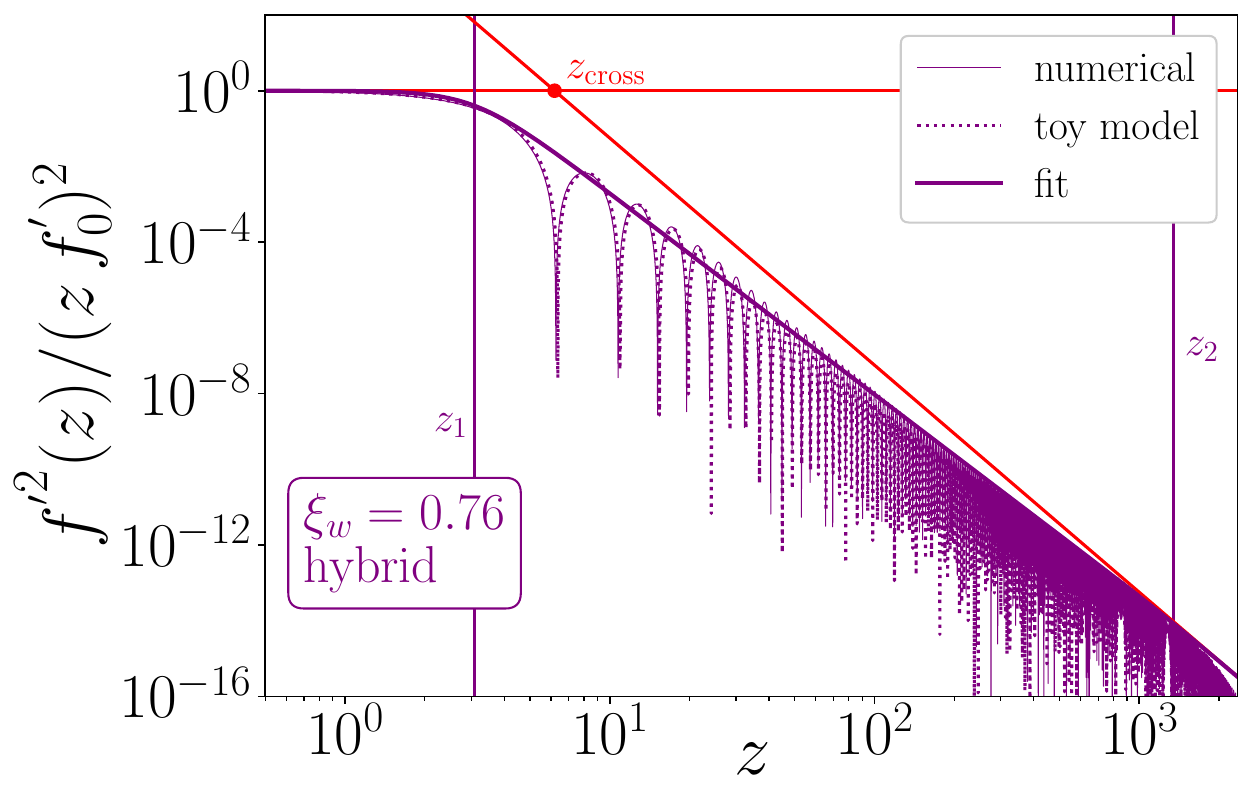}
    \caption{
    Same as \Fig{fp2_fp20} for hybrids
    with $\alpha=0.1$ and a bubble wall
    velocity approaching $\vCJ (\alpha) \simeq 0.78$.
    We find that, as we approach $\xi_w \to \vCJ^-$, the
    second scale $z_2$ becomes progressively larger:
    this is due to the fact that
    it is characterized by the inverse distance between
    discontinuities, $\tilde \Delta \xi^{-1}$ (see \Eq{fit_z2}
    and the right panel in
    \Fig{fig:comparison_z1z2}
    for a comparison between $\Delta \xi^{-1}$ and $\tilde{\Delta} \xi^{-1}$).
    A second scale determined by the inverse fluid shell thickness,
    $\Delta \xi^{-1}$, would not capture well the $z$ dependence of ${f'}^2(z)$, see \Tab{tabFITS}.
    }
    \label{hybrids_closeto_vcj}
\end{figure}

After averaging over bubble nucleation locations and times
(see \Secss{kinetic_sp_bubbles}{FL_template}), $z_2$ determines the second break in the velocity power spectrum. 
Ultimately, this then enters in the GW power spectrum. 
We therefore find that, 
contrary to what 
proposed in previous literature, where the sound shell thickness has been identified as the relevant scale entering the GW spectrum
\cite{Hindmarsh:2013xza,Hindmarsh:2016lnk,Hindmarsh:2019phv}, 
it is the 
difference between the discontinuities for deflagrations and the
fluid shell thickness for detonations the relevant scale.
This difference is particularly relevant for hybrids.
As an illustrative example, the confined
hybrid with $\xi_w = 0.7$ shown in \Fig{fp2_fp20}
has $\tilde \Delta \xi \simeq 0.01$ and $\Delta \xi \simeq 0.14$, hence larger $\Delta \xi$ but
smaller $\tilde \Delta \xi$ than the hybrid with $\xi_w = 0.6$,
for which $\tilde{\Delta} \xi \simeq 0.05$ and $\Delta \xi \simeq 0.07$
(see \Fig{fig_ip} and \Tab{tab:scales_fpz}).
Indeed, from \Fig{fp2_fp20}, it appears that the position of the second break is clearly found at larger $z$ for $\xi_w = 0.7$.
The fact that the second scale for the hybrids has to be computed from $\tilde{\Delta} \xi$
is even more clear
in \Fig{hybrids_closeto_vcj}, where $\fpsq$ is shown for three
values of $\xi_w$ that are approaching $\vCJ$ from below.
It can be seen that as $\xi_w \to \vCJ^{-}$, $z_2$ increases and, hence, it is not well
characterized by $\Delta \xi$.
    
\begin{table}[b!]
    \centering
    \begin{tabular}{ccccccccc} \hline
    type & $\xi_w$ 
    & $\fpsqz$
    & $\fpenvsq$ & $\xi_+$ & $\tilde \Delta \xi$ &
    ${\Delta} \xi$ &
    $\xi_{\rm sh}$ & $\alpha_+$ \\ \hline
    def & 0.1 & 1.09$ \times 10^{-8}$ & 1.43$ \times 10^{-3}$ & 0.68 & 0.48 & -- & 0.58 & 0.10 \\
    def & 0.2 & 6.60$ \times 10^{-7}$ & 2.40$ \times 10^{-2}$ & 0.78 & 0.38 & -- & 0.58 & 0.09 \\
    def & 0.3 & 6.78$ \times 10^{-6}$ & 1.36$ \times 10^{-1}$ & 0.88 & 0.28 & -- & 0.58 & 0.09 \\
    def & 0.4 & 3.17$ \times 10^{-5}$ & 5.72$ \times 10^{-1}$ & 0.98 & 0.18 & -- & 0.58 & 0.08 \\
    def & 0.5 & 8.68$ \times 10^{-5}$ & 3.11 & 1.10 & 0.10 & -- & 0.60 & 0.07 \\
    hyb & 0.6 & 1.43$ \times 10^{-4}$ & 12.9 & 1.25 & 0.05 & 0.07 & 0.65 & 0.05 \\
    hyb & 0.7 & 2.88$ \times 10^{-4}$ & 26.8 & 1.41 & 0.01 & 0.14 & 0.71 & 0.04 \\
    det & 0.8 & 1.41$ \times 10^{-3}$ & 6.59 & 1.38 & 0.22 & -- & 0.80 & 0.10 \\
    det & 0.9 & 3.60$ \times 10^{-3}$ & 3.52 & 1.48 & 0.32 & -- & 0.90 & 0.10 \\
    det & 1.0 & $7.45 \times 10^{-3}$ & 2.71 & 1.58 & 0.42 & -- & 1.00 & 0.10 \\
    \end{tabular}
    \caption{
    Relevant set of parameters to reconstruct the single and double broken power law
    fits in \Eqs{eq:f_templateSBPL}{eq:f_templateDBPL}
    of the envelope of $\fpsq$ for $\alpha = 0.1$ and different
    values of $\xi_w$.
    $\fpsqz$ provides the amplitude of $\fpsq$ in the
    $z \to 0$ limit.
    Together with the envelope in the $z \to \infty$ limit, $\fpenvsq$, the crossing 
    scale $z_{\rm cross}$ and
    the intermediate slope $\gamma$ can be reconstructed
    using \Eq{slope_z}.
    The two scales $z_1$ and $z_2$ are summarized in \Tab{tab:scales_fpz}
    and depend either
    on the positions of the front and the back (detonations) or the positions
    of the discontinuities (deflagrations): $z_1$ depends
    on their sum ($\xi_+$), while $z_2$ depends on their difference ($\tilde \Delta \xi$).
    For comparison, we give
    the value of $\Delta \xi \equiv \xif - \xib$ when it is different
    than $\tilde \Delta \xi$, i.e.,
    for hybrids.
    The amplitude $\fpenvsq$ can be reconstructed from the values
    of the velocity
    jumps at $\xi_w$ and at $\xi_\sh$, which can be computed
    from $\alpha_+$ and
    $\xi_\sh$
    following the procedure described in \App{1d_profiles}.
    }
    \label{tabFITS}
\end{table}

We summarize in \Tab{tabFITS} the values $\fpsqz$ and $\fpenvsq$ that
determine the asymptotic limits of $\fpsq$,
and the values $\xi_1$ and $\xi_2$ that characterize the two scales
($\xi_+$ and $\tilde \Delta \xi$)
for benchmark phase transitions with $\alpha = 0.1$ and different
wall velocities.
We also provide $\xi_\sh$ 
and $\alpha_+ \leq \alpha$,
where
$\alpha_+$ can be
used to determine the values of the velocity jump at $\xi_w$ (see \Sec{matching})
and, hence, to determine $\fpenvsq$ together with $\xi_\sh$
and $v_\sh$, using \Eq{fpenv_inf}.

\begin{figure}
    \centering
    \includegraphics[width=0.49\linewidth]{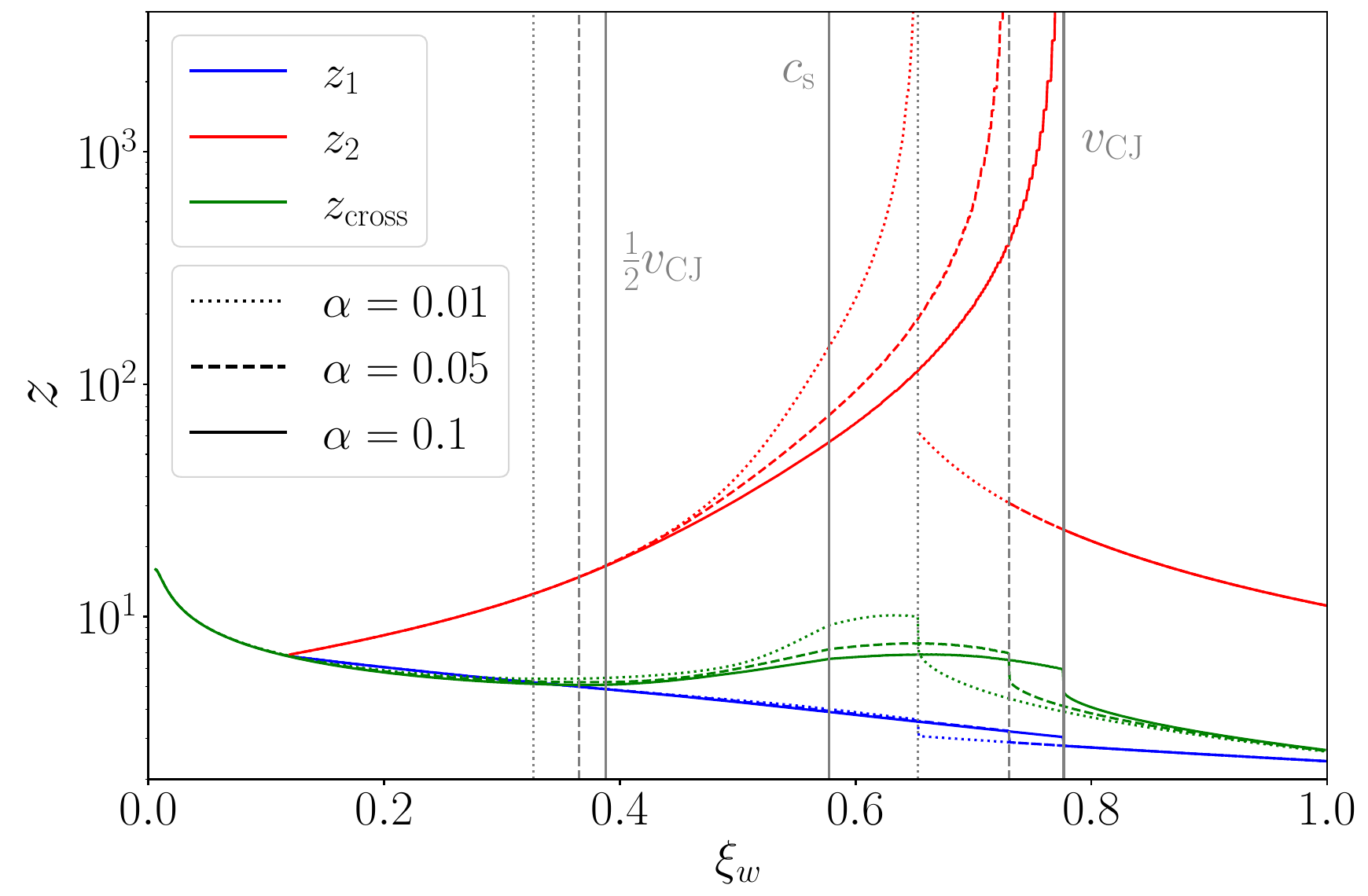}
    \includegraphics[width=0.49\linewidth]{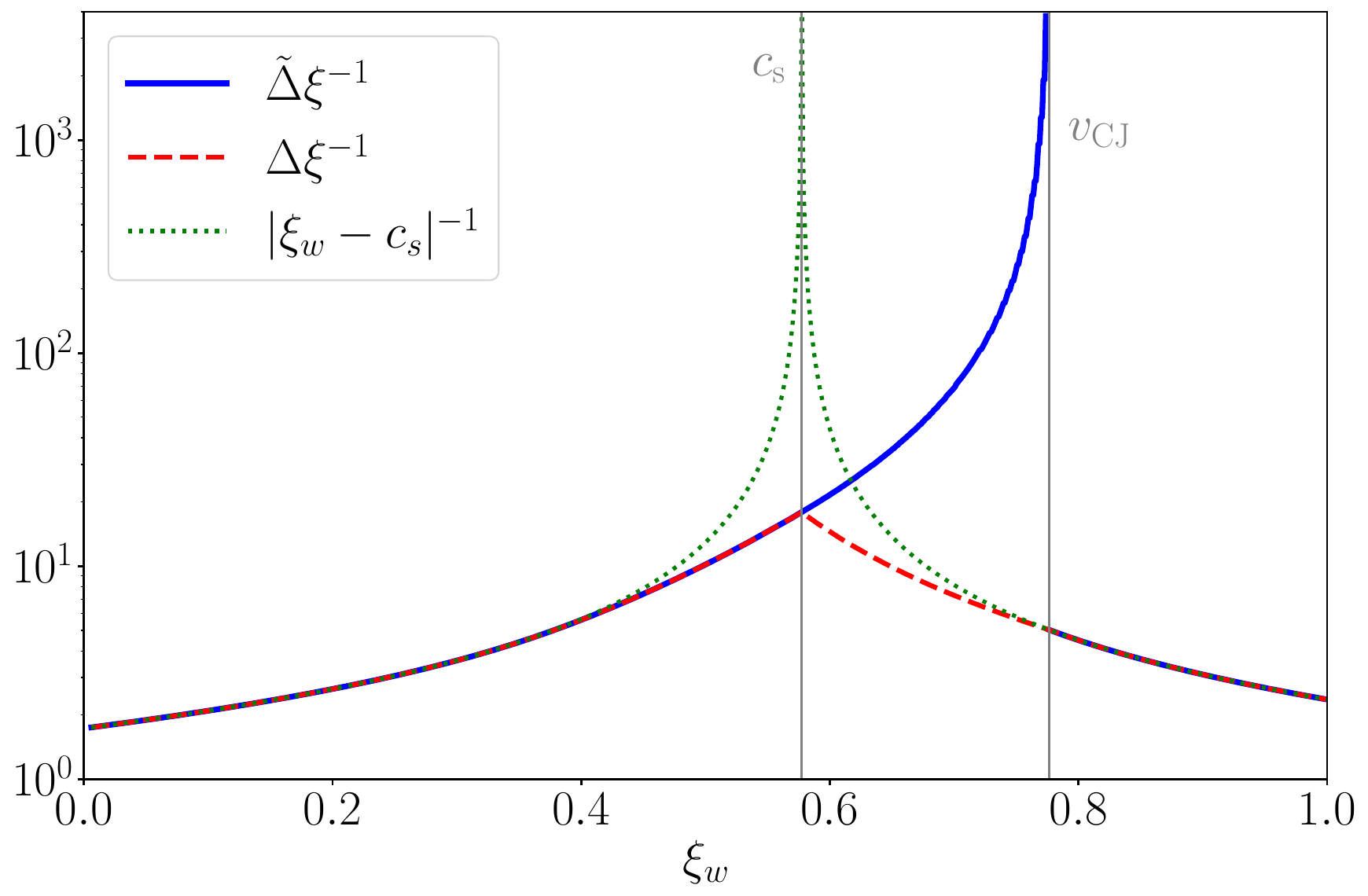}
    \caption{
    {\em Left panel:} Dependence on the bubble wall velocity, $\xi_w$, of the parameters $z_1$ (blue) and $z_2$ (red),
    given in \Eqs{fit_z1}{fit_z2}, that characterize the shape of $\fpsq$.
    The double broken power law structure in the envelope of $\fpsq$  
    arises when $\xi_w \gtrsim \tfrac{1}{2} \vCJ (\alpha)$,
    corresponding to the value at which $z_\cross$ (green),
    given in \Eq{slope_z},
    starts to deviate from $z_1$.
    The Chapman-Jouget speed, $\vCJ (\alpha)$,
    the speed of sound, $\cs$, and $\tfrac{1}{2} \vCJ (\alpha)$,
    are indicated by vertical gray lines.
    We show $z_1$, $z_2$, and $z_\cross$ for $\alpha = 0.01$ (dotted),
    $0.05$ (dashed), and $0.1$ (solid).
    {\em Right panel:} Comparison of $\tilde{\Delta} \xi^{-1}$ (blue),
    $\Delta \xi^{-1}$ (red), and $|\xi_w-\cs|^{-1}$ (green) for $\alpha = 0.1$.
    The choices $\Delta \xi^{-1}$ \cite{Hindmarsh:2013xza,Hindmarsh:2016lnk,Hindmarsh:2019phv,Jinno:2020eqg,Jinno:2022mie,Caprini:2024gyk,Caprini:2024hue,Caprini:2024ofd} and
    $|\xi_w-\cs|^{-1}$ \cite{Caprini:2015zlo,Caprini:2019egz,RoperPol:2023bqa}
    have been previously considered in the literature to determine the
    position of the second
    spectral break.
    We show how previous choices deviate from $\tilde \Delta \xi^{-1}$ for
    hybrids.
    The largest separation between the two scales occurs
    when $\xi_w \lesssim \vCJ (\alpha)$, differing from the usual
    assumption that it occurs when the fluid shell thickness
    reaches its minimum at $\xi_w \approx \cs$ \cite{Caprini:2019egz,Hindmarsh:2019phv,Hindmarsh:2020hop,RoperPol:2023bqa}.
    }
\label{fig:comparison_z1z2}
\end{figure}

\subsection{Single and double broken power law templates for $\fpsq$}
\label{sec:f_template}

As we have seen in \Sec{subsec_RiemannL}, the asymptotic limits of $\fpsq$
for large and small scales are respectively reached at $z_1 = 2 \pi/\xi_1$ and $z_2 = 2\pi/\xi_2$,
summarized in \Tab{tab:scales_fpz}.

The dependence of $z_1$ and $z_2$ on $\xi_w$ for various values of $\alpha$ is shown in \Fig{fig:comparison_z1z2}. From the numerical results, shown in \Fig{fp2_fp20}, we find that the function $\fpsq$ takes the form of a 
double broken power law modulated by oscillations.
When $\xi_w \simeq \tfrac{1}{2} \vCJ$, we see that the scales $z_1$ and $z_2$ become very
close to each other and,
for $\xi_w \lesssim \half \vCJ$,
the envelope of $\fpsq$ takes the form of a single broken power law (SBPL) given by
\begin{equation}
    {f'}^2_{\! \! \rm SBPL} (z) = {f_0'}^2 \, z^2 \,
    \biggl[1 + \biggl(\frac{z}{z_1} \biggr)^{a_1}\biggr]^{-\frac{6}{a_1}}
    \qquad \text{for \ } \xi_w \lesssim \tfrac{1}{2} \, \vCJ (\alpha)\,,
\label{eq:f_templateSBPL}
\end{equation}
where the value of $a_1$ determines the smoothness of the transition of $\fpsq$ around $z_1$ and has been set to $4$ in \Fig{fp2_fp20}.
This value of $a_1$ provides a good fit for all the values of $\xi_w$
and $\alpha$ and it is large enough to reproduce the sharp
transition of $f'^2(z)$ from the large- to the intermediate-scales
regime.
Even though we only show results for $\alpha = 0.1$ in \Fig{fp2_fp20}, we have
confirmed that the presented fits are also valid for other values of $\alpha$.

\begin{figure}[t]
    \centering
    \includegraphics[width=0.9\linewidth]{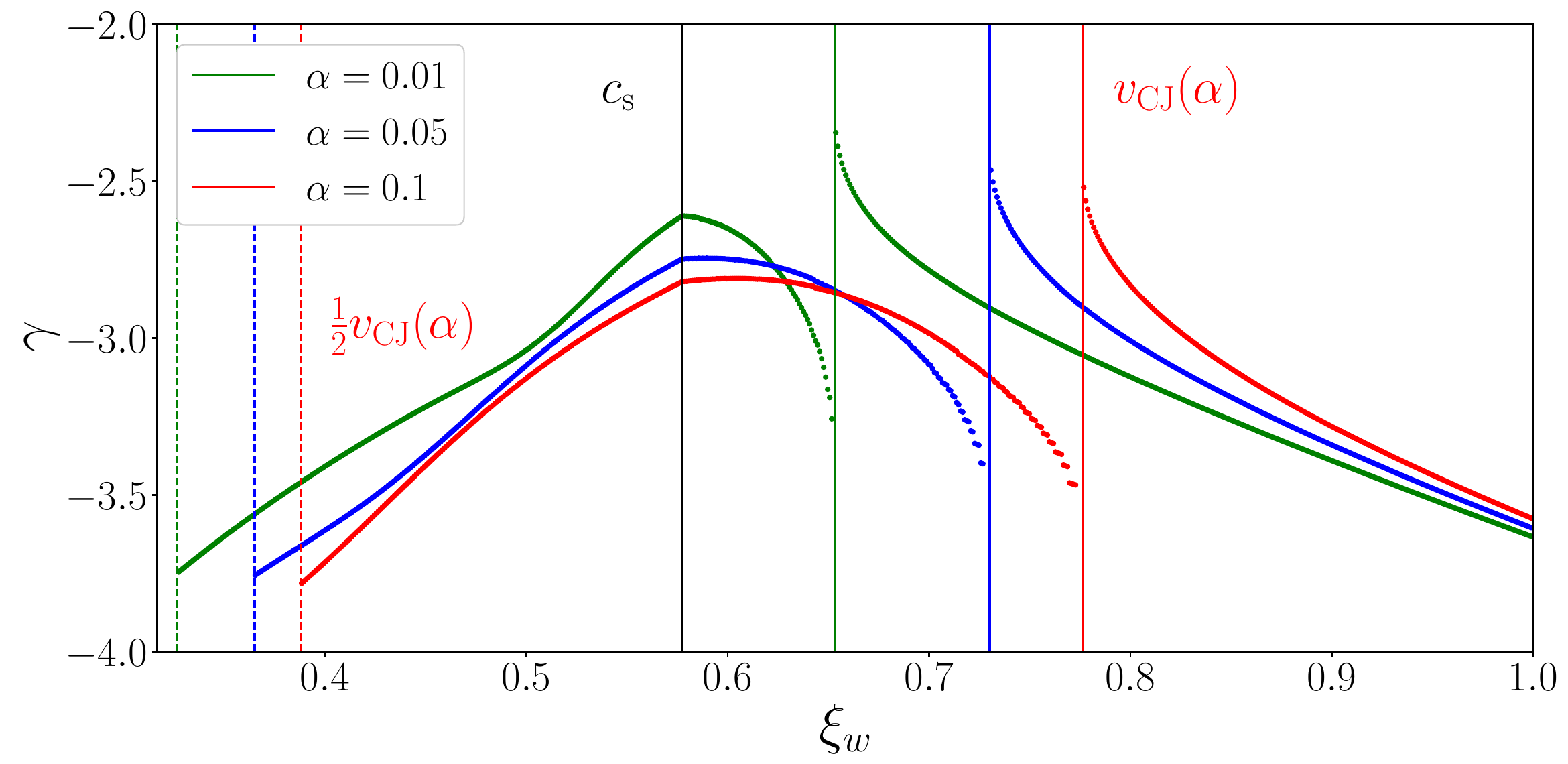}
    \caption{
    Intermediate slope $\gamma$, defined in \Eq{slope_z}, as a function of $\xi_w$ for different values of $\alpha= 0.01$ (green), 0.05 (blue), and 0.1 (red).
    For each value of $\alpha$,
    we plot $\gamma$
    in the range $\xi_w \geq \tfrac{1}{2} \vCJ (\alpha)$ (indicated
    by vertical, colored dashed lines),
    in which we fit the envelope of $\fpsq$
    with a double broken power law [see \Eq{eq:f_templateDBPL}].
    The Chapman-Jouget speed, $\vCJ (\alpha)$, 
    and the speed of sound, $\cs$, are indicated with
    vertical solid lines.
    }
\label{fig:gamma_vs_vw}
\end{figure}

On the other hand, when $\xi_w \gtrsim \tfrac{1}{2}\,
\vCJ$, the second scale $z_2$ becomes
clearly separated from $z_1$.
This can be interpreted as the appearance of a second discontinuity,
since it is roughly for $\xi_w \gtrsim \tfrac{1}{2}\,
\vCJ$ that the velocity jump at $\xi_\sh$ starts to be non-negligible.
Therefore,
we see a clear intermediate slope in the envelope of $\fpsq$,
which can be fit with a double broken power law (DBPL),
\begin{equation}
    {f'}^2_{\! \! \rm DBPL} (z) = {f_0'}^2 \, z^2 \, \biggl[1 + \biggl(\frac{z}{z_1}\biggr)^{a_1}\biggr]^{\frac{\gamma - 2}{a_1}}
    \, \biggl[1 + \biggl(\frac{z}{z_2}\biggr)^{a_2}
    \biggr]^{\frac{-\gamma - 4}{a_2}}
    \qquad \text{for \ } \xi_w \gtrsim \tfrac{1}{2}
    \vCJ (\alpha) \,,
    \label{eq:f_templateDBPL}
\end{equation}
where $a_1$ and $a_2$ respectively 
determine the smoothness
of the transitions around $z_1$ and $z_2$.
Both parameters are set to $4$ in \Fig{fp2_fp20},
which
are found to give the best fit to the numerical results.
The intermediate slope
can be computed as
\begin{equation}
    \gamma = 2 \left[ 1-3\,\frac{\log(z_2/z_\cross)}{\log(z_2/z_1)} \right]
    \in [-4,2]
    \,, \qquad \text{with \ } z_\cross = \left|\frac{f_\env'}{f'_0}\right|^{1/3}\,. \label{slope_z}
\end{equation}
We have defined $z_\cross$ (see \Fig{fig:comparison_z1z2}) as the scale at which the two asymptotic
branches of the envelope of $\fpsq$ cross.
We show the intermediate slope $\gamma$ for different values of $\xi_w$
and $\alpha$ in \Fig{fig:gamma_vs_vw}.
Notice that, when all scales collapse to one $z_2\simeq z_1\simeq z_\cross$, 
then $\gamma \to -4$,
and we recover the single broken power law fit, as in \Eq{eq:f_templateSBPL}.
The expressions for $z_1$ and $z_2$ in \Eqs{fit_z1}{fit_z2} have been tested for values of $\xi_w\gtrsim 0.1$. 
For smaller values of $\xi_w$, we propose to use the single broken power law fit in \Eq{eq:f_templateSBPL} with $z_1$ substituted with $z_\cross$ as the latter is more accurate. 
The necessary quantities to reconstruct the
values of the fitting parameters $z_1$, $z_2,$ and $\gamma$
are reported in \Tab{tabFITS} for a range of wall velocities
and the benchmark value $\alpha = 0.1$.
As mentioned above, the deviations of $z_{\rm cross}$ from $z_1$ happen
for $\xi_w \gtrsim \tfrac{1}{2} \vCJ$ (see \Fig{fig:comparison_z1z2}
for $\alpha = 0.1$) and
justify the need for a second scale $z_2$
and a double broken power law fit for $\xi_w \gtrsim \tfrac{1}{2} \vCJ$.
For these deflagrations, the jump of the velocity profile at
$\xif = \xi_\sh$ is not negligible, as can be seen in \Fig{fig_ip},
while for subsonic deflagrations with $\xi_w \lesssim \tfrac{1}{2}
\vCJ$ the velocity jump is close to zero.

\subsection{Toy models for the velocity spectral shape}
\label{sec:toymodel}

In this section, we propose simplified analytical toy
models that can be used to mimic 
the velocity profiles.
These toy models are constructed using the minimal amount of information from the velocity profiles that is relevant to provide an accurate velocity power spectrum and, hence,
the resulting GW spectrum.
In particular, to mimic $\fpsq$ for
subsonic deflagrations, we consider
a quadratic velocity profile with a discontinuity at $\xib=\xi_w$.
When a shock is formed for $\xi_w \gtrsim \tfrac{1}{2} \, \vCJ$,
we
add a second discontinuity at $\xif = \xi_\sh$.
For the case of hybrids, we consider a piecewise
linear-constant profile that continuously goes to zero at $\cs$ and
presents discontinuities at $\xi_w$ and at $\xif=\xi_{\rm sh}$.
Finally, to mimic detonations,
we use again a quadratic profile with a
single discontinuity at $\xif=\xi_w$.
The parameters of our toy models
are fixed by reproducing the
constants $\fpsqz$ and $\fpenvsq$
using \Eqs{asymptotic_fpz}{fpenv_inf}, which
determine the asymptotic limits
of $\fpsq$.
These toy models are illustrated in \Fig{toy_velocity}
and described in \Secs{quadratic}{lin_const_toy}.
In \App{appendix_toy}, we compare the proposed quadratic and linear-constant toy models with even simpler
ones (e.g., constant
and linear toy models), which allow us to exactly reproduce the locations $z_1$
and $z_2$ and one of the two asymptotic limits (e.g., $\fpenvsq$)
but they fail to reproduce
the opposite asymptotic limit (e.g., $\fpsqz$) and, hence, the intermediate slope $\gamma$ for cases that follow a double broken power law.
Therefore, the quadratic
and linear-constant toy models provide the simplest analytical
profiles to reproduce both asymptotic limits of $\fpsq$,
as we illustrate in \Figs{fp2_fp20}{hybrids_closeto_vcj}.
Throughout the rest of the paper, we study the deviation of the results computed
with the full numerical profiles $v_\ip (\xi)$ (see \Fig{fig_ip})
with respect to the results obtained using the toy models.
In particular, \Fig{fp2_fp20} compares the results for the function
$\fpsq/(zf_0')^2$.
We find great agreement between the full profiles and the
toy models of \Fig{toy_velocity}.

\begin{figure}
    \centering
    \includegraphics[width=.8\textwidth]{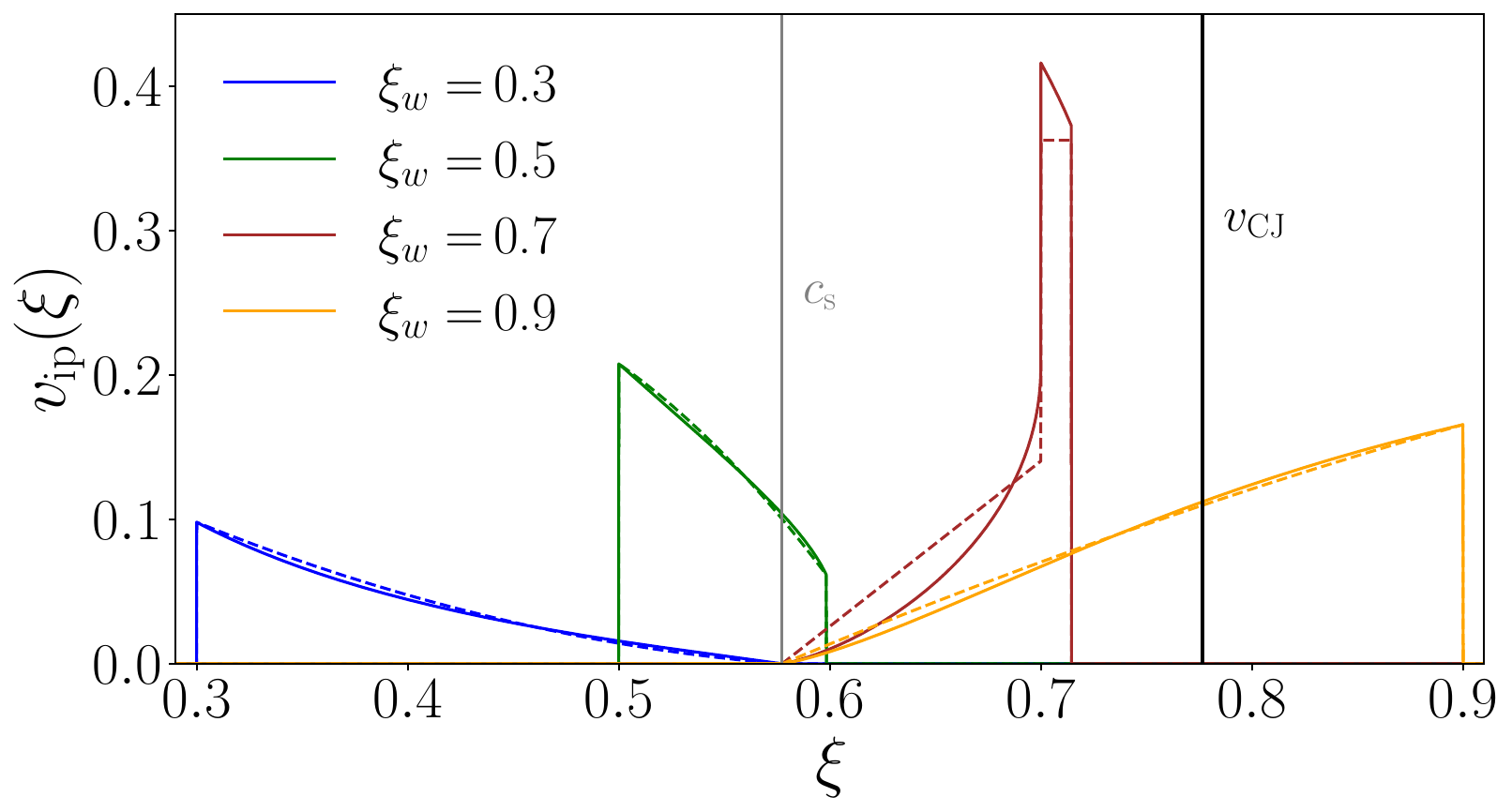}
    \caption{
    Numerical fluid profiles (solid lines) compared to the analytical toy models (dashed lines).
    The quadratic profiles of \Eq{general_quad_toy_model} with one
    or two discontinuities
    are used as toy models
    for subsonic deflagrations (see blue and green lines)
    and detonations (see orange lines).
    For hybrids, the linear-constant profile of \Eq{vtoy_lin_const} is used (see brown lines).
    }
    \label{toy_velocity}
\end{figure}

\subsubsection{Quadratic toy model for subsonic deflagrations and detonations}
\label{quadratic}

Let us consider a generic quadratic velocity profile,
continuous in the range $\xi \in (\xib, \xif)$,
\begin{align}
    v_{\rm quad}(\xi) = (C_2 \, \xi^2 + C_1 \, \xi + C_0) \mathcal{X}_{(\xib, \xif)}(\xi) \,,
    \label{general_quad_toy_model}
\end{align}
with $\mathcal{X}_{(\xib, \xif)}(\xi) \equiv \Theta(\xi-\xib)-\Theta(\xi-\xif)$
being the characteristic function of the interval $(\xib, \xif)$, and where $C_i \, (i=0,1,2)$ are free parameters.
Note that constant and linear velocity profiles can be recovered by taking $C_2 = 0$
and further taking $C_1 = 0$ for the constant case.
The corresponding $f'(z)$ has an analytical expression that can be obtained
from \Eq{def_fpz} as
\begin{align}
    \frac{f'_{\rm quad} (z)}{4\pi} = &\, C_2 \biggl[ \frac{\xi^3 \sin (z \xi)}{z^2} +
    \frac{4 \xi^2 \cos (z \xi)}{z^3} - \frac{8\xi \sin (z\xi)}{z^4}
    - \frac{8 \cos (z \xi)}{z^5}
     \biggr]_{\xib}^{\xif}  \nonumber \\
    &\, +   C_1 \biggl[ \frac{\xi^2 \sin (z \xi)}{z^2} + \frac{3 \xi \cos (z \xi)}{z^3} 
    - \frac{3 \sin (z \xi)}{z^4}\biggr]_{\xib}^{\xif} + C_0 \biggl[
    \frac{\xi \sin (z \xi)}{z^2} + \frac{2 \cos (z \xi)}{z^3} \biggr]_{\xib}^{\xif} \,,
    \label{general_fprime_toy}
\end{align}
where each term in brackets corresponds to subtracting its value evaluated
at $\xib$ from its value evaluated at $\xif$.
From \Eq{general_fprime_toy}, we
have the following limits for $z \to 0$ and $z \to \infty$,
\begin{subequations}
\begin{align}
    f'_{\rm quad}(z) &= - 4 \pi \,\biggl[ \frac{\xi^4}{3}
    \bigl( \tfrac{1}{6} C_2\, \xi^2 + \tfrac{1}{5} C_1 \, \xi + \tfrac{1}{4} C_0 \bigr)  \biggr]_{\xib}^{\xif}  z  + \mathcal{O}(z^3)  \quad \text{for } z \to 0 \,,
    \label{lim_z0_toy}
    \\
    f'_{\rm quad}(z) &=  - \frac{4 \pi}{z^{2}} \Bigl[-\xi \sin (z \xi) \, v_{\rm quad} (\xi) \Bigr]_{\xib}^{\xif} + \mathcal{O}(z^{-3})
     \quad \text{for } z \to \infty \,,
    \label{lim_zinf_toy}
\end{align} 
\end{subequations}
which, as expected, follow the same power laws as the ones of $f'(z)$
[see \Eqs{asymptotic_fpz}{eq:f1_zInfty_dependence}].
Note that in the expansion around $z \to 0$ of \Eq{lim_z0_toy},
all divergent terms in \Eq{general_fprime_toy}
of order $z^{-n}$ with $n > 0$ cancel out.

Since we want to match the $z \to 0$ and $z \to \infty$ limits
to the numerical
velocity profile $f'(z)$, using \Eqs{asymptotic_fpz}{eq:f1_zInfty},
we require the following conditions,
\begin{subequations}
\label{conditions_quad_toy}
\begin{align}
     & \tfrac{1}{6} C_2\, (\xif^6 - \xib^6) + \tfrac{1}{5} C_1 \, (\xif^5 - \xib^5)
    + \tfrac{1}{4} C_0 (\xif^4 - \xib^4) = \int_{\xib}^{\xif} \xi^3 \, v_{\rm ip} (\xi) \dd \xi \,, 
    \label{toy_z_to_zero_condition}\\
     & \xib \sin (z \xib)
    \, v_{\rm quad} (\xib^+) 
    - \xif \sin (z\xif) \, v_{\rm quad} (\xif^-) 
    \nonumber \\  & \hspace{60mm} =  \xi_w \sin (z \xi_w)
    \Delta v_{\rm ip}(\xi_w) + \xi_{\rm sh}
    \sin (z \xi_{\rm sh}) \Delta v_{\rm ip}(\xi_{\rm sh}) \,.
    \label{toy_z_to_inf_condition}
\end{align}
\end{subequations}
A quadratic toy model, characterized by
the constants $C_i$
and by the positions of $\xib$ and $\xif$
can be defined to meet the conditions in \Eqq{conditions_quad_toy} for a 
generic profile that is continuous in $\xi \in (\xib, \xif)$ (i.e.,
subsonic deflagrations and detonations).
In particular, we can determine two of the three constants by imposing
that $v_{\rm quad} (\xib^+) = v_\ip (\xib^+) = \vb$ and
$v_{\rm quad} (\xif^-) = v_\ip (\xif^-) = \vf$,
where $\vb$ and $\vf$ are the
values of the velocity profile at
$\xib$ and $\xif$, respectively
(see \App{1d_profiles}).
One possible choice respecting such conditions is
\begin{align}
    v_{\rm quad}(\xi) = \biggl[ \vf - (\vf - \vb)
    \frac{(\xif-\xi) (\xi-\xiA)}
    {(\xif-\xib) (\xib-\xiA)} \biggr] \mathcal{X}_{(\xib, \xif)}(\xi) \,,
    \label{vtoy_quad_b_2}
\end{align}
where we have introduced a third constant $\xiA$ as one of the two zeros
(together with $\xif$) of the polynomial $v_{\rm quad} (\xi) - v_f$.
\EEq{vtoy_quad_b_2} corresponds to \Eq{general_quad_toy_model} with the choice  
\begin{equation}
     C_2 = \frac{\vf - \vb}{(\xif - \xib)
     (\xib - \xiA)}\,, \qquad C_1 = - (\xif + \xiA) \, C_2\,,
     \qquad C_0 = \vf + \xif \xiA \, C_2\,.
\end{equation}
This automatically satisfies \Eq{toy_z_to_inf_condition}.
Then, we
use \Eq{toy_z_to_zero_condition} to determine the remaining constant, $\xiA$,
\begin{align}
    \label{choice_quad}
     \xiA & = \frac{\xib \, \Delta \xi \biggl[ 
    \displaystyle\int_{\xib}^{\xif} \xi^3 \, v_{\rm ip} (\xi) \dd \xi - \tfrac{1}{4} (\xif^4-\xib^4)
    \vf \biggr] + (\vb - \vf) \bigl[ \tfrac{1}{6}
    (\xif^6-\xib^6) - \tfrac{1}{5} \xif (\xif^5 - \xib^5) \bigr]}{\Delta \xi \biggl[
    \displaystyle\int_{\xib}^{\xif} \xi^3 \, v_{\rm ip} (\xi) \dd \xi - \tfrac{1}{4} (\xif^4-\xib^4)
    \vf \biggr] + (\vb - \vf) \bigl[ \tfrac{1}{5} 
    (\xif^5-\xib^5) - \tfrac{1}{4} \xif (\xif^4-\xib^4) \bigr]}\,.
\end{align}

The toy model of \Eq{vtoy_quad_b_2} applies to subsonic deflagrations, which have in general
two discontinuities at $\xib$ and $ \xif$.
For $\xi_w \lesssim \frac{1}{2} v_{\rm CJ}$, the velocity profile
continuously goes to zero at $\xif$, hence
$\vf \to 0$ as $\xif \to \cs$, while still presenting
a discontinuity at $\xib$.
This model also applies to detonations, which take a value $\vf$ at $\xif = \xi_w$,
and continuously go to zero at $\xib = \cs$.
In \Figs{fp2_fp20}{hybrids_closeto_vcj}, we show that the quadratic profiles provide an accurate modeling
of the numerical results for the
asymptotic limits and also
captures the oscillations
in $z$.
This is true for all detonations and subsonic deflagrations with $\xi_w \gtrsim \tfrac{1}{4}
\vCJ (\alpha)$.
For smaller values of $\xi_w$ (e.g.,
see $\xi_w = 0.1$ in \Fig{fp2_fp20}), we find slight
discrepancies around $z_1$ between the numerical and the toy model results.
This is potentially due to 
the fact that the
quadratic profile becomes
negative in this range of $\xi_w$.
This could be improved using a cubic profile.
However, this does not seem to considerably affect our results, so we restrict
our analysis to the use of quadratic toy models for simplicity.

\subsubsection{Linear-constant toy model for hybrids}
\label{lin_const_toy}

For hybrids, the velocity profile has two discontinuities: one
at $\xif = \xi_\sh$ and another one at $\xi_w \in (\xib, \xif)$,
where $\xib = \cs < \xi_w$.
Hence, hybrids are discontinuous within the fluid shell and
we cannot use the profile in \Eq{general_quad_toy_model}.
As we show in \App{appendix_toy},
the simplest alternative is
a linear-constant (lc) toy model,
\begin{align}
    v_{\rm lc}(\xi) = 
    \vA \frac{\xi-\xib}{\xi_w - \xib} \mathcal{X}_{(\xib, \xi_w)}(\xi) +
    \vB \mathcal{X}_{( \xi_w, \xif)}(\xi) \,,
    \label{vtoy_lin_const}
\end{align}
where $\xib$ and $\xif$ are now chosen to be $\cs$ and $\xi_{\rm sh}$ of the full profile.
This profile corresponds to the sum of
a linear profile defined in $\xi \in (\xib, \xi_w)$
with $C_2=0$, $C_1=\vA/(\xi_w-\xib)$, and $C_0 = - \xib C_1$, and 
a constant profile with $C_0 = \vB$ in $\xi \in (\xi_w, \xif)$,
presenting hence a discontinuity at $\xi_w$.
Using \Eq{general_fprime_toy},
we find the following function $f'(z)$ for the linear-constant profile,
\begin{align}
    \frac{f'_{\rm lc}(z)}{4\pi} = &\,
    \vA \biggl[ \frac{ \xi_w \,
    \sin (z \xi_w)}{z^2} + \frac{(3 \xi_w - 2 \xib) \cos (z \xi_w) - \xib \cos (z \xib)}{z^3 (\xi_w-\xib)}
    + 3\,\frac{\sin (z \xib )- \sin (z \xi_w)}{z^4 (\xi_w-\xib)} \biggr] \nonumber \\
    &\, + \vB
    \biggl[\frac{\xi_{\rm sh} \sin (z \xi_{\rm sh}) - \xi_w \sin (z \xi_w)}{z^2} + \frac{2}{z^3}
    \bigl[\cos (z \xi_{\rm sh}) - \cos (z \xi_{w}) \bigr] \biggr]\,.
\end{align}
In the $z \to 0$ and $z \to \infty$ limits, it becomes
\begin{subequations}
\begin{align}
    f'_{\rm lc}(z) = &\, \pi \biggl[
    \frac{\vA}{15} \,
    (\xib^4+\xib^3 \xi_w + \xib^2 \xi_w^2 + \xi_w^3 \xib - 4 \xi_w^4) -
    \frac{\vB}{3} ( \xi_{\rm sh}^4 - \xi_w^4 ) \biggr] \, z + \mathcal{O}(z^3) \quad \text{for $z \to 0$}\,, \\
    f'_{\rm lc}(z) = &\, \frac{4\pi}{z^2}  \, \bigl(\vA \, \xi_w \sin(z \xi_w) +   \vB
    [\xi_{\rm sh} \sin (z \xi_{\rm sh}) - \xi_w \sin (z \xi_w)]\bigr) + \mathcal{O}(z^{-3}) \qquad \,
    \text{for $z \to \infty$}\,.
\end{align}
\end{subequations}
Hence, in order to reproduce the $z \to 0$ limit in \Eq{asymptotic_fpz}
and the $z \to \infty$ limit in \Eq{eq:f1_zInfty},
we have to simultaneously satisfy the conditions
\begin{subequations}
\begin{align}
     & \vB  ( \xi_{\rm sh}^4 - \xi_w^4 ) -  \frac{\vA}{5} \,
    (\xib^4+\xib^3 \xi_w + \xib^2 \xi_w^2 + \xi_w^3 \xib - 4 \xi_w^4) = 4\int_0^\infty
    \xi^3 \, v_{\rm ip} (\xi)  \dd \xi \,,
    \label{condition_linconst_z_to_zero} \\
    &    (\vB-\vA) \, \xi_w \sin(z \xi_w) -  \vB
    \xi_\sh \sin (z \xi_\sh)  =
    \xi_w \, \sin(z\xi_w)\,  \Delta v_\ip(\xi_w) + \xi_\sh \, \sin(z\xi_\sh) \,
    \Delta v_\ip(\xi_\sh)\,.
    \label{condition_linconst_z_to_inf}
\end{align}
\end{subequations}
If we want to reproduce the full oscillations
of $\fpsq$, we have to choose $\vB=- \Delta v_\ip(\xi_\sh) = v_\sh$
and $\vA = - \Delta v_\ip(\xi_\sh) - \Delta v_\ip(\xi_w) = v_\sh - (v_+ - v_-)$.
However, this choice does not allow to reproduce the $z \to 0$ limit of $\fpsq$
since \Eq{condition_linconst_z_to_zero} is then not generally satisfied.
Hence, we choose to reproduce only the envelope of $\fpsq$ in the $z \to \infty$ limit, for which, using \Eq{fpenv_inf},
the condition in \Eq{condition_linconst_z_to_inf} becomes
\begin{align}
   |\vB-\vA| \xi_w + \vB \xi_{\rm sh} = \xi_{\rm sh} |\Delta v_\ip(\xi_\sh)|
    +
    \xi_w \,  |\Delta v_\ip(\xi_w)|\,.
    \label{condition_linconst_z_to_inf_env}
\end{align}
We then find the following choice for $\vB$ and $\vA$ (assuming
that $\vB > \vA$),
\begin{subequations}
\begin{align}
    \vA & = \frac{5 (\xi_{\rm sh} - \xi_w) (\xi_{\rm sh}^2+\xi_w^2)
     [\xi_{\rm sh} v_\sh
    +
    \xi_w \,  (v_+ - v_-)]   -  20 \displaystyle\int_0^\infty \xi^3 v_\ip(\xi) \dd \xi}{\xib^4+\xib^3 \xi_w + \xib^2 \xi_w^2 + \xi_w^3 \xib - 4 \xi_w^4 -
    \xi_w  \xi_\sh^3 - \xi_w^3 \xi_\sh + \xi_w^2 \xi_\sh^2 + \xi_w^4}\,, \\
    \vB & = \frac{\xi_{\rm sh} v_\sh
    +
    \xi_w \,  (v_+ - v_-) + \vA \xi_w}{\xi_{\rm sh}+ \xi_w}  \,,
\end{align}
\end{subequations}
which satisfies both \Eqs{condition_linconst_z_to_zero}{condition_linconst_z_to_inf_env}
at the same time.
Adding a linear profile in the range $\xi \in (\xi_w, \xif)$ would
allow us to recover the same oscillations at large $z$ than the
original profile if we set the same values of the velocity jumps
at the discontinuities.
In this case, we would have
three free parameters: $\vA = v (\xi_w^{-})$, $\vB = v (\xi_w^{+})$,
and $\vC = v(\xif^{-})$.
Hence, to reproduce the full $z \to \infty$ limit,
we would need to impose $\vC = v_{\rm ip}(\xif^{-})$
and $\vB-\vA = v_+-v_-$.
We would then still have
the freedom to choose $\vB+\vA$ such that the $z \to 0$ limit is reproduced.
This is possible, but it would involve a more complicated determination of the
toy model parameters.
Since for the final results in \Sec{FL_template}, recovering the exact
oscillations at large $z$ is not essential
to reproduce the slopes and the breaks
of the velocity spectrum, we use the simpler linear-constant toy model.

\section{Two-point correlation of the velocity field before collisions}
\label{kinetic_sp_bubbles}

\subsection{In coordinate space}
\label{bij_realspace}

The total velocity field is the linear superposition of the field induced by each of the $N_b$ expanding bubbles in the volume $V$, as described in \Eq{super_vel_w}.
Its two-point correlation function
$B_{ij} (t, \xx, \yy) = \bra{v_i (t, \xx) \, v_j (t, \yy)}$ 
can thus be written as,
\begin{align}
    B_{ij}  (t, \xx , \yy) 
    \equiv
    \Biggl\langle \,
    \sum_{n, m = 1}^{N_b}
    v_i^{(n)} \Bigl(t-t_0^{(n)}, \xx-\xx_0^{(n)}\Bigr) \,
    v_j^{(m)} \Bigl(t-t_0^{(m)}, \yy-\yy_0^{(m)}\Bigr) \!
    \Biggr\rangle \,,
    \label{eq:Bij_def}
\end{align}
where the brackets indicate an ensemble average over the nucleation locations
$\xx_0^{(n)}$ and $\yy_0^{(m)}$, and the nucleation times $t_0^{(n,m)}$.
Since we
assume a homogeneous statistical distribution of the nucleation locations,
the resultant velocity field at any time follows a
homogeneous and isotropic distribution.
Indeed, we note that, even though each realization of 
the velocity field produced by a distribution of bubbles is localized and has a preferred direction at any given point,
it becomes homogeneous and isotropic
after averaging over all possible nucleation locations
when they are homogeneously distributed in space.
We present general properties of statistically homogeneous and isotropic fields in \App{appendix_correlation}. 

The velocity fields induced by bubbles nucleated
at different
locations are uncorrelated, and the ensemble average over $\xx_0^{(n)}$ and $\yy_0^{(m)}$
is non-trivial only if $n = m$. 
Therefore, in the rest of the paper, we use the notation
\begin{equation}
    B_{ij} (t, \xx, \yy) = 
    n_b(t)\,
    \bra{V \tilde B_{ij} (t-t_0, \xx, \yy)}_{t_0}\,,
    \label{Bij_nb}
\end{equation}
where the sum in \Eq{eq:Bij_def} is replaced by the average number density of 
bubbles, $n_b(t)$, defined in \Eq{nb_t}, and
$\tilde B_{ij}$ corresponds to the integral over nucleation locations in
\Eq{eq:Bij_def} divided by the volume $V$ where bubbles can be nucleated,
\begin{equation}
    \tilde B_{ij} (t - t_0, \xx, \yy) = \frac{1}{V} \int v_i^{(n)}
    \Bigl(t - t_0^{(n)}, \xx - \xx_0^{(n)}\Bigr)
    \, v_j^{(n)}
    \Bigl(t - t_0^{(n)}, \yy - \xx_0^{(n)} \Bigr) \dd^3 \xx_0 \,.
    \label{tilde_Bij}
\end{equation}
Since $\tilde B_{ij}$ is the same for all bubbles, we omit from now on the superscript $(n)$ in the velocity $\vv$,
and in the nucleation location and time, $\xx_0$ and $t_0$.
The average over nucleation times $t_0$ 
is studied in \Sec{ensemble_times}, while in \App{time_dist}
we review the basic ingredients of the
exponential and simultaneous
distributions
of nucleation times that are 
commonly chosen in the literature.

\subsubsection{Tensor structure}

Due to statistical homogeneity and isotropy,
the two-point correlation function of the velocity field at
the points $\xx$ and $\yy \equiv \xx + \rr$,
introduced in \Eq{Bij_nb},
has the following tensor structure,
\begin{equation}
    B_{ij} (t, \rr) 
    = (\delta_{ij} - \hat r_i \hat r_j) B_{\rm N} (t, r) + \hat r_i \hat r_j B_{\rm L} (t, r)\,,
    \label{general_Bij}
\end{equation}
where $B_{\rm L} (r)$ and $B_{\rm N} (r)$ are the longitudinal (L) and normal (N) components of the correlation function [see \Eq{Bij_corr}],
which depend uniquely on the radial distance $r = |\rr|=|\xx - \yy|$.
We note that the decomposition of any homogeneous
and isotropic field is identical, e.g., $\tilde B_{ij}$ can also be decomposed
into $\tilde B_{\rm L} (r)$ and $\tilde B_{\rm N}(r)$ using \Eq{general_Bij}.

\subsubsection{Normal and longitudinal components}

We now compute the correlation function $\tilde B_{ij}$ explicitly for a generic velocity profile by averaging over all possible nucleation locations $\xx_0$.
The radial velocity field induced by an expanding bubble is
$v_i (t - t_0, \xx - \xx_0) = v_\ip (\xi_x)\, \hat \xi_{x, i}$ 
with 
$\xxi_{x} = (\xx - \xx_0)/(t - t_0)$, $\hat \xxi_x = \xxi_x/\xi_x$, and $\xi_x = |\xxi_x|$,
where the self-similar velocity profile induced by an expanding bubble is $v_\ip(\xi)$.
For a generic profile, $\tilde B_{ij}$, given in \Eq{tilde_Bij},
can be expressed as
\begin{equation}
    \tilde b_{ij} (\xxi_r)
    \equiv
    \frac{V \tilde B_{ij} (t - t_0, \rr)}{(t - t_0)^3}  
    = \int
    v_\ip (\xi_x) \, v_\ip (\xi_y) \, \hat \xi_{x, i} \, \hat \xi_{y, j} \dd^3 \xxi_0 \,,
    \label{average_x0}
\end{equation}
where we have defined the additional self-similar variables
$\xxi_{y} \equiv (\yy - \xx_0)/(t - t_0)$,  $\xxi_r \equiv \rr/(t - t_0)$, and $\xxi_0 \equiv \xx_0/(t - t_0)$.
A similar calculation was carried out in ref.~\cite{Caprini:2007xq} for
a linear velocity profile.
We extend their analysis to a generic radial profile $v_\ip (\xi)$.

To identify the relevant integration domain, we observe that the correlation
between the velocity fields at the points $\xx$ and $\yy$ is non-zero only if, at the time $t$, both
past lightcones of $\xx$ and $\yy$ intersect with the compact support of the $v_\ip$
profile sourced by the bubble nucleated at $(t_{0}, \xx_{0})$.
Hence, the integral
over nucleation locations is computed
over this intersection volume, $V_i$, and normalized by the total volume $V \supset V_i$. 
The volume $V$ in the normalization does not affect
the final result,
$B_{ij}$, given that
$n_b (t) = N_b(t)/V$ in \Eq{Bij_nb}.
Reference~\cite{Caprini:2007xq} normalized the integral in \Eq{average_x0} by the intersection volume $V_i$ instead of the total volume.
However, as we show below, if the normalizing volume depends on the distance $r$, the normal spectral density $\tilde F_{\rm N}$ before the average over nucleation times is non-zero,
which is
in contradiction with the result expected for irrotational fields (see \Sec{potential} and ref.~\cite{MY75}).
This is resolved if one would then correct the volume $V \to V_i (r)$
in \Eq{Bij_nb}, used to define $n_b(t)$, getting rid of the additional
$r$ dependence such that the final spectral density $F_{\rm N}$
would become zero even if $\tilde F_{\rm N}$ is not (see discussion in \Sec{FL_momentum}).

\begin{figure}
\begin{center}
    \includegraphics[scale=0.5]{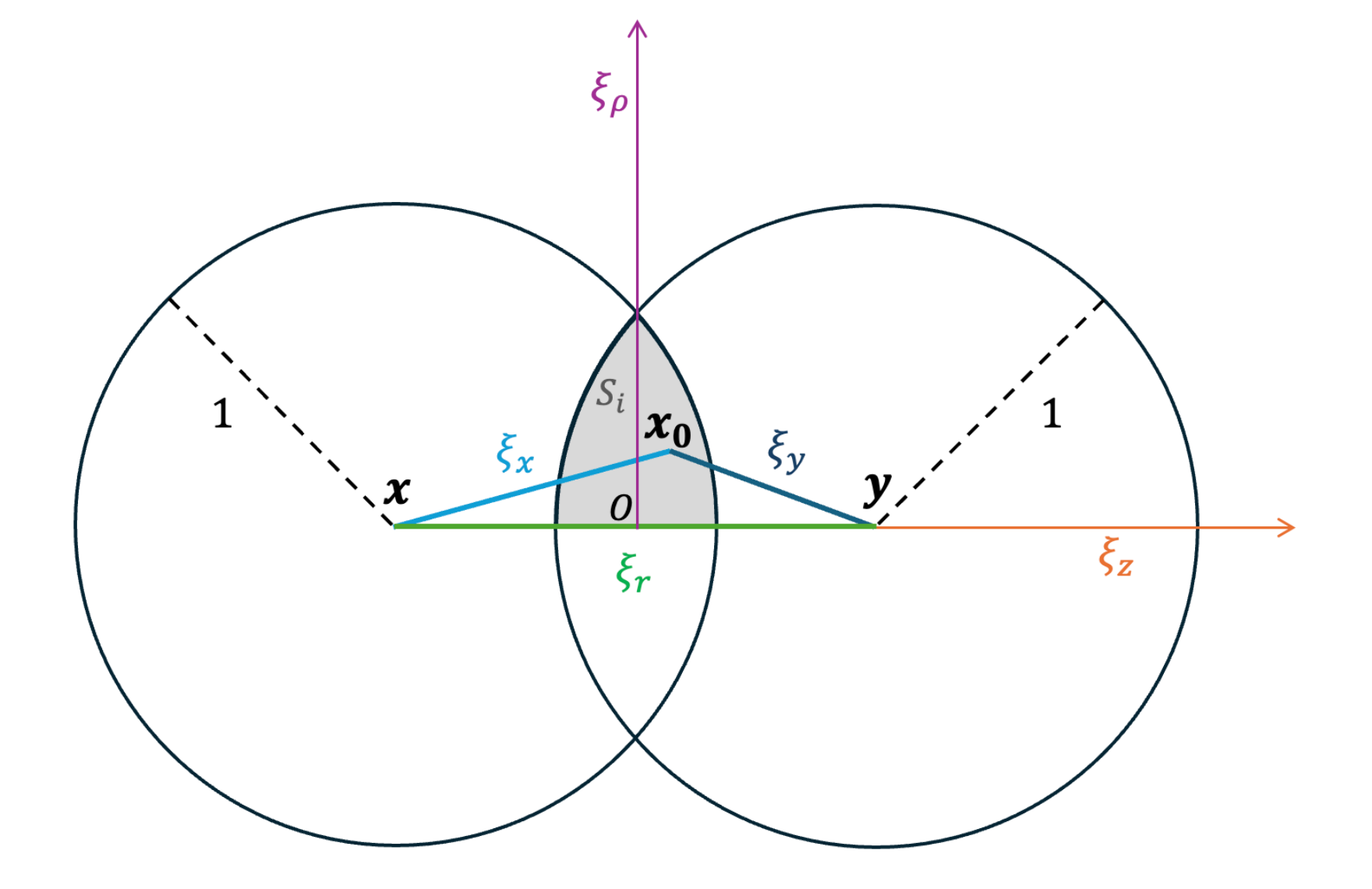}
    \caption{
    Example of contributing intersection of two light-cones, centered respectively at
    $\xx$ and $\yy$, for the evaluation of the integral in \Eq{average_x0}.
    We use self-similar variables
    $\xxi_x \equiv (\xx -\xx_0)/(t-t_0)$,
    $\xxi_y \equiv (\yy -\xx_0)/(t-t_0)$,
    $\xxi_r \equiv \rr/(t - t_0)$,
    and $\xxi_0 = \xx_0/(t - t_0)$.
    We choose a cylindrical coordinate system
    such that a point $\xxi_0$
    can be identified with coordinates 
    $(x\equiv \xi_\rho \cos \theta, y \equiv \xi_z, z \equiv \xi_\rho \sin \theta)$.
    The intersection
    volume $V_i$ is found by rotating the surface
    of revolution $S_i$ (gray shaded area)
    an angle $\theta \in (0, 2\pi)$ around the $\xi_z$ axis.
    The distance of $\xxi_0$ from the origin is $\sqrt{\xi_\rho^2+\xi_z^2}$.
    }
    \label{lcs}
\end{center}
\end{figure}

The system of coordinates used to parametrize $V_i$ and integrate \Eq{average_x0}
analytically is sketched in \Fig{lcs}.
Specifically, we introduce self-similar cylindrical coordinates
$(\xi_\rho, \xi_z, \theta)$ and set the origin
at the intermediate point between $\xx$ and $\yy$,
being $\xi_\rho$ and $\xi_z$ the distance to the origin
projected, respectively, in the perpendicular and parallel
directions to the vector $\xxi_r$, as indicated in \Fig{lcs}, 
while $\theta \in (0, 2\pi)$
is the azimuthal angle
in the plane perpendicular to ${\pmb \xi}_r$.
In this system of coordinates, we write
\begin{subequations}
\begin{align}
    \xxi_0 = (\xi_\rho \cos \theta, \xi_z, \xi_\rho \sin \theta)\,,
    & \quad \dd^3 \xxi_0 = \, \xi_\rho \dd \xi_\rho \dd \xi_z \dd \theta\,, \\
    {\pmb \xi}_x =
    - (\xi_\rho \cos \theta, \xi_z + \half \xi_r, 
    \xi_\rho \sin \theta )\,, & \quad  \
    {\pmb \xi}_y =
    - (\xi_\rho \cos \theta, \xi_z - \half \xi_r, \xi_\rho \sin \theta 
    ) \,.
\end{align}
\end{subequations}

Since the volume $V_i$ is axisymmetric, we can take the integral over $\theta \in (0, 2\pi)$,
and obtain
\begin{equation}
    \tilde b_{ij} (\xxi_r) = 2\pi \int_{S_i} \!\!
     \dd \xi_\rho \dd \xi_z \, \xi_\rho 
    \frac{v_\ip(\xi_x) \, v_\ip(\xi_y)}{\xi_x \, \xi_y} \left(\begin{array}{ccc}
        \half \xi_\rho^2   & 0 & 0 \\
        0 & \xi_z^2 - \fourth \xi_r^2 & 0 \\ 
        0 & 0 & \half \xi_\rho^2 
    \end{array} \right) \,,
    \label{eq_bij1}
\end{equation}
where $S_i$ is the surface of revolution, corresponding to the gray shaded area
in \Fig{lcs},
that produces the volume $V_i$ by a full rotation in the azimuthal direction.
The contributing matrix components of $\tilde b_{ij}$
are only diagonals because
the off-diagonal terms drop
thanks to the integral over $\theta$. They can be identified as 
$\tilde b_{11} = \tilde b_{33} = \tilde b_{\rm N}$
and 
$\tilde b_{22} = \tilde b_{\rm L}$.
The symmetry of the integrand for $\xi_z \to - \xi_z$
allows us to express it as
twice the integral over the region $\xi_z \in [0, 1 - \half \xir]$,
$\xi_\rho^2 \in [0, 1 - (\half \xir + \xi_z)^2 ]$.

\begin{figure}
    \centering
    \includegraphics[width=.49\textwidth]{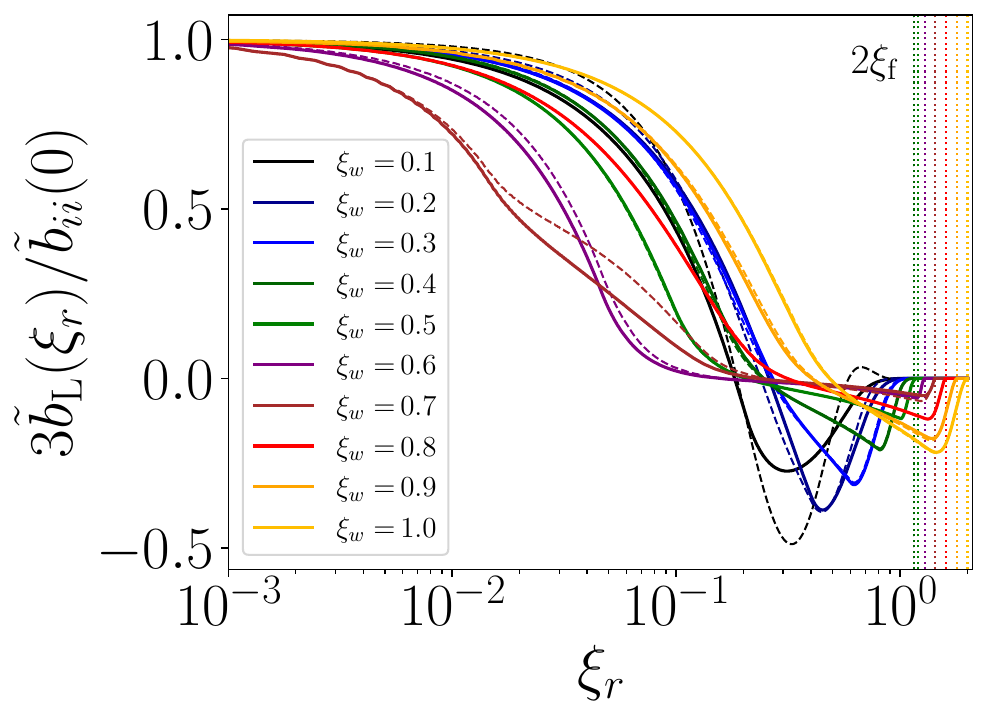}
    \includegraphics[width=.49\textwidth]{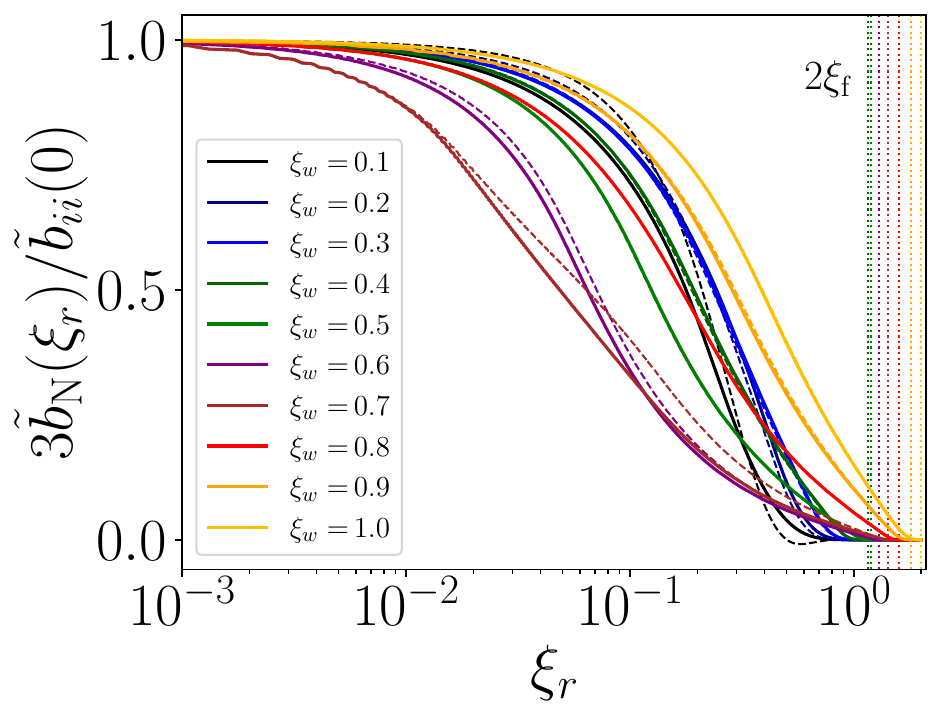}
    \caption{
    Longitudinal (left) and normal (right) components of the correlation
    function $\tilde b_{\rm N, L} (\xi_r)$ normalized by their values at $\xi_r = 0$, 
    $\onethird \tilde b_{ii} (0)$.
    As a benchmark,
    we have used $\alpha = 0.1$.
    The normal component,
    $\tilde b_{\rm N}$, is always positive,
    but the longitudinal one, $\tilde b_{\rm L}$,
    can become negative.
    Both correlation functions 
    vanish at the distance $\xi_r = 2 \, \xif \leq 2$ (vertical dotted lines),
    as expected from causality.
    The correlation functions found
    from $v_\ip (\xi)$ (solid lines) are compared to those found
    using the toy models (dashed lines) presented in \Sec{sec:toymodel}
    (i.e., quadratic for subsonic deflagrations and detonations,
    and linear-constant for hybrids).
    }
    \label{Bii_vs_xi}
\end{figure}

Finally, we change integration variables to $\xi_x$ and $\xi_y$ using 
the following relations,
\begin{equation}
    \xi_\rho^2 (\xi_x, \xi_y) = \half
    (\xi_x^2 + \xi_y^2 - \half \xi_r^2 -
    2 \xi_z^2 )\,, \qquad \xi_z (\xi_x, \xi_y) = \frac{\xi_x^2 - \xi_y^2}{2 \xir} \,,
    \label{coordinates_change}
\end{equation}
such that 
$\xir \, \xi_\rho \dd \xi_\rho \dd \xi_z = \xi_x \, \xi_y \dd \xi_x \dd \xi_y$.
Then,
the region of integration corresponds to
$\xi_x \in \bigl[\half \xir, 1\bigr]$, $\xi_y \in [|\xi_r - \xi_x|, \xi_x]$,
where the lower bound of $\xi_y$ can be understood
from the triangular inequality applied to the triangle
formed by the vectors $\xxi_x$, $\xxi_y$, and $\xxi_r$
(see \Fig{lcs}),
and \Eq{eq_bij1} can be written, using the decomposition in \Eq{general_Bij}, as
\begin{align}
    &\tilde b_{\rm N, L} (\xir) = \frac{2 \pi}{\xir}
    \int_{\half \xir}^1 \!\! \dd \xi_x \, v_\ip(\xi_x)
    \int_{|\xi_r - \xi_x|}^{\xi_x}\!\!  \dd \xi_y \, v_\ip(\xi_y)  \,
    I_{\rm N, L} (\xi_x, \xi_y) \,, \label{BNL_generic}
\end{align}
with $I_{\rm N} (\xix, \xiy) = \xi_\rho^2$ and $I_{\rm L} (\xix, \xiy) = 2 \xi_z^2 - \half\xi_r^2$.
Since the change of variables is applicable under the condition $\xi_r\neq0$,
we stress that \Eq{BNL_generic} is not
valid in the limit $\xi_r\to0$, where
instead we find
$\tilde b_{\rm N} (0) = \tilde b_{\rm L} (0) = \onethird \tilde b_{ii} (0)$ [see \Eq{bii0_irrot}].
We note that the results of \App{appendix_correlation} apply to any
statistically homogeneous and isotropic field.
\FFig{Bii_vs_xi} shows the normalized
$\tilde b_{\rm N, L} (\xi_r)$ for benchmark phase transitions with
$\alpha = 0.1$, compared to the correlation functions
found using the toy models of \Sec{sec:toymodel}.
The causal decay to zero at distances $\xir  \geq 2 \, \xif$ is clearly visible.

\subsubsection{Components relation for irrotational fields}
\label{diff_eq_bnl}

If the velocity field is purely irrotational, i.e., $\nab \times \vv = {\pmb 0}$,
the components of the corresponding correlation function satisfy
the following differential equation (see \Sec{potential} and ref.~\cite{MY75}),
\begin{equation}
    \tilde b_{\rm N} (\xi_r) + \xi_r \, \partial_{\xi_r} \tilde b_{\rm N} (\xi_r) = \tilde b_{\rm L} (\xi_r)\,.
    \label{diff_eq_pot_text}
\end{equation}
With \Eq{BNL_generic} at hand, we can verify if this is the case computing
\begin{equation}
    \xir \, \partial_{\xir} \tilde b_{\rm N}
    (\xir) =
    2 \pi \, \partial_{\xir} \Biggl[
    \int_{\half \xir}^1  \!\! \dd \xi_x \, v_\ip(\xi_x)  
    \int_{|\xi_r - \xi_x|}^{\xi_x} \!\! \dd \xi_y \, v_\ip(\xi_y) \, 
    I_{\rm N}
    (\xix, \xiy) \Biggr]
    - \tilde b_{\rm N}
    (\xir) \,. \label{aux_diff}
\end{equation}
\EEq{aux_diff} can be rearranged into
\begin{align} \label{aux3}
    \tilde b_{\rm N} (\xir) 
    + \xir \, \partial_{\xir} \tilde b_{\rm N}
    (\xir) 
    = - 2 \pi
    \int_{\half \xir}^1  
    \!\! \dd \xi_x\, v_\ip(\xi_x) \Biggl[ &v_\ip( 
    |\xir - \xi_x|)\, I_{\rm N}
    (\xi_x,
    |\xir - \xi_x|)\, {\rm sgn} (\xir - \xi_x) \nonumber \\ 
    &-  
    \int_{|\xi_r - \xi_x|}^{\xi_x} 
    \!\!\dd \xi_y \, v_\ip(\xi_y) \, \partial_\xir I_{\rm N}
    (\xix, \xiy) \Biggr]\,.
\end{align}
Setting $I_{\rm N} = \xi_\rho^2$ we find that the boundary term vanishes
since $\xi_y=|\xi_r-\xi_x|$ implies $\xi_\rho=0$,
while for the
second term we can use $\xir \partial_\xir I_{\rm N}
= 2 \xi_z^2 - \half \xi_r^2 = I_{\rm L}$.
Hence, using these results in \Eq{aux3}, we find that indeed
\Eq{diff_eq_pot_text} is satisfied.
\EEqs{aux_diff}{aux3} hold also after changing N$\,\leftrightarrow\,$L.

As discussed in the paragraph after \Eq{average_x0},
using $V_i (\xi_r)$ as the normalization volume in \Eq{average_x0}
would spoil this result, since then
    $\xir \, \partial_\xir \tilde b_{\rm N} \to V_i^{-1}\, 
    \xir \, \partial_\xir \tilde b_{\rm N} -
    V_i^{-2} \, \xir \, \tilde b_{\rm N} \, 
    \partial_\xir V_i $.
This is what is found in ref.~\cite{Caprini:2007xq},
where \Eq{diff_eq_pot_text} is not satisfied.
However, it would then be required to accordingly define
$n_b (t) = N_b(t)/V_i(\xi_r)$ in \Eq{Bij_nb},
such that
the resulting correlation function $B_{ij}$
is independent of
$V_i$ and the analogous of \Eq{diff_eq_pot_text}
is satisfied for the $B_{\rm N} (r)$ and $B_{\rm L} (r)$ components.
Here, we rather use the generic volume $V \supset V_i$, independent
of $\xi_r$, leading to the expected relation between the longitudinal and
normal components of $\tilde b_{ij}$ as given in \Eq{diff_eq_pot_text}.

\subsection{In momentum space}
\label{FL_momentum}

Because of the statistical homogeneity and isotropy
of the velocity field,
its spectral density
${\cal F}_{ij}$, defined from the correlation function
in Fourier space as
\begin{align}
    \Bra{v_i (t, \kk) \, v_j^{*} (t, \kk')}
    & = (2 \pi)^3 \, \delta^3 (\kk - \kk') \,
    {\cal F}_{ij} (t, \kk)\,,
    \label{Fij_two}
\end{align}
has the following tensor structure \cite{MY75}
\begin{equation}
    {\cal F}_{ij} (t, \kk) = (\delta_{ij} - \hat k_i \hat k_j ) F_{\rm N} (t, k) + \hat k_i \hat k_j F_{\rm L} (t, k)\,,
    \label{decomp_Fij2}
\end{equation}
where the spectral density is decomposed into longitudinal ($\FL$)
and normal ($\FN$) components [see \Eq{decomp_Fij}].
In analogy to \Eq{average_x0}, we denote the two-point function averaged
over nucleation locations with a tilde,
$\tilde {\cal F}_{ij}$,
\begin{equation}
    {\cal F}_{ij} (t, \kk) = n_b(t) \Bra{V \tilde {\cal F}_{ij} (t-t_0, \kk)}_{t_0} \,.
    \label{def_tildeFij}
\end{equation}

The decomposition of $\tilde {\cal F}_{ij}$
into $\tilde F_{\rm L}$ and $\tilde F_{\rm N}$
follows again directly from \Eq{decomp_Fij2}.
As we consider
statistically homogeneous and isotropic irrotational
fields, the normal component of the spectral density vanishes \cite{MY75}.
Indeed, re-expressing $\tilde {\cal F}_{ij}$ using \Eq{velo_fz}, we find

\begin{align} \label{corrected}
    \bra{v_i (t, \kk) \, v_j^\ast (t, \kk')}_{\xx_0}
    = &\, \frac{1}{V}
    \int v_i(t - t_0, \kk) \,
    v_j^\ast (t - t_0, \kk') \dd^3 \xx_0 \nonumber \\ = &\,
    \frac{(2 \pi)^3}{V} \, \delta^3 (\kk - \kk')
    \, \hat k_i \,
    \hat k_j (t - t_0)^6 \fpsq\,,
\end{align}
where $z = k (t - t_0)$. 
Comparing \Eq{corrected} with the general decomposition in 
\Eqs{Fij_two}{decomp_Fij2} applied to $\tilde{\cal F}_{ij}$, we directly
find $\tilde f_{\rm N} (z) = 0$ and
\begin{equation} \label{fL_fpz}
    \tilde f_{\rm L} (z) 
    \equiv
    \frac{V \, \tilde F_{\rm L} (t - t_0, k)}{(t - t_0)^6}
    = \fpsq \,.
\end{equation}
This longitudinal spectral density was directly computed in momentum
space in refs.~\cite{Hindmarsh:2016lnk,Hindmarsh:2019phv}.
Alternatively, it can be computed from $\tilde b_{\rm N}$ and $\tilde b_{\rm L}$ using \Eq{FL_from_BNBL},
\begin{equation}
    \tilde f_{\rm L} (z)
    = 4 \pi \int_0^\infty
    \!\! \dd \xir \, \xi_r^2 \,
    j_0 (z\xir) \bigl[2 \, \tilde b_{\rm N} (\xir) + \tilde b_{\rm L} (\xir) \bigr] \,.
    \label{fL_vs_bNL}
\end{equation}
Using \Eq{BNL_generic},
\begin{align}
    \tilde f_{\rm L} (z) 
    =  \frac{8 \pi^2}{z} 
    \int_0^\infty \!\!\dd \xir\, \xir \sin (z \xir)  
    \int_{\half \xir}^1\!\!\dd \xi_x\, v_\ip (\xi_x)  
    \int_{|\xi_r - \xi_x|}^{\xi_x}\!\!\dd \xi_y\, v_\ip (\xi_y)
    (\xi_x^2 + \xi_y^2 - \xi_r^2 ) \,.
    \label{fl_int_1}
\end{align}
The region of integration $\xi_r \in (0, \infty)$, $\xi_x \in \bigl[\half \xir, 1\bigr]$,
$\xi_y \in [|\xi_r - \xi_x|, \xi_x]$, extended back to the region of points with $\xi_z < 0$,
is equivalent to
$\xi_x$, $\xi_y \in (0, \infty)$, $\xir \in [|\xi_x - \xi_y|, \xi_x + \xi_y]$,
\begin{equation}
    \tilde f_{\rm L} (z) = 
    \frac{4 \pi^2}{z} 
    \int_{0}^\infty\!\!\dd \xi_x\, v_\ip (\xi_x)  
    \int_{0}^{\infty}\!\!\dd \xi_y\, v_\ip (\xi_y) 
    \int_{|\xi_x - \xi_y|}^{\xi_x + \xi_y} \!\!\dd \xir\,
    (\xi_x^2 + \xi_y^2 - \xi_r^2 ) \sin (z \xir)  \,.
    \label{FL_2}
\end{equation}
The integral over $\xir$ yields
\begin{align}
    \int_{|\xi_x - \xi_y|}^{\xi_x + \xi_y} \!\!\dd \xir\, \sin (z \xir)
    (\xi_x^2 + \xi_y^2 - \xi_r^2 )  
    &=   
    \frac{4}{z^3} \Bigl[ \sin (z \xi_x) \sin (z \xi_y) + z^2 \,
    \xi_x \, \xi_y \cos (z \xi_x) \cos (z \xi_y) \nonumber \\
    \label{integral_dxi}
    &\hphantom{=   
    \frac{4}{z^3} \Bigl[} - z \,
    \xi_x \cos (z \xi_x) \sin (z \xi_y) - z \, \xi_y \sin (z \xi_x) \cos (z \xi_y) \Bigr] \,.
\end{align}
Substituting \Eq{integral_dxi} into \Eq{FL_2}, we find that the integral
becomes symmetric in $\xi_x$ and $\xi_y$, and
separable, leading to
$\tilde f_{\rm L} (z) = \fpsq$ [see \Eq{def_fpz}].
Therefore, after correcting the normalization volume used for the average in physical space, the approach of ref.~\cite{Caprini:2007xq}
leads to a correlation function that is compatible with the
longitudinal spectral density $\tilde f_{\rm L} (z) = \fpsq$ found
in ref.~\cite{Hindmarsh:2016lnk}, as expected.

The exact $k$ dependence of
$F_{\rm L} (t, k) = n_b (t) \bra{ V \tilde F_{\rm L} (t - t_0, k)}_{t_0}$
requires a description of the ensemble average over nucleation times $t_0$,
which is addressed in \Sec{ensemble_times}.
However, the asymptotic limits of $\fpsq$ in $z \to 0$ and $z \to \infty$ become simple power laws in $z=k(t-t_0)$ [see \Eqs{asymptotic_fpz}{eq:f1_zInfty_dependence_a}].
Therefore, after performing the average over nucleation times $t_0$ at fixed $t$, one finds again the same power laws for  the asymptotic $k \to 0$ and $k \to \infty$ of $\FL(k)$ and
$\tilde F_{\rm L} (k)$: they
scale proportionally to $k^2$ and $k^{-4}$, respectively.

The non-zero component of the spectral density,
$\FL (t, k)$, decays as $k^2$ or steeper in the $k \to 0$ limit,
due to the velocity field
being irrotational and causal \cite{MY75,Hindmarsh:2019phv} (see 
discussion in \Sec{potential}).
In our case, the causality condition
is verified (see \Fig{Bii_vs_xi}) and, hence, \Eq{asymptotic_fpz} yields
\begin{equation}
    \tilde f_{\rm L} (z) =
    \fpsqz\, z^2 + \mathcal{O} (z^4) \quad \text{for } z \to 0 \,,
    \label{eq:zto0_fL}
\end{equation}
as shown in \Fig{fp2_fp20}.
As $z \to \infty$, we find $\fpsq \sim z^{-4}$ numerically (see \Fig{fp2_fp20}),
and analytically for a generic velocity profile
[see \Eqq{eq:f1_zInfty_dependence}].
Hence, $\FL (k) \to k^{-4}$ in the $k \to \infty$ limit.
This is a direct consequence of squaring the function $f'(z)$,
whose asymptotic limit $z \to \infty$ is determined by the
discontinuities
present in the velocity profile $v_\ip (\xi)$.
Additionally, since $\fpsq = \tilde f_{\rm L} (z)$ is related
to the correlation functions $\tilde b_{\rm N, L} (\xir)$
via \Eq{fL_vs_bNL},
this asymptotic limit should
be a consequence of the smoothness of $\tilde b_{\rm L}$.
Based on preliminary work,
we expect that $\tilde f_{\rm L} \sim z^{-4}$ is a consequence
of $\tilde b_{\rm L}$ ($\tilde b_{\rm N}$)  being a function of class
${\cal C}^1$ (${\cal C}^2$), which
presents discontinuities in its second (third) derivative
with respect to $\xi_r$.

\subsection{Ensemble average over nucleation times}
\label{ensemble_times}

Physical observables for the bubble ensemble, $\mathcal{O}$, are obtained after averaging
over the bubble nucleation history,
\begin{equation}
    \mathcal{O} (t) = \bra{\mathcal{\tilde{O}}(t, t_0)}_{t_0}
    \equiv
    \frac{1}{n_b(t)}
    \int_{t_c}^t
    \Gamma(t_0)\,\mathcal{\tilde{O}}(t, t_0) \dd t_0
    \,,\qquad \text{with \ } n_b(t)
    \equiv
    \int_{t_c}^t \, \Gamma(t_0) \dd t_0\,,
    \label{def_averages}
\end{equation}
where the time $t_c$ marks the moment at which the critical temperature is reached,
while
\begin{equation}
    \Gamma(t) = p(t) \, h(t)\,,
\end{equation}
is the probability distribution in time, per unit volume, of nucleating a bubble,
with $p(t)$ the probability per unit volume in the
symmetric phase,
and $h(t)\in [0, 1]$ the volume fraction in
the symmetric phase,
which can be computed from $p(t)$ as described in \App{time_dist} using \Eq{h_t}.

Hence, the correlation function in physical space becomes
[see \Eqs{Bij_nb}{average_x0}]
\begin{equation}
    B_{ij} (t, \rr) = \int_{t_c}^t  V \tilde B_{ij} \, \Gamma (t_0) \dd t_0  =
    \int_{t_c}^{t} \Gamma (t_0) (t - t_0)^3  \dd t_0
    \int v_i (\xxi_x) \, v_j (\xxi_y) \dd^3 \xxi_0\,. 
    \label{Bij_nucl_times}
\end{equation}
Similarly, in momentum space the components of the velocity
correlator
${\cal F}_{ij} (t, \kk) = \hat k_i \hat k_j F_{\rm L} (t,k)$,
given in
\Eqss{decomp_Fij2}{fL_fpz}, can be evaluated from $\tilde f_{\rm L} (z) = \fpsq$
by computing an average of single-bubble profiles over different nucleation times,
\begin{align}
    \FL (t, k) = n_b(t) \, \Bra{V \tilde F_{\rm L} (t - t_0, k)}_{t_0}
    =  \int_{t_c}^t
    \Gamma(t_0) (t - t_0)^6 \,
    \tilde f_{\rm L} \bigl[(t-t_0) \,k\bigr] \dd t_0\,.
    \label{eq_FL1}
\end{align}

We describe in \App{time_dist} how $p(t)$ and $h(t)$ are obtained in 
two particular choices for expanding
the Euclidean action, $S(t)$,
around a characteristic time $t_*$.
In particular,
when the action is a decreasing function of time,
a linear expansion around
the characteristic time $t_\ast$
yields an exponentially increasing nucleation probability
distribution per unit volume,
\begin{equation}
    p_{\rm exp} (t > t_c) = p_\ast \exp[\beta(t - t_\ast)]\,,
    \label{eq:pexp}
\end{equation}
where $p_{*}$ is given in \Eq{pt_exp_explicit} and $\beta \equiv - S'(t_\ast)$
is the rate
that characterizes the inverse
time scale of the phase transition
(see \Sec{exp_nucl}).
On the other hand,
a quadratic expansion
around a minimum of the Euclidean action
leads to a Gaussian probability distribution
in time (see \Sec{sim_nucl}),
\begin{equation}
    p_{\rm Gauss} (t > t_c) = p_\ast \exp
    \bigl[-\half \beta_2^2 (t - t_\ast)^2\bigr]\,,
\end{equation}
where $p_{*}$ depends on $\beta_2 \equiv \sqrt{S''(t_\ast)}$
[see \Eq{pt_sim_explicit}].
In the limit of simultaneous nucleation,
$p_{\rm Gauss}$ can be approximated with a delta distribution,
as described in \Sec{sim_nucl},
\begin{equation}
    p_{\rm sim} (t) = 
    \lim_{\beta_2\to \infty}
    p_{\rm Gauss} (t)
    =
    \bar n_b^{\rm sim}
    \delta(t - t_\ast)\,.
    \label{eq:psim}
\end{equation}
Note that we can define a finite
$\beta_{\rm eff}$, such that the asymptotic number of bubbles $\bar n_b$
takes the same expression for both exponential [see \Eq{nb_exp2}]
and simultaneous [see \Eq{nb_sims}] nucleations,
\begin{equation} \label{asympt_nb}
    \lim_{t \to \infty} \bar n_b^{\rm exp}(t)=
    \bar n_b^{\rm exp} = \frac{\beta^3}{8 \pi \xi_w^3}\,, \qquad
    \lim_{t \to \infty} \bar n_b^{\rm sim}(t)=
    \bar n_b^{\rm sim} =
    \frac{\beta_{\rm eff}^3}{8 \pi \xi_w^3}\,.
\end{equation}
A bar over
$n_b$ indicates that it is computed in the limit
$\beta(t_\ast - t_c) \gg 1$ or $\beta_2 (t_\ast - t_c) \gg 1$, which implies that nucleation is assumed to occur far from the critical time.

The two cases presented above are commonly known as exponential and simultaneous nucleation, 
and the timescale associated with the
phase transition is parametrized by the inverse of $\beta$ and $\beta_{\rm eff}$,
respectively.
To streamline the notation, in the following we only use
$\beta$ for both cases.
Since the dependence on the time of nucleation
in $p(t)$ and, hence, $\Gamma(t)$,
appears only in the dimensionless combination $\tilde t = \beta \, t$,
the result of the integral in \Eq{eq_FL1}
is a function of $\tilde t$ and $\tilde k = k/\beta$.
The dependence of $p_{\rm sim} (t)$
on $\tilde t = \beta_{\rm eff} \, t$ in \Eq{eq:psim} becomes apparent taking into
account that $\beta_2/\beta_{\rm eff}$ is a constant, given in \Eq{ratio_betaeff},
and that $\delta (\tilde t) = \delta (t)/\beta_{\rm eff}$.

For the exponential and Gaussian models,
the function $\Gamma(\tilde t)$
depends thus on the choice of the characteristic time scale $\tilde{t}_\ast$,
which can be fixed by choosing the volume of the symmetric phase at this time,
$h(\tilde t_*) = h_\ast$, and
on the time interval $\tilde t_*-\tilde{t}_c$.
 \FFigs{hhs}{hs_sims} in \App{time_dist}
show numerical results for $\gamma$,
which is the normalized nucleation rate,
\begin{equation}
    \gamma(\tilde{t}) \equiv \frac{\Gamma({t})}{\beta \, n_b}\,,
    \qquad \text{ with \ } \int_{-\infty}^{\infty}
    \gamma(\tilde t) \dd \tilde t = 1\,,
    \label{def_gamma_of_t}
\end{equation}
for different values of $\tilde t_\ast - \tilde t_c$ and $h_\ast$.
The distribution function $\gamma$ is normalized with the asymptotic number density of bubbles,
$n_b = n_b({\tilde t \to \infty})$.

\begin{figure}[t]
    \centering
    \includegraphics[width=0.8\linewidth]{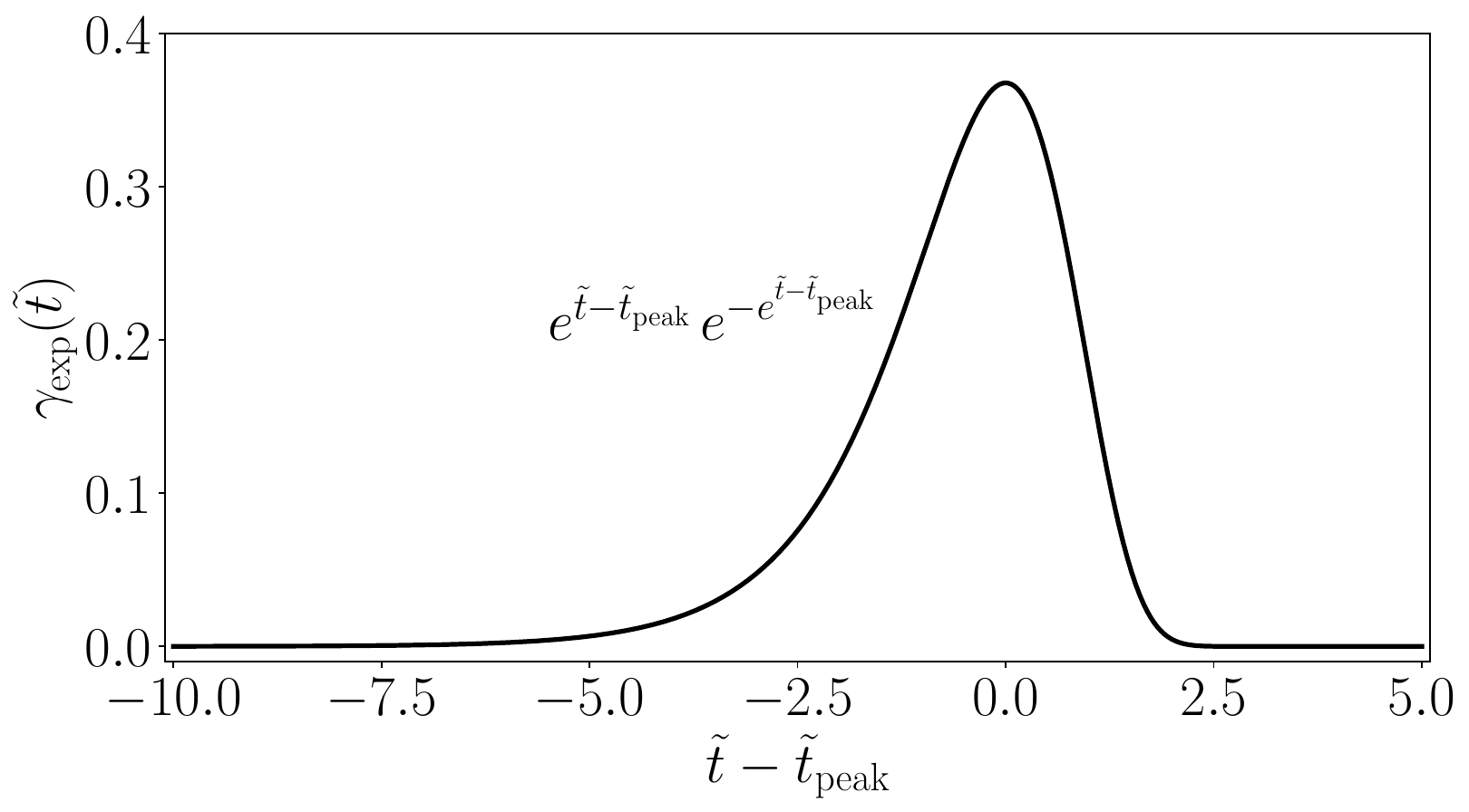}
    \caption{
    Normalized probability distribution for exponential nucleation,
    $\gamma_{\rm exp} (\tilde{t})$, defined in \Eq{def_gamma_of_t},
    as a function of $\tilde{t}-\tilde{t}_{\rm peak}$, with $\tilde{t}_{\rm peak}=\tilde{t}_\ast+C_{\rm exp}$ and $C_{\rm exp}=-\ln(-\ln h_\ast)$.
    The plot shows the asymptotic
    limit $\tilde{t}_\ast - \tilde{t}_c \gg 1$,
    for which the functional dependence of $\gamma_{\rm exp}$ with $\tilde t - \tilde t_{\rm peak}$
    does not depend on the choice of $h_{\ast}$
    (see \Sec{exp_nucl} for details).
    }
    \label{fig:gammaexp_as}
\end{figure}

In the following, we will restrict our
analysis to the case where $\tilde t_\ast - \tilde t_c \gg 1$,
such that we can effectively set $\tilde t_c \to -\infty$
\cite{Ignatius:1993qn,Hindmarsh:2019phv},
and, hence, we will omit the bar on $\bar n_b$ from now on.
We will consider two choices of the characteristic time for
exponential nucleation, $h_\ast = 1/e$ and $h_\ast \simeq 0.71$.
The former choice takes $t_\ast$ as the time when the rate of change
of the symmetric volume reaches its maximum \cite{Enqvist:1991xw,Hindmarsh:2019phv,Hindmarsh:2020hop},
while the latter choice
takes $t_\ast$ as the percolation time
\cite{Athron:2023xlk,Caprini:2024ofd} (see details in \Sec{exp_nucl}).
However, taking the limit $\tilde{t}_\ast - \tilde{t}_c \gg 1$,
the function $\gamma_{\rm exp} (\tilde{t})$ only depends on $\tilde{t}-\tilde{t}_{\rm peak}$,
where $\tilde{t}_{\rm peak} = \tilde{t}_\ast + C_{\rm exp}$ indicates the time at
which $\gamma_{\rm exp} $ reaches its peak value, with $C_{\rm exp}=-\ln(-\ln h_\ast)$,
as we show in \App{time_dist} and in \Fig{fig:gammaexp_as}.
Therefore, our results, when described
as functions of $\tilde t - \tilde t_{\rm peak}$, apply to any generic choice of $h_\ast$.
For Gaussian nucleation, we will restrict our analysis to the simultaneous case,
which effectively assumes $h_\ast \to 1$ and $\beta_2 (t_\ast - t_c) \gg 1$ (see details in \Sec{sim_nucl}).

To single out the effects of the time dependence introduced by
$\tilde f_{\rm L} (z) = \fpsq$ from the specific choice of the nucleation history in \Eq{eq_FL1},
we introduce
\begin{align}
    A(\tilde t) \equiv 
    \int_{ t_c}^t \Gamma(t_0) (t - t_0)^6
    \dd t_0 = \frac{n_b}{\beta^6} \int_{\tilde t_c}^{\tilde{t}}
    \gamma(\tilde t_0) (\tilde t - \tilde t_0)^6 \dd \tilde t_0 \,,
    \label{eq:nucl-time-average-norm}    
\end{align}
and use it to define a normalized ${\tt F}_{\rm L}$ from $\FL$,
\begin{equation}
    \FL (\tilde t, \tilde k) = A(\tilde t) \,
    {\tt F}_{\rm L}(\tilde t, \tilde k)\,.
    \label{normalized_FL}
\end{equation}

\begin{figure}[t]
   \centering
    \includegraphics[width=0.45\textwidth]{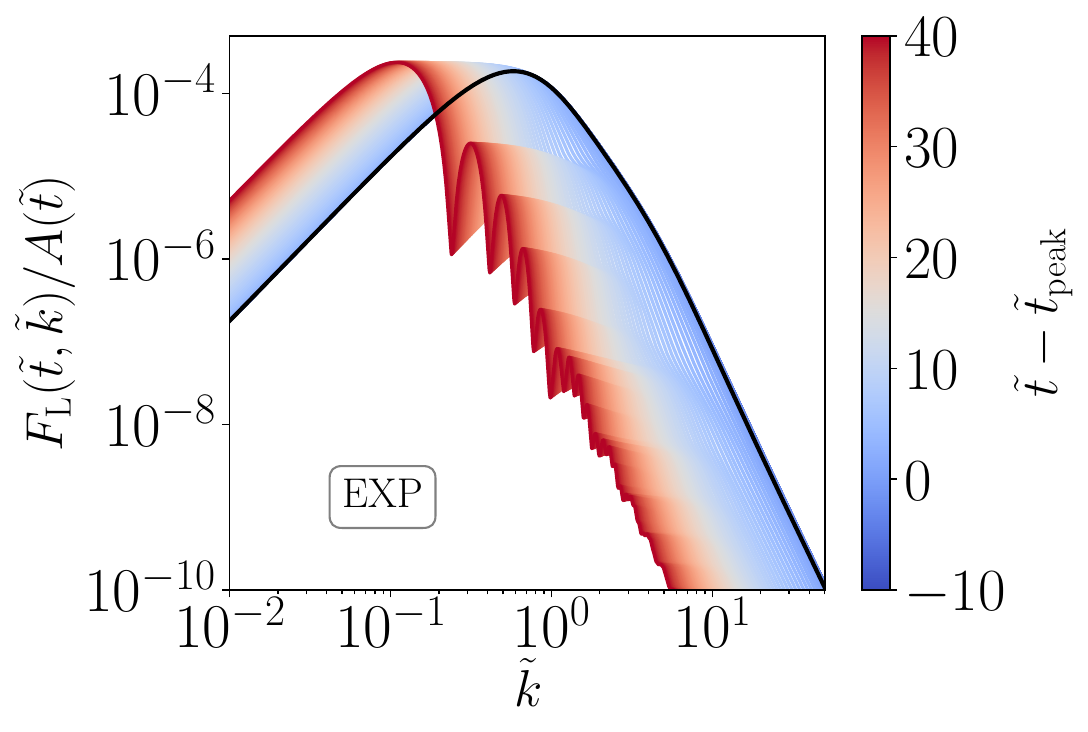}
    \includegraphics[width=0.45\textwidth]{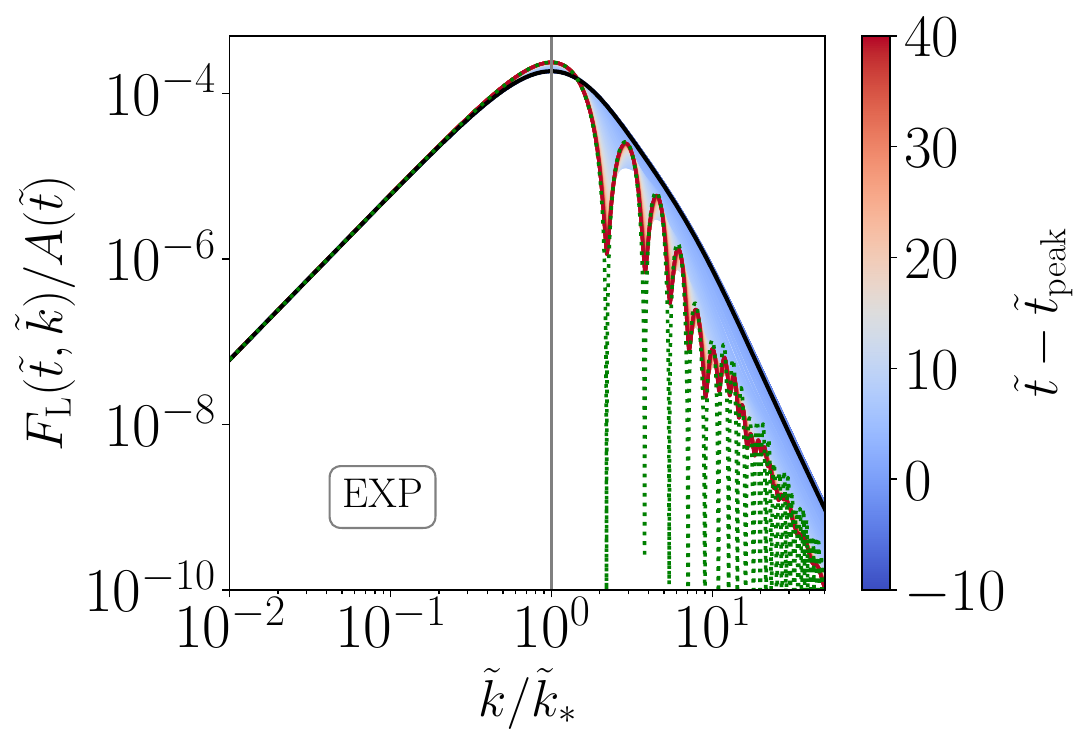}
    \includegraphics[width=0.45\textwidth]{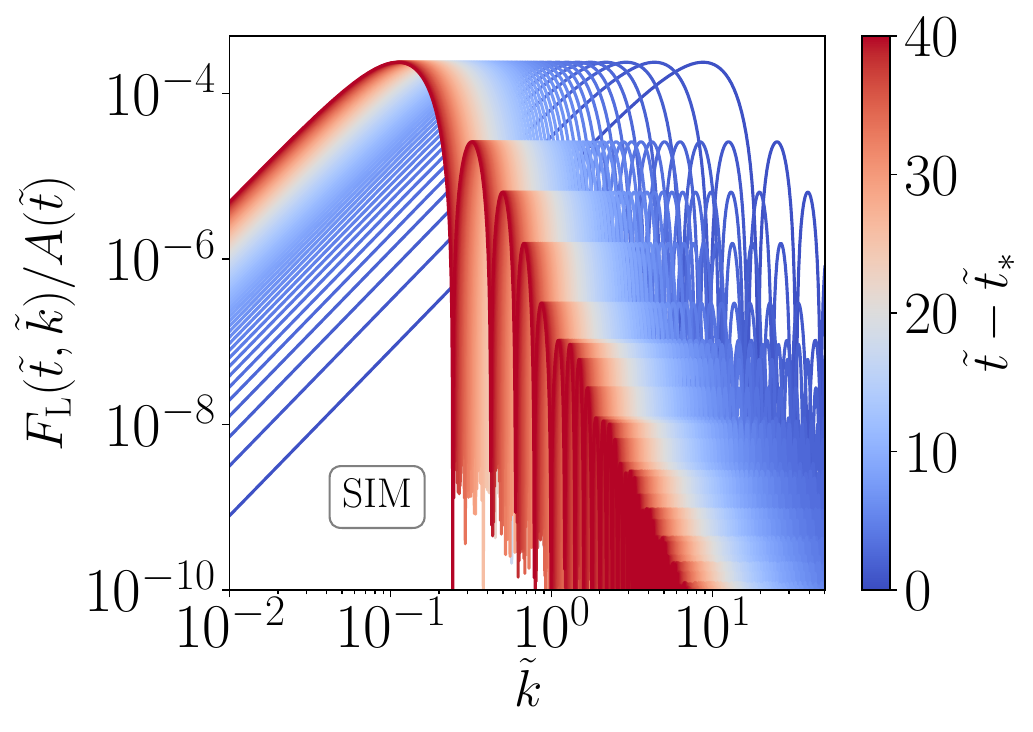}
    \includegraphics[width=0.45\textwidth]{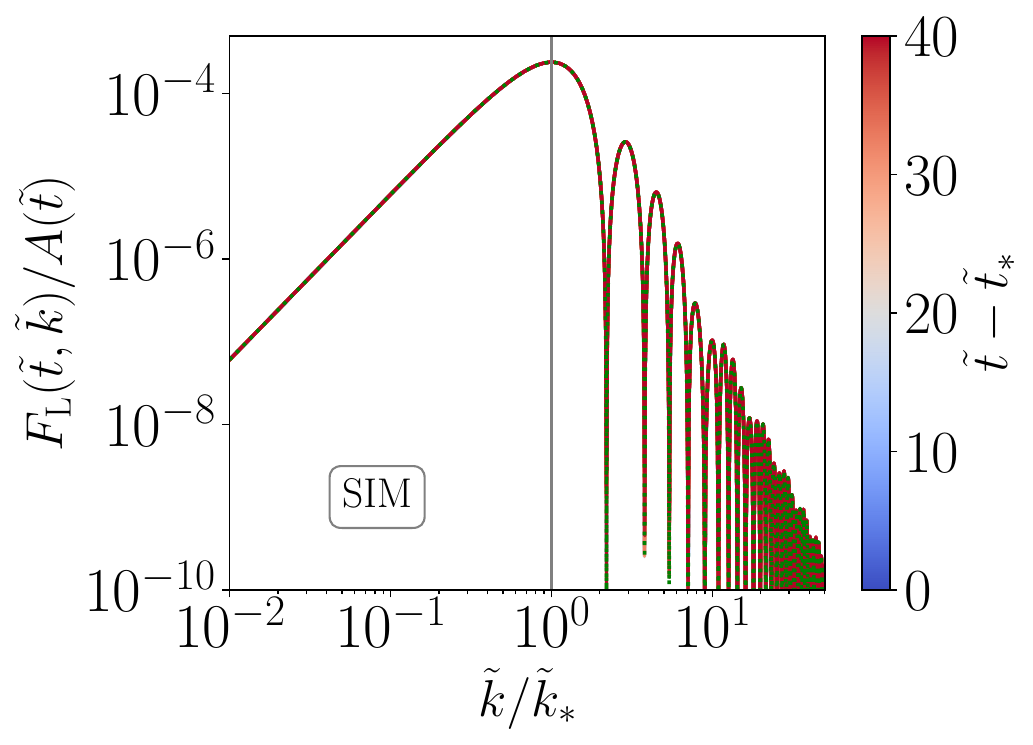}
    \caption{
    Normalized power spectral density of the velocity field
    in the bubble expansion phase, ${\tt F}_{\rm L} (\tilde{t}, \tilde k)
    \equiv \FL (\tilde{t}, \tilde k)/A(\tilde{t})$,
    as a function of time $\tilde t$ for the cases of
    exponential (EXP, top panels) and simultaneous (SIM, bottom panels) 
    nucleation.
    The power spectral density is shown as a function of $\tilde k$ (left panels)
    and the normalized $\tilde k/\tilde k_\ast$ (right panels).
    The evolution with time of the normalizing
    function $A(\tilde t)$ [see \Eq{eq:nucl-time-average-norm}] and the position of the
    peak $k_\ast (\tilde t)$ are shown in \Fig{kin_spec_av_nucl2}.
    For illustration, we take a benchmark phase transition with
    $\xi_w = 0.4$ and $\alpha = 0.1$.
    Green dotted lines in the right panels show ${f'}^2 (z/z_\ast)$,
    with $z_* \simeq z_1$ being the peak position of $\fpsq$.
    We see that in the exponential case,
    the function ${\tt F}_{\rm L} (\tilde k/\tilde k_\ast)$
    tends to ${f'}^2 (z/z_\ast)$ in the $\tilde t - \tilde t_\ast \gg 1$ limit
    and at all times in the simultaneous case.
    The black solid lines indicate
    $\tilde{t}=\tilde{t}_{\rm peak}=\tilde t_*+C_{\rm exp}$ in the case of exponential nucleation.
    }
    \label{kin_spec_av_nucl}
\end{figure}

\begin{figure}[t]
   \centering
    \includegraphics[width=0.49\textwidth]{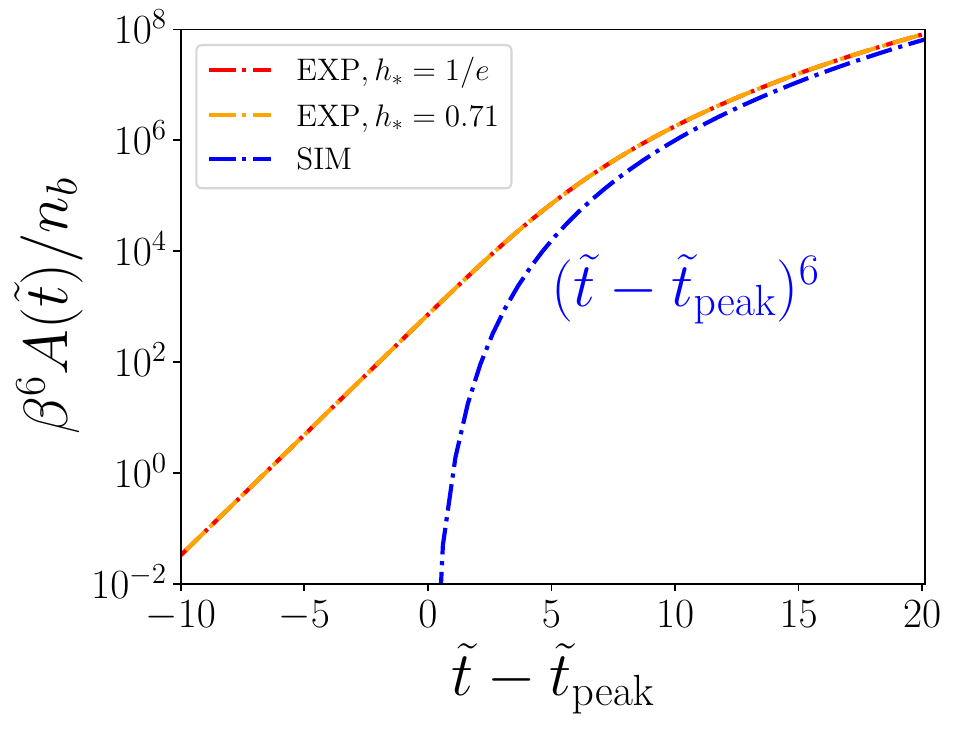}
    \includegraphics[width=0.47\textwidth]{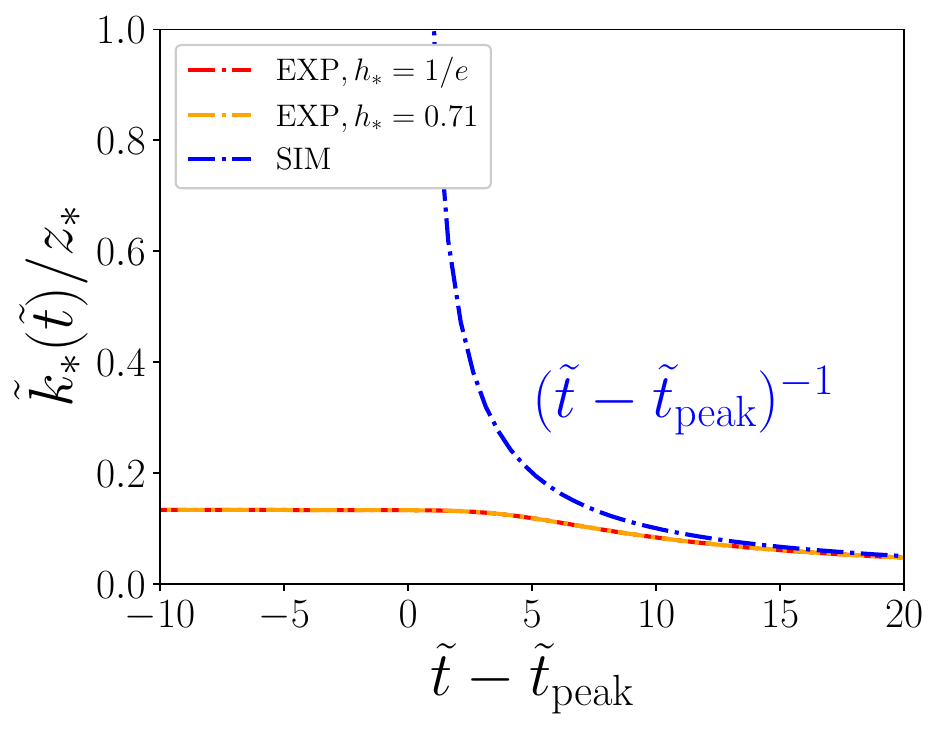}
    \caption{
    Time evolution of the function $A (\tilde t)$ [see \Eq{eq:nucl-time-average-norm}] (left panel)
    and the spectral peak $\tilde k_\ast$ (right panel) that characterize
    the velocity power spectral density, $\FL(\tilde{t}, \tilde{k})$,
    presented in \Fig{kin_spec_av_nucl},
    for exponential nucleation with $h_\ast = 1/e$ (red lines) and
    $h_\ast =0.71$ (orange lines),
    and for simultaneous nucleation
    (blue lines).
    Both $A$ and $\tilde k_\ast$
    are shown as a function of $\tilde t - \tilde t_{\rm peak}$, where
    $\tilde t_{\rm peak} = \tilde t_\ast + C$ with $C_{\rm exp} = - \ln (-\ln h_\ast)$
    and $C_{\rm sim} = 0$.
    In the simultaneous case, we find 
     $A = n_b(\tilde t - \tilde t_{\rm peak})^6 \Theta(\tilde t -\tilde t_{\rm peak})$ and $\tilde{k}_\ast
= z_\ast/(\tilde t - \tilde t_{\rm peak})$, (note that $\tilde k_*$ is defined only for $\tilde t\geq \tilde t_{\rm peak}$).
In the exponential case, instead, the aforementioned behavior with $\tilde t -\tilde t_{\rm peak}$ is only reached asymptotically   at $\tilde t - \tilde t_{\rm peak} \gtrsim 15$. The spectral peak of the $\fpsq$ function $z_\ast$ is approximately
    the value of $z_1$ investigated in \Sec{subsec_RiemannL}.}
    \label{kin_spec_av_nucl2}
\end{figure}

The time evolution of the normalized power spectral density
${\tt F}_{\rm L} (\tilde{t},\tilde{k})$ is shown in \Fig{kin_spec_av_nucl}, while in \Fig{kin_spec_av_nucl2},
we provide the time evolution of the
normalization amplitude $A(\tilde{t})$ (left panel) and the position
of the spectral peak
$\tilde{k}_*(\tilde{t})$ (right panel).
Since we consider the $\tilde t_\ast - \tilde t_c \gg 1$ limit,
the resulting
functions depend solely on the time difference $\tilde t - \tilde t_\ast$.
We find that for both exponential
and simultaneous nucleation, $A(\tilde t)$ accurately captures the growth
of $\FL$ with time, such that the amplitude of ${\tt F}_{\rm L} \equiv \FL/A$
remains roughly constant
in all panels of \Fig{kin_spec_av_nucl}.
For the simultaneous case, when we compensate $\tilde k$ with $\tilde k_\ast$, we find
that the spectrum collapses exactly into ${f'}^2(z/z_\ast)$ at all times, 
meaning that the remaining time dependence is only due to the shift of $\tilde k=z/(\tilde t-\tilde t_*)$ with time. 
On the other hand, 
for the exponential case, we observe some residual time dependence
on the normalized spectral shape ${\tt F}_{\rm L} (k/k_\ast)$, which also eventually
reaches ${f'}^2(z/z_\ast)$ at late times.
In general, the ensemble average over nucleation times introduces an additional
scale $1/\beta$, embedded in our normalization $\tilde k \equiv k/\beta$, which
enters in determining the resulting length scales of the velocity spectrum, as we study in \Sec{FL_template}.

In the simultaneous case,
the normalized distribution function becomes $\gamma(\tilde t) = \delta (\tilde t - \tilde t_\ast)$,
hence,
\begin{align}
    A_{\rm sim}(t) =
    n_b (t - t_\ast)^6 \, \Theta(t-t_\ast) \,.
    \label{a_of_t_sim_analytical}
\end{align}
Moreover,
we find the following analytical solution for
$\FL$ (shown in the bottom panels of \Fig{kin_spec_av_nucl}),
\begin{align}
    \FL^{\rm sim}(\tilde{t}, \tilde{k}) =
    A_{\rm sim} (t) \,
    {f'}^2\bigl[\tilde{k}(\tilde{t}-\tilde{t}_*)\bigr] \,.
    \label{fl_sim_analytical}
\end{align}
Hence,
the evolution of the spectral peak of $\FL^{\rm sim}$ is
$\tilde{k}_*^{\rm sim} (\tilde t) = z_*/(\tilde{t}-\tilde{t}_\ast)$,
where $z_*$ is the spectral peak of $\fpsq$.
The value of $z_\ast$ is approximately $z_1$, which has been computed
for all phase transitions in \Sec{subsec_RiemannL}.
In particular, for the single broken power law used in \Eq{eq:f_templateSBPL}, $z_\ast/z_1 =
2^{-a_1^{-1}} \simeq 0.84$.
The appearance of a second scale $z_2$ in the double broken power law
can slightly modify the relation between $z_1$ and $z_*$,
but in general we find $z_\ast/z_1 \approx (0.85-1.05)$.
Both $A_{\rm sim}(t)/n_b = (t - t_\ast)^6 \, \Theta(t-t_\ast)$
and $\tilde k_\ast^{\rm sim}$
are shown in \Fig{kin_spec_av_nucl2}, where $C_{\rm sim} = 0$ indicates
that the position of the peak of $\gamma_{\rm sim}$ is
at $\tilde t_{\rm peak} = \tilde t_\ast$.

For the case of exponential nucleation instead, these results
are only approximately valid
in the asymptotic limit $\tilde{t}-\tilde{t}_{\rm peak} \gg 1$.
In this limit,
since $\gamma_{\rm exp} (\tilde{t})$ is
peaked at $\tilde t_{\rm peak} = \tilde{t}_*+C_{\rm exp}$, 
with
$C_{\rm exp} = - \ln(-\ln h_\ast)$,
as shown in \Fig{fig:gammaexp_as},
we approximately have
\begin{equation}
    \lim_{\tilde t - \tilde t_{\rm peak} \gg 1} A_{\rm exp}(t) \simeq
    n_b (t-t_{\rm peak})^6\,.
\end{equation}
Again in this asymptotic limit,
a crude estimate of the longitudinal correlator
is therefore found by approximating
$\gamma_{\rm exp} (\tilde{t})\approx \delta(\tilde t  - \tilde t_{\rm peak})$,
such that $F_{\rm L}$ becomes
a function only of the time difference
$\tilde t - \tilde t_{\rm peak}$,
\begin{equation}
    \lim_{\tilde t - \tilde t_{\rm peak} \gg 1}
    \FL^{\rm exp}
    (\tilde t, \tilde k)  
    \simeq
    A_{\rm exp} (t) \,
    {f'}^2 \bigl[\tilde k\,
    (\tilde t - \tilde t_{\rm peak})
    \bigr]\,.
    \label{spectrum_exp}
\end{equation}
Indeed,
we observe in \Figs{kin_spec_av_nucl}{kin_spec_av_nucl2}
that the functional form at late times becomes the same
for both exponential and simultaneous nucleation
(for which $C_{\rm sim} = 0$ and $\tilde t_{\rm peak} = \tilde t_\ast$),
\begin{equation}
    \lim_{\tilde t - \tilde t_{\rm peak} \gg 1} \FL
    (\tilde t, \tilde k) = n_b\, (\tilde t - \tilde t_{\rm peak})^6\,
    {f'}^2\, \bigl[\tilde k \,
    (\tilde t - \tilde t_{\rm peak})\bigr]\,.
\end{equation}
The main consequence of
$\gamma_{\rm exp}$ being highly peaked
around $\tilde t_{\rm peak}$ is that
the oscillations of $\fpsq$ are not smoothed out at late time,
since the shape of $F_{\rm L}$ eventually inherits the shape of $\fpsq$.
For the cases of exponential and Gaussian nucleation,
the phase transition might complete before this
limit is reached.
Therefore, in general, we shall use \Eq{eq_FL1} for numerical evaluation
of $\FL(\tilde t, \tilde k)$.

In general, we show in \Sec{exp_nucl} that for any choice of $h_\ast$
in the exponential case,
the distribution function $\gamma_{\rm exp}$ can be uniquely expressed as a function
of
$\tilde t - \tilde t_{\rm peak}$ (see also \Fig{fig:gammaexp_as}).
We indeed observe in \Fig{kin_spec_av_nucl2} that both $A(\tilde t)$
and $\tilde{k}_\ast (\tilde t)$ collapse at all times when shown as a
function of $\tilde t - \tilde t_{\rm peak}$
for the two choices of $h_\ast = 1/e$ and 0.71.
At early times, we find that
the ratio $\tilde k_\ast/z_\ast \simeq 0.13$
remains roughly constant for $\tilde t - \tilde t_{\rm peak} \lesssim 5$.
In the following sections, we will be interested in computing
$\FL$ at the beginning of the sound-wave regime, $\tilde t_{\rm sw}$,
which we model in \Sec{across_collisions}.
Since for exponential nucleation
$\tilde t_{\rm sw}$ might occur before the $\tilde t - \tilde t_\ast \gg 1$
asymptotic limit is reached, we will have
some dependence on this time of the resulting spectral shape and the position
of the spectral scales, as we investigate in \Sec{FL_template}.

\subsection{Anisotropic stresses in the phase
of expanding bubbles}
\label{anis_stress1}

To conclude the treatment of the bubble expansion phase,
in this section we show that the transverse
and traceless (TT) projection of
the energy-momentum tensor of a superposition of expanding bubbles
is identically zero and,
hence, does not give rise to GWs. 
This is the extension to momentum space of the calculation
given in Appendix A of ref.~\cite{Caprini:2007xq}.
The TT-projected energy-momentum tensor in momentum space is
$\Pi_{ij}(t,\kk) = \Lambda_{ijkl}(\hat\kk) \, T_{kl}(t,\kk)$,
where $\Lambda_{ijkl}(\hat\kk)$ is
the projection tensor,
\begin{align}
    \Lambda_{ijkl} (\hat \kk) \equiv
    \half (P_{ik}P_{jl} +P_{il}P_{jk} -  P_{ij}P_{kl}) \,, \qquad
    \text{with \ } P_{ik}  (\hat \kk) = \delta_{ij} - \hat k_i 
    \hat k_j\,.
\end{align}
The energy-momentum tensor, given in \Eq{perf_fluid} for fluid
perturbations, can be expressed as
$T_{ij}(t, {\xx}) = X_i(t, \xx) X_j(t, \xx)+ p(t, \xx)\, \delta_{ij}$,
where we have defined
$X_i \equiv \sqrt{w} \gamma v_i$ \cite{vorticity}.
The TT projection of the diagonal terms 
vanishes.
Using the Fourier transform of the irrotational
velocity field induced by a single bubble, given in \Eq{velo_fz},
generalized to
$X_i (t, \pp)=\hat{p}_i X_\ip (t, p) \, e^{i \pp \cdot \xx_0}$,
the anisotropic stress tensor in momentum space becomes
\begin{align}
    \Pi_{ij}(t,\kk)
    = &\, e^{i \kk \cdot \xx_0}
    \int \frac{\dd^3 \pp}{(2\pi)^3}\, 
    \Lambda_{ijkl}(\hat k) \, \hat p_k \, \hat{\tilde p}_l \,
    X_\ip (t,p) \, X_\ip(t,\tilde p)
    \nonumber \\ 
    = & \frac{1}{4 \pi^2}
    e^{i \kk \cdot \xx_0}
    \int_{-1}^1 \dd \vartheta
    \int_0^\infty X_\ip (t,p) \, X_\ip(t,\tilde p) \,
    \frac{p^3}{\tilde p} \dd p
    \nonumber\\
    & \quad \times 
    \biggl\lbrace 
    \frac{1}{2} \bigl[
    \delta_{ij}(1-\vartheta^2) - \hat k_i \hat k_j (1+\vartheta^2) \bigr]
    - \int_0^{2\pi}\!\!\frac{\dd \phi}{2\pi} \,
    \bigl[\hat p_i \hat p_j
    -(\hat p_i \hat k_j + \hat p_j \hat k_i) \vartheta \bigr]
    \biggr\rbrace
    \,, \label{eq:anisotropic:stress_spher}
\end{align}
where we have defined $\tilde \pp = \kk - \pp$
and expressed the integral over $\pp$ in
spherical coordinates $(p,\theta,\phi)$, such that $\pp\cdot\kk = p k \vartheta$
and $\tilde p^2 = p^2 + k^2 - 2pk\vartheta$, being $\vartheta \equiv \cos \theta$.
Since the integral over $\phi$ cancels out with
the first term in brackets,
\begin{equation}
\label{integral_phi}
    \int_0^{2\pi} \frac{\dd \phi}{2\pi} \bigl[\hat p_i \hat p_j 
    -(\hat p_i    \hat k_j + \hat p_j \hat k_i) \vartheta \bigr]
    = \frac{1}{2}\bigl[
    \delta_{ij}(1-\vartheta^2) - \hat k_i \hat k_j (1+\vartheta^2) \bigr]\,,
\end{equation}
one finds indeed $ \Pi_{ij}(t,\kk)=0$.

This result shows that in general the energy-momentum tensor of
a radial velocity profile, i.e.,~induced by the same bubble $n$ \cite{Caprini:2007xq}, has identically zero traceless and transverse projection, $\Pi_{ij} = 0$. 
The velocity field resultant from many expanding and uncollided bubbles
also does not produce anisotropic stresses $\Pi_{ij}$.
Therefore, there is no GW production from the bubble expansion phase.
GWs start to be sourced by the collisions of fluid
shells, when the energy-momentum tensor $T_{ij}(t,\xx)$
contains also the velocity field arising from the collided bubbles. 
Indeed, the product of velocities induced by different
bubbles $n$ and $m$
in \Eq{eq:anisotropic:stress_spher} adds the
term $e^{i \pp \cdot \xx_0^{(n)}} e^{i \tilde \pp \cdot \xx_0^{(m)}}
= e^{i \pp \cdot (\xx_0^{(n)} - \xx_0^{(m)})} e^{i \kk \cdot \xx_0^{(m)}}$, leading to a non-vanishing
integral \cite{Jinno:2016vai,Jinno:2017fby,Cai:2023guc}.
Furthermore, at the end of the phase transition, the fluid perturbations
in the linear regime
can be described as a superposition of sound
waves, induced by different bubbles, leading to the
production of GWs as described in the sound shell model 
\cite{Hindmarsh:2016lnk,Hindmarsh:2019phv,RoperPol:2023dzg,Sharma:2023mao,part2}.

In general, one aims at calculating the normalized GW energy density power
spectrum $\OmGW  (t, k)$.
Solving the GW equation
with the Wronskian method, one finds
\cite{Caprini:2018mtu,RoperPol:2023dzg,Caprini:2024ofd}
\begin{align}
    &\OmGW  (t, k) 
    = \frac{3 k^3}{4 \pi^2} F_{\rm GW}^0 
    \int_{t_\ast}^{t}\! \frac{\dd t_1^{ }}{t_1}
    \int_{t_\ast}^{t} \!\! \frac{\dd t_2^{ }}{t_2} \,
    F_\Pi^{ } (t_1^{ }, t_2^{ }, k)  
    \cos k(t - t_1^{ }) \cos k(t - t_2^{ }) \,, \label{eq:omGW}
\end{align}
where $F_{\rm GW}^0 \simeq 1.6 \times 10^{-5}$ is the redshift of the GW energy
density to present time, assuming all Standard Model particles
are relativistic at the time of GW formation
\cite{Caprini:2019egz,RoperPol:2023dzg,Caprini:2024hue},
and $F_\Pi$ is the unequal-time correlator (UETC) of 
the anisotropic stresses,
\begin{equation}
    \bra{ \Pi_{ij}^{ } (t_1^{ },\kk^{ })  \,
          \Pi^*_{ij} (t_2^{ },\kk')}  
    \equiv (2 \pi)^3 \, \delta^3 (\kk^{ } - \kk') \,
    F_\Pi^{ }(t_1^{ }, t_2^{ }, k) \,. \label{eq:UETC:PiPi}
\end{equation}
The two-point function of the anisotropic stress on the left-hand
side of the above equation is a four-point function of the field
$X_i(t,\kk)$ that is in general unknown.
The calculation is then tackled by assuming that
the fields entering the four-point function in \Eq{eq:UETC:PiPi} are normally 
distributed.
Wick's (Isserlis') theorem then applies, and the correlator can be decomposed in terms of 
two-point functions of the velocity field \cite{Isserlis:1916}.

However, in the present case, applying Wick's theorem
leads to a non-vanishing
$F_\Pi^{ }(t_1^{ }, t_2^{ }, k)$ and, therefore, to GW production
[see \Eq{eq:omGW}], despite the fact that the velocity field before collisions
is spherically symmetric. 
To clarify this point, let us
apply Wick's theorem to the two-point correlator of $\Pi_{ij}$, similarly
to the calculation of ref.~\cite{Caprini:2007xq}.
Assuming, as is usually done in the literature, that the velocity and enthalpy perturbations
are small,
we can express $X_i \simeq \bar w \, v_i$,
where $\bar w$ is the average enthalpy.
Then, one obtains
\cite{Caprini:2007xq,RoperPol:2023dzg,Caprini:2024ofd}
\begin{align}
    F_\Pi^{ }(t_1^{ }, t_2^{ }, k)
    &\, = 
    \label{eq:UETC:PiPi:Isserlis}
    \frac{\bar w^2}{4 \pi^2} \,
    \int_0^\infty \FL (t_1, t_2, p) \, p^4 \dd p
    \int_{-1}^1
    \frac{\FL (t_1, t_2, \tilde p)}{\tilde p^2}  (1 - \vartheta^2)^2  \dd \vartheta
    \neq 0  \,,
\end{align}
where $\tilde \qq = \kk - \qq$.
The spectral densities $\FL$ in \Eq{eq:UETC:PiPi:Isserlis}
are non-zero even when the velocity field from the same bubble
is considered (see \Sec{kinetic_sp_bubbles}).
The angular integral on $\vartheta$ is also not trivial, yielding a
non-vanishing $F_\Pi$.

Applying Wick's theorem leads to a non-vanishing GW signal,
in contradiction with $\Pi_{ij} = 0$ demonstrated previously.
This indicates that 
the velocity field before collisions is non-Gaussian.
Indeed, the statistic of the velocity field given by the presence of many uncollided bubbles
is far from Gaussian, and the velocity field distribution becomes
normal only in the sound-wave regime, once it corresponds to the superposition of
velocity shells from many collided bubbles, characterized as sound waves.
Therefore, applying Wick's theorem in the uncollided regime is not justified, and a
non-zero 
$F_\Pi^{ }(t_1^{ }, t_2^{ }, k)$ 
cannot describe
the bubble expansion phase. 
However, it might be
regarded as an estimate of the anisotropic stresses induced by the collisions of the
fluid shells surrounding the bubbles.
This was indeed
used in ref.~\cite{Caprini:2007xq} to estimate  the GW signal.
Simplifying the four-point correlation function
of the velocity field with Wick's theorem renders it non-zero also for points
$\xx$ and $\yy$ belonging to different bubbles,
as it contains the product of the velocity field two-point correlation
functions of the two bubbles, which is non-zero. 
The approach could therefore be justified
if one further assumes that the velocity
field is not too much perturbed during the collision and sound-wave phases.
However, although the UETC of \Eq{eq:UETC:PiPi:Isserlis} describes it at the initial time
of the sound-wave phase, the velocity
field evolves following the sound-wave
solution during the sound-wave regime.
We will compare this result with the one obtained following
the sound shell model in ref.~\cite{part2}.

\section{Fluid perturbations across collisions}
\label{across_collisions}

As the bubbles grow, the perturbed fluid shells start to collide.
Within the sound shell model, the velocity perturbations
at this moment provide the initial conditions to
study the perturbative dynamics after the collisions.
The regime after the period of collisions
corresponds to the acoustic or sound-wave
regime and is commonly described following
the sound shell model \cite{Hindmarsh:2016lnk,Hindmarsh:2019phv}.
In the present section,
we  study the evolution of the kinetic energy with time. 
In particular, in \Sec{vrms_single} we review the
energy conservation argument to estimate the conversion
of the false vacuum energy
into thermal and kinetic energy for a single
bubble
\cite{Espinosa:2010hh,Hindmarsh:2019phv}.
We then extend this argument to the phase of
bubble expansion, where the velocity field
is described by the superposition of $N_b$ bubbles.
Since we consider the linear regime of fluid
perturbations, we assume that the average
kinetic energy density is $\bra{\rhokin}
= \bra{w \gamma^2 v^2} \simeq \bar w \bra{v^2} = \bar w
\, v_\rms^2$, where $\bar w$ is the average enthalpy.
Hence, we focus on studying the evolution of the rms
velocity $v_\rms = \sqrt{\bra{v^2}}$.
In \Sec{velocity_rms_mult}, we evaluate the time evolution
of $v_\rms^2$ during the expansion phase of multiple uncollided bubbles.
We then point out in \Sec{EffCollTime} that
an implicit assumption of the sound shell model is that the
kinetic energy density is conserved across collisions.
Then, imposing this conservation of kinetic
energy, we estimate the
effective deterministic time $\tsw$
as the time at which the kinetic energy density
around expanding bubbles matches the
result at the beginning of the sound-wave regime.

\subsection{Velocity and energy conservation for a single bubble}
\label{vrms_single}

\begin{figure}[t]
    \centering
    \includegraphics[width=.6\textwidth]{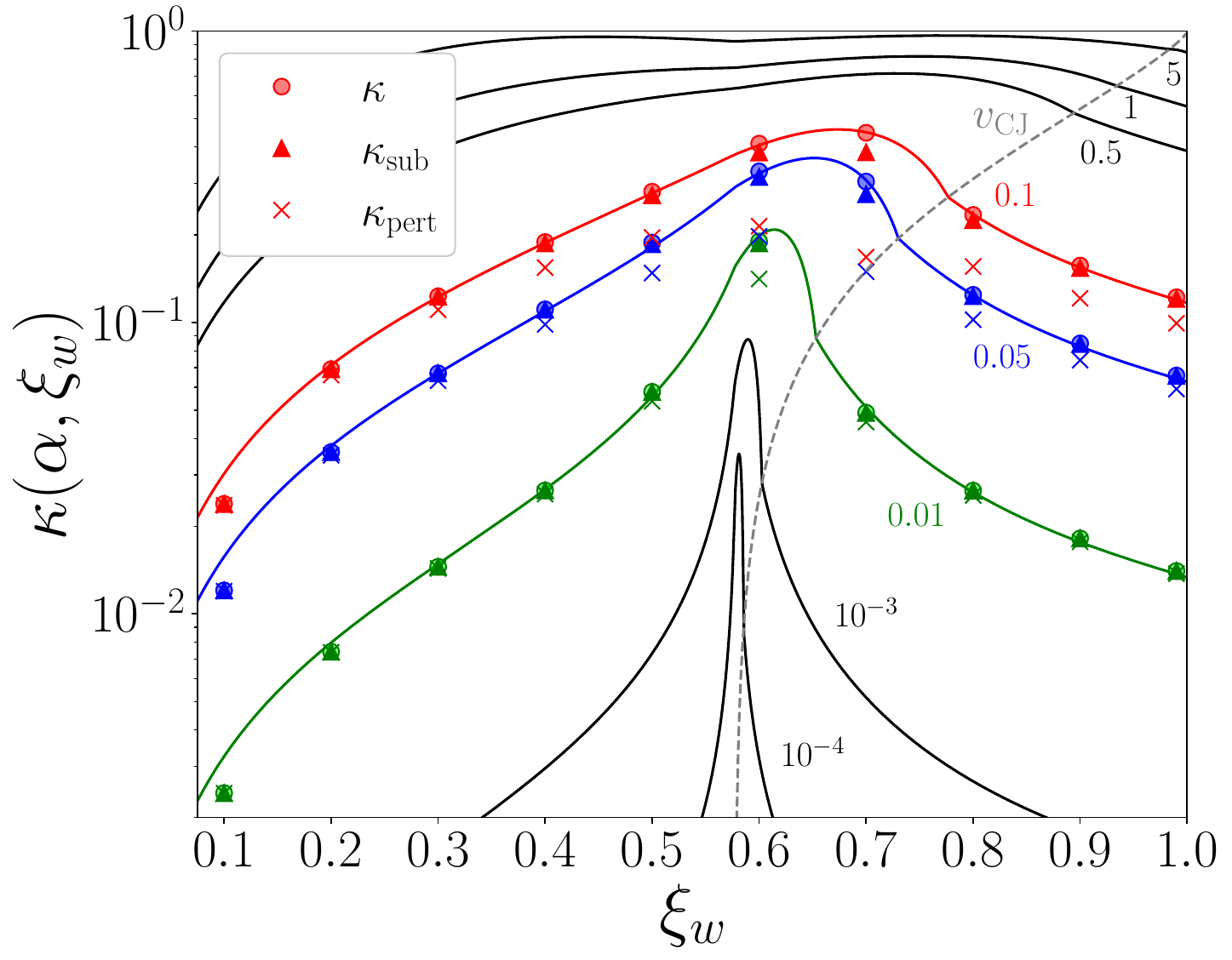}
    \caption{
    Kinetic efficiency factor $\kappa$
    as a function of the wall velocity $\xi_w$ and phase transition
    strength $\alpha$.
    The solid lines
    are evaluated using the semi-analytical fits of ref.~\cite{Espinosa:2010hh}
    for different values of $\alpha$, with the colored lines
    indicating the benchmark values shown in previous figures,
    $\alpha = 0.01$ (green), 0.05 (blue), and 0.1 (red).
    The colored dots indicate the values of $\kappa$,
    defined in \Eq{eq_kappa},
    the triangles indicate $\kappa$ when we take
    subrelativistic $\gamma \simeq 1$
    and still allow for large enthalpy perturbations ($\kappa_{\rm sub}$), and crosses 
    indicate $\kappa$ when we treat perturbatively both velocity
    and enthalpy perturbations ($\kappa_{\rm pert}$), as done in \Eq{vrms2}.
    The dashed gray line indicates the Chapman-Jouget
    speed $\vCJ$ separating deflagrations from detonations.
    }
    \label{kappas_alpha}
\end{figure}

Let us first estimate the macroscopic energy density  conversion that sources perturbations
in the fluid velocity after
a broken-phase bubble has nucleated
\cite{Espinosa:2010hh,Hindmarsh:2019phv}.
Initially, the energy density of the primordial plasma is homogeneous and takes the value
$e_n$, evaluated at the nucleation time $t_n$, defined in \Eq{nucleation_time}.
In the bag equation of state,
the false vacuum contribution in the symmetric phase is
treated as a constant $\epsilon$ that only depends on $T_n$ [see \Eq{bag_eos_intro}], such that the total energy
density just before the
nucleation of the first bubble
is
\begin{equation}
T^{00} =  e_n + \epsilon \,.
\label{eq:T00_nucl}
\end{equation}
As bubbles of the broken phase are nucleated and expand,
the vacuum energy density is released into the plasma, producing fluctuations in the fluid
velocity, $v$, and enthalpy, $w  = w_n + \delta w$.
Once the transition is completed, the radiation energy takes the perturbed value
$e = e_n + \delta e$, and the total energy density is
\begin{equation}
    T^{00} = w \gamma^2 - p
     = \rhokin + e \,,
\end{equation}
where $\rhokin = w \gamma^2 v^2$ is the kinetic energy density 
and we have used
the identity $\gamma^2 v^2 = \gamma^2 - 1$.
Then, the conservation of energy applied to a single uncollided expanding
bubble implies,
\begin{equation}
    {\cal V}_w \epsilon = 
    \int (
     \rhokin + \delta e) \dd^3 \xxi
    = 4 \pi \int_0^1 \bigl[
    X_\ip^2 (\xi)
    + \delta e_\ip (\xi)
    \bigr] \, \xi^2 \dd \xi\,,\label{energy_budget}
\end{equation}
where $X_\ip (\xi) = [\sqrt{w_\ip} \, \gamma_\ip \, v_\ip] (\xi)$
is the square root of the kinetic energy density profile,
and ${\cal V}_w \equiv V_w/(t - t_0)^3 = \tfrac{4 \pi}{3}  \xi_w^3$
is the normalized
volume occupied by the bubble.
We note that $X = \sqrt{\bar w} \, U_{f}$, where $U_f$ is the enthalpy-weighted
velocity as defined in refs.~\cite{Hindmarsh:2013xza,Hindmarsh:2015qta,Hindmarsh:2019phv,Caprini:2019egz,Caprini:2024gyk}
and $\bar{w}$ is the average enthalpy defined in \Eq{mean_enthalpy}.
Dividing \Eq{energy_budget} by its left-hand side, we find the 
efficiency coefficients of kinetic and thermal
energy conversion, 
$\kappa$ and $\omega$
\cite{Kamionkowski:1993fg,Espinosa:2010hh,Hindmarsh:2019phv},
\begin{subequations}
\label{cons_kappa}
\begin{equation}
    1 = \kappa + \omega \,,
\end{equation}
with
\begin{equation}
    \label{eq_kappa}
    \kappa = \frac{4}
    {\xi_w^3 \alpha w_n} \int_0^1
    X_\ip^2 (\xi)\, \xi^2
    \dd \xi \qquad \text{and} \qquad \omega = \frac{3}
    {\xi_w^3 \alpha w_n} \int_0^1
    \delta w_\ip (\xi)\, \xi^2
    \dd \xi\,,
\end{equation}
\end{subequations}
where we have used $\epsilon = \tfrac{3}{4} \alpha w_n$
and $\delta e_\ip(\xi)/e_n = \delta w_\ip (\xi)/w_n$ in the bag equation
of state.
Reference~\cite{Espinosa:2010hh} gives a prescription to
determine analytically $\kappa (\alpha, \xi_w)$.
The result for the benchmark values of $\alpha = 0.01,$ $0.05$, and $0.1$
is illustrated by the colored
lines in \Fig{kappas_alpha}, and compared to numerical values from
\Eq{cons_kappa} (colorful dots).
We find numerical consistency of this fit
with a maximum relative error of 3\% (see numerical values in \Tab{tab:kappas}).

Since we are interested in studying
fluid perturbations in the linear regime,
we assume small velocity ($\gamma^2 \approx 1$)
and small enthalpy perturbations, such that $w = \bar w + \delta w$,
with $\delta w \ll \bar w$.
We define the average value $\bar w$ as
\begin{equation}
    \bar w = \bra{w}_{\rm sph \, 3d} =
    \frac{3}{4 \pi} \int w_\ip (\xi) \dd^3 \xxi =  3 \int_0^1
    w_\ip(\xi)
    \, \xi^2 \dd \xi\,,
    \label{mean_enthalpy}
\end{equation}
where the angle brackets indicate 3D spherical average over the
normalized volume of the sphere with radius $\xi = 1$. 
We show the numerical  values of $\bar w/w_n$ in \Tab{tab:kappas},
which are close to unity even for the
moderately large value of $\alpha = 0.1$.
In the limit of small perturbations, $\bar w \simeq w_n$.
We note that in previous literature (see, e.g., ref.~\cite{Hindmarsh:2019phv}),
a similar average is defined
for the rms enthalpy-weighted velocity field, normalizing by the bubble
volume ${\cal V}_w$, such that $(\bar U_f^{1 \rm d})^2 \equiv \bra{U_f^2}_{\rm sph\,3d}/\xi_w^3$.
The values of $\bar w_{1\rm d} \equiv \bar w/\xi_w^3
\simeq w_n/\xi_w^3$ computed including the ${\cal V}_w$ normalization
differ from $\bar w$ by a factor
of $\xi_w^3$.
Hence, $\bar w_{1\rm d}$ becomes
very different than $\bar w$ for small $\xi_w$,
even for small perturbations.

\begin{table}[b]
    \centering
    \begin{tabular}{ccccccc} \hline
       $\xi_w$
       & $ \bar w/w_n $
       & $w_{\rm ref}/w_n$ 
       & $\kappa$ & $\kappa_{\rm pert}$ & $\omega$ \\ \hline
        $0.1$ & $1.000$ & $1.012$ & $0.024$ & $0.024$ & $0.977$ \\
        $0.2$ & $1.001$ & $1.048$ & $0.069$ & $0.066$ & $0.931$ \\
        $0.3$ & $1.002$ & $1.113$ & $0.123$ & $0.111$ & $0.877$ \\
        $0.4$ & $1.005$ & $1.224$ & $0.189$ & $0.154$ & $0.811$ \\
        $0.5$ & $1.009$ & $1.436$ & $0.281$ & $0.196$ & $0.719$ \\
        $0.6$ & $1.013$ & $1.921$ & $0.411$ & $0.214$ & $0.588$ \\
        $0.7$ & $1.019$ & $2.674$ & $0.448$ & $0.168$ & $0.548$ \\
        $0.8$ & $1.039$ & $1.504$ & $0.234$ & $0.156$ & $0.766$ \\
        $0.9$ & $1.061$ & $1.295$ & $0.157$ & $0.121$ & $0.843$ \\
        $1.0$ & $1.085$ & $1.228$ & $0.122$ & $0.099$ & $0.878$ \\
    \end{tabular}
    \caption{
    Values of the mean enthalpy $\bar w$
    [see \Eq{mean_enthalpy}], the reference
    enthalpy $w_{\rm ref}$ [see \Eq{w_ref}], and the $\kappa$ and $\omega$
    coefficients [see \Eq{cons_kappa}]
    for benchmark phase
    transitions with $\alpha = 0.1$.
    We also present the
    efficiency coefficient $\kappa_{\rm pert}$ computed
    in the perturbative limit as in \Eq{vrms2}.
    }
    \label{tab:kappas}
\end{table}

In the perturbative
limit of small velocity and enthalpy perturbations,
$X = \sqrt{\bar w} \, U_f \simeq \sqrt{w_n} v$ since $U_f \simeq v$ and $\bar w \simeq w_n$, and the efficiency $\kappa$
can be approximated as
\begin{equation}
    \kappa \simeq \kappa_{\rm pert}
    \equiv \frac{4}{\xi_w^3 \alpha}
    \int_0^1 v^2_\ip (\xi) \, \xi^2 \dd \xi =
    \frac{4 }{3 \, \xi_w^3 \alpha}
    \bra{v^2}_{\rm sph \, 3d}
    \,.
    \label{vrms2}
\end{equation}
However, as velocity or enthalpy fluctuations become larger,
\Eq{vrms2}
is not an accurate approximation of the efficiency $\kappa$,
since the omission of $w_\ip (\xi)$
in the integral of \Eq{cons_kappa} leads to significant relative
errors, as shown in \Fig{kappas_alpha} and \Tab{tab:kappas}.
Alternatively, the enthalpy-weighted velocity $U_f = \sqrt{w_\ip/\bar w} \, \gamma_\ip v_\ip$
\cite{Hindmarsh:2017gnf,Caprini:2019egz,Cutting:2019zws,Caprini:2024hue,Correia:2025qif} or
the variable $X_\ip \equiv \sqrt{\omega_\ip}
\, \gamma_\ip \, v_\ip$ \cite{vorticity} are
used to study the
GW production in the limit of large enthalpy
perturbations, i.e., when $\kappa \neq \kappa_{\rm pert}$.
To estimate this difference, we define a reference
enthalpy $w_{\rm ref}$
\begin{equation}
    w_{\rm ref} 
    \equiv 
    \frac{\displaystyle\int_0^1 X_\ip^2 (\xi) \, \xi^2 \dd \xi}
    {\displaystyle\int_0^1
    v_\ip^2 (\xi) \, \xi^2 \dd \xi}\,, \qquad \text{such that \ }
    \kappa = \frac{w_{\rm ref}}{w_n} \kappa_{\rm pert} = \frac{4}{\xi_w^3 \alpha} \frac{w_{\rm ref}}{w_n}
    \int_0^1 v^2_\ip (\xi) \, \xi^2 \dd \xi\,.
    \label{w_ref}
\end{equation}
We provide numerical values of $\kappa_{\rm pert}$ and $w_{\rm ref}$
in \Tab{tab:kappas} for $\alpha = 0.1$, which allows us to quantify
the importance of nonlinear terms in the fluid dynamics.
As visible in \Fig{kappas_alpha},
the major source of error in determining $\kappa$ for $\alpha \lesssim 0.1$
is due to neglecting the enthalpy perturbations.
Indeed, one can see that the values of $\kappa_{\rm sub}$ (in triangles),
which is computed as in \Eq{eq_kappa}
assuming subrelativistic velocities (i.e., $\gamma \simeq 1$)
but still including enthalpy perturbations in the integral over $\xi$,
are much closer to $\kappa$ (in dots)
than $\kappa_{\rm pert}$ (in crosses).

On a separate note,
since the function $\fpsq$, studied in \Sec{sec_1bubble},
represents the
single-bubble velocity correlation function, its integral
over $z$ is related to the integral over $\xi$ and, hence,
to $\tilde b_{ii} (0)$,
via Plancherel's theorem,
\begin{align}
    \tilde b_{ii} (0) = \frac{1}{2 \pi^2}
    \int_0^\infty \fpsq  \, z^2
    \dd z =
    4 \pi \int_0^1 v_\ip^2 (\xi) \, \xi^2 \dd \xi =
    \frac{4\pi}{3}
    \bra{v^2}_{\rm sph \, 3d} 
    \,.
    \label{bL_bN_0}
\end{align}
Indeed, if we take ${\cal V} = \tfrac{4\pi}{3}$
as the fractional volume occupied by the spherical
shell with radius $\xi = 1$ in \Eq{average_x0}, and reinterpret
the integral over nucleation locations $\xxi_0$ as an integral over the
single-bubble volume, we find
\begin{equation}
    \frac{\tilde B_{ii} (t - t_0, 0)}{(t - t_0)^3} =
    \frac{3}{4 \pi} \tilde b_{ii} (0) =
    \bra{v^2}_{\rm sph\, 3d}\,.
\end{equation}
We note that $\tilde b_{ii} (0)$ is identified with ${\cal V}_w \, (\bar U_{f}^{\rm 1d})^2$
in previous work
\cite{Hindmarsh:2015qta,Hindmarsh:2017gnf,Hindmarsh:2019phv},
under the assumption of small perturbations.
Since the field is irrotational, the longitudinal
and normal components of the correlation function satisfy
$\tilde b_{\rm L} (0) = \tilde b_{\rm N} (0) = \frac{1}{3}
\tilde b_{ii}(0)$
[see \Eq{bii0_irrot}].

To prove the validity of the second equality in
\Eq{bL_bN_0}, we use the expression
of $\fpsq = \tilde \fL (z)$ given in \Eqs{FL_2}{integral_dxi},
\begin{align}
    \frac{1}{2 \pi^2} \int_0^\infty \fpsq \, z^2 &\, \dd z = 8
    \int_0^\infty v_\ip (\xi_x) \dd \xi_x \int_0^\infty v_\ip (\xi_y) \dd \xi_y \nonumber \\ \times 
    \int_0^\infty &\, \biggl[\frac{\sin (z \xi_x) \sin (z \xi_y)}{z^2} - \frac{2}{z} \xi_x \cos (z \xi_x) \sin (z \xi_y)
    + \xi_x\, \xi_y \cos (z \xi_x) \cos (z \xi_y) \biggr] \dd z \,,
    \label{fp2_z2_dz}
\end{align}
where the integral over $z$ can be split in three terms.
The  first two terms are
\begin{align}
    \int_0^\infty &\, \biggl[ \frac{\sin (z \xi_x) \sin (z \xi_y)}{z^2} -   \frac{2 \xi_x \cos (z \xi_x)
    \sin (z \xi_y)}{z}  \biggr] \dd z \nonumber \\ &\,
    = \frac{\pi}{4}
    (\xi_x + \xi_y - |\xi_x - \xi_y|) -
    \frac{\pi}{2} \xi_x \biggl(1 - \frac{\xi_x - \xi_y}{|\xi_x - \xi_y|} \biggr) =
    \frac{\pi}{2} \left\{ \begin{array}{cc}
       \xi_y \ , & \quad {\rm when \ } \xi_x > \xi_y \ ,  \\
       - \xi_x \ , & \quad {\rm when \ } \xi_x \leq \xi_y \ .
    \end{array} \right.
    \label{term1}
\end{align}
After integration over $\xi_x$ and $\xi_y$, the contribution of
\Eq{term1} to \Eq{fp2_z2_dz} vanishes.
The remaining term in \Eq{fp2_z2_dz} is
\begin{equation}
    \xi_x \, \xi_y \int_0^\infty \cos z \xi_x \cos z \xi_y \dd z = \frac{\pi}{2}  \xi_x \, \xi_y \bigl[ \delta(\xi_x + \xi_y) + \delta (\xi_x - \xi_y) \bigr]\,,
    \label{term2}
\end{equation}
which, substituted back into \Eq{fp2_z2_dz}, reduces to the
trace of the correlation function of the velocity field, $\tilde b_{ii} (0)$,
as we wanted to prove in \Eq{bL_bN_0}.

\subsection{Time evolution of the kinetic energy of multiple uncollided bubbles}
\label{velocity_rms_mult}

\begin{figure}[t]
    \centering
    \includegraphics[width=0.8\linewidth]{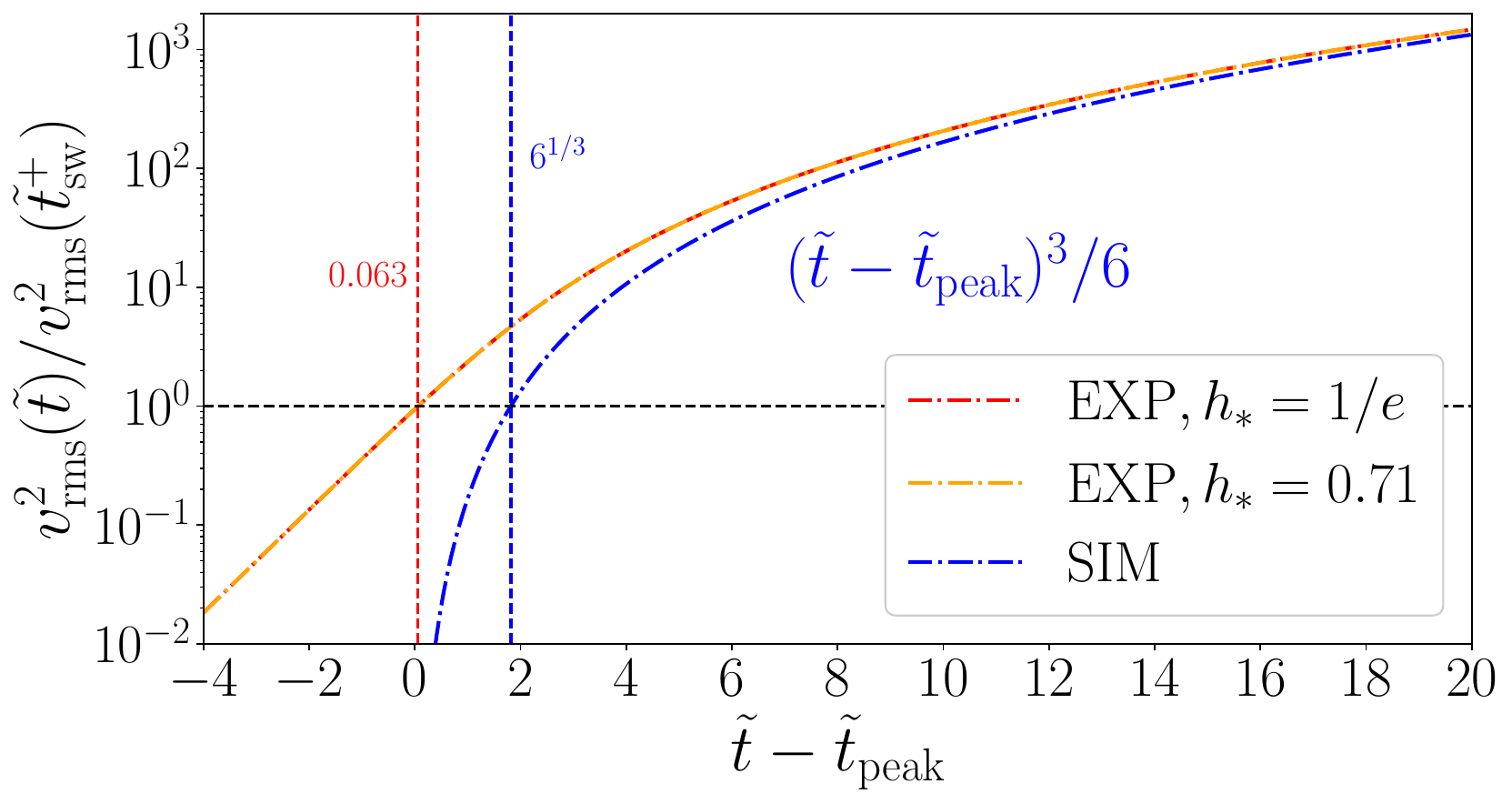}
    \caption{
    Evolution of the mean squared fluid velocity
    $v_\rms^2 (\tilde t) \equiv \bra{v^2(\tilde t)} = B_{ii} (\tilde t, 0)$, given in \Eq{vrms_expanding},
    during the phase of bubble expansion, normalized by
    $\tilde b_{ii} (0)/{\cal V}_w$,
    which corresponds to the value at the beginning of the sound-wave
    regime, $v_\rms^2(\tilde t_\sw^+)$,
    as we show in \Sec{EffCollTime} [see \Eq{eq:vrms2_coll+}].
    The red and orange lines correspond to the cases of 
    exponential nucleation with $h_\ast = 1/e$ and 0.71, respectively,
    while the blue line corresponds to the case of simultaneous nucleation.
    For the cases of exponential nucleation, $v^2_\rms$ is a universal
    function of $\tilde t - \tilde t_{\rm peak}$, independent of the choice
    of $h_\ast$.
    The simultaneous case follows the function $\tfrac{1}{6}
    (\tilde t - \tilde t_{\rm peak})^3$,
    which is also followed by the exponential case at late times,
    $\tilde t - \tilde t_{\rm peak} \gtrsim 20$.
    The curves cross unity at $\ttsw$, i.e.,
    when \Eq{eq:tcoll-condition} is satisfied, which happens at
    $\tilde{t}_{\rm sw}-\tilde t_{\rm peak} \simeq 0.063$
    for exponential nucleation (vertical red dashed line),
    and at $\tilde{t}_{\rm sw}-\tilde{t}_\ast = 6^{1/3} \simeq 1.82$ for simultaneous nucleation (vertical blue dashed line).
    }
    \label{fig:tcollnew}
\end{figure}

Extending the previous results to a superposition of multiple bubbles,
the evolution of the ensemble average of the
kinetic energy density, estimated as
$\bra{\rhokin} = X_\rms^2 = \bar w \, \bar U_{f}^2 \simeq
w_n v_\rms^2$ in the limit of
small fluid perturbations,
can be computed from
the power spectral density $\FL (t, k)$, studied in \Sec{ensemble_times}.
Integrating $\FL$ [see \Eq{eq_FL1}] over $k$, we can find the time evolution of
$v_\rms^2 (t) \equiv \bra{v^2 (t)} = B_{ii} (t, 0)$
via Plancherel's theorem [see \Eq{traces_relation}],
\begin{align}
    v_\rms^2 (t) = \bra{v^2(t)} = &\,  \frac{1}{2 \pi^2}
    \int_0^\infty \FL (t, k) \, k^2 \dd k \nonumber \\ = &\,
    \frac{1}{2 \pi^2}
    \int_0^\infty k^2 \dd k \int_{t_c}^t \Gamma(t_0) (t - t_0)^6
    \tilde \fL \bigl[k (t  - t_0)\bigr] \dd t_0\,.
    \label{vrms_from_FL}
\end{align}
Changing the integration variables from $t_0$
and $k$ to $t_0$ and $z \equiv k (t - t_0)$,
we find
\begin{align} \label{vrms_expanding}
    v_\rms^2 (\tilde t)
    = &\, \frac{1}{2 \pi^2}
    \int_0^\infty \tilde \fL(z) \, z^2 \dd z \int_{t_c}^t \Gamma(t_0) (t - t_0)^3
    \dd t_0 = \tilde b_{ii} (0) \int_{t_c}^t \Gamma(t_0) (t - t_0)^3
    \dd t_0
    \nonumber \\ = &\,
    \frac{\tilde b_{ii} (0)}{{\cal V}_w} \frac{1}{6}
    \int_{\tilde t_c}^{\tilde t}
    \gamma(\tilde t_0) (\tilde t - \tilde t_0)^3 \dd \tilde t_0
    = \frac{1}{8} \kappa_{\rm pert} \alpha \int_{\tilde t_c}^{\tilde t}
    \gamma(\tilde t_0) (\tilde t - \tilde t_0)^3 \dd \tilde t_0
    \,,
\end{align}
where we have used $\beta^3/n_b = 8 \pi \xi_w^3 = 6 {\cal V}_w$,
given in \Eq{asympt_nb},
and ${\cal V}_w = \tfrac{4}{3} \pi \xi_w^3$, such that their ratio is 6.
We have also used the relation $\tilde b_{ii} (0) = \pi \xi_w^3 \kappa_{\rm pert}\alpha$, which can be obtained combining \Eqs{vrms2}{bL_bN_0}.
This $v_{\rm rms}^2$ represents the evolution of the rms velocity
of a single bubble multiplied by
the expectation value of the fractional
volume occupied by the $N_b$ nucleated bubbles 
over time, shown in \Fig{fig:tcollnew}.

For simultaneous nucleation, $\gamma_{\rm sim}
(\tilde t) = \delta (\tilde t - \tilde t_\ast)$,
and $v_\rms^2$ becomes
\begin{equation} \label{vrms_sim}
    (v_\rms^{\rm sim})^2 (\tilde t) =
    \frac{\tilde b_{ii}(0)}{{\cal V}_w}
    \frac{1}{6} \, (\tilde t - \tilde t_\ast)^3 = 
    \frac{1}{8} \, \kappa_{\rm pert} \alpha \, (\tilde t - \tilde t_\ast)^3 \,.
\end{equation}
On the other hand,
$v_\rms^2$ in the exponential case follows
that of the simultaneous case
only in the asymptotic limit $\tilde t - \tilde t_{\rm peak}
\gg 1$,
where, as in \Sec{ensemble_times}, we can approximate the peaked distribution
function $\gamma(\tilde t) \simeq \delta (\tilde t - \tilde t_{\rm peak})$, and
find
\begin{equation}
    \lim_{\tilde t - \tilde t_\ast \gg 1}
    (v_\rms^{\rm exp})^2 (\tilde t) \simeq \frac{\tilde b_{ii}(0)}{{\cal V}_w}
    \frac{1}{6} \, (\tilde t - \tilde t_{\rm peak}^3)
    = \frac{1}{8} \, \kappa_{\rm pert}
    \alpha \, (\tilde t - \tilde t_{\rm peak})^3 \,.
\end{equation}
\FFig{fig:tcollnew} shows the evolution of $v_\rms^2$, normalized by
$\tilde b_{ii}(0)/{\cal V}_w$,
as a function of the time difference with respect to the time
when the peak of $\gamma$ is reached, $\tilde t - \tilde t_{\rm peak}$.
We note that $\tfrac{1}{6} (\tilde t - \tilde t_{\rm peak})^3$ (shown in green)
is followed at all times for the simultaneous case, while it is followed
by the exponential case only at late times $\tilde t - \tilde t_{\rm peak} \gtrsim 20$.

The result in \Eq{vrms_expanding} also applies for a 
deterministic realization of the nucleation history
of $N_b$ bubbles in a volume $V$.
In this case,
the distribution function $\Gamma(t_0)$ collapses to a delta at each
time of nucleation of a bubble $n$,
\begin{equation}
    \Gamma(t_0) = \frac{1}{V} \sum_{n = 1}^{N_b} \delta (t_0 - t_n)\,,
\end{equation}
and the resulting $v_\rms^2$ is
\begin{equation}
    v_\rms^2 (t) = \frac{\tilde b_{ii} (0)}{V}
    \sum_{n = 1}^{N_b} (t - t_n)^3 = \frac{\tilde b_{ii} (0)}{V}
    \frac{V_{\rm bp}}{{\cal V}_w} = \frac{3}{4}  \kappa_{\rm pert} \alpha \,
    {\cal V}_{\rm bp} (t)\,, \label{vrms_deterministic}
\end{equation}
where ${\cal V}_{\rm bp} \equiv V_{\rm bp}/V$ is the fraction of
the volume in the broken phase $V_{\rm bp} = {\cal V}_w \sum_i (t - t_i)^3$,
and we have
related $\tilde b_{ii} (0)$ to $\kappa$ combining \Eqs{vrms2}{bL_bN_0}.
An analogous approach was used in ref.~\cite{Caprini:2024gyk} (see their appendix B)
to track the evolution of
the kinetic energy fraction $K (t) \equiv \bra{\rhokin}/(e_n + \epsilon)$
during the bubble expansion phase, 
which can also be written as $K(t)=\tfrac{4}{3} \tfrac{v_{\rm rms}^2(t)}{(1+\alpha)}$,
using $\bra{\rhokin} = w_n v_{\rm rms}^2$, $e_n=3w_n/4$,
$\epsilon= \alpha \, e_n$.
From \Eq{vrms_deterministic}, we find that the evolution of the
deterministic kinetic energy fraction is
\begin{equation}
    K (t) = \frac{\kappa \alpha}
    {1 + \alpha} {\cal V}_{\rm bp} (t) = K_\xi \, {\cal V}_{\rm bp} (t)\,.
    \label{kinetic_energy_fraction_evolution}
\end{equation}
The authors in ref.~\cite{Caprini:2024gyk} compare the fraction
$K(t)/{\cal V}_{\rm bp} (t)$ to $K_\xi \equiv \kappa \alpha/(1 + \alpha)$
to test the convergence of their simulations in their figure~14.

\subsection{Conservation of kinetic energy and initial
time of the sound-wave phase}
\label{EffCollTime}

As the bubbles expand, eventually they start colliding and producing GWs.
After the period of collisions,
the initial conditions for the velocity field
at the beginning of the
sound-wave regime, following the sound shell model, 
are obtained by averaging
the function $\fpsq$ over the
bubble lifetime $T$
(see \Sec{distribution_lifetimes} for details about the
bubble lifetime distribution $\nu$) \cite{Hindmarsh:2019phv,RoperPol:2023dzg} 
\begin{align}
    \FL(\tilde t_\sw^+, \tilde k) =
    \frac{n_b}{\beta^6} \int_0^{\infty}
    \nu(\tilde{T}) \, \tilde{T}^6 {f'}^2 (\tilde{k} \tilde{T})
    \dd \tilde T \,,
    \label{eq:FLsw}
\end{align}
where $\tilde T = T \beta$ and we define $\tilde t_\sw$ as the
starting time of the sound-wave regime.
The first question we want to address is what is the corresponding
kinetic energy density at the beginning of the sound-wave regime,
which is determined by $\rhokin \simeq w_n v_\rms^2 (\ttsw^+)$ in the limit of small fluid 
perturbations.
Then, we want to understand at
which time,
$v_\rms^2 (\tilde t)$ in the uncollided expanding phase would reach the value of $v_\rms^2 (\ttsw^+)$
if one ignores the potential impact of collisions in the kinetic
energy, i.e., assuming that kinetic
energy is conserved.
To answer these questions, we
insert \Eq{eq:FLsw} in \Eq{vrms_from_FL} and change the first integration
variable to $z=\tilde k \tilde T$.
Then, the two integrals become independent, and $v_\rms^2$ becomes
\begin{align}
    v_\rms^2 (\tsw^+)
    = \frac{1}{2 \pi^2} \int_0^\infty \FL (\tilde t_\sw^+, 
    \tilde k) \, k^2 \dd k = 
    \frac{n_b}{\beta^3} \frac{1}{2 \pi^2}
    \int_0^\infty \fpsq \, z^2 \dd z
    \int_0^\infty
    \nu(\tilde T) \, \tilde T^3 \dd\tilde T \,.
    \label{eq:vrms_tcol+}
\end{align}

For any choice of the nucleation distribution,
the third moment of $\nu(\tilde{T})$, $\nu_3$,
has a unique interpretation as the normalized
volume in the broken phase at $\tsw$, which takes a universal value
of 6, as we prove in \Eq{thirdmoment_nu}.
It was  already pointed out in ref.~\cite{Hindmarsh:2019phv} that $\nu_3 = 6$
for exponential
and simultaneous nucleation.
We demonstrate this for any $\nu(T)$ by starting from its definition
[see \Eq{lifetimes_ssm}],
using the modeling of the collisions proposed in
the sound shell model (see section 4.2 of
ref.~\cite{Hindmarsh:2019phv} and \Sec{distribution_lifetimes}),
\begin{align}
   \nu_3 = 
    - \frac{1}{\beta \,n_b}
    \int_{\tilde t_{c}}^{\infty} p( \tilde t)  \dd \tilde t 
    \int_0^{\infty}
    \frac{\dd h( \tilde t + \tilde T)}{\dd  \tilde T}
    \, \tilde T^3 \dd \tilde T =
   \frac{3}{\beta \,n_b}
   \int_{\tilde t_c}^{\infty}
    h(x) \dd x
   \int_{\tilde t_c}^{x}
   p(\tilde t) \, (x-\tilde t)^2  \dd \tilde t \,,
   \label{eq:nuT3is6}
\end{align}
where we have integrated by parts the first expression
and defined the variable $x \equiv \tilde T + \Tilde t$.
Using the relation between $p$ and $h$ from \Eq{h_t},
the following property is satisfied,
\begin{equation} \label{proof_aux}
   3\, h(x) \int_{\tilde t_c}^{x}
   p(\tilde t) \, (x-\tilde t)^2  \dd \tilde t
   = - \frac{\beta^4}{{\cal V}_w}
   \frac{\dd h(x)}{\dd x}   \,, 
\end{equation}
and \Eq{eq:nuT3is6} becomes
\begin{align}
    \nu_3 \equiv \int_0^{\infty}
    \nu (\tilde T) \, \tilde T^3  \dd \tilde T 
    & = 
     - \frac{\beta^3}{n_b} \frac{1}{{\cal V}_w}
    \int_{\tilde t_c}^{\infty} \dd h(x)
    = \frac{\beta^3}{n_b} \frac{1}{{\cal V}_w}
    = 6 \,.
    \label{thirdmoment_nu}
\end{align}
All in all, \Eq{eq:vrms_tcol+} becomes
\begin{align}
    v_\rms^2 (\tsw^+)
    = \frac{\tilde b_{ii}(0)}{{\cal V}_w}
    = \frac{3}{4}
    \kappa_{\rm pert} \, \alpha \simeq \bar{U}_f^2 \,,
    \label{eq:vrms2_coll+}
\end{align}
where the rms enthalpy-weighted velocity $\bar{U}_f$
reduces to $v_\rms (t_{\rm sw}^+)$
in the limit of small perturbations.
Remarkably, the result in \Eq{eq:vrms2_coll+} 
depends only on the single-bubble
kinetic energy, characterized by $\kappa_{\rm pert} \alpha$ [see \Eq{vrms2}].
As discussed in \Sec{vrms_single}, when the enthalpy perturbations
become large, $\kappa_{\rm pert}$ is no longer a good approximation
of $\kappa$.

We note that our analysis during the expansion phase can be generalized
to larger perturbations.
In fact, the derivation carried out in this section, leading to \Eq{eq:vrms2_coll+}, can be applied to any
fluid variable, e.g., to $X_\rms^2$,
implying that all squared averaged variables from the fluid are assumed
to be conserved across collisions.
This implies that given the collision modeling
of the sound shell model, the kinetic energy
density is conserved upon collisions also in the
regime of large fluid perturbations.
However, vorticity can be produced during collisions
and at later times \cite{Pen:2015qta,Dahl:2024eup,RoperPol:2025lgc,vorticity},
and the fluid perturbations after collisions might not, in general,
follow those described by the sound shell model \cite{Cutting:2019zws,RoperPol:2019wvy,Caprini:2024gyk,Correia:2025qif}.
Hence, since the sound shell model applies in the
linear regime of perturbations, we restrict our analysis
to small perturbations, such that $X_i = \sqrt{w_n} \, {U_f}_i
\approx \sqrt{w_n}\, v_i$, $\kappa \approx \kappa_{\rm pert}$, and $w_{\rm ref}
\approx w_n$.

We now estimate an effective time setting the
beginning of the sound-wave regime,
$\ttsw$, by imposing that $v_\rms^2$ 
is left unchanged from $\ttsw^-$ to $\ttsw^+$.
Then, equality between \Eqs{vrms_expanding}{eq:vrms2_coll+}
leads to the following condition to determine $\ttsw$,
\begin{equation}
    \frac{v_\rms^2(\ttsw^-)}{v_\rms^2(\ttsw^+)}
    =
    \frac{1}{6}
    \int_{\tilde t_c}^{\ttsw}
    \gamma(\tilde t_0)
    (\ttsw - \tilde t_0)^3 \dd \tilde t_0
    = 1 \,.
    \label{eq:tcoll-condition}
\end{equation}
The values of $\tilde t_{\rm sw}$ predicted for simultaneous and exponential
nucleations are indicated by the vertical dashed lines in \Fig{fig:tcollnew}.
In particular, for simultaneous nucleation, we find $\tilde t_{\rm sw} - \tilde t_\ast = 6^{1/3}$
using \Eq{vrms_sim}, while for exponential nucleation, we find numerically
$\tilde t_{\rm sw} - \tilde t_{\rm peak} \simeq 0.063$.

In summary,
we have defined an effective  starting time
of the sound-wave regime $\ttsw$,
such that the kinetic energy at this time
is identical to the kinetic energy found
at the beginning of the sound-wave
regime, computed by
averaging over the bubble lifetime distribution $\nu(T)$.

In general, for a stochastic distribution of the time $t_1$ at which the sound-wave regime
starts after collisions,
the velocity power spectral density in the aftermath of collisions is 
\begin{equation} \label{FL_joint_dist}
    \FL (k) = \int_{t_c}^{\infty} \dd t_1 \int_{t_c}^{t_1}
    P(t_0, t_1) (t_1 - t_0)^6 \tilde f_{\rm L}
    \bigl[(t_1 - t_0) k \bigr] \dd t_0\,,
\end{equation}
where $P(t_0, t_1)$ is  the joint distribution function of nucleation
and collision times $t_0$ and $t_1$.
In ref.~\cite{Hindmarsh:2019phv}, by assuming that the fluid perturbation around
each bubble exactly becomes the initial condition of the sound wave originating
from that bubble, they argue that $\FL$
does not depend separately on the nucleation and collision times,
but only on their difference.
Hence, they turn
$P(t_0, t_1)$ into a distribution over the bubble lifetimes $T = t_1 - t_0$. 
In our approach, based on the definition of a deterministic effective 
time $\ttsw$ satisfying \Eq{eq:tcoll-condition}, the joint distribution function becomes $P(t_0, t_1) = \Gamma(t_0) \, \delta(t_1 - \tsw)$,
leading to
\begin{equation} \label{FL_endtsw}
    \FL (k) = \int_{t_c}^{t_\sw} \Gamma(t_0)
     (t_\sw - t_0)^6 \tilde f_{\rm L}
    \bigl[(t_1 - t_0) k \bigr] \dd t_0\,,
\end{equation}
which corresponds to \Eq{eq_FL1} evaluated at $t_\sw$.
A better modeling of the joint distribution function during the period
of collisions, incorporating
a stochastic distribution of $t_1$ around $t_{\rm sw}$, could improve the modeling
presented in our work.

\section{Spectrum of the velocity field at the beginning of the
sound-wave phase}
\label{FL_template}

\begin{figure}
    \centering
    \includegraphics[width=0.6\linewidth]{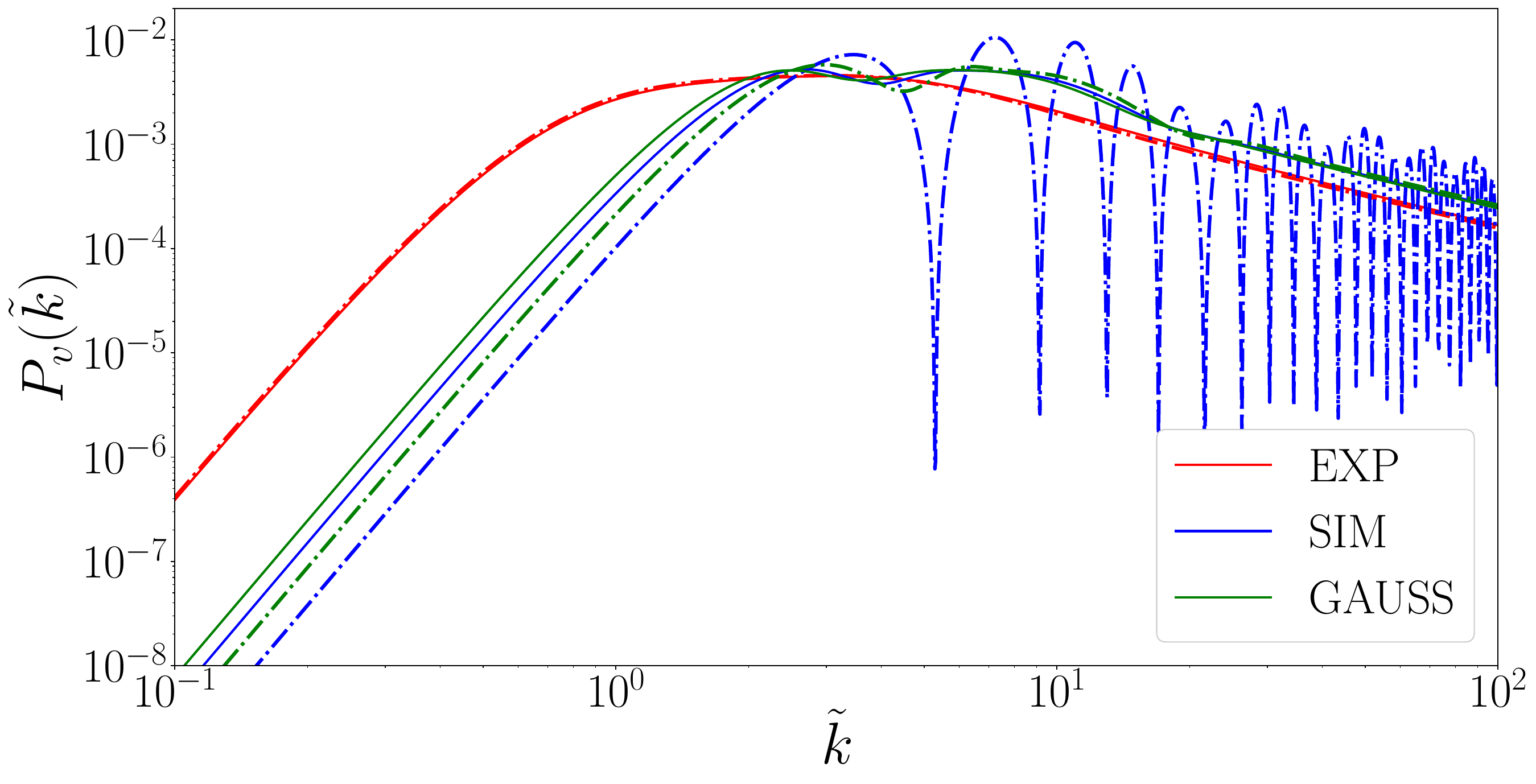}
    \includegraphics[width=0.6\linewidth]{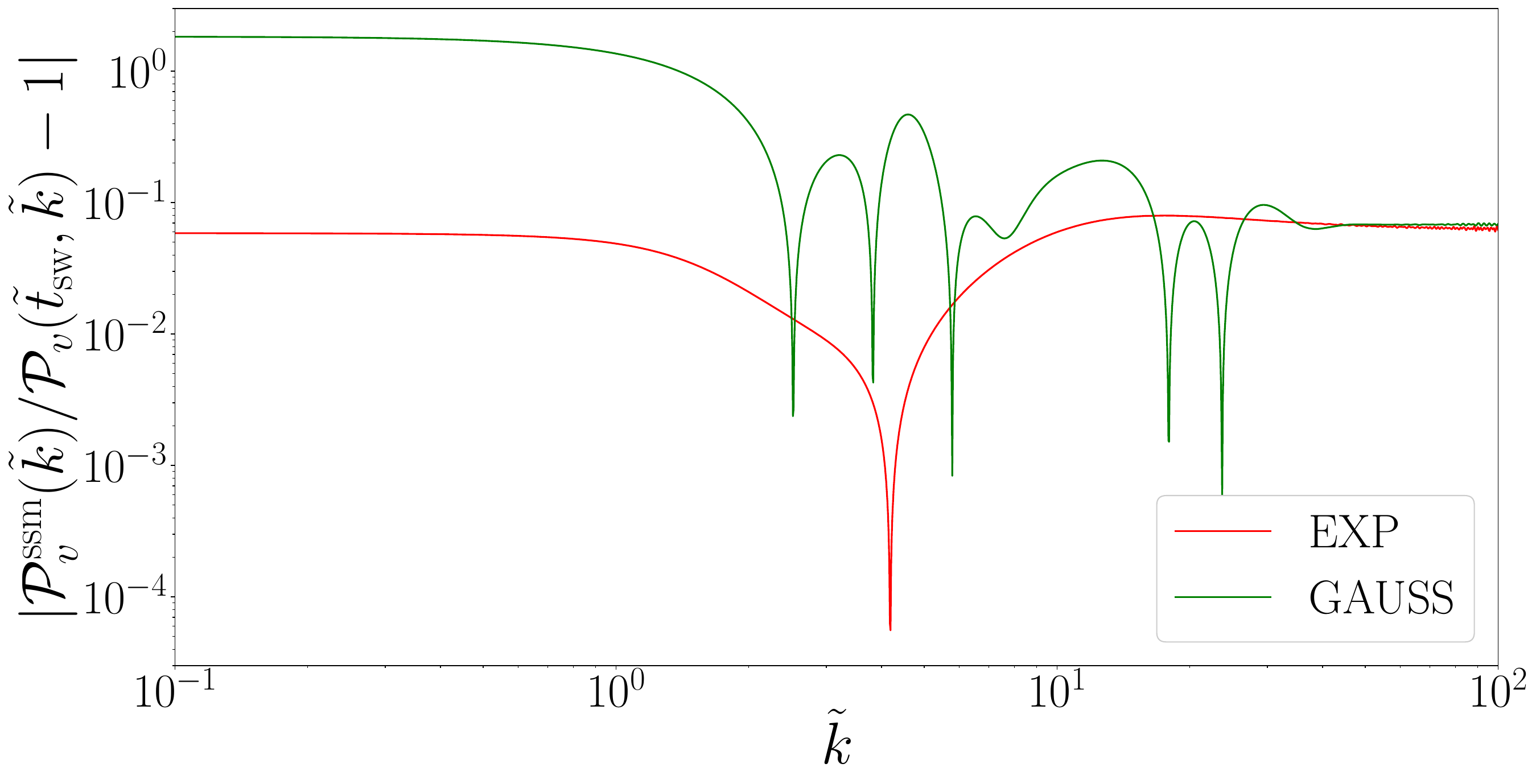}
    \caption{
    Velocity spectrum, ${\cal P}_v (\tilde k) \equiv
    k^3 \FL(\tilde k)/(2 \pi^2)$, defined such that
    $v_\rms^2 = \int {\cal P}_v (\tilde k) \dd \ln \tilde k $
    at the effective sound-wave time
    $\tilde t_\sw$
    [see \Eq{eq_FL1}] for
    the case of exponential nucleation (red lines),
    the case of simultaneous nucleation (blue lines), and
    the case of Gaussian nucleation with $\ln h_\ast = -0.01$
    (green lines),
    which approaches the
    simultaneous case but its $\gamma(t)$ is not a delta function,
    as shown in \Fig{hs_sims}.
    The respective velocity spectra ${\cal P}_v^{\rm ssm} (\tilde k)$
    evaluated according to the sound shell model using \Eq{def_FL_ssm}
    are shown for comparison in solid lines.
    The choice of phase transition parameters is $\alpha=0.1$ and $\xi_w =0.4$
    for illustration, as in \Fig{kin_spec_av_nucl}.
    We find that only for the case of exponential nucleation the velocity spectra
    following both approaches almost
    coincide, with a relative difference below 8\% (lower plot).
    In the case of Gaussian nucleation, both approaches lead to comparable spectra but
    differences are larger
    at small $\tilde k$,
    with relative differences up to 180\%, while for the simultaneous nucleation, the appearance
    of oscillations in $\PP_v$ makes the differences between the
    two approaches large at all scales.
    However, we note that the integrated quantities are still leading to the same
    $v_\rms^2$ in both approaches, and that we expect these oscillations
    to smooth out once we compute the GW spectrum, as we investigate
    in ref.~\cite{part2}.
    }
    \label{fig:FL_ev}
\end{figure}

In \Sec{sec_1bubble}, we studied the function $\fpsq$, which describes the power spectral
density of the velocity field resulting from a superposition of expanding
bubbles randomly nucleated in space.
We showed that the envelope of this function is characterized by the amplitudes in two
asymptotic limits, $\fpsqz$ when $z \to 0$ and $\fpenvsq$ when $z \to \infty$,
where $\fpsqz$ is related to the integral of the velocity profile over $\xi$
[see \Eq{asymptotic_fpz}] and $\fpenvsq$ is related to its
discontinuities [see \Eq{fpenv_inf}].
Two scales appear naturally, $z_1$ and $z_2$, determining the transitions to each
of these asymptotic limits,
potentially allowing for an intermediate regime
with slope $\gamma$ when both scales are sufficiently separated [see \Eq{slope_z}].
Then, in \Sec{ensemble_times}, we described the time evolution of the power spectral
density $\FL(t, k)$, resulting from a stochastic distribution of the
nucleation times, considering both exponential and simultaneous nucleation in time.
Finally,
we showed in \Sec{across_collisions} that the modeling of collisions in the sound
shell model implies that the kinetic energy density
is conserved
from the uncollided to the sound-wave phase,
allowing us to define an effective time $t_\sw$
that describes the initial time of the sound-wave regime.
In the following, we propose to use $\ttsw$ to evaluate $\FL$  [see \Eq{FL_endtsw}]
as an estimate of the velocity spectrum at the initial time of the sound-wave regime.
For exponential nucleation,
we show that our approach leads to velocity spectra
that are almost equal to the ones obtained in the sound shell model \cite{Hindmarsh:2016lnk,Hindmarsh:2019phv,RoperPol:2023dzg},
using the bubble lifetime distribution function $\nu$
(see \Sec{distribution_lifetimes}), as indicated in \Eq{eq:FLsw},
in the limit $\tilde t_\ast - \tilde t_c \gg 1$.
In particular, we find maximum relative differences of up to 8\% between the two approaches
(see \Fig{fig:FL_ev}).
However, for the case of simultaneous nucleation, we find that the resulting
velocity spectrum using our approach differs from the one found in the sound shell model.
In particular, this is mostly due to the fact that oscillations from $\fpsq$
are not smoothed out in the simultaneous case in our approach.
For a Gaussian distribution, strongly peaked around $t_\ast$, we find good
agreement at large $\tilde k$ (see \Fig{fig:FL_ev}).
The advantage of defining this effective time
is that it does not rely on the determination of the distribution of bubble lifetimes,
and therefore the only distribution function involved is the one for the times
of nucleation, $\Gamma(t) = p(t) \, h(t)$, through \Eq{FL_endtsw},
reducing the number of manipulations needed to generalize the model.
On the other hand, in our approach, the main assumption is that all collisions
effectively occur at a deterministic time $t_{\rm sw}$,
which is the same for all bubbles.
Effectively, this is a way to bypass our ignorance of the details of the collision phase. 
As discussed above, 
a better modeling of $\FL$ could be achieved if we knew the exact distribution function of the collision time.

We show in \Sec{asymptotic_FL}
the asymptotic limits of the velocity spectrum and
how the original scales $z_1$ and $z_2$ convert in the two scales that characterize
the velocity spectrum $\FL(\tsw, k)$ at the starting time of the sound-wave regime.
Then, we provide in \Sec{template_FL}
a template for the function $\FL(\tsw, k)$
using the two scales, $k_1$ and $k_2$,
and evaluate its intermediate
slope $\sigma$.
We show that the prefactor $\fpsqz$ that appears in the template can be related to the coefficient $\kappa$, defined in \Eq{vrms2}.
Therefore, our template of $\FL (\tsw, k)$ is expressed 
only in terms of the simplest known quantities
that can be evaluated from the 1D profiles (see \Fig{fig_ip}), with the minimal
required amount of information, while still improving its accuracy with respect to previous
estimates used in the literature.

We already introduced the velocity power spectral density
at the beginning of the sound-wave phase in \Eq{eq:FLsw}, 
following the modeling of collisions of the sound shell model (ssm)
\cite{Hindmarsh:2019phv,RoperPol:2023dzg},
\begin{align}
    \FL^{\rm ssm} (\tilde{k})=\frac{n_b}{\beta^6}
    \int_0^{\infty} \nu(\tilde{T}) \, \tilde{T}^6 \,
    {f'}^2 (\tilde{k} \, \tilde{T})\dd \tilde{T} \,,
    \label{def_FL_ssm}
\end{align}
where for the cases of exponential and simultaneous nucleation,
$\nu_{\rm exp} (\tilde T) = e^{-\tilde T}$ and $\nu_{\rm sim} (\tilde T)
= \half \,
\tilde T^2 \exp(-\tfrac{1}{6} \tilde T^3)$ (see \Sec{distribution_lifetimes}
and refs.~\cite{Hindmarsh:2016lnk,Hindmarsh:2019phv}).
As shown in \Sec{EffCollTime} [see \Eq{FL_endtsw}],
one can alternatively define an effective
time $\ttsw$ at which the velocity
spectrum in the bubble expansion phase gives the same
kinetic energy density, $\rhokin \simeq w_n \, v_\rms^2 = \tfrac{3}{4} \, w_n \,
\kappa_{\rm pert}\alpha$,
as the one at the beginning of the sound-wave regime,
\begin{align}
    \FL (\tilde k) = \FL (\ttsw, \tilde{k}) = \frac{n_b}{\beta^6}
    \int_{\tilde{t}_c}^{\ttsw}
    \gamma(\tilde{t}_0) \, (\ttsw-\tilde{t}_0)^{6 } \, {f'}^2
    \bigl[\tilde{k} \, (\ttsw-\tilde{t}_0)\bigr]
    \dd \tilde{t}_0\,.
\end{align}

We show in \Fig{fig:FL_ev} the velocity power spectrum
$\PP_v (\tilde k) = k^3 \FL (\tilde k) /(2 \pi^2)$, 
defined such that $v_\rms^2 (\tilde t_{\rm sw})
\equiv \int \PP_v (\tilde k) \dd \ln \tilde k$.
We find that both computations
are almost equivalent
for the exponential nucleation case,
with less than $8\%$ relative difference at all $\tilde k$
in the $\tilde t_\ast - \tilde t_c \gg 1$ limit
that we focus on.
For finite values of $\tilde t_\ast - \tilde t_c$, the two approaches give 
larger relative differences in the power
spectrum.

For the case of simultaneous nucleation, the two approaches lead to  velocity spectra that are  different {in particular for what concerns the oscillatory behaviour}.
This is mostly due to the fact that in our approach, the sound-wave time is treated deterministically,
such that the $\delta$ function in the nucleation time
distribution,
$\gamma_{\rm sim} = \delta(\tilde t - \tilde t_\ast)$, leads to $\FL (\tilde k)
= n_b (\tsw - t_\ast)^6
\, {f'}^2 \bigl[\tilde k (\tilde t_\sw - \tilde t_\ast)\bigr] \, \Theta(t - t_\ast)$,
as shown in \Eqs{a_of_t_sim_analytical}{fl_sim_analytical}.
{$\FL$ therefore inherits the oscillatory behaviour of ${f'}^2$. }
However, if one introduces a distribution around $t_{\rm sw}$ in \Eq{FL_joint_dist},
as effectively considered in the sound shell model
via the introduction of a lifetime distribution $\nu(T)$,
then $\fpsq$ is averaged over
different collision times.
As a result, its oscillations are no longer in phase and are therefore smoothed out.
In any case, after taking the convolution of \Eq{eq:UETC:PiPi:Isserlis}
to compute the anisotropic stresses,
the oscillations that appear in our approach should smooth out, leading
to a resulting GW spectrum that should be similar for both approaches.
We investigate the potential discrepancy between
the two approaches in ref.~\cite{part2}.

We also consider the case of Gaussian nucleation with $\ln h_\ast \simeq -0.01$,
whose distribution function is shown in \Fig{hs_sims} in brown.
The distribution $\gamma (\tilde t)$ in this case
approaches the simultaneous one,
but it is still spread around $\tilde t_\ast$.
We still observe some differences in the velocity spectra resulting from the two approaches,
but the oscillations are smoother
and the spectra are similar
at large $\tilde k$.
At small $\tilde k$, the two approaches yield a power law in $\tilde k$ that differs by a factor of 2.

In all cases, we remind the reader that the integral of $\PP_v$ over $\ln k$ yields
the same result, $v_\rms^2 \equiv \tilde b_{ii}(0)/{\cal V}_w = \tfrac{3}{4}
\, \kappa_{\rm pert} \alpha$,
and we expect that the differences between the two choices
in the resulting GW spectrum are suppressed.
In the following, we focus on studying the asymptotic limits of
$\FL (\tilde k) \equiv \FL^{\rm ssm} (\tilde k)$ and providing a template based on the minimal
amount of information from the original
velocity self-similar profiles $v_\ip (\xi)$ induced in a first-order phase
transition.

\subsection{Asymptotic limits of the velocity spectrum}
\label{asymptotic_FL}

We show in \App{appendix_correlation} that the causality condition for an irrotational field
implies that $F_{\rm L} \to k^2$ in the $k \to 0$ limit \cite{MY75,Hindmarsh:2019phv}.
We indeed find
that, in this limit, $\FL(\tilde{k})$ inherits the large-scale behavior of $\fpsq$, i.e.,$\propto z^2=\tilde k^2\tilde T^2$.
Taking this limit in \Eq{def_FL_ssm}, we find
$\FL(\tilde k) = \FLz \, \tilde k^2 + {\cal O} (\tilde k^4)$, where
\begin{equation}
   \FLz =
    \frac{n_b}{
    \beta^6} \, \fpsqz A_{0}\,, \qquad \text{with \ }
    A_0
    = \int_0^\infty \nu(\tilde T)\, \tilde T^8 \dd \tilde T\,.
    \label{rel_FL0_fp0}
\end{equation}
Using the $\nu(\tilde T)$ of exponential
and simultaneous nucleation [see \Eqs{nu_exp_as}{nu_sim_as}],
we find
\begin{align}
    A_{0}^{\rm exp} = \int_0^\infty e^{-\tilde T}\, \tilde T^8
    \dd \tilde T = 8! = 40320\,, \qquad
    A_0 ^{\rm sim} = \frac{1}{2}
    \int_0^\infty \exp\bigl(- \tfrac{1}{6} \tilde T^3
    \bigr) \, \tilde T^{10}  \dd \tilde T = 6^{8/3} \, \Gamma\bigl(\tfrac{11}{3}
    \bigr) \,.
    \label{A0_expsim}
\end{align}

We know that in the $z \to \infty$ limit,
$\fpsq = z^{-4} \fpinfsq (z) + {\cal O} (z^{-5})$,
where $f'_{\infty} (z)$,
given in \Eq{eq:f1_zInfty},
depends on the discontinuity jumps and positions of the velocity profile.
Hence, in the limit $k \to \infty$, we define
$\FL^\infty$, such that $\FL (\tilde k) = \FL^\infty \, \tilde k^{-4}
+ {\cal O} (\tilde k^{-5})$,
\begin{align} \label{FL_infinity}
    \FL^\infty = \frac{n_b}{\beta^6} \, \fpenvsq \, A_\infty\,,
    \qquad \text{with \ } A_\infty =
    \lim_{\tilde k \to \infty} \int_0^\infty \frac{\fpinfsq
    (\tilde k \tilde T)}{\fpenvsq} \nu(\tilde T)\, \tilde T^2
    \dd \tilde T\,.
\end{align}
Since the envelope of $\fpinfsq (z)$
over the oscillations at large $z = \tilde k \tilde T$ becomes
$\fpenvsq$, we have introduced it in the definition of $A_\infty$.
Then, the coefficient $A_\infty$ takes the following values
\begin{align}
    A_{\infty} =
    \frac{1}{2}
    \int_0^{\infty} \nu(\tilde{T}) \, \tilde{T}^2 \dd \tilde{T}\,,
    \label{fl_env}
\end{align}
where we find the $\half$ factor numerically.
This factor appears due to the fast oscillations
of the $\sin^2 (z)$ and $\cos^2 (z)$ functions that appear in $\fpinfsq$,
which are effectively averaged out when we take the integral over $\tilde T$.
For exponential and simultaneous nucleations, the values of $A_\infty$ become
\begin{align} \label{Ainfty_expsim}
    A_{\infty}^{\rm exp}
    = &\,
    \int_0^\infty \half\, \tilde T^2 \, e^{-\tilde T}
    \dd \tilde T = 1 \,,  \qquad
    A_{\infty}^{\rm sim} = 
    \int_0^\infty \half\, \tilde T^{4} \exp\bigl(- \tfrac{1}{6} \tilde T^3
    \bigr) \dd \tilde T = \half \,
    6^{2/3} \, \Gamma\bigl(\tfrac{5}{3} \bigr) \,.
\end{align}

\subsection{Template of the velocity spectrum}
\label{template_FL}

We found in the previous section that, similarly to $\fpsq$, the function
$\FL(\tilde k)$ takes the asymptotic limit $\FL (\tilde k) = \FLz \tilde k^2$ when
$k \to 0$ and $\FL (\tilde k) = \FL^{\infty} \tilde k^{-4}$ when $k \to \infty$.
Therefore, we can relate the asymptotic limits of $\FL$ with the integral over
the self-similar velocity profiles,
$\FL^0$, and with the velocity jumps of
the velocity profiles, $\FL^\infty$.

In analogy to $z_\cross$, defined in \Eq{slope_z}, we can define $\tilde{k}_\cross$ based on
these two asymptotic limits.
Using \Eqs{rel_FL0_fp0}{FL_infinity}, $\tilde k_\cross$ becomes
\begin{align}
    \tilde k_\cross = \biggl( \frac{\FL^{\infty}}{\FLz}\biggr)^{1/6}
    = 
    \left|\frac{f_\env'}{f'_0}\right|^{1/3}
    \biggl(\frac{A_\infty}{A_0} \biggr)^{1/6}
    = z_\cross \biggl(\frac{A_\infty}{A_0} \biggr)^{1/6}\,.
    \label{def_kcross}
\end{align}

Then, introducing \Eqs{A0_expsim}{Ainfty_expsim} into \Eq{def_kcross},
we can find the ratio $\tilde k_\cross/z_\cross$ for exponential
and simultaneous nucleation,
which allows us to compute the overall shift in the scales that occurs due to the
bubble nucleation history and only depends on the distribution
function of nucleation times,
\begin{align} \label{kcross_expsim}
    \frac{\tilde{k}_\cross^{\rm exp}}{z_\cross}
    = (8!)^{-1/6} \simeq 0.17 \,, \qquad
    \frac{\tilde{k}_\cross^{\rm sim}}{z_\cross}
    = 320^{-1/6} \simeq 0.38\,.
\end{align}
This result already pinpoints an important remark.
The spectral peak of the power spectrum $\FL$ can be determined
using the scale $z_1$
found in \Sec{subsec_RiemannL}, which only depends on the self-similar profiles, and
the constant factor $\tilde k_\cross/z_\cross \equiv (A_\infty/A_0)^{1/6}$
that only depends on the nucleation history.
Indeed, since we find that the ratio {$\tilde k_\cross/z_\cross$}
does not depend on $\xi_w$
and $z_1 = z_\cross$ at $\xi_w \lesssim \half \, \vCJ$, we can then
use \Eq{def_kcross} to determine $\tilde k_1/z_1 = \tilde k_\cross/z_\cross$.
This result allows us to predict
{the positions of the spectral
peak,}
and compare among different nucleation
histories.
{Furthermore,}
it
{provides an explanation
to the location of the spectral
peak,}
{which has been found to be at larger scales for the
case of
exponential nucleation than
for the simultaneous case in previous work}
\cite{Hindmarsh:2019phv,RoperPol:2023dzg}.

We can now compare the value of $\tilde k_1/z_1$ with $\tilde k_*/z_\ast$ studied in \Sec{ensemble_times}.
We found there that $\tilde k_\ast/z_\ast$ for exponential
nucleation remained constant and equal
to $\sim 0.13$ at times $\tilde t - \tilde t_{\rm peak}
\lesssim 3$ (see \Fig{kin_spec_av_nucl2}), which is the range
where $\tilde t_{\rm sw} - \tilde t_{\rm peak} \simeq 0.063$ lays  (see \Fig{fig:tcollnew}).
Since $z_\ast\simeq z_1$, we consistently find the very similar values 0.13 and 0.17. 
For simultaneous nucleation, on the contrary,
we found in \Sec{ensemble_times} that $\tilde k_\ast/z_\ast = (\tilde t - \tilde t_\ast)^{-1}$,
which, evaluated at $\tilde t_\sw - \tilde t_\ast = 6^{1/3}$  (see \Fig{fig:tcollnew}), gives us the
value $\tilde k_\ast/z_\ast \simeq 0.55$.
This value significantly differs from the value of
0.38 found in \Eq{kcross_expsim} using $\nu(T)$ to compute $\FL$,
as it can also be seen in \Fig{fig:FL_ev} comparing the blue
solid and dash-dotted lines.
This is not surprising, since in this case, the resulting 
power spectrum ${\cal P}_v$ is different in the two approaches.

Analogously to the templates that we have presented for $\fpsq$
in \Sec{sec:f_template},
we provide single and double broken power laws for $\FL$
in the following.
When the two scales $\tilde k_1$ and $\tilde k_2$ are close enough,
such that there is no evident intermediate slope,
$\FL$ can be fit
with a single broken power law,
\begin{align}
    \FL^{\rm SBPL} (\tilde k) = \FLz \,
    \tilde k^2 \Biggl[ 1+ \biggl(\frac{\tilde k}{\tilde k_1} \biggr)^{b_1} \Biggr]^{-6/b_1}
    \qquad \text{for \ } \xi_w \lesssim \tfrac{1}{2}
    \, \vCJ (\alpha)\,.
    \label{FL_SBPL}
\end{align}

In the case in which the separation
between
the two scales is sufficiently large, an intermediate slope $\sigma$ appears,
and we need to fit $\FL$ using a double broken power law,
\begin{align}
    \FL^{\rm DBPL} (\tilde k) = \FLz \, \tilde  k^2 \Biggl[ 1+ \biggl(\frac{\tilde k}{\tilde k_1} \biggr)^{b_1} 
    \Biggr]^{\frac{\sigma-2}{b_1}}\,
    \Biggl[ 1+ \biggl(\frac{\tilde k}{\tilde k_2} \biggr)^{b_2} \Biggr]^{\frac{-\sigma-4}{b_2}}
    \qquad \text{for \ } \xi_w \gtrsim \tfrac{1}{2}\,  \vCJ (\alpha)\,,
    \label{FL_DBPL}
\end{align}
where 
\begin{align}
    \sigma =  2 \left[ 1-3\,\frac{\log(\tilde k_2/\tilde k_\cross)}{\log(\tilde k_2/\tilde k_1)} \right]
    \in [-4,2]
    \,,
    \qquad \text{with } \tilde k_\cross = \biggl|\frac{f_{\rm env}'}{f_0'}\biggr|^{1/3}
    \biggl(\frac{A_\infty}{A_0}\biggr)^{1/6}\,.
    \label{slope_k}
\end{align}

\begin{figure}[t]
    \centering
    \includegraphics[width=.32\textwidth]{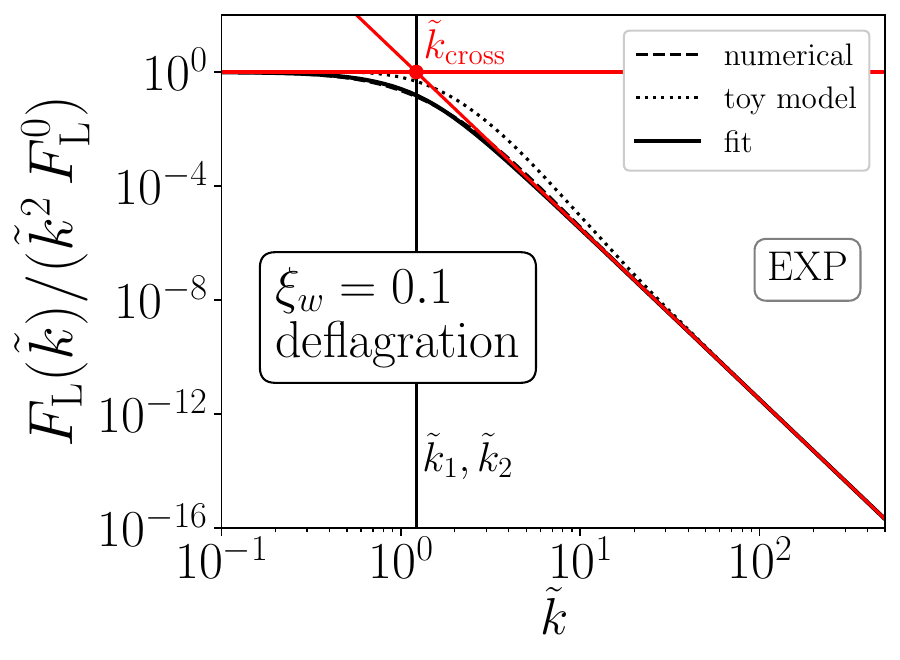}
    \includegraphics[width=.32\textwidth]{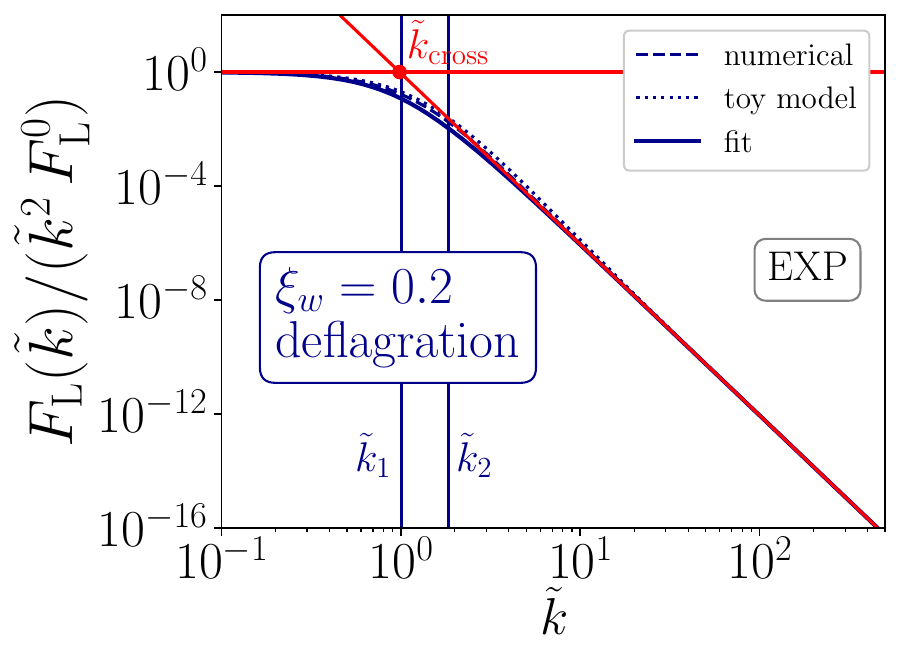}
    \includegraphics[width=.32\textwidth]{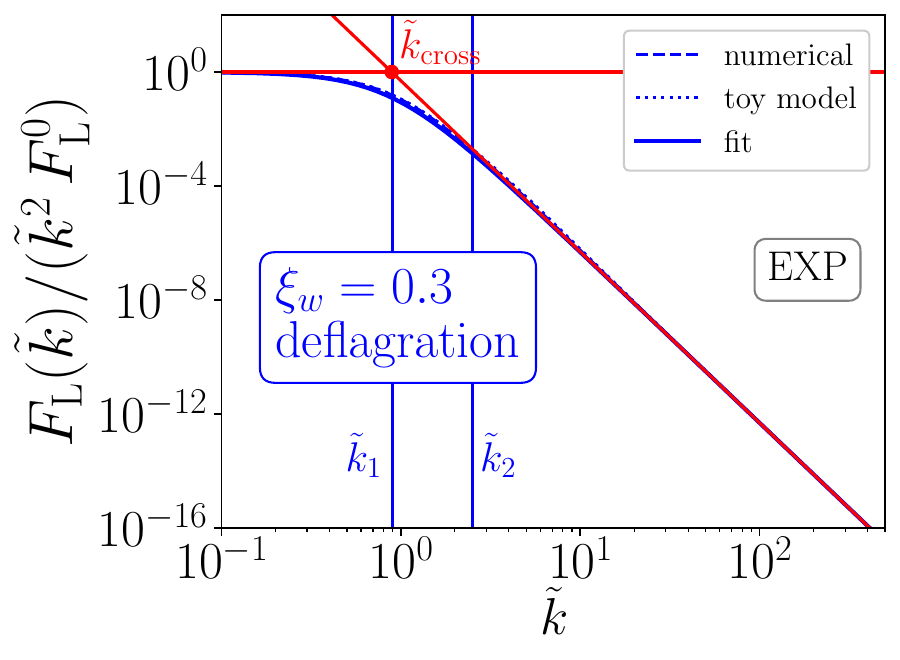}
    \includegraphics[width=.32\textwidth]{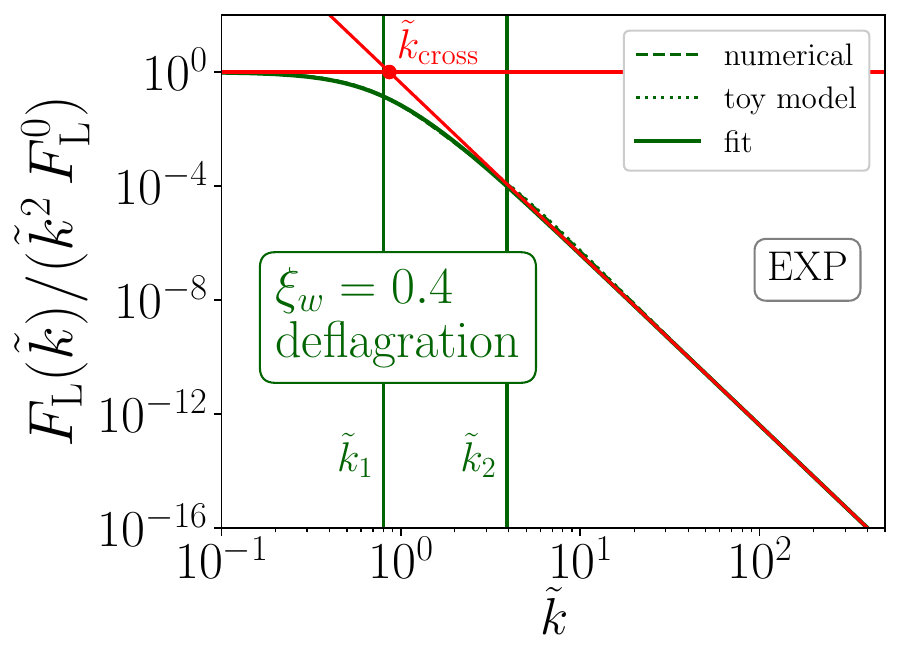}
    \includegraphics[width=.32\textwidth]{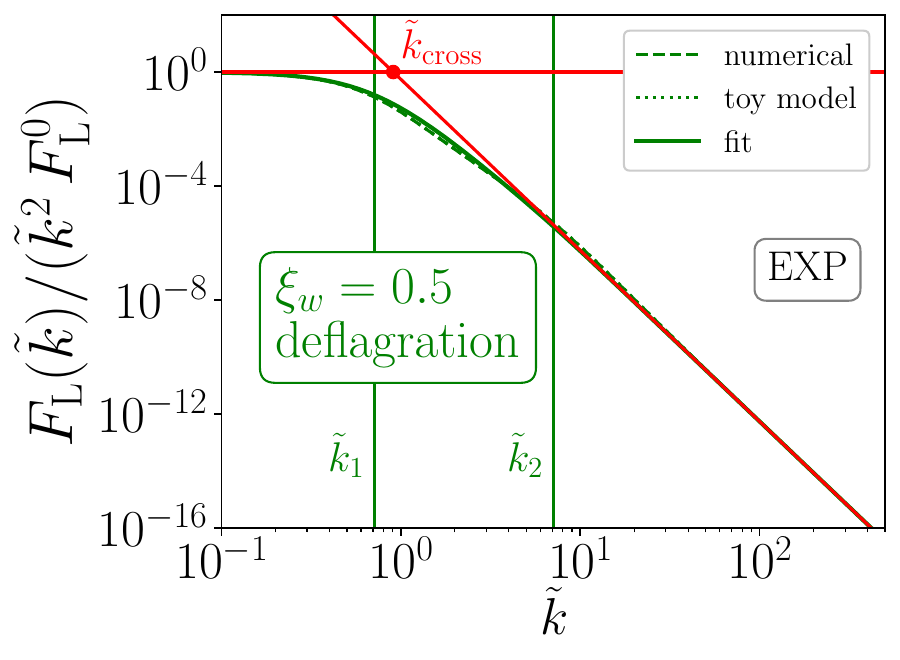}
    \includegraphics[width=.32\textwidth]{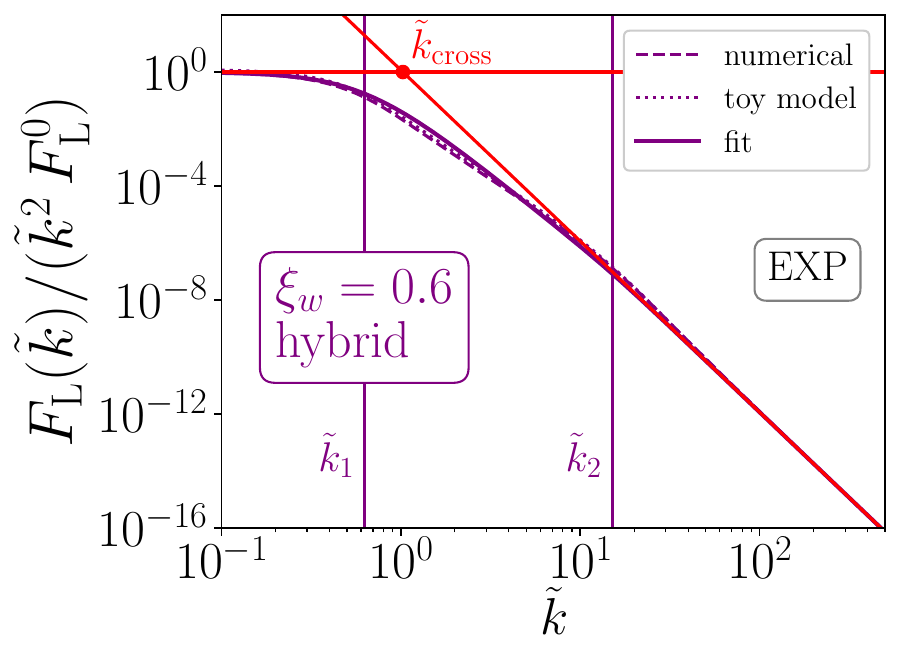}
    \includegraphics[width=.32\textwidth]{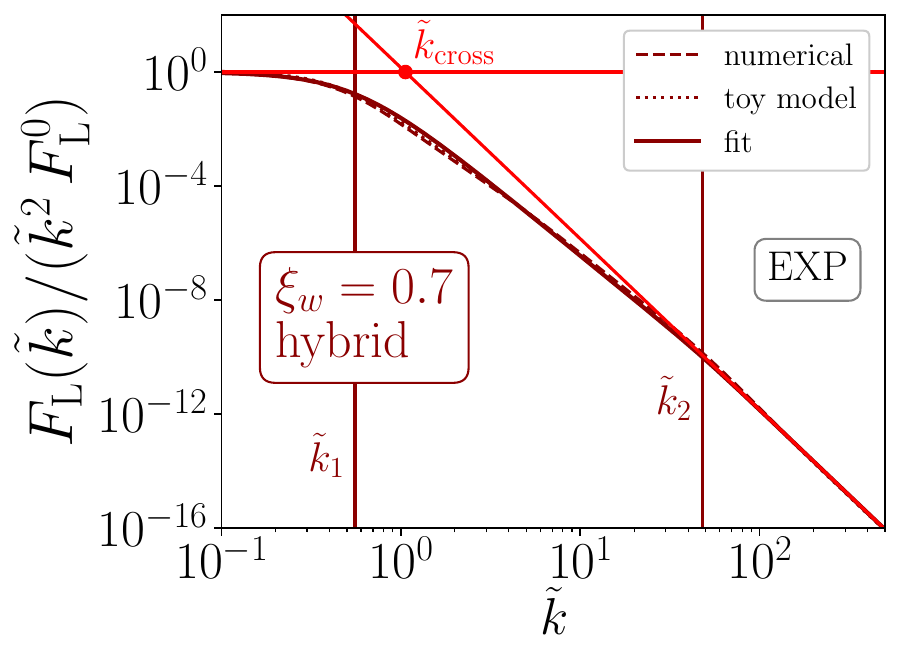}
    \includegraphics[width=.32\textwidth]{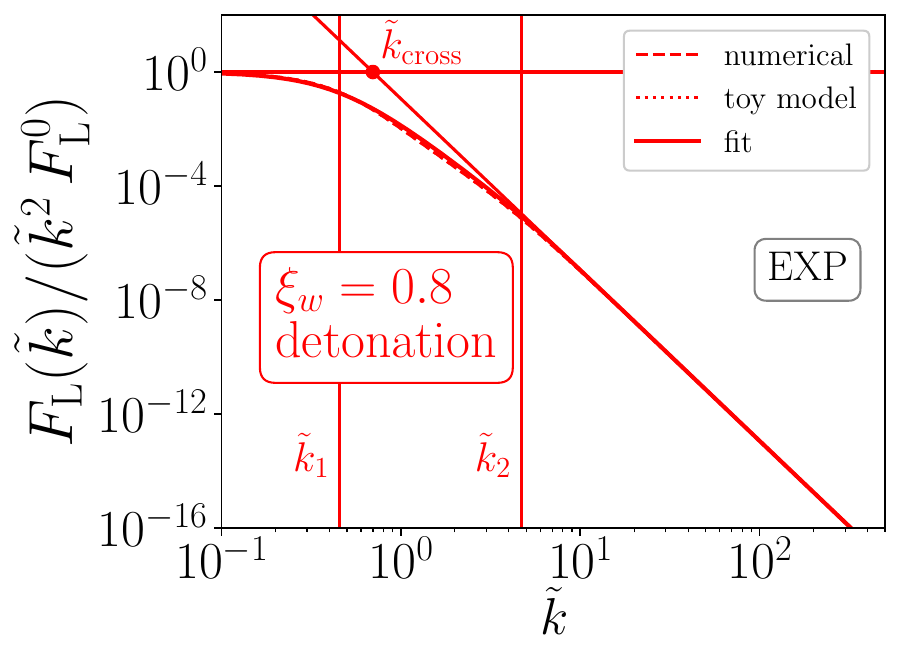}
    \includegraphics[width=.32\textwidth]{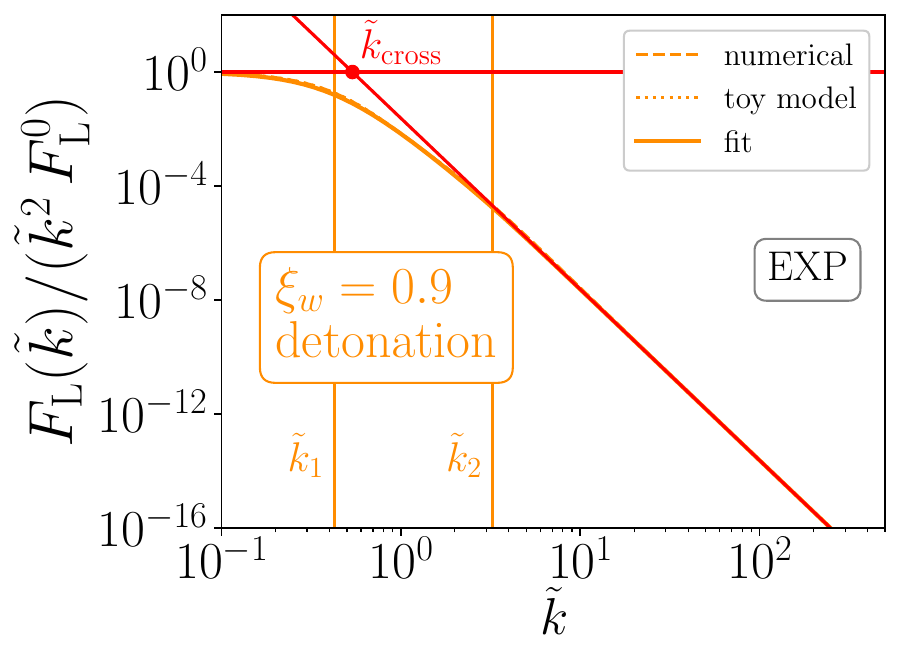}
    \caption{
    Numerical results for the power spectral
    density of the velocity field, $\FL(k)$, computed from \Eq{def_FL_ssm} for an exponential (EXP)
    nucleation history in the limit $\tilde{t}_*-\tilde{t}_c \to \infty$
    (see details in \Sec{exp_nucl}), considering benchmark phase transitions
    with $\alpha=0.1$ and a range of $\xi_w$.
    $\FL (\tilde k)$ is normalized by its asymptotic $k \to 0$ branch, $\tilde k^2 \FLz$.
    The plot also shows the single or double broken power law fit presented in \Sec{template_FL} and the velocity spectral
    density computed from the toy models for the velocity profiles presented in \Sec{sec:toymodel}.
    The vertical lines indicate the locations of $\tilde k_1$ and $\tilde k_2$,
    given in \Eq{scales_relation}.
    The red dot is at the location of $\tilde k_\cross$ [see \Eq{slope_k}], which is defined
    as the location of the intersection between the two asymptotic limits
    $\FLz \, \tilde k^2$ and $\FL^\infty \tilde k^{-4}$.
    }
    \label{newfits_FL_EXP}
\end{figure}

The power spectral density of the velocity, $\FL (\tilde k)$,
found numerically for phase transitions with
$\alpha = 0.1$ and different wall velocities, is shown in \Figs{newfits_FL_EXP}{newfits_FL_SIM},
respectively for the exponential and simultaneous nucleation cases,
compared with the single broken and double broken power law fits.
We have fixed $b_1=2$ and $b_2=4$ in the exponential case,
and $b_1=4$ and $b_2=4$ in the simultaneous case for all values of $\xi_w$.
The smaller value of $b_1$ for the exponential case allows us to reproduce, as shown in \Fig{newfits_FL_EXP}, a smoother transition from
the large- to the intermediate-scale
regimes.
Moreover, we compare the results of the full profiles with those
corresponding to
the toy models described in
\Sec{sec:toymodel}, which are quadratic profiles for subsonic
deflagrations and detonations, and linear-constant profiles for hybrids.
We can observe how the resulting $\FL$ obtained using the toy models accurately
describes the numerical results, exactly reproducing the asymptotic limits
and the intermediate slope by construction.
This is true even for hybrids, for which the envelope of the function
$\fpsq$ in the $z \to \infty$ was fixed, but not the oscillations in $z$ (see discussion in
\Sec{sec:toymodel}), showing that $\FL$ is not strongly affected by the exact
oscillations of $\fpsq$ at large $z$.

\begin{figure}[t]
    \centering
    \includegraphics[width=.32\textwidth]{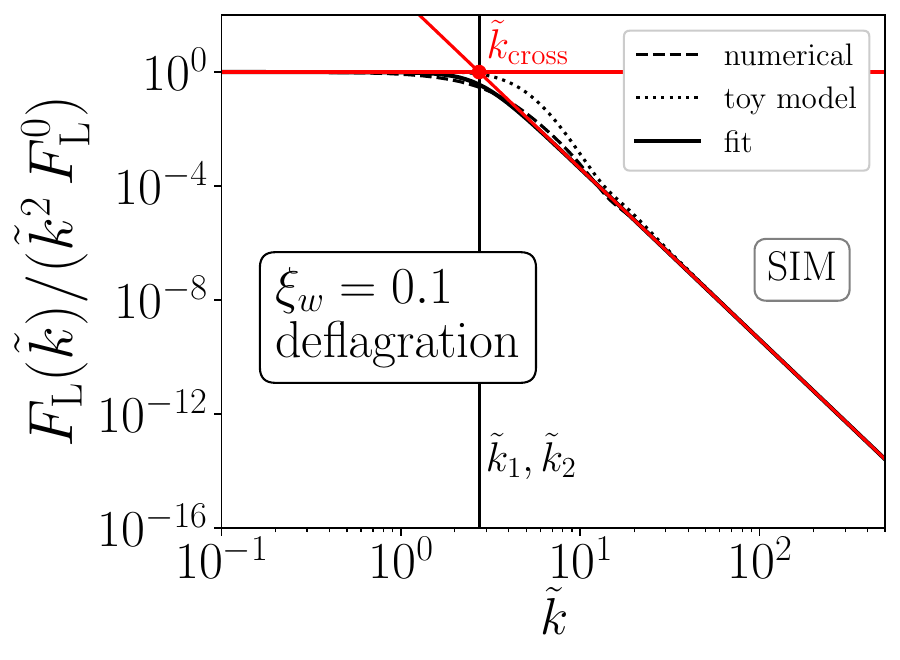}
    \includegraphics[width=.32\textwidth]{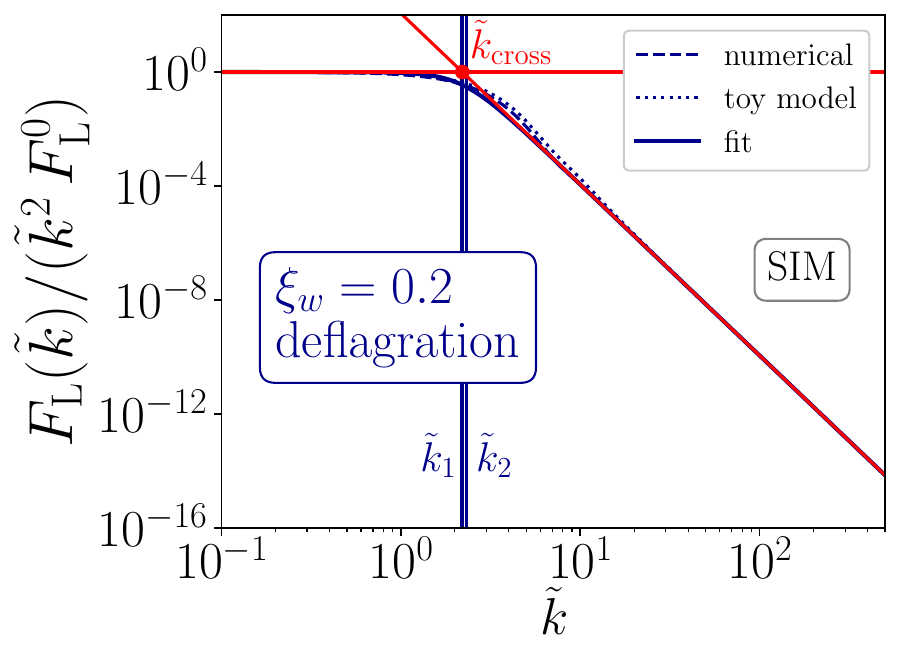}
    \includegraphics[width=.32\textwidth]{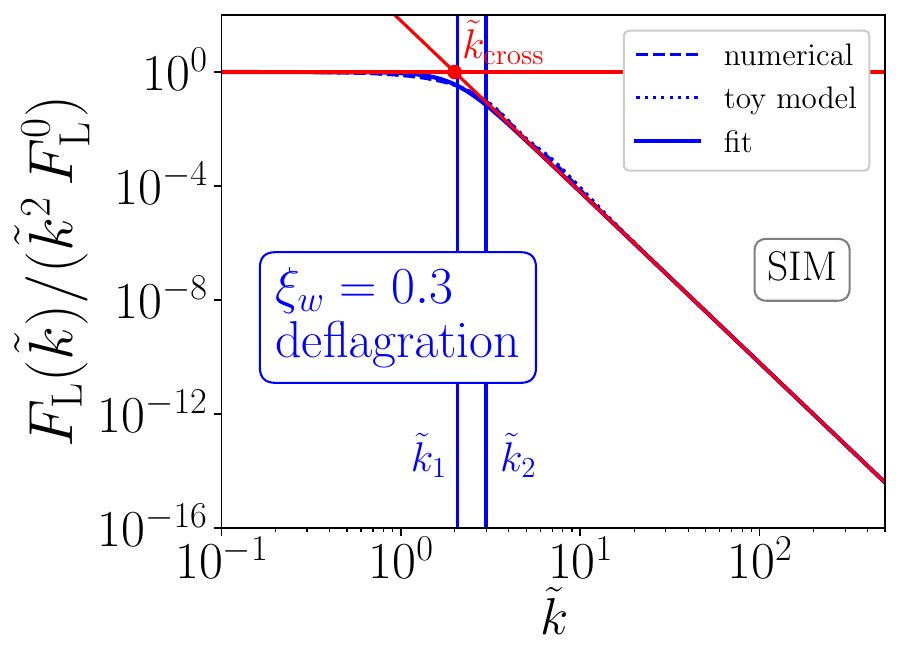}
    \includegraphics[width=.32\textwidth]{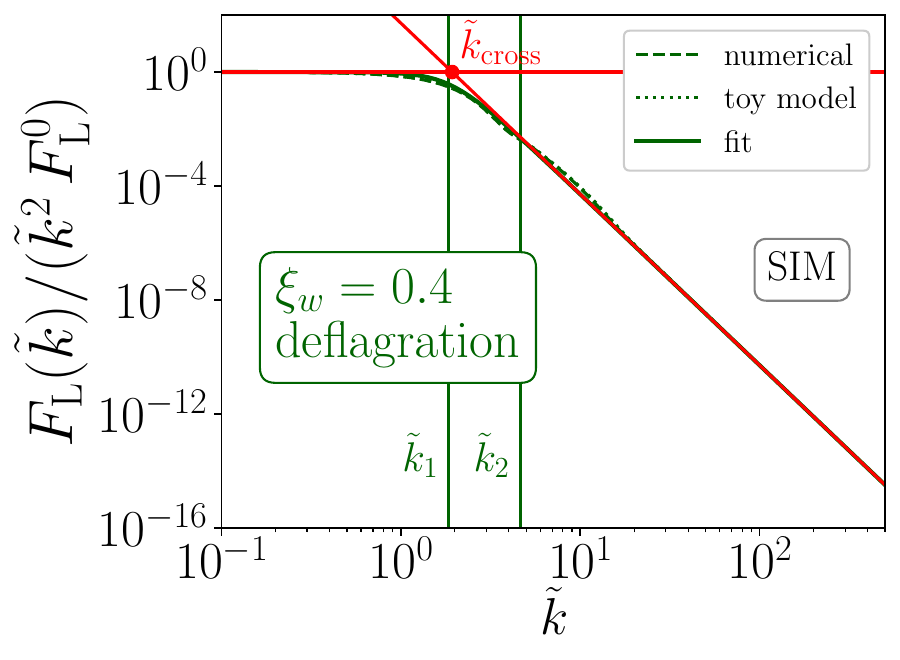}
    \includegraphics[width=.32\textwidth]{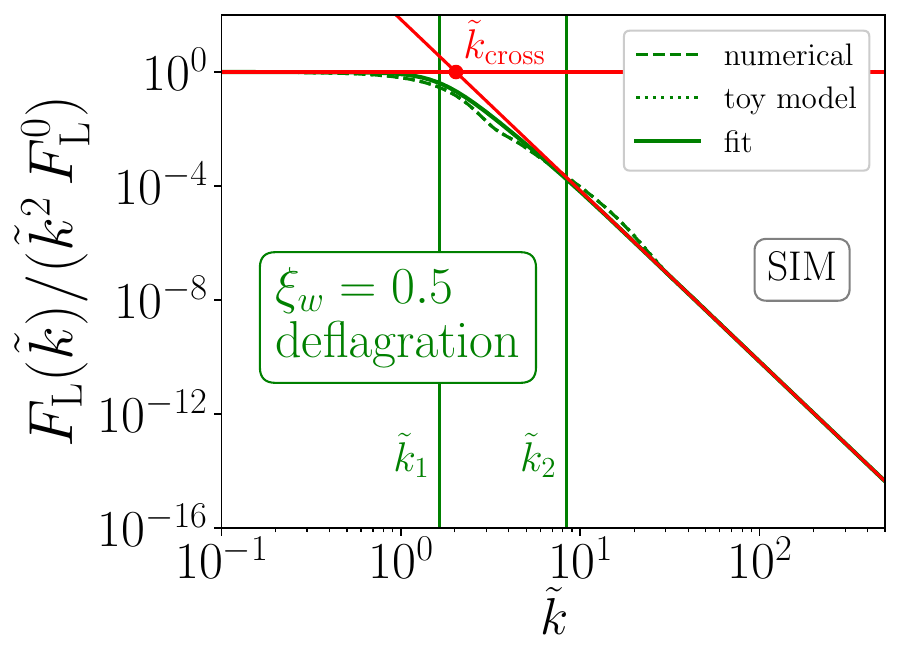}
    \includegraphics[width=.32\textwidth]{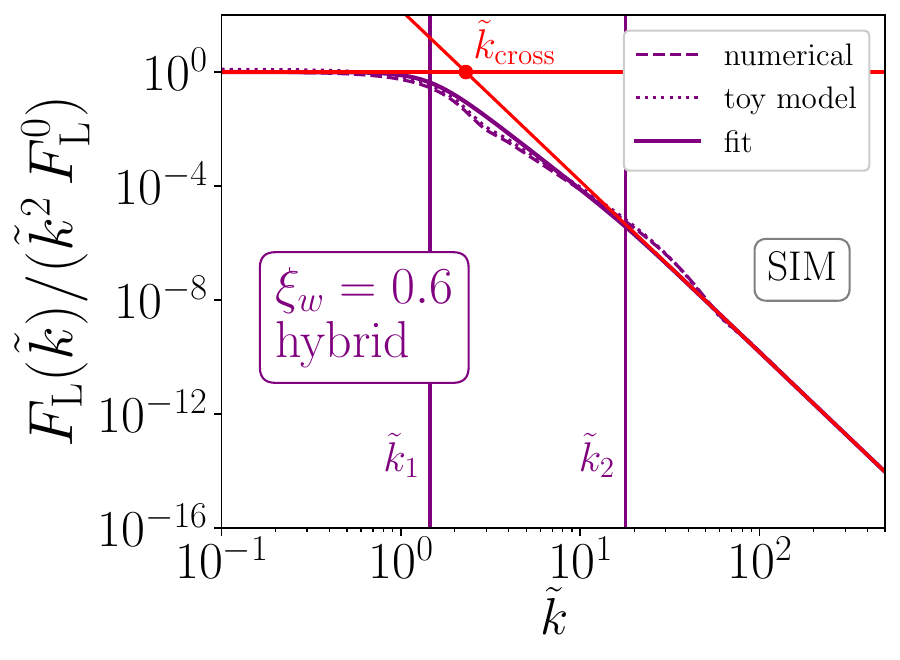}
    \includegraphics[width=.32\textwidth]{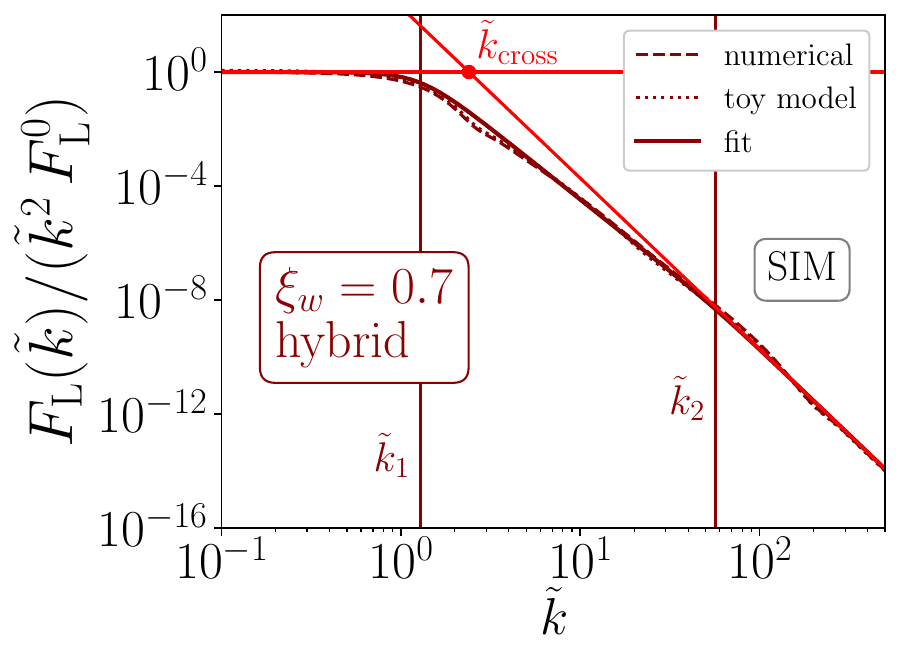}
    \includegraphics[width=.32\textwidth]{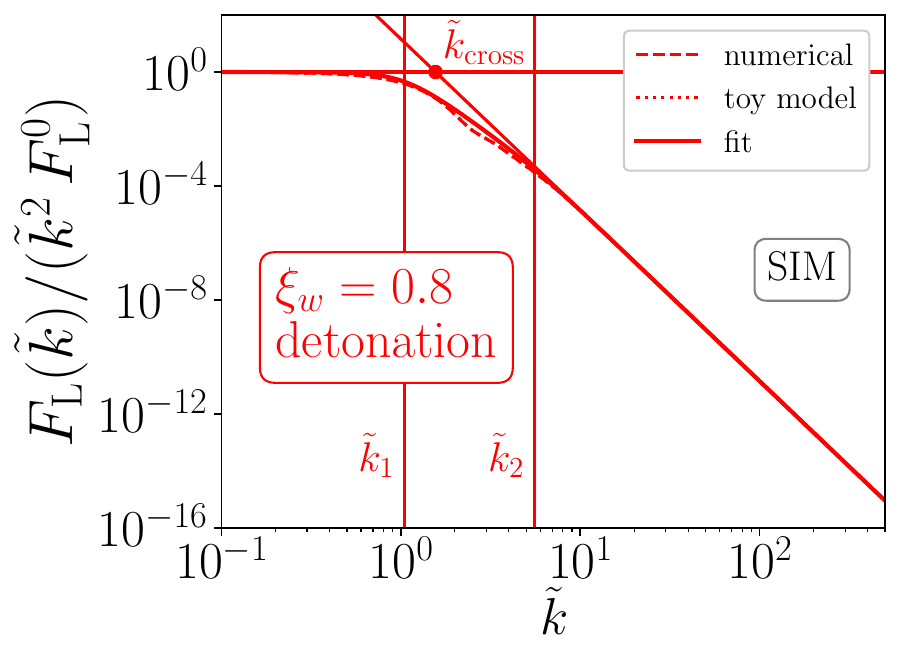}
    \includegraphics[width=.32\textwidth]{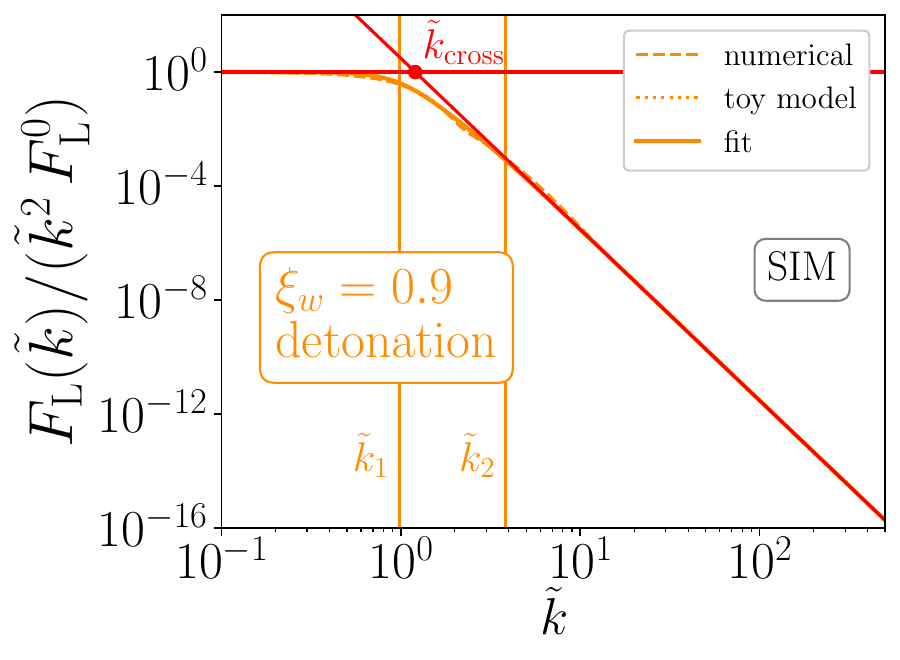}
    \caption{
    Same as \Fig{newfits_FL_EXP} for the case of simultaneous (SIM) nucleation.
    }
    \label{newfits_FL_SIM}
\end{figure}

We now proceed to compute the ratios $\tilde k_1/z_1$ and   $\tilde k_2/z_2$ numerically, by performing the integration over $\nu (T)$
in \Eq{def_FL_ssm}.
We find that the factors $\tilde k_1/z_1$ 
and $\tilde k_2/z_2$ are different from each other, and that they only depend on
the type of nucleation probability distribution,
but not on the strength $\alpha$ of the phase transition neither on the bubble
wall velocity $\xi_w$,
\begin{equation}
\label{scales_relation}
    \frac{\tilde k_{1, 2}^{\rm exp}}{z_{1, 2}} \simeq \{0.17, 0.21\}\,, \qquad
    \frac{\tilde k_{1, 2}^{\rm sim}}{z_{1, 2}} \simeq \{0.38, 0.25\}\,.
\end{equation}
These values accurately describe the positions of the spectral peaks for
the single and double broken power laws that can be used to reproduce the velocity
spectrum for different phase transitions, as can be seen in \Figs{newfits_FL_EXP}{newfits_FL_SIM}.
Furthermore, as argued after \Eq{kcross_expsim}, $\tilde k_1/z_1$ is equal
to $\tilde k_\cross/z_\cross$. 

\begin{figure}[t]
    \centering
    \includegraphics[width=0.9\linewidth]{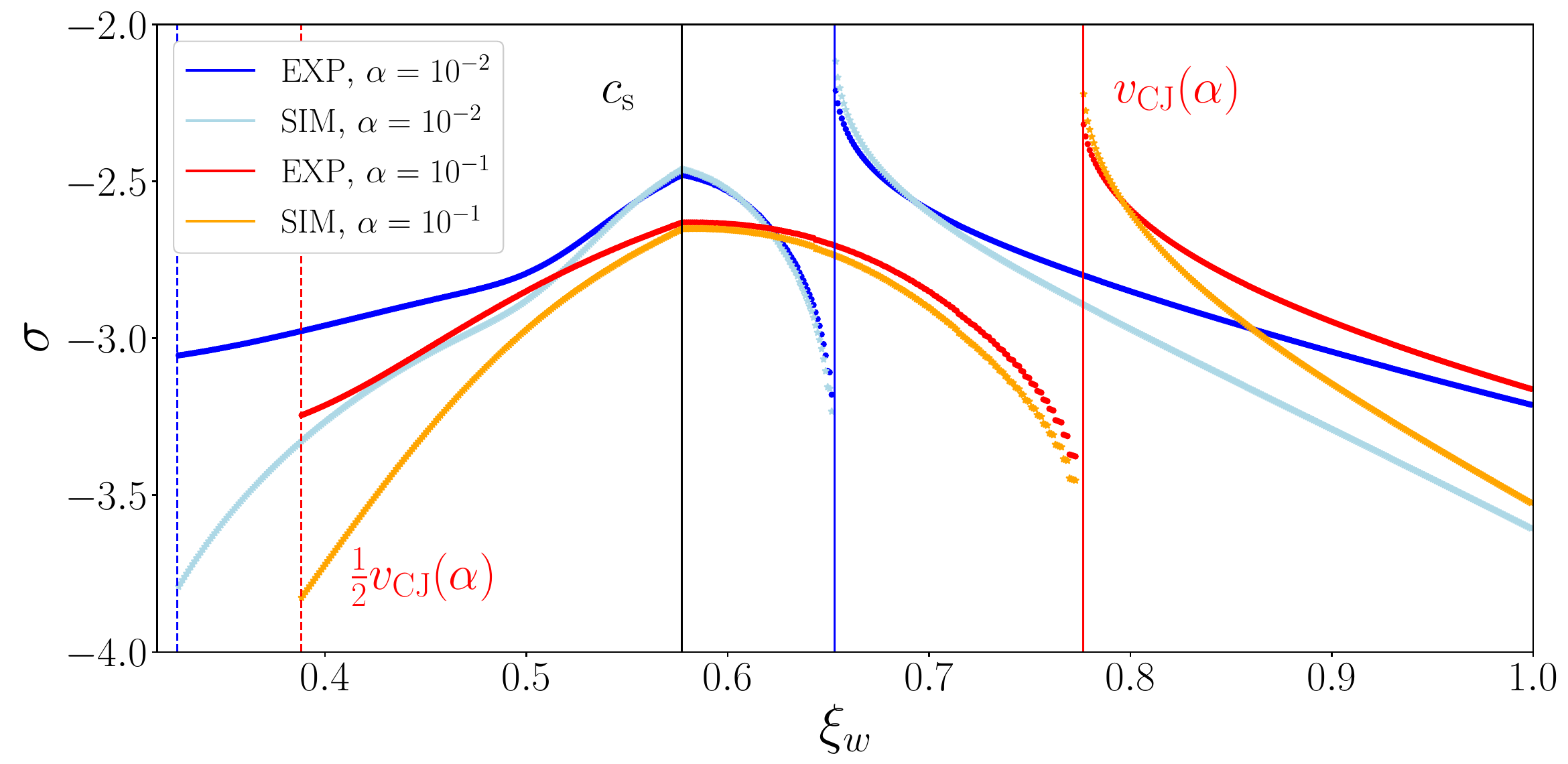}
    \caption{Intermediate slope $\sigma$, defined in \Eq{slope_k},
    as a function of $\xi_w$ for different values of $\alpha = 0.01$ and $0.1$.
    For each value of $\vCJ(\alpha)$,
    we plot $\sigma$ only
    in the range $\xi_w \geq \tfrac{1}{2} \vCJ (\alpha)$ (indicated
    by vertical dashed lines),
    in which we fit the envelope of $\FL$
    with a double broken power law [see \Eq{FL_DBPL}].
    The sound ($\cs$) and the Chapman-Jouget ($\vCJ$) speeds,
    determining the transition between sub and supersonic deflagrations
    and between deflagrations and detonations, are indicated with
    vertical solid lines.
    We show the results for both exponential (EXP) and simultaneous (SIM) nucleation histories.
    }
\label{fig:sigma_vs_vw}
\end{figure}

We note that the different coefficients in the two ratios of \Eq{scales_relation} modify the slope
$\sigma$ (see \Fig{fig:sigma_vs_vw}) with respect to $\gamma$, i.e., the intermediate
slope found for $\fpsq$ (see \Fig{fig:gamma_vs_vw}), not to be confused with the
normalized nucleation rate.
Indeed, defining $\tilde{k}_{\rm cross}/z_{\rm cross} = \tilde k_1/z_1 = a$
and $\tilde k_2/z_2 = b$, 
the slope $\sigma$ in \Eq{slope_k} becomes
\begin{equation}
    \sigma = 2 \biggl[1 - 3 \frac{\log(z_2/z_\cross) + \log(b/a)}{\log(z_2/z_1) + \log(b/a)}
    \biggr] = \gamma - 6 \log(b/a) \frac{\log(z_\cross/z_1)/\log(z_2/z_1)}
    {\log(z_2/z_1) + \log(b/a)}\,.
\end{equation}
Hence, the value of $\log(b/a)$ quantifies how different the slope $\sigma$ is to the
slope $\gamma$ computed in \Sec{sec:f_template},
and $\sigma$ reduces to $\gamma$
in the limit $b = a$, where the slope is preserved.
Of course, $\sigma$ also reduces to $-4$ as
$\gamma$ does, in the limit $z_2 \to z_\cross \to z_1$,
independently of the ratio $b/a$.
We note that the value of $\sigma \to -3$ as $z_\cross \to z_1$ in \Fig{fig:sigma_vs_vw}
(see $\sigma$ at $\xi_w = \half \, \vCJ$), which
seems in contradiction with the appearance of only one power law, is just
a numerical artifact due
to the value of the smoothness parameter
$b_1 = 2$ used for exponential nucleation, which makes
the transition smoother than when we use $b_1 = 4$ for $\fpsq$ in \Fig{fig:gamma_vs_vw}
and for simultaneous nucleation.

For both the single and the double broken power law fits,
the amplitude of the velocity spectrum in the $k \to 0$ limit,
$\FLz$, independently 
of the nucleation history,
is proportional to $\fpsqz$
[see \Eq{rel_FL0_fp0}].
The latter is related to the third moment of the single-bubble
velocity profile $v_\ip (\xi)$
[see \Eq{asymptotic_fpz}].
However, as shown in \Sec{velocity_rms_mult}, the integral over $k$ of ${\cal P}_v/k \equiv k^2 \FL (k)/(2 \pi^2)$
should be proportional to $v_\rms^2 \equiv \tfrac{3}{4} \kappa_{\rm pert} \, \alpha$
[see \Eq{eq:vrms2_coll+}].
This allows us to write
\begin{equation} \label{final_temps}
    \PP_v (\tilde k) = v_\rms^2 \, {\cal N} \,
    S (\tilde k) = \tfrac{3}{4} \, \kappa_{\rm pert} \, \alpha
    \, {\cal N} \, S (\tilde k) \,,
\end{equation}
where ${\cal N}$ is a normalization constant, defined
such that
$\int S(\tilde k) \, \dd \ln \tilde k = {\cal N}^{-1}$
and $S$ represents either the single or the double broken power law,
based on the appropriate fit for each case,
\begin{subequations}
\label{SS_funcs}
\begin{align} \label{S_SBPL}
    S_{\rm SBPL} (\tilde k) = &\,
    \biggl(\frac{\tilde k}{\tilde k_1}\biggr)^5
    \Biggl[ 1+ \biggl(\frac{\tilde k}{\tilde k_1} \biggr)^{b_1} \Biggr]^{-6/b_1}\,,
    \\
    \label{S_DBPL}
    S_{\rm DBPL} (\tilde k) = &\,
    \biggl(\frac{\tilde k}{\tilde k_2}\biggr)^5
    \Biggl[ 1+ \biggl(\frac{\tilde k}{\tilde k_1} \biggr)^{b_1} \Biggr]^{\frac{\sigma-2}{b_1}}\,
    \Biggl[ 1+ \biggl(\frac{\tilde k}{\tilde k_2} \biggr)^{b_2} \Biggr]^{\frac{-\sigma-4}{b_2}}\,.
\end{align}
\end{subequations}
The normalization constants ${\cal N}$ relate the amplitude $\FLz$
and $v_\rms^2$ {for both the SBPL and the DBPL}, and can be expressed analytically (see \App{analytical_comp} for a detailed derivation of these
expressions, in particular \Eq{kappav_to_FL_1} for the single broken power law,
and \Eqs{kappav_to_FL_2}{kappav_to_FL_3} for the double broken power law with $b_1 = b_2$
and $b_1 \neq b_2$ respectively).

Combining \Eqs{final_temps}{S_SBPL},
and introducing ${\cal N}_{\rm SBPL}$ from \Eq{kappav_to_FL_1}, we find
\begin{mymathbox}
\begin{align} \label{final_temp_SBPL}
    \PP_{v, \, {\rm EXP/SIM}}^{\rm SBPL} (\tilde k)
    &= 
    \frac{b_1 \Gamma \bigl( \tfrac{6}{b_1} \bigr)}{\Gamma \bigl( \tfrac{5}{b_1} \bigr)\Gamma
    \bigl( \tfrac{1}{b_1} \bigr)} \, v_\rms^2
    \,
    S_{\rm SBPL} (\tilde k)
    \quad \text{for \ } \xi_w \lesssim \half \, \vCJ\,,
\end{align}
\end{mymathbox}
\noindent
for the velocity spectra that can be approximated as a single broken power law.
The prefactor $b_1 \Gamma(6/b_1) \Gamma(5/b_1)^{-1} \Gamma(1/b_1)^{-1}$ takes the values
1.09 and 1.70 for $b_1 = 4$ and $b_1 = 2$, respectively.

For the velocity spectra described by double broken power laws,
the integral of $S_{\rm DBPL}$ can be expressed via the hypergeometric function
when $b_1 = b_2 = b$, which we find to represent $\FL$ for simultaneous nucleation
($b = 4$).
Introducing ${\cal N}_{\rm DBPL}$ from \Eq{kappav_to_FL_2}, we write
\begin{mymathbox}
\vspace{-2mm}
\begin{align} \label{FL_from_kappa_sim}
    \PP_{v, \, {\rm SIM}}^{\rm DBPL} (\tilde k) = 
    \frac{\Gamma\bigl(\tfrac{6}{b}\bigr)}
    {\Gamma\bigl(\tfrac{5}{b}\bigr) \Gamma\bigl(\tfrac{1}{b}\bigr)} &\,
    \!\,_2F_1^{-1} \Bigl(\tfrac{2-\sigma}{b}, \tfrac{5}{b}; \tfrac{6}{b};
    1 - \bigl[\tfrac{k_2}{k_1}\bigr]^b \Bigr) \, v_\rms^2
    \, S_{\rm DBPL} (\tilde k)
    \quad \text{for \ } \xi_w \gtrsim \half \, \vCJ\,.
    \end{align}
\end{mymathbox}
\noindent
In this case, the amplitude of the velocity spectrum depends on the slope $\sigma$
and the ratio $k_2/k_1$.

For the case of exponential nucleation, we find that different smoothness
parameters ($b_1 = 2$ and $b_2 = 4$) are required to fit the numerical
results.
In this case, ${\cal N}_{\rm DBPL}$ can be expressed using the Fox $H$-function
in \Eq{kappav_to_FL_3},
\begin{mymathbox}
\begin{align} \label{FL_from_kappa_exp}
    \PP_{v, \, {\rm EXP}}^{\rm DBPL} (\tilde k)   =  &\,
    \, b_2 
    \Gamma\bigl(\tfrac{2 - \sigma}{b_1}\bigr) \, \Gamma\bigl(\tfrac{4 + \sigma}{b_2}\bigr)
    \, v_\rms^2
    \Biggl\{ H_{2,2}^{2,2}\! \left[
    \left(\frac{k_2}{k_1}\right)^{b_1}
    \;\middle|\;
    \begin{array}{l}
    \bigl(-4,\frac{b_1}{b_2}\bigr),\;
    \bigl(1-\frac{2-\sigma}{b_1},1 \bigr)
    \\
    (0,1),\;
    \bigl(\frac{\sigma+4}{b_2}-5,\frac{b_1}{b_2} \bigr)
    \end{array}
    \right] \Biggr\}^{-1} 
    \nonumber \\  &\, \times 
    S_{\rm DBPL} (\tilde k)
    \hspace{70mm} \text{for \ } \xi_w \gtrsim \half \, \vCJ\,.
    \end{align}
\end{mymathbox}

These templates provide accurate fits to the velocity power spectral density
with the minimum amount of information required from the original
single-bubble fluid velocity profiles
and the distribution function of the nucleation times.
The scales
$\tilde{k}_1$ and $\tilde{k}_2$ for exponential and simulteanous
nucleation are computed using \Eq{scales_relation}
from the scales $z_1$ and $z_2$ [see \Eqs{fit_z1}{fit_z2}],
which characterize the power spectral density of the velocity field
averaged over nucleation locations, $\fpsq$.
Then, the intermediate slope $\sigma$ is defined in \Eq{slope_k}
and can be computed from $\tilde{k}_1$,
$\tilde{k}_2$, and $\tilde{k}_\cross$, with $\tilde k_\cross/z_\cross$
given in \Eq{kcross_expsim}.
$z_\cross$ depends on the ratio $f'_0/f'_{\rm env}$,
where $f'_0$ is given by the integral over the 
self-similar velocity profile in \Eq{asymptotic_fpz},
and $f'_{\rm env}$ by the discontinuities of the velocity profile in \Eq{fpenv_inf}.
Finally, the amplitude can either be expressed directly
using $\FLz$ [see
{\Eqs{FL_SBPL}{FL_DBPL}}], or using the integrated amplitude
$v_\rms^2 = \tfrac{3}{4} \kappa_{\rm pert} \alpha$
{[see \Eqss{final_temp_SBPL}{FL_from_kappa_exp}]}
where $\kappa_{\rm pert}$ is
given by \Eq{vrms2} and
characterizes the kinetic energy density in the perturbative
limit of small fluid perturbations.
Hence, we have clearly identified the origin of each of the scales, asymptotic limits,
and intermediate slopes, and our results can be {directly} applied for all different
types of fluid profiles and {two common choices of} distribution functions of nucleation times.

\section{Conclusions}
\label{concls}

The gravitational wave (GW) background produced during a first-order phase
transition in the early Universe is one of the most promising
cosmological backgrounds that can be potentially observed with
current and future GW experiments.
It is of particular relevance to explore BSM physics leading
to a first-order electroweak phase transition that can be probed
by the Laser Interferometer Space Antenna.

In this work, we investigate the fluid perturbations associated
to first-order phase transitions, paying particular attention
to the origin of the scales that are commonly assumed to describe the
GW spectrum in the literature based on fits to numerical simulations
of the fluid and scalar field dynamics during first-order
phase transitions \cite{Hindmarsh:2013xza,Hindmarsh:2015qta,Hindmarsh:2017gnf,Cutting:2019zws,Jinno:2020eqg,Jinno:2022mie,Caprini:2024gyk,Correia:2025qif},
and on theoretical modeling based on the sound shell model \cite{Hindmarsh:2016lnk,Hindmarsh:2019phv,RoperPol:2023dzg,Sharma:2023mao}.
To describe the fluid perturbations,
we restrict our analysis to the bag equation of state \cite{Espinosa:2010hh}.
The results presented in our work provide the initial conditions
for the sound-wave regime that some of the authors
considered in ref.~\cite{RoperPol:2023dzg}.
The study of the fluid perturbations during the sound-wave regime
and the impact of our results to determine the GW spectrum, in
particular the relevant scales and the different slopes, is presented
in a separate publication \cite{part2}.

We study the statistical properties and spectral shape
of the velocity field, whose power spectral
density is shown to 
{be fully characterized by its} longitudinal
component, as expected for fields with zero vorticity (irrotational),
which is given in \Eq{eq_FL1},
\begin{equation} \label{FL_concl}
    \FL (\tilde t, \tilde k) = \frac{n_b}{\beta^6}
    \int_{t_c}^t \gamma(\tilde t_0) (\tilde t - \tilde t_0)^6 \, 
    {f'}^2
    \bigl[(\tilde t - \tilde t_0)\, \tilde k \bigr] \dd \tilde t_0\,,
\end{equation}
where $\gamma$ describes the normalized nucleation rate of broken-phase
bubbles, described in detail in \App{time_dist},
$n_b \equiv \beta^3/(8 \pi \xi_w^3)$ is the expected
number density of bubbles at the end of the phase transition, $\beta$ is
the nucleation rate parameter, $\xi_w$ the wall velocity, and $\tilde k = k/\beta$
and $\tilde t = t \beta$ are normalized wave numbers and times.
The power spectrum in \Eq{FL_concl} describes the velocity field
during the phase of expansion of uncollided bubbles.

The spectral function ${f'}^2 (z)$,
where $z = (\tilde t - \tilde t_0) \, \tilde k$, can be traced back
to the Fourier transform of the single-bubble velocity profile
$v_\ip (\xi)$,
as given by \Eq{velo_fz}, where $\xi = r/(t - t_0)$, $r$ being the distance to the
center of the bubble.
For a benchmark phase transition with $\alpha = 0.1$, we show
the velocity profiles in \Fig{fig_ip}, following the description provided
in \App{1d_profiles}.
The relation between $v_\ip (\xi)$ and $f'(z)$, provides us with key
information on the spectral shape of the resulting velocity
spectrum.
Indeed, we found 
in \Sec{subsec_RiemannL} that the two asymptotic limits of
${f'} (z)$ when $z \to 0$ (large scales) and $z \to \infty$ (small scales)
can be determined
analytically from specific properties of $v_\ip (\xi)$.
In particular, the third moment of the velocity profile $v_\ip (\xi)$
determines the value of ${f'}(z)$
at large scales [see \Eq{asymptotic_fpz}],
\begin{equation}
    \lim_{z \to 0} f'(z) = -
    \frac{4 \pi}{3} z \int_0^\infty
    \xi^3 \, v_{\rm ip} (\xi)  \dd \xi\,,
\end{equation}
while the locations of the discontinuities of $v_{\rm ip} (\xi)$ and
the values of the velocity jumps $\Delta v_\ip$, determine exactly
the asymptotic limit of ${f'} (z)$ at small scales [see \Eqq{eq:f1_zInfty_dependence}],
\begin{align} \label{fp_inf_concl}
    \lim_{z \to \infty} f' (z) =  - \frac{4\pi}{z^2}
    \bigl[
    \xi_\sh \, \sin(z\xi_\sh) \,
    \Delta v_\ip(\xi_\sh)
    + \xi_w \, \sin(z\xi_w)\,  \Delta v_\ip(\xi_w)
    \bigr] \,.
\end{align}
The above equation holds for deflagrations,
which present two discontinuities: one at $\xi_w$ and one at $\xi_\sh$.
For detonations, with only one discontinuity at $\xi_w$, the asymptotic limit of $f'(z)$ is
\begin{align} 
    \lim_{z \to \infty} f' (z) =  - \frac{4\pi}{z^2}\,
    \xi_w \, \sin(z\xi_w)\,  \Delta v_\ip(\xi_w) \,.
\end{align}

Furthermore, we note that $\fpsq
\sim z^2$ and $\sim z^{-4}$
respectively in the two asymptotic limits $z \to 0$ and $z \to \infty$.
This behavior can be directly inferred from the statistical
properties of discontinuous, causal, and
irrotational fields that follow
statistically homogeneous and isotropic distributions, as we discuss
in \Sec{kinetic_sp_bubbles} and \App{appendix_correlation}.
We
found that the scales $z_1$ and $z_2$ at which both
asymptotic limits are reached, depend precisely on the
locations of the two discontinuities that
appear in the velocity profiles of deflagrations:
one discontinuity is always present at the bubble wall ($\xi_w$),
and a second discontinuity is due to the formation of a shock in front
of the wall at $\xi_\sh$.
For detonations, only one discontinuity is present.
In this case, we found
that the two scales $z_1$ and $z_2$ are determined by the front and the back
of the fluid shell, as previously argued in the literature.
Also for subsonic deflagrations, the front and the back locations
of the fluid shell coincide with the discontinuity locations.
For hybrids, however, one needs to correct the
estimate of the separation between the two scales $z_1$ and $z_2$
in previous literature.
To summarize, for a broad range of $\xi_w$ and $\alpha$,
the locations of the two scales determining the breaks
of the function $\fpsq$ are
the following [see \Eqs{fit_z1}{fit_z2}],
\begin{equation}
    \xi_1 = \frac{2 \pi }{z_1} \simeq
    {\frac{1}{3} \, \xi_+} \times \left\{ \begin{array}{l}
    4\,, \\ 5\,,
    \end{array}\right. \qquad
    \xi_2 = \frac{2 \pi }{z_2} \simeq
    {\frac{1}{3} \,
    \tilde \Delta \xi} \times
    \left\{ \begin{array}{ll}
    6\,, & \qquad \rm{for \ deflagrations}\,,   \\
    4\,, & \qquad
    \rm{for \ detonations}\,.
    \end{array}\right. 
\end{equation}
We identified two scales, $\xi_+$ and $\tilde \Delta \xi$, where the former
is $\xi_+ = \xif + \xi_w {= \xi_\sh + \xi_w}$ for deflagrations and $\xif + \xib = {\xi_w + \cs}$ for detonations,
and the latter is $\tilde \Delta \xi = \xif - \xi_w = {\xi_\sh - \xi_w}$ for deflagrations
and $\tilde \Delta \xi = \Delta \xi = \xif - \xib = {\xi_w - \cs}$ for detonations.
The scale $\tilde \Delta \xi$ coincides with the
fluid shell thickness $\Delta \xi = \xif - \xib$ for subsonic deflagrations
and detonations, but not for hybrids.
For subsonic deflagrations with $\xi_w \lesssim \half \, 
\vCJ (\alpha)$, the velocity jump becomes negligibly small at $\xi_\sh
\to \cs$, where $\cs = \tfrac{1}{\sqrt{3}}$ is the speed
of sound and $\vCJ$ is the Chapman-Jouget speed.
In these cases, the second scale does not appear in the velocity
spectral density, which is then only characterized by $\xi_1$.
{This is a consequence of a vanishing
discontinuity jump at $\xi_\sh$ as $\xi_\sh$ approaches $\cs$.}

The resulting $\fpsq$ is, in general, a highly oscillatory function at
large values of $z$, as can be seen in \Eq{fp_inf_concl}.
Its envelope can be determined by a single or a double broken power law,
depending on whether we have one or two relevant scales.
We provided in \Sec{sec:f_template} templates that accurately reproduce the envelope
that we find from the numerical velocity profiles (see \Fig{fp2_fp20}).
These fits reproduce accurately the locations $z_1$ and $z_2$, the asymptotic limits,
and the intermediate slopes in the cases described by a double broken
power law.
The intermediate slope in $\fpsq$ can be found using \Eq{slope_z},
and it only depends on the asymptotic amplitudes and locations of the
spectral breaks, which can be traced back to the properties of the
single-bubble velocity profiles.

To test the robustness of our results and provide analytical results, we studied toy models in \Sec{sec:toymodel}. 
We found that the simplest toy models that accurately reproduce
the velocity profiles and the spectral shape of the envelope of $\fpsq$
are quadratic (for subsonic deflagrations and detonations) and piecewise
linear-constant with a discontinuity at $\xi_w$ (for hybrids).
The proposed
toy models can be used to compute $\fpsq$ analytically,
providing a useful tool to study the resulting velocity
spectrum at the beginning of the sound-wave regime and, hence,
the resulting anisotropic stresses and the GW spectrum.

In \Sec{kinetic_sp_bubbles}, we studied the correlation function
of the velocity field as a function of the separation $\rr$ between
two points in coordinate space, $B_{ij} (\rr)$,
and related it to the
power spectral density $\fpsq$ in momentum space.
We revised previous studies in
the literature, and provided new results regarding the relation between
the longitudinal and normal components of the correlation function
[see \Eq{diff_eq_pot_text}],
which is a consequence of the curl-free condition of the
spherical velocity profiles.

In \Sec{ensemble_times}, we presented how the scales and amplitude of the
velocity spectrum $\FL$, obtained from $\fpsq$ after taking the average
over nucleation times, as indicated in \Eq{FL_concl}, evolve with time
during the phase of expanding uncollided bubbles.
We focus our study on two types of nucleation distributions that are
commonly considered for phase transitions in the literature
(exponential and simultaneous), which we review
in \App{time_dist}.
To conclude our analysis of the fluid perturbations during the phase
of bubble expansion, we reviewed in \Sec{anis_stress1}
the implications of applying
Wick's theorem during this phase, which leads to a non-vanishing
two-point correlator of the anisotropic stress tensor, in contradiction with the spherical symmetry
of the system.
We explicitly indicate how this is a consequence of the non-Gaussianities
of the field during this period and, hence, the inapplicability
of Wick's theorem.

We then evaluate the conservation of kinetic energy density during
the expansion phase of multiple uncollided bubbles, showing how 
the kinetic energy density of the system can be described by the efficiency factor
$\kappa$ \cite{Kosowsky:1991ua,Espinosa:2010hh}, obtained from the single-bubble fluid profiles, and used
in the literature to determine the amplitude of the GW spectrum.
Furthermore, we showed that the modeling of collisions via
the introduction of a bubble lifetime distribution in the sound
shell model implicitly implies that the mean squared fluid fields
(i.e., velocity and enthalpy) are conserved during the period of
collisions.
Based on this finding, we proposed an alternative method to determine
the velocity power spectrum at the initial time of the sound-wave
phase that is instead only
based on the conservation of kinetic energy density during the
initial period of collisions.
Following this approach, we defined the characteristic,
deterministic time at which the sound-wave regime starts, $\tilde t_{\rm sw}$,
and computed it for the cases of exponential
and simultaneous nucleation in \Sec{EffCollTime}.
The proposed approach leads to the same velocity power spectrum
that the original sound shell model in the case of exponential
nucleation (with relative errors in the spectrum below 8\%),
as can be seen in \Fig{fig:FL_ev},
but it leads to different results as the
distribution function becomes closer to the simultaneous case,
i.e., as the dispersion of the distribution function around
the peak time $\tilde t_{\rm peak}$ diminishes.
This is a consequence of the fast oscillations in $\fpsq$, which
are not suppressed when we compute $\FL$ in the simultaneous case
following our approach.
When we introduce an average over the bubble lifetimes, as done
in the sound shell model, these oscillations are smoothed out.
However, we expect that the resulting anisotropic stresses and GW
spectrum will not be strongly affected by the presence of the oscillations,
since they are computed from a convolution of the velocity power
spectrum, as indicated in \Eq{eq:UETC:PiPi:Isserlis}.
We investigate the resulting GW spectrum following both
approaches in a separate publication \cite{part2}.

Finally, in \Sec{FL_template}, we provide a template of the
velocity power spectrum at the initial time of the sound-wave regime,
which is the key ingredient to determine the GW
spectrum from sound waves.
We evaluated the asymptotic limits of $\FL$ at large and small
scales, which depend on the asymptotic limits of the original
${f'}^2 (z)$ function, and on {the eighth (large-scales limit) and
second (small-scale limit) moments}
of the distribution
function of the bubble lifetimes [see \Eqs{rel_FL0_fp0}{fl_env}].
We predict the positions of the velocity power spectrum $\FL$,
$\tilde k_1$ and $\tilde k_2$, which are related to $z_1$
and $z_2$, as indicated in \Eq{scales_relation},
\begin{equation}
    \frac{\tilde k_{1, 2}^{\rm exp}}{z_{1, 2}} \simeq \{0.17, 0.21\}\,, \qquad
    \frac{\tilde k_{1, 2}^{\rm sim}}{z_{1, 2}} \simeq \{0.38, 0.25\}\,.
\end{equation}
The resulting $\tilde k_1 \equiv k_1/\beta$ and $\tilde k_2 \equiv k_2/\beta$
determine the characteristic 
scales of $\FL$, given as a fraction of the nucleation
rate parameter $\beta$, which can be related to the mean separation of the bubbles
$R_\ast$ using \Eqs{asympt_nb}{rstar_as}.
The two scales $\tilde k_1$ and $\tilde k_2$ have different
ratios to the original scales $z_1$ and $z_2$, hence implying that
the particular nucleation distribution can impact not only
the position of the resulting scales, but also the intermediate slope
of the velocity spectrum, which can be computed using \Eq{slope_k}.
{However, we find that the ratios $\tilde k_{1,2}/z_{1,2}$ are independent
of $\xi_w$ and $\alpha$ and, hence,
only depend on the distribution function
of nucleation times.}
We provided a single and double broken power law template
for the velocity spectrum $\FL$, which only depends on the two scales
$\tilde k_1$ and $\tilde k_2$, on the intermediate slope $\sigma$,
and on the amplitude in the $k \to 0$ limit, $\FLz$, given in \Eq{rel_FL0_fp0}.
We also showed that, instead of using $\FLz$, we can use the integral
of $\FL$ over wave numbers, which is determined by the efficiency factor
$\kappa$ [see \Eq{vrms2} and \Fig{kappas_alpha}]
\begin{equation}
    \kappa \simeq \kappa_{\rm pert}
    \equiv \frac{4}{\xi_w^3 \alpha}
    \int_0^1 v_\ip^2 (\xi)
    \, \xi^2 \dd \xi\,,
\end{equation}
where $\kappa_{\rm pert}$ approximates
$\kappa$ in the linear regime
of small fluid perturbations.
Based on analytical relations between the amplitude and the
integral of the single and double broken power law templates presented
in \App{analytical_comp}, we provided in \Sec{template_FL} a template for the velocity spectrum
in which the amplitude is determined by $\kappa$.

The main result of our work is a generic double broken
power law template of the velocity spectrum at the beginning
of the sound-wave regime, given
in \Eqss{final_temp_SBPL}{FL_from_kappa_exp}, that contains the
minimal required information from the original velocity profiles,
{described by:}
\begin{enumerate}
    \item The two scales $\tilde k_1$ and $\tilde k_2$ of the
    velocity power spectral density $\FL$, which are characterized by
    either the positions of the discontinuities of the
    single-bubble profiles $v_\ip (\xi)$
    or the positions
    of the front and the back of the fluid shells, and a constant
    factor that depends on the nucleation distribution function;
    \item The overall amplitude, which can be recast in terms of the
    integral over of the kinetic
    energy density fraction of the single-bubble profile, $\kappa$,
    given in \Eq{cons_kappa};
    \item The intermediate slope $\sigma$, which is computed using the
    two asymptotic limits of the function ${f'}(z)$ and integrals over
    the bubble lifetime distributions, as given in \Eq{slope_k}, and shown
    in \Fig{fig:sigma_vs_vw}.
\end{enumerate}

The spectral breaks and slopes found in the velocity spectrum
determine the power
spectrum of the anisotropic stresses and of the GW spectrum produced in the sound-wave
phase.
Hence, the two scales that we identified in the velocity spectrum indicate the
origin of the two scales that are usually considered in the literature
for GW backgrounds from first-order phase transitions sourced by
sound waves, based on theoretical modeling following
the sound shell model and on numerical simulations.
These findings are crucial because they predict the geometric parameters of the GW spectrum arising from first-order
phase transitions, which can then be used to reconstruct the thermodynamic
and fundamental parameters. These respectively describe the dynamics
of first-order phase transitions and the underlying fundamental BSM theories leading to the first-order phase transition \cite{Caprini:2024hue}.
Providing simple analytical formulas for the geometric parameters can
significantly improve the efficiency
of our studies of parameter reconstruction from
the GW signals, allowing us to explore a broader range of phase
transition parameters efficiently, and allowing us to explore new physics
from GW observables in the early Universe.

\acknowledgments

ARP is grateful to Thomas Konstandin and Isak Stomberg for useful discussions.
SP is grateful to Jani Dahl, Jonas El Gammal, Lorenzo Giombi, Mark Hindmarsh, Mikko Laine,
Kari Rummukainen and Jorinde van de Vis for useful discussions.
ARP and ASM are supported
through the Swiss National Science Foundation
SNSF Ambizione grant \href{https://data.snf.ch/grants/grant/208807}{208807},
and CP and SP through the
SNSF Project Funding grant \href{https://data.snf.ch/grants/grant/212125}{212125}.
ARP, SP and ASM acknowledge the hospitality of CERN
where part of this work was conducted.
The authors are grateful for the support
provided by Nordita during the program
``{\em Numerical Simulations of Early Universe Sources of Gravitational Waves},'' by the Bernoulli Center during
the program ``{\em Generation, evolution, and observations of cosmological magnetic fields},''
by the Centro de Ciencias de Benasque
Pedro Pascual during ``{\em The Dawn of Gravitational Wave Cosmology}'' conference,
and by CERN during the workshop
``{\em Advancing gravitational wave predictions from cosmological first-order phase transitions},''
where preliminary results of this work were presented.
ARP and ASM acknowledge the support of CERN during the 
2025 Pencil Code school
and user meeting,
where part of this work was conducted.
ARP acknowledges the hospitality of the Yukawa Institute for Theoretical Physics during the Gravity and Cosmology 2024
workshop.

\newpage

\appendix

\section{Hydrodynamics of an expanding bubble}
\label{1d_profiles}

In this section, following refs.~\cite{Espinosa:2010hh,Hindmarsh:2016lnk}, we solve the hydrodynamic conservation laws,
assuming spherical symmetry for a single broken-phase
bubble expanding at a terminal
velocity $\xi_w$, 
such that the resulting fluid velocity and enthalpy profiles,
$v_\ip (\xi)$ and $w_\ip (\xi)$,
can be described in terms of the reduced parameter $\xi \equiv r/t$,
where $r$ is the distance to the center of the bubble and $t$ is the time since its nucleation.
We provide a Python library, {\tt hydro\_bubbles},
as part of the {\sc CosmoGW} public code \cite{CosmoGW_GH,cosmogw_manual}
that computes
the self-similar profiles for different wall velocities and strengths of the phase transition,
following the description given in this appendix.

The hydrodynamic equations can be found by projecting the fluid conservation laws,
\begin{equation}
    \partial_\mu T^{\munu} = 0 \,,
    \label{hydro_app}
\end{equation}
in the parallel
and perpendicular directions of the fluid
\cite{Espinosa:2010hh,Rezzolla:2013dea,RoperPol:2025lgc},
where $T^\munu \equiv w \, U^\mu U^\nu + p\, \eta^\munu$ models
a perfect fluid [see \Eq{perf_fluid}], i.e., a fluid in local thermal equilibrium,
described in Minkowski space-time with metric tensor $\eta^{\mu\nu}$.
The conservation laws obtained from \Eq{hydro_app} are exact when
we are describing processes much shorter than one Hubble time, such
that the expansion of the Universe can be neglected.
As mentioned in \Sec{sec_1bubble}, in our work we assume short phase transition
durations, such that $\beta/H_\ast \gg 1$.
Moreover, we consider the bag equation of state to describe
the scalar field such that, out of the bubbles, $T^\munu$ is
only described by the fluid, as discussed at the beginning of \Sec{sec_1bubble}.

\subsection{Spherically symmetric fluid solutions}
\label{spherical_1d}

Spherical symmetry implies that
$U^\mu = \gamma (1, v \,\hat{r}^i)$, being $v$ the radial velocity,
$\gamma = (1 - v^2)^{-1/2}$ the corresponding Lorentz factor,
and $\hat r^i$ the unitary radial vector.
Moreover, we express space-time derivatives with respect to the self-similar
variable $\xi$,
 \begin{equation}
    t \partial_t = - \xi \partial_\xi\,, \qquad 
    r \partial_r =   \xi \partial_\xi \,,
    \label{ders_xi}
\end{equation}
where we omit any residual time dependence not incorporated in $\xi$,
due to the assumption
that the fields have reached stationary self-similar profiles.
Projection of \Eq{hydro_app} in the parallel direction
to the fluid velocity
leads to the equation of energy conservation,
\begin{equation}
    U_\nu \partial_\mu T^{\munu} = 0 
    \quad \Rightarrow \quad
    U^\mu \partial_\mu e + w \partial_\mu U^\mu = 0 \,,
    \label{cont_eq}
\end{equation}
where we have used the properties $U^\mu U_\mu = -1$ and
$U_\nu \partial_\mu U^\nu = \half \partial_\mu (U^\nu U_\nu) = 0$.
Using \Eq{ders_xi}, the spherically symmetric version of the
continuity equation \eqref{cont_eq} becomes
\begin{align}
    \frac{\dd \ln w_\ip (\xi)}{\dd \xi} = \frac{1+c_s^2}{\xi -v_\ip (\xi)}
    \biggl(2\frac{v_\ip (\xi)}{\xi} - \bigl[1 - \xi\, v_\ip (\xi)\bigr]
    \frac{\dd v_\ip (\xi)}{\dd \xi}
    \biggr) \,. \label{eq:dwdxi1}
\end{align} 

On the other hand, the projection of \Eq{hydro_app} in the orthogonal direction using the projection
tensor $h^\munu = \eta^\munu + U^\mu U^\nu$, with the property
$h^\mu_{\ \, \nu} \, U^\nu = 0$, yields the relativistic Euler equation,
\begin{align}
    h^\alpha_{\ \, \nu} \, \partial_\mu T^\munu = 0
    \quad \Rightarrow \quad
    w U^\mu \partial_\mu U^\alpha + \partial^\alpha p + U^\alpha U^\mu \partial_\mu p = 0\,,
    \label{Euler}
\end{align}
which in the direction  parallel to $\hat r^i$ becomes
\begin{align}
    \frac{\dd \ln w_\ip (\xi)}{\dd \xi} =
    \gamma^2_\ip (\xi) (1+c_s^{-2})  \,
    \mu(\xi, v_\ip)
    \frac{\dd v_\ip (\xi)}{\dd \xi} \,. \label{eq_hydro_ip_w}
\end{align}
The Lorentz transformation of the velocity is
\begin{equation}
    \mu(\xi,v) \equiv \frac{\xi-v}{1-\xi v}\,.
    \label{mu_xi}
\end{equation}
Combining \Eqs{eq:dwdxi1}{eq_hydro_ip_w}, we obtain the differential equation for
the radial velocity field \cite{Espinosa:2010hh},
\begin{align}
    \frac{\dd v_\ip (\xi)}{\dd \xi} &= 2 \, \frac{v_\ip (\xi)}{\xi}
    \frac{1-v_\ip^2 (\xi)}{1-\xi\, v_\ip (\xi)} \biggl(\frac{\mu^2(\xi,v_\ip)}{c_s^2} - 1 
    \biggr)^{-1} \,.
    \label{eq_hydro_ip_v}
\end{align}

We note that \Eq{eq_hydro_ip_v} has a singularity point at $\mu(\xi, v) = \cs$, i.e., when $v = \mu(\xi, \cs)$.
Hence, since
physical solutions are not allowed to cross this singular point,
the velocity profile
$v_\ip (\xi)$ remains monotonic at each side of the bubble wall (it can either increase or decrease,
but the sign of $\dd v_\ip/\dd \xi$ can only change across a discontinuity).
To find the fixed points of the trajectory expressed by \Eq{eq_hydro_ip_v}, we write it
in terms of a path parameter $q$, such that
\begin{subequations}
\begin{empheq}[left={ \displaystyle\frac{\dd v_\ip}{\dd \xi} = \frac{\dd v_\ip}{\dd q}
\frac{\dd q}{\dd \xi} \Rightarrow\empheqlbrace\,}]{align}
    & \frac{\dd v_\ip}{\dd q} = 2 \, \cs^2 \, v_{\rm ip} \, (1 - v_{\rm ip})^2
    \, (1 - \xi v_{\rm ip}) \ , \\
    & \frac{\dd \xi}{\dd q} = \xi \bigl[
    (\xi - v_{\rm ip})^2 - \cs^2 \, (1 - \xi v_{\rm ip})^2
    \bigr] \, .
\end{empheq}
\end{subequations}
Fixed points (fp)
are then found by imposing $\dd v_{\rm ip}/\dd q = \dd \xi/\dd q = 0$
\cite{Espinosa:2010hh},
\begin{equation}
    (\xi, v_\ip)_{\rm fp} = \lbrace (0, 0), (\cs, 0), (1, 1) \rbrace\,.
\end{equation}

\EEq{eq_hydro_ip_v} for $v_\ip (\xi)$ is decoupled from \Eq{eq_hydro_ip_w}, so it can be solved first, and then \Eq{eq_hydro_ip_w} can be
integrated to find the enthalpy profile $w_\ip (\xi)$,
\begin{equation}
    w_\ip (\xi) = w_0 \exp\biggl[\bigl(1 + \cs^{-2}\bigr)
    \int_{v_0}^{v_\ip (\xi)} \gamma^2 \mu (\xi, v) \, \dd v \biggr]\,,
    \label{sol_wip}
\end{equation}
where $v_0$ is a reference velocity for
which the value of the enthalpy, $w_0$, is known.

The enthalpy $w$ can alternatively be described by the
energy density fluctuations
$\lambda \equiv (e - e_{\rm sym})/w_n$,
where $e_{\rm sym} = e_n + \epsilon$ refers to the radiation energy density
in the region  where the fluid is not perturbed outside the bubble
in the symmetric phase \cite{Hindmarsh:2017gnf,Giese:2021dnw},
being $e_n$ and $w_n$ the background
energy density and enthalpy evaluated
at the nucleation temperature $T_n$.
This field is useful to describe the fluid perturbations during the sound-wave
regime in refs.~\cite{Hindmarsh:2016lnk,Hindmarsh:2019phv,RoperPol:2023dzg,RoperPol:2025lgc},
as we discuss in ref.~\cite{part2}.
As mentioned in \Sec{sec_1bubble}, we restrict the description of the
thermal potential from the phase transition to the bag constant $\epsilon$,
such that $\epsilon_+ = \epsilon$ and $\epsilon_- = 0$,
and to a radiation equation of state with
$\cs^2 = \partial p/\partial e = \tfrac{1}{3}$
in both the symmetric and broken phases.
This choice corresponds to the bag equation of state
\cite{Chodos:1974je,Espinosa:2010hh},
and the density fluctuations
become
\begin{equation}
    \lambda_\ip (\xi) = \tfrac{3}{4} \Bigl(\tfrac{w_\ip(\xi)}{w_n} - 1\Bigr) -
    \left\{\begin{array}{cl}
    \threefourth \alpha  & {\rm \ when \ } \xi \leq \xi_w\,, \\
    0 &  {\rm \ when \ } \xi > \xi_w \,,
    \end{array} \right.
    \label{lambda_ip}
\end{equation}
where we have defined $\alpha = \fourthird \epsilon/w_n$ as the parameter
that describes the strength of
the phase transition.

\subsection{Matching conditions across the bubble wall}
\label{matching}

The fluid across the bubble wall can be modeled as a
discontinuity in the limit of infinitesimally small bubble wall
thickness.
In this limit, the dynamics of the scalar field are restricted  to the
bubble wall. Out of the bubbles, we assume that $\partial_\xi \varphi = 0$ and that
the fluid conservation laws describe the full dynamical system.
In the frame where the bubble wall is at rest, the equations of momentum and energy
conservation, \Eq{hydro_app}, can be rewritten as $\partial_r T_{0r} = \partial_r T_{rr} = 0$. 
Integrating them across the wall leads to the following
conditions for the enthalpy, the pressure, and the velocity field in the wall frame,
$\tilde v$, expressed as \cite{Espinosa:2010hh,Hindmarsh:2019phv}
\begin{gather}
    w_+ \tilde \gamma_+^2 \tilde v_+^2 + p_+ = w_- \tilde \gamma_-^2 \tilde v_-^2 + p_-\,, \qquad
    w_+ \tilde \gamma_+^2 \tilde v_+ = 
    w_- \tilde \gamma_-^2 \tilde v_- \,.
    \label{matching_2}
\end{gather}
These equations can be rearranged in the following way \cite{Espinosa:2010hh}
\begin{equation}
    \tilde v_+ \tilde v_- = \frac{p_+ - p_-}{e_+ - e_-}\,,
    \qquad \frac{\tilde v_+}{\tilde v_-} = \frac{e_- + p_+}{e_+ + p_-}\,.
    \label{general_shock}
\end{equation}
The pressure and energy density in and outside the bubbles are given by
\begin{equation}
    p_\pm = \tfrac{1}{3} a_\pm T_\pm^4 - \epsilon_\pm\,, \qquad e_\pm = a_\pm T_\pm^4 + \epsilon_\pm\,,\label{p_pm_e_pm}
\end{equation}
with $\epsilon_-=0$ and $\epsilon_+=\epsilon$ in the bag equation of state.
The relations in
\Eq{general_shock} then become \cite{Espinosa:2010hh,Hindmarsh:2019phv,Giese:2020rtr}
\begin{subequations}
\begin{gather}
    \tilde v_+ \tilde v_- = \frac{1 - (1 - 3 \alpha_+)\, r}{3 - 3 (1 + \alpha_+)\, r}\,,
    \qquad \label{vplus_vminus}
    \frac{\tilde v_+}{\tilde v_-} = \frac{3 + (1 - 3 \alpha_+) \, r}{1 + 3 (1 + \alpha_+)\, r}\,,
\end{gather}
where
\begin{equation}
    \alpha_+ = \frac{4}{3} \frac{\epsilon}{w_+} \quad \text{and} \quad
    r = \frac{w_+}{w_-}
    = \frac{\tilde \gamma_{-}^{2} \tilde v_-}{\tilde \gamma_+^2 \tilde v_+}\,.
    \label{alpha_plus}
\end{equation}
\end{subequations}

The two matching conditions across the wall in
\Eq{vplus_vminus} can be combined to obtain the velocities $\tilde v_{\pm}$
as a function of the parameter $\alpha_+$,
\begin{subequations}
\label{vplus_from_vminus2}
\begin{align}
    \tilde v_+ = &\, \frac{1}{2(1+\alpha_+)} \left[
    \tilde v_- + \frac{1}{3\tilde v_-}  \pm \sqrt{\biggl(\tilde v_- - \frac{1}{3\tilde v_-}\biggr)^2 + 4\alpha_+^2 + \frac{8}{3} \alpha_+} \ 
    \right]\,, \label{vplus_from_vminus} \\
    \tilde v_- = &\, \frac{1}{2} \left[
    (1 + \alpha_+)\tilde v_+ + \frac{1 - 3\alpha_+}{3\tilde v_+} \pm
    \sqrt{\biggl[(1 + \alpha_+) \tilde v_+ +
    \frac{1 - 3\alpha_+}{3\tilde v_+}\biggr]^2 - \frac{4}{3}}\ \right] \,.
    \label{vminus_from_vplus}
\end{align}
\end{subequations}
Equivalently, we can also express $\alp$ as
\begin{equation}
    \alp = \frac{(1 - \vp/\vm) (1 - 3 \vp \vm)}
    {3 \, (\vp^2 - 1)}\,. \label{alppl_vs_vs}
\end{equation}

The strength of the phase transition is described by the parameter $\alpha \equiv \fourthird \, \Delta \theta/w_n$, which reduces to $\alpha = \epsilon/e_n$ in the
bag equation of state.
The relation between $\alpha$ and $\alpha_+$ 
is straightforward if $w_+/w_n$ is known: $\alpha\, w_n = \alpha_+\, w_+$.

\subsection{Classification of solutions}
\label{class_sols}

To find the boundary conditions to solve \Eq{eq_hydro_ip_v}, we take into
account that the fluid is at rest, $v= 0$, at $\xi = 0$ due to radial symmetry
and at $\xi \geq 1$ due to causality.
Furthermore,
depending on the bubble wall speed $\xi_w$ and the phase transition strength $\alpha$,
we distinguish three types of solutions: detonations, subsonic deflagrations,
and supersonic deflagrations (also known as hybrids) \cite{Espinosa:2010hh}.
In the following subsections,
we describe the boundary conditions across the
bubble wall leading to each of these types of solutions.
The radial profiles of the velocity, $v_{\rm ip} (\xi)$,
and the enthalpy, $w_\ip (\xi)$, obtained
in the bag equation of state for the different types of solutions
are shown in \Fig{fig_ip}
for a benchmark value of
$\alpha = 0.1$ and a range of wall velocities $\xi_w$ from 0.1 to 1.

\begin{figure}[t]
    \centering
    \includegraphics[width=.49\textwidth]{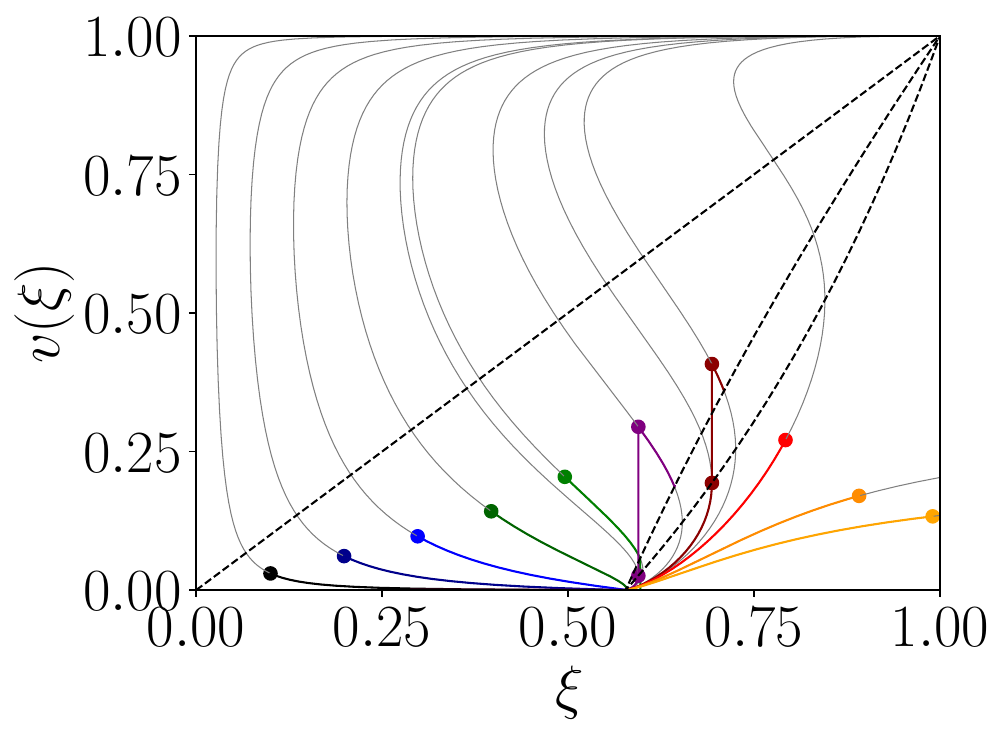}
    \includegraphics[width=.49\textwidth]{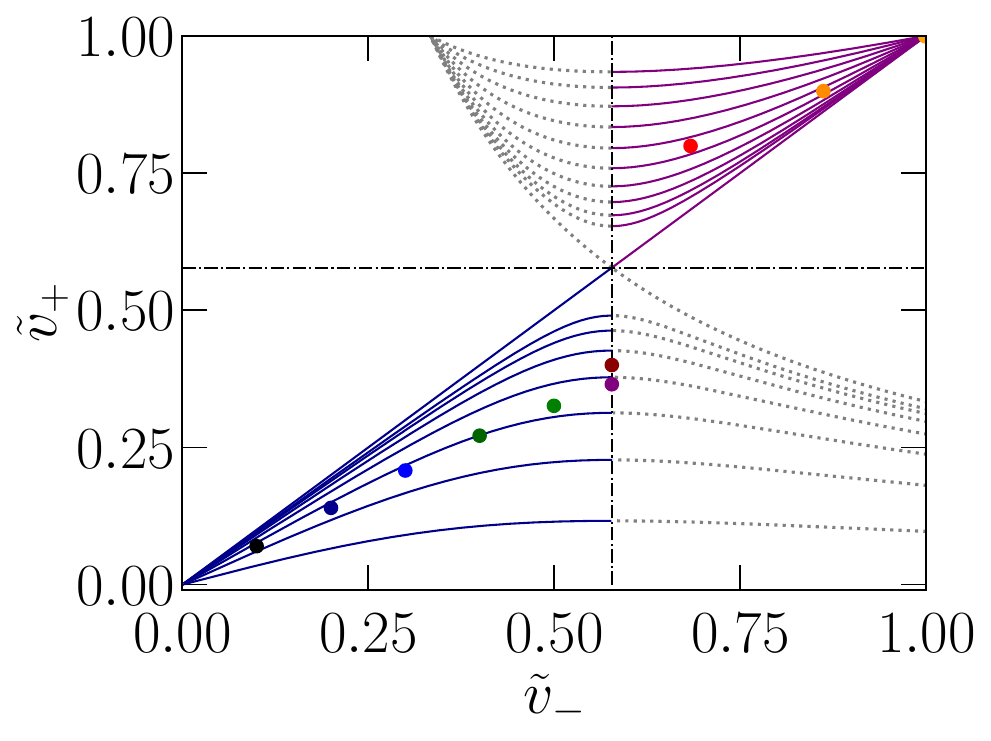}
    \caption{{\em Left panel:} Generic solutions $v(\xi)$ (gray lines) to \Eq{eq_hydro_ip_v} that
    start at the boundary conditions (colored dots) of each solution $v_\ip (\xi)$
    shown in \Fig{fig_ip} for $\alpha = 0.1$.
    Physical solutions occur below the diagonal dashed line $v = \xi$
    to ensure $\tilde v_+ \geq 0$.
    The other two black dashed lines indicate respectively
    the position of the shock formation
    $(\xi_\sh, v_\sh)$ [see \Eq{cond_shock}] and the
    singularity point $\mu (\xi, v) = \cs$, where $\dd v/\dd \xi \to - \infty$ [see \Eq{mu_xi}].
    {\em Right panel:} Possible branches of $\tilde v_+$ and $\tilde v_-$ for different values of $\alpha_+$
    (10 uniformly log-spaced values between 0.01 and 1) are shown by solid blue and purple lines,
    being $\vp = \vm$ and $\vp \vm = \tfrac{1}{3}$
    the limiting branches when $\alpha = 0$.
    For $\alpha_+ > \tfrac{1}{3}$, $\tilde v_+ < 0$ so no deflagration
    solutions exist (see blue lines).
    The lower left branch indicates deflagrations with $\tilde v_{\pm} \leq \cs$, while
    the upper right branch, $\tilde v_{\pm} > \cs$,
    indicates detonations.
    For hybrids, $\tilde v_- = \cs$ (see vertical dashed line).
    The values corresponding to the solutions in \Fig{fig_ip}
    are shown as colorful dots, where we note that $\alpha_+ (\xi_w) < \alpha$
    for deflagrations, so the dots are not aligned with 
    any constant $\alpha_+$ curves. Colors of dots correspond to the
    values of $\xi_w$ indicated in \Fig{fig_ip}.
    }
    \label{types_sol}
\end{figure}

The solutions shown in \Fig{fig_ip} are a subset of
the generic dynamical solutions $v(\xi)$ to \Eq{eq_hydro_ip_v},
represented by the gray curves in the left panel of \Fig{types_sol}. 
This subset of solutions is obtained by integrating the equations of motion starting
at the $(\xi, v)$ point determined
by the boundary conditions and ending either at a fixed point or when a fluid
shock is formed.
The resulting fluid velocity is then perturbed within a fluid shell
$\xi \in (\xib, \xif)$ and becomes zero outside the shell.
The right panel of \Fig{types_sol} shows the velocity at both sides of the wall
in the wall reference frame, $\tilde v_\pm$, for different
values of $\alpha_+$.
The relation $\tilde v_+$ vs $\tilde v_-$ can be computed
using \Eqq{vplus_from_vminus2}.

\subsubsection*{Detonations}
\label{deto}

A detonation occurs when the bubble wall propagates faster than any perturbation
in the fluid, such that the fluid is at rest
outside the bubble, $\tilde v_+ = \xi_w > \cs$.
The relative velocity $\tilde v_-$ can then be computed using \Eq{vminus_from_vplus},
and converted to a velocity in the global reference frame by
its Lorentz transform, $v_- = \mu(\xi_w, \tilde v_-)$, using \Eq{mu_xi}.
The solution $v_{\rm ip} (\xi)$ for $\cs \leq \xi \leq \xi_w$ is obtained by integrating \Eq{eq_hydro_ip_v}
from the bubble wall at $(\xi, v) = (\xi_w, v_-)$ inwards
until the stationary
point $(\cs, 0)$ is reached \cite{Espinosa:2010hh}.
Hence, at $\xi$ in the range $\cs \leq \xi \leq \xi_w$, we require that
the velocity increases with $\xi$, $\dd v_{\rm ip}/\dd \xi > 0$.
This condition implies that $\tilde v_- \geq \cs$, as it can be inferred from \Eq{eq_hydro_ip_v},
which can only be obtained by taking the positive argument in \Eq{vminus_from_vplus}.
In detonations, the front and back positions of the fluid shell
are then $\xif = \xi_w$ and $\xib = \cs$,
and only one discontinuity is present at $\xi_w$.
In addition, detonation
solutions exist only if the integrand in \Eq{vminus_from_vplus} is positive.
This condition provides the minimum value of $\xi_w$ that leads to a detonation,
\begin{equation}
    \xi_w = \tilde v_+ \geq \vCJ (\alpha_+) = \frac{1 +
    \sqrt{\alpha_+(2 + 3 \alpha_+)}}{\sqrt{3}(1 + \alpha_+)}
    \geq \frac{1}{\sqrt{3}}\,,
    \label{Chapman}
\end{equation}
known as the Chapman-Jouget speed $\vCJ$.
At smaller wall velocities $\xi_w$, a deflagration solution will develop.
When $\xi_w = \vCJ$, we find the limiting case $\tilde v_- = \cs$.
Finally, \Eq{sol_wip} is integrated in the same range of $\xi \in (\xib, \xif)$
to obtain the enthalpy profile $w_{\rm ip} (\xi)$.
We start the integration at $\xif = \xi_w$, where $v_\ip (\xif^-) = v_-$
and the value $w_\ip (\xif^-) = w_-$ is found from the matching condition in \Eq{matching_2},
\begin{equation}
    \frac{w_-}{w_n} = \frac{w_-}{w_+} =
    \frac{\tilde \gamma_+^2 \tilde v_+}
    {\tilde \gamma_-^2 \tilde v_-}\,.
\end{equation}
Since the fluid is unperturbed in the outer side of the wall,
we have $w_+ = w_n$ and $\alpha_+ = \alpha$.

\subsubsection*{Subsonic deflagrations}
\label{subs_defl}

Subsonic deflagrations take place when the fluid perturbations can propagate
faster than the wall, $\xi_w \leq \cs$.
In this case,
the fluid is at rest
inside the bubble so $\tilde v_- = \xi_w$ and $\tilde v_+$ is computed using \Eq{vplus_from_vminus}.
The value of the velocity just outside the  wall is given by its Lorentz transform,
$v_+ = \mu(\xi_w, \tilde v_+) = (\xi_w-\tilde v_+)/(1-\xi_w\tilde v_+)$,
and positivity of $v_+ > 0$ requires taking
the negative argument in \Eq{vplus_from_vminus}.
In addition, to ensure that the fluid velocity is always smaller
than the wall velocity, we require that $\tilde v_+ \geq 0$.
This can only be achieved when $\alpha_+ \leq {1\over 3}$, indicating
that subsonic deflagrations are not possible for larger values of $\alpha_+$.
This upper bound on $\alpha_+$ leads to an upper bound
$\alpha \leq \alpha_{\rm max} \simeq \onethird (1 - \xi_w)^{-13/10}$, found empirically in
ref.~\cite{Espinosa:2010hh}, where $\xi_w \leq \cs$.
Hence, subsonic deflagrations can only be found for
$\alpha \lesssim 1.02$.
\FFig{limit_alphas} shows the value of $\alpha$ found numerically
for $\alpha_+ = \tfrac{1}{3}$ for deflagrations, compared to the analytical fit
given in ref.~\cite{Espinosa:2010hh}.

The fluid velocity outside the bubble is then computed solving \Eq{eq_hydro_ip_v}, and it decreases
until a shock front is encountered and the velocity field drops to zero.
The appearance of the shock does not allow the solution to
cross the singularity in $\dd v_\ip/\dd \xi$
at $\mu (\xi, v) = \cs$ that would
lead to an unphysical double-valued function $\xi(v)$.
Since the region where the shock occurs is in the symmetric phase, the trace
anomaly across the shock is zero, i.e., $\alpha_+^\sh = 0$ and, hence,
the matching condition in \Eq{vplus_vminus} yields $\tilde v_+^\sh\, \tilde v_-^\sh = {1 \over 3}$,
where now $\pm$ denote the values at the left and right of the shock. 
At $\xi \geq \xi_\sh$, the fluid is at rest, with
$\tilde v_+^\sh = \xi_\sh$ and
\begin{equation}
    v_-^\sh = \mu(\xi_\sh, \tilde v_-^\sh) =
    \frac{\xi_\sh - \tilde v_-^\sh}{1 - \tilde v_-^\sh \, \xi_\sh} = \frac{3 \xi_\sh^2 - 1}{2 \xi_\sh}\,.
    \label{cond_shock}
\end{equation}
It follows that this condition is equivalent to $\xi_\sh \, \mu(\xi_\sh, v_-^\sh) = {1 \over 3}$.
In this case, the positions of the front and the back of the fluid shell are
respectively $\xif = \xi_\sh \geq \cs$ and $\xib = \xi_w$,
and a discontinuity is present at both locations.
We note that even though a shock always forms, as $\xi_w$ decreases,
the position of the shock becomes closer
to $\cs$ and, hence, the velocity jump across the shock becomes close
to zero for small $\xi_w$.

The enthalpy is obtained by integrating \Eq{sol_wip} once $v_\ip (\xi)$ is known.
We start by the value at the inner side of the shock,
which, using \Eq{matching_2}, is
\begin{equation}
    \frac{w_-^\sh}{w_n} = \frac{w_-^\sh}{w_+^\sh} = \frac{\tilde \gamma_-^{\rm sh\,^2}
    \tilde v_-^\sh}{{\tilde \gamma_+^{\rm sh\,^2}} \tilde v_+^\sh} =
    \frac{9 \xi_\sh^2 - 1}{3(1 - \xi_\sh^2)}\,,
    \label{cond_shock_w}
\end{equation}
and integrate inwards in the range $\xi_w \leq \xi \leq \xi_\sh$ until $w_+/w_n$.
As mentioned above, for small values of $\xi_w$,
the presence of a shock is negligible, and we can
integrate inwards from $\xi_\sh \simeq \cs$, where $w_-^\sh \simeq 
w_n$, until $w_+/w_n$.
Inside the bubble at $\xi < \xi_w$,
we use the matching condition in \Eq{matching_2}
to find $w_-$ from $w_+$.

For subsonic deflagrations, the fluid outside the bubble
is perturbed up to $\xi_\sh > \xi_w$.
Hence, the enthalpy on the outer side of the bubble wall is different from its unperturbed
value at the
nucleation temperature, $w_+ > w_n$, such that $\alpha_+<\alpha$ \cite{Espinosa:2010hh,Hindmarsh:2019phv,Giese:2020znk}.
Since we are in general
interested in computing the velocity and enthalpy profiles for
a given value of $\alpha$ at $T_n$, instead of for $\alpha_+$,
we need to use a
shooting algorithm based on the Newton-Raphson method to iteratively
solve the fluid profiles.
First, 
we start with the guess $\alpha_+^{(0)} =\alpha$.
Then, we compute
$w_+^{(0)}/w_n$ and $\alpha^{(0)}$ from the solution and correct the guess
$\alpha_+^{(1)} \to \alpha^{(0)} w_n/w_+^{(0)}$.
We proceed iteratively until the correct value of $\alpha_+$ that leads to
the desired $\alpha$ outside the bubble is found.

\subsubsection*{Supersonic deflagrations (hybrids)}
\label{hybr}

Supersonic deflagrations, also known as hybrids,
take place at intermediate wall speeds, $\cs \leq \xi_w \leq \vCJ (\alpha)$ [see \Eq{Chapman}].
At larger $\xi_w$, the solution becomes a detonation, such that
$\alpha_+ = \alpha$ and hence $\vCJ$ denoting the boundary
between detonations and hybrids can be evaluated from $\alpha$, as done in \Eq{chapman2}.
In this case, the fluid is perturbed at both sides of the bubble wall.
Outside the bubble wall there is a 
shock wave, as in a subsonic deflagration; however, in this case  $\xi_w > \tilde v_- = \cs$ because there is also
a rarefaction wave propagating inwards until the
stationary point $(\xi, v) = (\cs, 0)$ is reached inside the bubble.
One thus solves \Eq{eq_hydro_ip_v} inwards, starting at the wall where
$v_- = \mu(\xi_w, \cs)$, to find the fluid velocity inside the bubble.
Like in a subsonic deflagration, outside the bubble wall the velocity decreases down to the formation of a shock front.
To find the fluid perturbations outside the bubble,
we now integrate outwards with $v_+ = \mu(\xi_w, \tilde v_+)$, where $\tilde v_+$ is computed
from $\tilde v_- = \cs$ using  \Eq{vplus_from_vminus}, until the shock front is reached [see \Eq{cond_shock}].
In this case, the fluid shell is confined between $\xif = \xi_\sh > \xi_w > \cs$ and $\xib = \cs$,
and it presents a discontinuity inside the fluid shell, at $\xi_w$, and a second
discontinuity at $\xif$.
As for subsonic deflagrations, to avoid negative $v_+$, we take the negative argument in \Eq{vplus_from_vminus},
and to avoid negative $\tilde v_+$, we require $\alpha_+ \leq {1 \over 3}$,
again corresponding to $\alpha_{\rm max} \simeq \onethird (1 - \xi_w)^{-13/10}$ \cite{Espinosa:2010hh}.
Finally, we integrate \Eq{sol_wip} to obtain the enthalpy profile $w_{\rm ip} (\xi)$
and use $w_-^\sh$ and $v_-^\sh$ for the reference values, as done for subsonic
deflagrations, until $\xi_w$ is reached.
Then, $w_-$ is computed from $w_+$ using the matching condition in \Eq{matching_2}, and the enthalpy
profile inside the bubble can be found integrating inwards from $\xi_w$ to $\cs$.

The resulting $\alpha$ is again larger than $\alpha_+$ and,
as in subsonic deflagrations, we need to perform a shooting algorithm to find the value of $\alpha_+$ that yields
the desired value of $\alpha$.
We note that when $\alpha_+ < \alpha$, the condition
$\vCJ (\alpha) > \xi_w > \vCJ (\alpha_+)$ could be reached, naively
indicating a detonation solution.
This can be seen for example
in \Fig{limit_alphas}, where for $\alpha_+ = \tfrac{1}{3}$, one
would naively find $\vCJ (\tfrac{1}{3}) = \tfrac{\sqrt{3}}{2}$.
However, notice that the condition in \Eq{Chapman} for the solution to
be a detonation is on
$\tilde v_+$, not on $\xi_w$, and the two are different for hybrids, where
$\tilde v_+$ is found from the matching conditions
with $\tilde v_- = \cs$.
From $\tilde v_- = \cs$ one finds an upper bound on $\tilde v_+ \leq \tfrac{1}{\sqrt{3}}$,
which is then always smaller than $\vCJ \geq \tfrac{1}{\sqrt{3}}$,
allowing for supersonic deflagrations to always exist as long as
$\alpha \lesssim \onethird (1 - \xi_w)^{-13/10}$ (see \Fig{limit_alphas}).

\section{Constant and linear toy models for the velocity profiles}
\label{appendix_toy}

In \Sec{sec:toymodel}, the quadratic
and linear-constant toy models for the velocity profiles have been introduced.
The former model is used
to mimic subsonic deflagrations and detonations,
while the latter model is used for hybrids.
We have argued that such profiles represent the simplest choice
that accurately reproduces the function $\fpsq$ from the full velocity profiles.
In this section, we explore the simpler constant and linear toy models
and compare them to the ones presented in \Sec{sec:toymodel},
as well as to the numerical profiles.

\subsection{Toy models for subsonic deflagrations}

Subsonic deflagrations are, in general, characterized
by two discontinuities: one at $\xib = \xi_w$,
and another one at $\xif = \xi_\sh$.
The corresponding velocity profiles are continuous in the
range $\xi \in (\xib, \xif)$.
The simplest choice to mimic such profiles is the constant toy model,
\begin{align}
    v_{\rm const}(\xi) = v_{\rm const} \, \mathcal{X}_{(\xib, \xif)}(\xi) \,,
    \label{const_toy_model} 
\end{align}
which corresponds to the general quadratic profile in \Eq{general_quad_toy_model}
with $C_2 = C_1 = 0$ and $C_0 = v_{\rm const}$.
Using this toy model, we can either reproduce the $z \to 0$
or the $z \to \infty$ limit of the envelope of the $f'(z)$ function,
but not both limits.
In particular, to reproduce the latter,
the condition in \Eq{toy_z_to_inf_condition} reduces
to the one obtained from \Eq{fpenv_inf}, which is less restrictive since
it does not depend on $z$, 
\begin{align}
    v_{\rm const} (\xif + \xib) =
    \xi_\sh v_\sh + \xi_w |v_+ - v_-|\,.
    \label{toy_z_to_inf_env_condition}
\end{align}
From \Eq{toy_z_to_zero_condition} and \Eq{toy_z_to_inf_env_condition},
we then find the following choices for $v_{\rm const}$,
\begin{align}
    \label{choices_vconst}
    v_{\rm const} = \left\lbrace  \begin{array}{ll}
    v_0 = \bra{v}_{{\rm sph} \, 4 \rm d}\,, & \
    \quad \text{to reproduce $f' (z)$ in the $z \to 0$ limit}\,,  \\
    v_\infty = \displaystyle\frac{\xib \, v_\ip (\xib^+) + \xif \, v_\ip(\xif^{-})}{\xi_+} \,,  & \
    \quad \text{to reproduce $f'_{\rm env}$ in the $z \to \infty$ limit}\,,
    \end{array} \right.
\end{align}
where $\xi_+  = \xi_w + \xi_\sh$,
and we have defined the spherical-shell average in $N$ dimensions as
\begin{equation}
    \bra{v}_{{\rm sph} \, N \rm d} = \frac{N}{\xif^N - \xib^N} \int_{\xib}^{\xif} \xi^{N - 1}
    v_\ip (\xi) \dd \xi\,.
\end{equation}
We note that in the case of subsonic deflagrations with $\xi_w \lesssim \half \, \vCJ$,
the discontinuity at $\xif = \xi_\sh$
becomes negligible and $v_\ip (\xif^-) = v_\sh \simeq 0$.

If we are also interested in reproducing the oscillations of $f'(z)$ at large $z$
(not just its envelope), the simplest choice to do so is
a linearly decreasing toy model,
\begin{align}
    v_{\rm lin, dec}(\xi) =
    \frac{(\xif - \xi) \vb + (\xi - \xib) \vf}
    {\Delta \xi} \mathcal{X}_{(\xib, \xif)}(\xi) \,, \qquad
    \text{with \ } \vb > \vf\,,
    \label{vtoy_linear_decreasing}
\end{align}
which corresponds to choosing $C_2=0$,
$C_1 = (\vf - \vb)/\Delta \xi$, and
$C_0 = \vf - \xif \, C_1 = (\vb \xif - \vf \xib)/\Delta \xi$ in \Eq{general_quad_toy_model}.
This toy model is characterized by
two constants, $\vb$ and $\vf$.
Fixing $\vf = v_\ip (\xif^-) = v_\sh$, the constant $\vb$ can be chosen to exactly match the amplitude of
either of the two independent $z \to 0$ and
$z \to \infty$ limits of $f'(z)$ [see \Eqs{asymptotic_fpz}{eq:f1_zInfty}],
\begin{align}
    \label{choices_vlin}
    \vb = \left\lbrace  \begin{array}{ll}
    v_0 = {\cal A}_f
    \displaystyle\int_{\xib}^{\xif} \xi^3 v_{\rm ip}(\xi) \dd \xi
    - \vf {\cal A}_f/{\cal A}_b
    \,, & \ \quad \text{to
    reproduce $f'(z)$ in the $z \to 0$ limit}\,,  \\
    v_\infty =
    v_\ip (\xib^+) = v_+ \,,  & \ \quad  \text{to
    reproduce $f'(z)$ in the $z \to \infty$ limit}\,,
    \end{array} \right.
\end{align}
where
\begin{equation}
    {\cal A}_f = \frac{20 \, \Delta \xi}{\xif^5+4 \xib^5 - 5 \xif \xib^4}\,,
    \qquad
    {\cal A}_b = \frac{20 \, \Delta \xi}{\xib^5+4 \xif^5 - 5 \xib \xif^4}\,.
\end{equation}

\begin{figure}
    \centering
    \includegraphics[width=.46\textwidth]{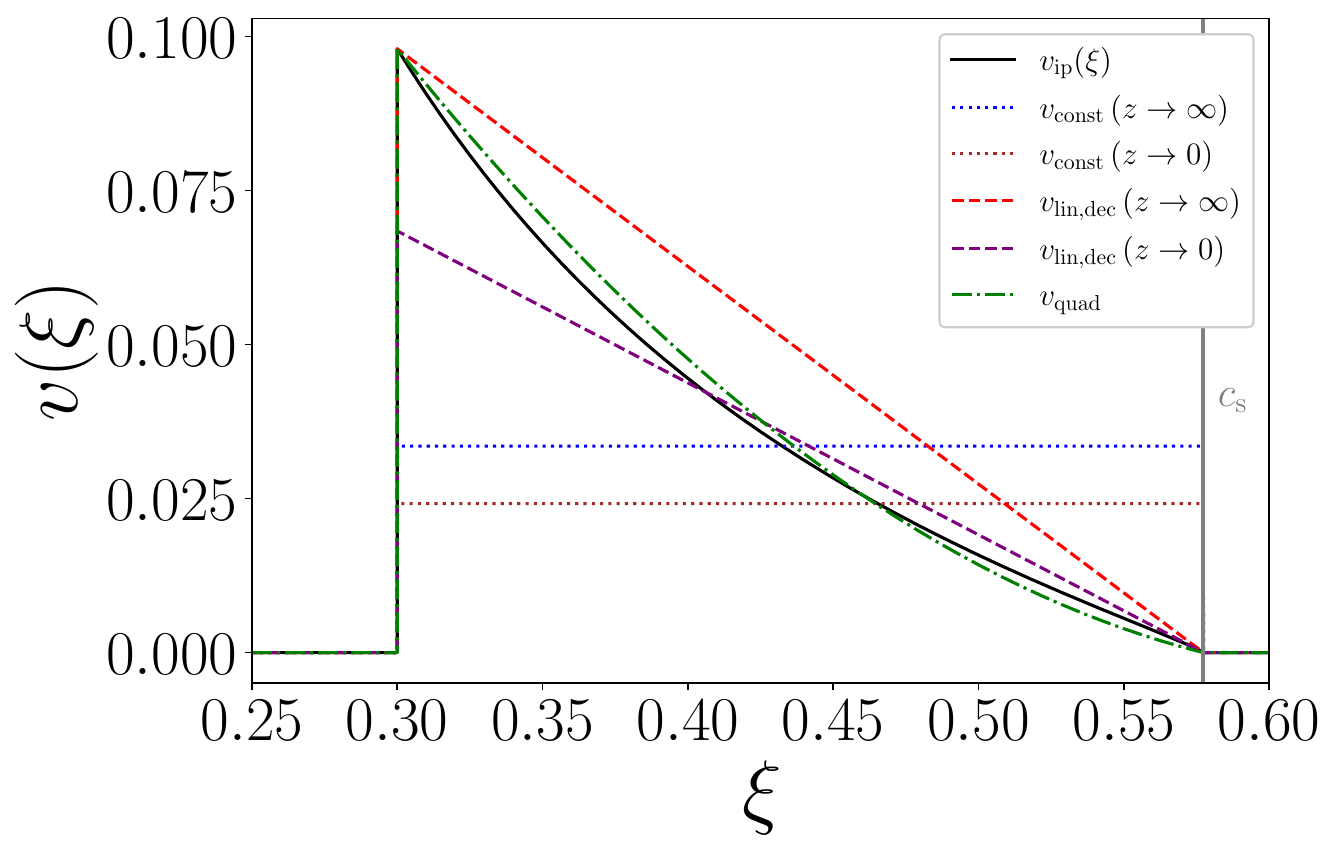}
    \includegraphics[width=.45\textwidth]{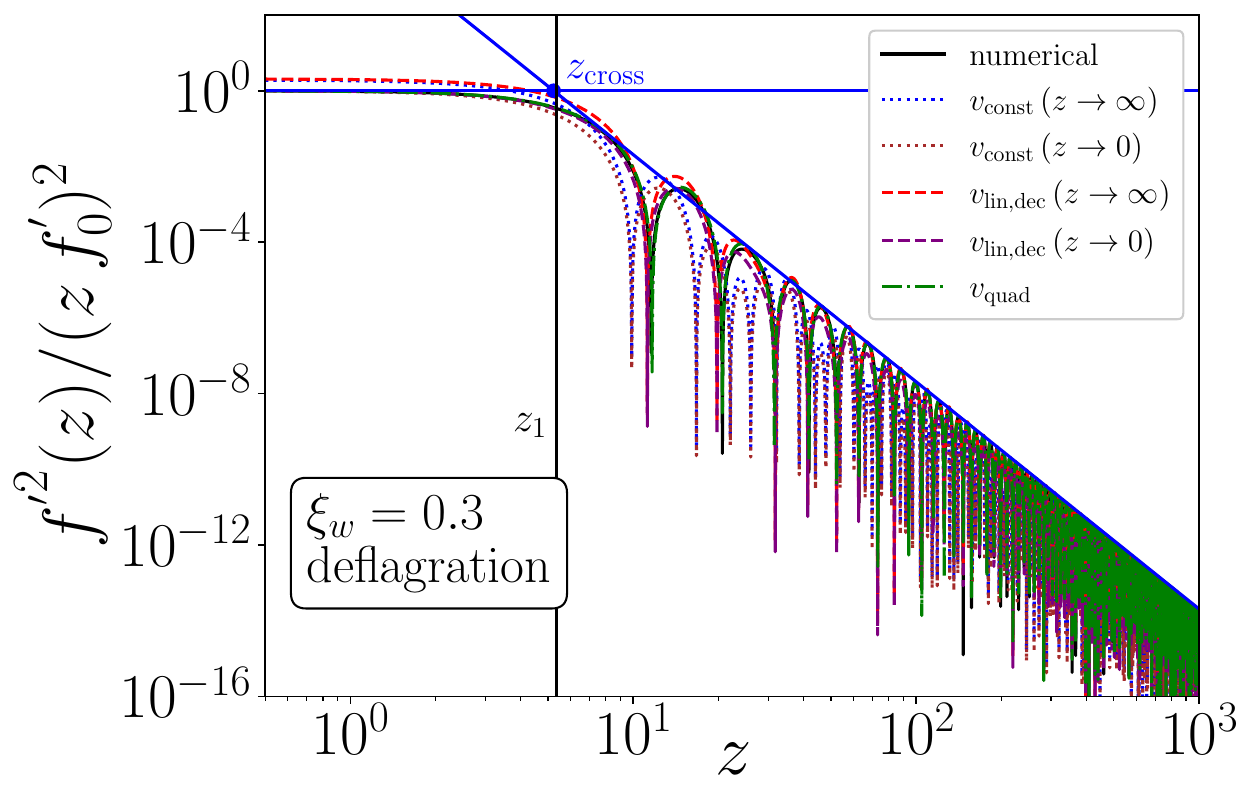}
    \includegraphics[width=.46\textwidth]{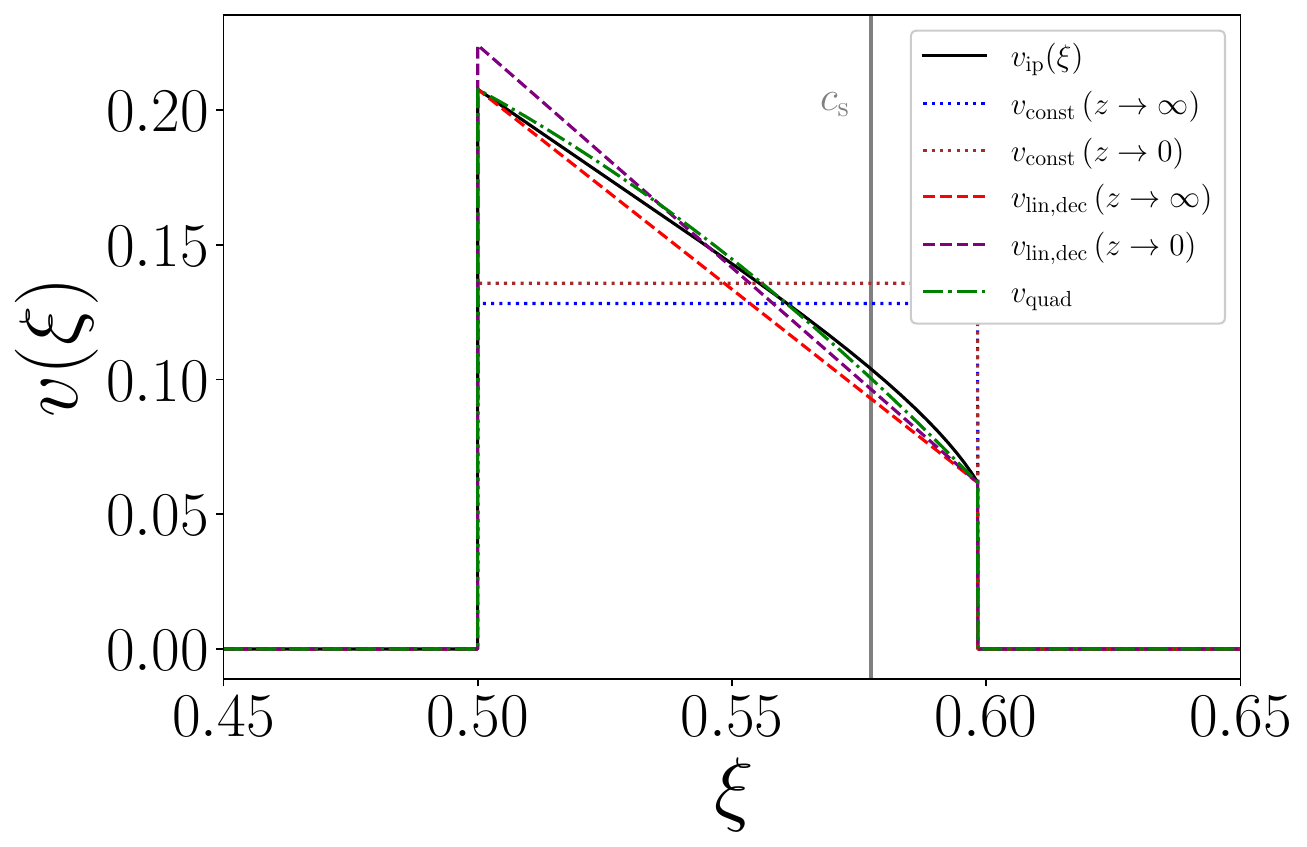}
    \includegraphics[width=.45\textwidth]{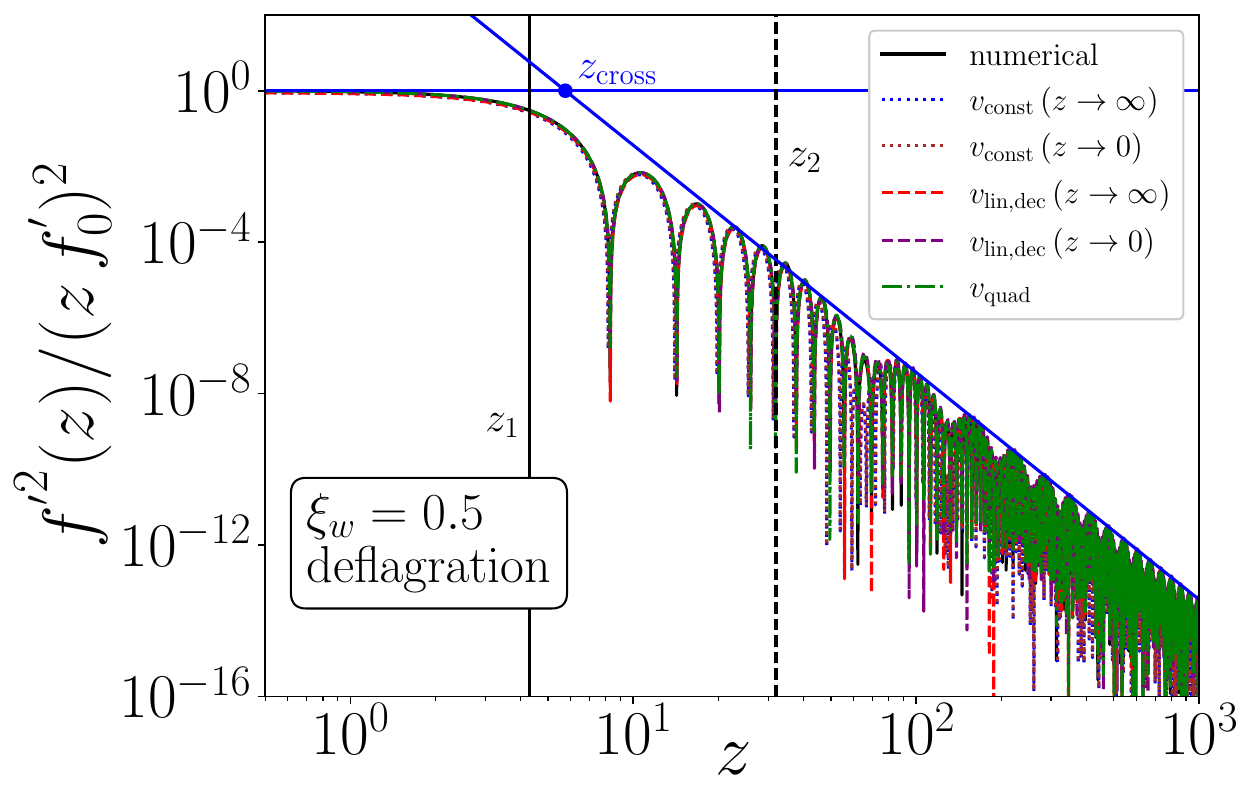}
    \caption{
    Comparison of the velocity profiles (left panels) and the $\fpsq$ functions (right panels),
    for $\alpha = 0.1$, and two values of $\xi_w = 0.3$ and 0.5,
    corresponding both to subsonic deflagrations.
    The deflagration with $\xi_w = 0.3$ does
    not present a visible shock at $\xif \to \cs$ (upper panels),
    while the deflagration with $\xi_w = 0.5$ presents
    a distinguishable shock at $\xif = \xi_\sh > \cs$
    (lower panels).
    The results for the full numerical profiles are shown
    in black lines and compared to
    the constant (dotted), linearly decreasing (dashed), and quadratic (dash-dotted) toy models.
    The constant and linear toy models can only reproduce either the $z \to 0$ or the $z \to \infty$
    limits of $\fpsq$ based on the choice of the model parameters
    [see \Eqs{choices_vconst}{choices_vlin}].
    On the other hand,
    the quadratic toy model
    can accurately reproduce both limits for the choice
    of parameters given in \Eq{choice_quad}.
    We show black vertical lines
    at $z_1$ and $z_2$, defined in \Eqs{fit_z1}{fit_z2} for
    deflagrations, which accurately predict
    the locations of the
    $\fpsq$ breaks for all cases (numerical profiles
    and toy models).}
    \label{full_vs_toy_1}
\end{figure}

In \Fig{full_vs_toy_1},
we compare the numerical profiles to the toy models for two choices of the wall velocity,
corresponding to subsonic deflagrations
where, for one case, the velocity jump at the shock is negligible
($\xi_w = 0.3$ with $\xi_\sh \to \cs)$
and, for the other case, the shock presents a clear
discontinuity $v_\sh$ at $\xi_\sh$ ($\xi_w = 0.5$).
We compare the toy models presented here, the quadratic model
presented in \Sec{quadratic}, and show how the quadratic toy model
is the only one capable of accurately reproducing both asymptotic limits
of $\fpsq$ function, including the oscillations at large
values of $z$.
The constant and linear toy models, even though simpler,
fail to reproduce both asymptotic limits at the same time,
since they do not have enough parameters to satisfy
both constraints in \Eqq{conditions_quad_toy}.
Furthermore, while the linear toy model can reproduce
the oscillations at large $z$ for the choice of
$\vb = v_\ip (\xib^+)$ and $\vf = v_\ip (\xif^-)$,
the constant toy model can only capture the
envelope of $\fpenvsq$ using $v_{\rm const} = v_\infty$ as
indicated in \Eq{choices_vconst}.
We note that for
all the toy models presented, we have chosen
$\xif$ and $\xib$ to be those of the numerical
profiles, and we find that $z_1$ and $z_2$ are
accurately predicting the positions of the breaks for
all cases (see \Fig{full_vs_toy_1}).
This result further reinforces the finding in \Sec{subsec_RiemannL}
that these scales only depend on $\xif$ and $\xib$.
Furthermore, since $z_{1,2}$ and the limit $z \to 0$
and $z\to\infty$ of $f'(z)$ are reproduced, the quadratic toy
models predict the same slope $\gamma$ at intermediate $z$
as the original profiles.

\subsection{Toy models for detonations}

Detonation solutions for the velocity profile are characterized by a single discontinuity at
$\xif = \xi_w$, and vanish at $\xib = \cs$. 
Hence, as before, the simplest way to model detonations
is to choose a constant model as in \Eq{const_toy_model}.
We can then reproduce either the $z \to 0$ or the $z \to \infty$
limits of the envelope of $\fpsq$ with the choices
already presented in \Eq{choices_vconst}, where in this case,
we set $v_\ip (\xib^+) \to 0$ and $v_\ip (\xif^-) = v_-$.
If we also want to reproduce the oscillations in the $z \to \infty$ limit, then the simplest way is to model
the velocity profile with
a linearly increasing toy model
that vanishes at $\xib$ and is smooth in the range $\xi \in (\xib, \xif)$,
\begin{align}
    v_{\rm lin, inc}(\xi) = \vf \frac{\xi - \xib}{\xif-\xib} \mathcal{X}_{(\xib, \xif)}(\xi) \,, 
    \label{vtoy_linear_increasing}
\end{align}
which corresponds to \Eq{vtoy_linear_decreasing} setting $\vb=0$.
Here, the constant $\vf = v_{\rm lin, inc}(\xif^{-})$ can be chosen to reproduce either of the $z \to 0$ or $z \to \infty$ limits of $f'(z)$.
Similar choices to those presented in \Eq{choices_vlin}
can be selected for $\vf$, such that they
reproduce one of the two limits,
\begin{align}
    \label{linear_choice}
    \vf = \left\lbrace  \begin{array}{ll}
    v_0 = \displaystyle {\cal A}_b
    \int_{\xib}^{\xif} \xi^3 v_{\rm ip}(\xi) \dd \xi
    \,, & \ \quad \text{to
    reproduce $f'(z)$ in the $z \to 0$ limit}\,,  \\
    v_\infty = v_\ip (\xif^-) = v_- \,,  & \ \quad \text{to
    reproduce $f'(z)$ in the $z \to \infty$ limit}\,.
    \end{array} \right.
\end{align}

\begin{figure}
    \centering
    \includegraphics[width=.46\textwidth]{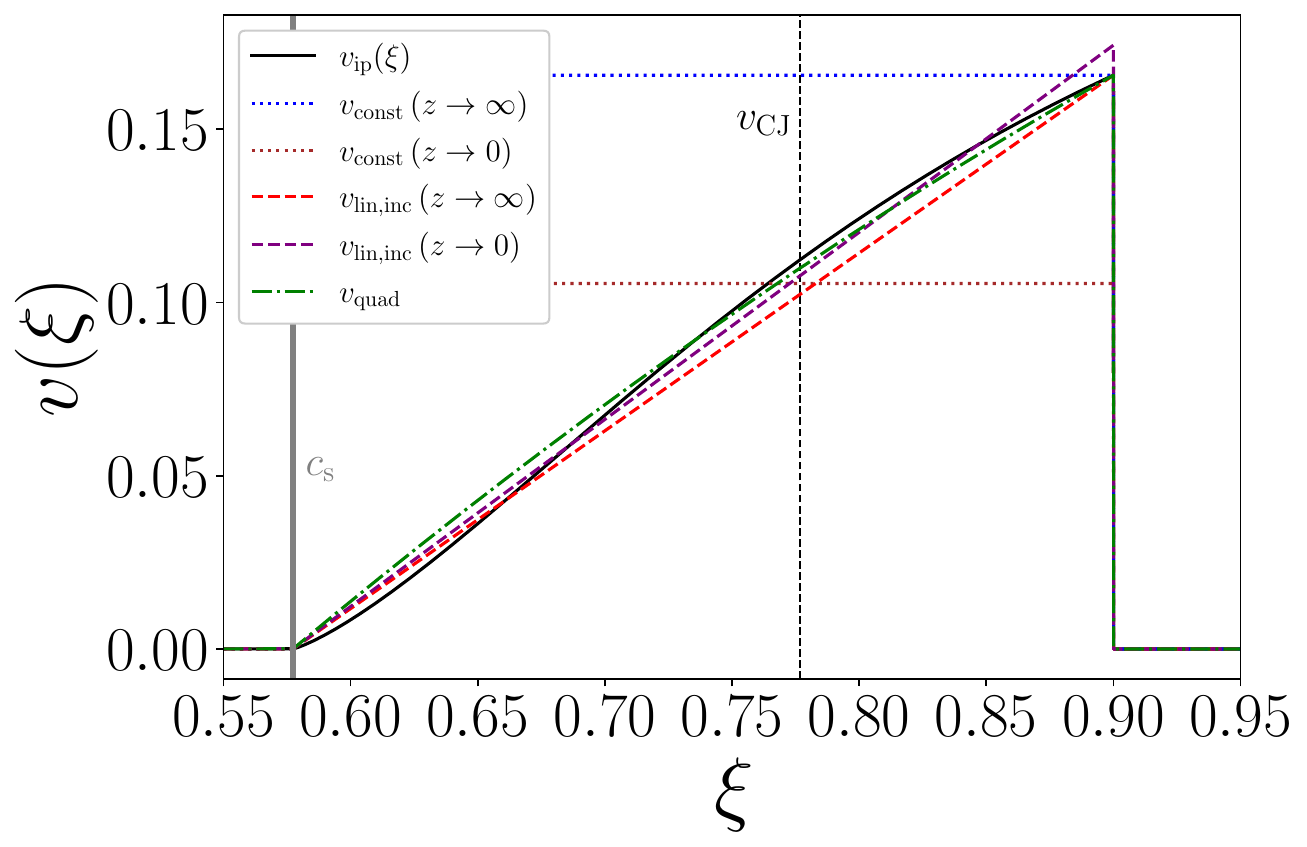}
    \includegraphics[width=.45\textwidth]{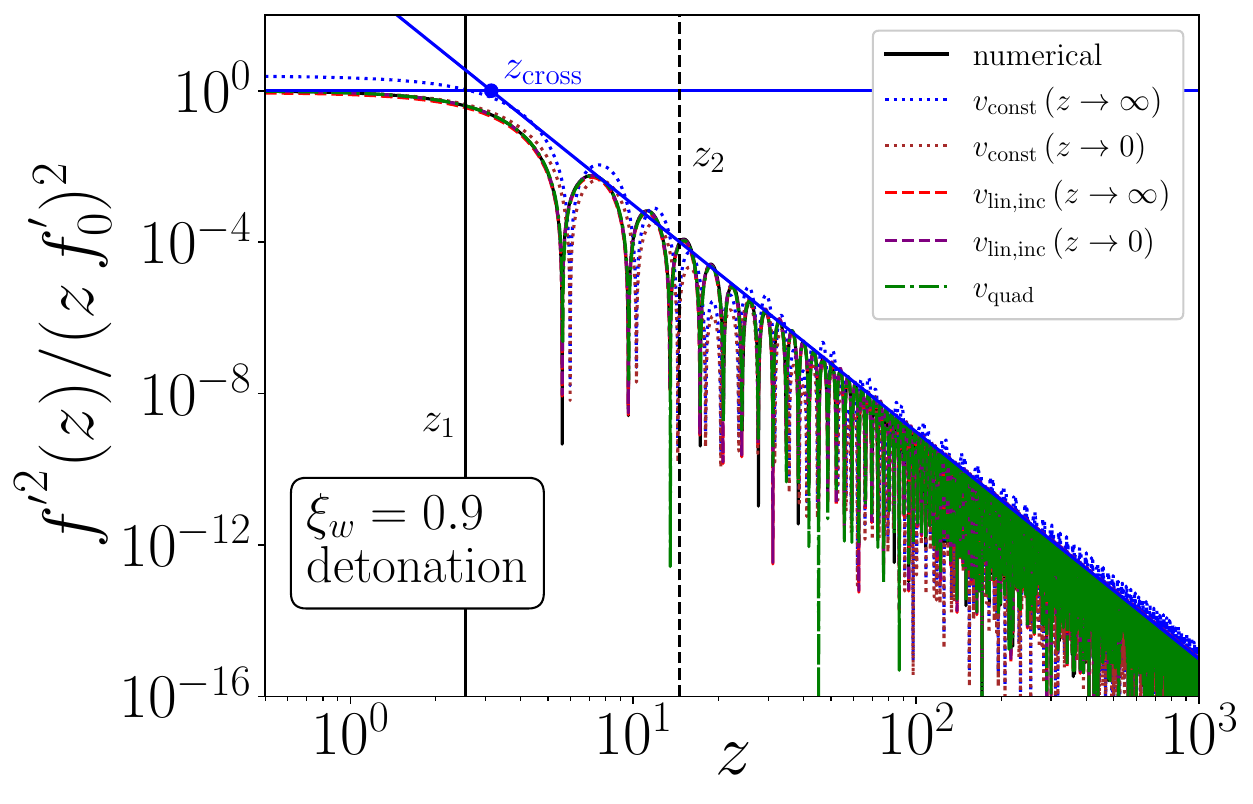}
    \caption{
    Comparison of the velocity profile (left)  and the
    function $\fpsq$ (right) from the numerical profile with those
    obtained using the toy models
    for $\alpha = 0.1$ and $\xi_w = 0.9$ 
    (corresponding to a detonation).
    We find that the constant and the linearly increasing toy models can reproduce either the
    $z \to 0$ or the $z \to \infty$ limits of $\fpsq$ for the choice of parameters indicated in the text,
    but not both of them, as expected.
    Although for this case, the difference between the toy models
    and the numerical profiles is not large,
    the quadratic toy model is more accurate to reproduce
    the numerical profile, being
    able to exactly reproduce both limits.}
    \label{full_vs_toy_4}
\end{figure}

In \Fig{full_vs_toy_4}, we compare the full numerical profiles to the toy models for a choice of the wall velocity corresponding to a detonation
($\xi_w =0.9$). Also in this case we see that, even if $z_1$
and $z_2$ are captured by all the presented toy models,
only the quadratic toy model can accurately reproduce
both $z \to 0$ and $z \to \infty$ limits at the same time,
and hence, also the intermediate slope of $\fpsq$.

\subsection{Toy models for hybrids}

Hybrid velocity profiles are characterized by two discontinuities: one at 
$\xi_w > \xib = \cs$,
and another one at $\xif = \xi_{\rm sh}$.
Therefore, they present a discontinuity in the range $\xi \in (\xib, \xif)$.
The simplest way to mimic such a profile with two discontinuities is
using the constant toy model of \Eq{const_toy_model}.
However, in this case, since we found in \Sec{subsec_RiemannL} that the scales
$z_1$ and $z_2$ are characterized by the positions of the discontinuities instead of by $\xib$ and $\xif$,
we set $\xib \to \xi_w$ instead of $\xib = \cs$ for the
constant toy model.
Since the constant toy model is characterized by a single parameter
$v_{\rm const}$, it can only reproduce either the $z \to 0$ or the
envelope of $f'(z)$ in the $z \to \infty$ limit.
Furthermore,
setting $v_{\rm const}$ to the value in \Eq{toy_z_to_inf_env_condition},
the $z \to \infty$ limit
cannot be satisfied for all $z$ using this toy model, 
which can only reproduce its envelope over oscillations.
If we want to alternatively reproduce the $z \to 0$ limit of $f'(z)$, 
we can instead choose $v_{\rm const} = v_0$, where $v_0$ is given
in \Eq{choices_vconst}.

\begin{figure}
    \centering
    \includegraphics[width=.4\textwidth]{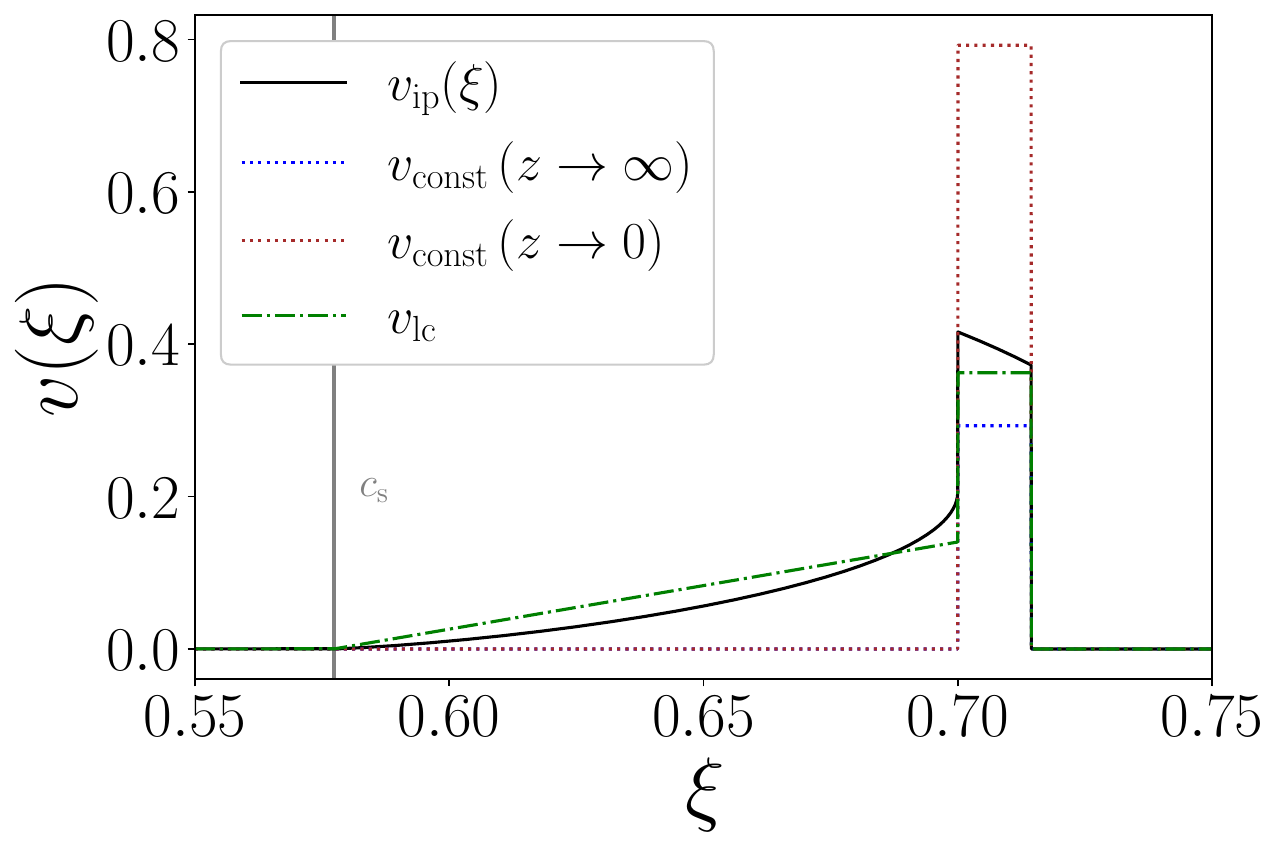}
    \includegraphics[width=.45\textwidth]{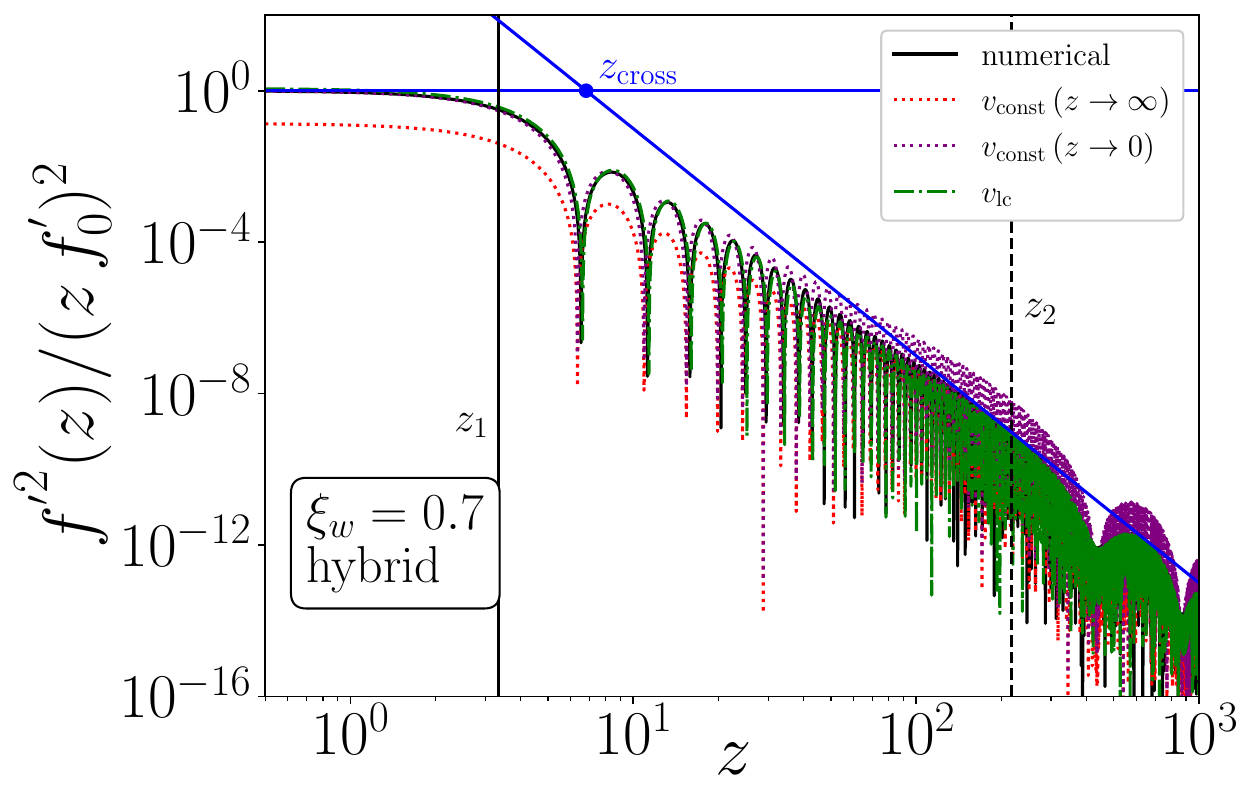}
    \caption{
    Comparison,
    for $\alpha = 0.1$ and $\xi_w = 0.7$ (corresponding to a hybrid),
    of the velocity profile (left) and the function $\fpsq$ (right)
    obtained from the full profiles and
    from the toy models.
    The constant toy models
    reproduce either the $z \to 0$ or $z \to \infty$ limit of $f'(z)$.
    The linear-constant (lc) toy model of \Eq{vtoy_lin_const}
    exactly reproduces the $z \to 0$ limit as well as the envelope of $f'(z)$
    in the $z \to \infty$ limit.}
    \label{full_vs_toy_3}
\end{figure}

In \Fig{full_vs_toy_3}, we compare the full numerical profiles to the constant
and linear-constant toy models for a choice of the wall velocity corresponding to a hybrid
solution ($\xi_w=0.7$).
We clearly see that a constant profile, contrary to the linear-constant toy model,
cannot accurately reproduce both limits $z \to 0$ and the envelope
in the $z \to \infty$ limit of $\fpsq$ at the same time.

\section{Correlations of statistically homogeneous and isotropic fields}
\label{appendix_correlation}

The velocity field presented in \Sec{kinetic_sp_bubbles},
which is produced by multiple expanding bubbles with
nucleation locations distributed homogeneously in space,
is statistically homogeneous and isotropic.
In the present section, we review the statistical description of
this type of fields.
The generic results are applied to irrotational fields
in \Sec{potential}.

\subsection{Two-point correlation function}

The two-point correlation function of a statistically homogeneous and
isotropic
field $\vv$ only depends on the separation vector $\rr$
between two points $\xx$ and $\yy = \xx + \rr$. 
Due to homogeneity and isotropy,
its tensor structure is given entirely by a linear combination
of the tensors $\delta_{ij}$ and $\hat r_i \hat r_j$, invariant under spatial
rotations and translations, yielding the unique\footnote{We note that helical
fields violate invariance under mirror reflections,
which leads to an additional independent component
$B_{\rm H} (r)$ in \Eq{Bij_corr}, proportional to $\hat r_l \varepsilon_{lij}$,
to generically describe $B_{ij} (\rr)$
(see, e.g., ref.~\cite{Brandenburg:2018ptt}).
In the present work, we only consider non-helical fields with $B_{\rm H} = 0$,
which implies that $\bra{v_i (\xx) \, \omega_i (\xx)} = 0$,
with $\oom = \nab \times \vv$.
}
decomposition \cite{MY75}
\begin{equation}
    B_{ij} (\rr) 
    = \bra{v_i (\xx) \, v_j(\xx + \rr)} 
    = (\delta_{ij} - \hat r_i \hat r_j) \, B_{\rm N} (r) +
    \hat r_i \hat r_j \,  B_{\rm L} (r) \,, \label{Bij_corr}
\end{equation}
where the brackets indicate an average over ensemble realizations.
The results of this section are independent of the time
at which the fields in \Eq{Bij_corr} are evaluated. Hence,
we omit any time dependence in the following.
$B_{\rm N} (r)$ and $B_{\rm L} (r)$ are the normal (N) and longitudinal (L)
components of the correlation function,
computed respectively from the
perpendicular and the parallel components of the
field $\vv$ with respect to $\rr$,
$v_{\rm N} (r)$ and $v_{\rm L} (r)$.
The normal and longitudinal correlation functions only depend on the
distance between the two points, $r = |\rr|$.

Because of the invariance under translations in coordinate space,
momenta are conserved in Fourier space,
\begin{align}
     \bra{v_i (\kk) \, v_j^*(\kk')} &=
    \int
      \dd^3 \yy
      \int   \,  e^{i (\kk \cdot \yy - \kk' \cdot \xx)}
      B_{ij} (\yy-\xx) \dd^3 \xx 
        \nonumber \\ 
      &=   (2 \pi)^3 \, \delta^3  (\kk - \kk') \int e^{i \kk \cdot \rr} B_{ij} (\rr) \dd^3 \rr 
      =   (2 \pi)^3 \, \delta^3 (\kk - \kk') \, {\cal F}_{ij} (\kk) \,,
      \label{two_point_fourier}
\end{align}
where the spectral density ${\cal F}_{ij} (\kk)$ has been defined as
the Fourier transform of $B_{ij} (\rr)$,
and it can be decomposed analogously in terms of
its longitudinal and normal components \cite{MY75},
\begin{align}
    {\cal F}_{ij} (\kk) \equiv \int e^{i\kk \cdot \rr} B_{ij} (\rr) \dd^3 \rr
    = P_{ij} (\hat \kk) \, F_{\rm N} (k)
    + \hat k_i \hat k_j F_{\rm L} (k) \,,
    \label{decomp_Fij}
\end{align}
where $P_{ij} (\hat \kk) = \delta_{ij} - \hat k_i \hat k_j$ is the projection
tensor perpendicular to $\hat{k}$.
The relation of the components of ${\cal F}_{ij}$
to the longitudinal and normal components of $B_{ij}$ is
\cite{MY75}
\begin{subequations}
\label{FNL_BNL}
\begin{align}
    F_{\rm L} (k) = &\,  4 \pi \int_0^\infty
    \Bigl(j_0 (kr) B_{\rm L} (r) + \frac{2 j_1 (kr)}{kr}
    \bigl[B_{\rm N} (r) - B_{\rm L} (r) \bigr] \Bigr)\, r^2 \dd r\,,
    \label{FL_BL_BN}
    \\
    F_{\rm N} (k) = &\,  4 \pi \int_0^\infty
    \Bigl(j_0 (kr) B_{\rm N} (r) + \frac{j_1 (kr)}{kr}
    \bigl[B_{\rm L} (r) - B_{\rm N} (r) \bigr] \Bigr) \, r^2 \dd r  \,,
    \label{FN_BL_BN}
\end{align}
\end{subequations}
where $j_n(x)$ is the $n$-th order spherical Bessel function
of the first kind.
In particular, $j_0 (x) = \sin x/x$ and $j_1 (x) = \sin x/x^2 - \cos x/x$.
We note that the value of $B_{ii}$ at $r = 0$ corresponds to the 
average value of the squared field $v_\rms^2 = \bra{v^2 (\xx)}$,
\begin{equation}
    v_\rms^2 = B_{ii} (0) = 2 B_{\rm N} (0) + B_{\rm L} (0) =
    \frac{1}{2\pi^2} \int_0^\infty \mathcal{F}_{ii} (k) \,
    k^2 \dd k =  \frac{1}{2\pi^2} \int_0^\infty 
    \bigl[2 F_{\rm N} (k) + F_{\rm L} (k) \bigr] k^2 \dd k\,.
    \label{traces_relation}
\end{equation}
The spectrum of the velocity field $\vv$, $E_v (k)$, is defined from the trace of
the two-point correlation function in momentum space [see \Eq{two_point_fourier}]
\begin{align}
    \bra{v_i (\kk) \, v_i^* (\kk')} =  
    (2 \pi)^6 \, \delta^3 (\kk - \kk') \, \frac{E_v (k)}{4 \pi k^2}
    \quad \Rightarrow \quad
    E_v (k) =
    \frac{k^2}{2 \pi^2} \bigl[2 \FN(k) + \FL (k) \bigr] \,.
    \label{EK_FK}
\end{align}
The spectrum is defined such that its integral over momenta
yields the average value of the squared field in physical space,
\begin{align}
    v_\rms^2 = \bra{v^2(\xx)} = &\, 
    \int \frac{\dd^3 k}{(2\pi)^3}\int \frac{\dd^3 k'}{(2\pi)^3} e^{-i\xx \cdot(\kk-\kk')}
    \bra{v_i (\kk) \, v_i^* (\kk')}  \nonumber \\
    = &\, \int_0^\infty E_v (k) \dd k = \int_{-\infty}^\infty \PP_v (k) \dd \ln k \,,
    \label{Parseval}
\end{align}
where we have also defined the spectrum $\PP_v (k) = k\,E_v(k)$, normalized
per unit of $\ln k$.
In the case of an irrotational velocity field with $F_{\rm N} = 0$, as the
one we find in the bubble expansion phase, the relation
$E_v(k) = k^2\FL(k)/(2\pi)^2$ is often given as a definition of the spectrum,
see, e.g., refs.~\cite{Caprini:2009fx,Hindmarsh:2019phv}.
Similarly, solenoidal fields with $\nab \cdot \vv = 0$ (e.g., purely vortical fluids or magnetic fields \cite{Caprini:2009fx,RoperPol:2022iel,Auclair:2022jod})
have $\FL = 0$ and it is common to identify the
spectrum with the normal component, $\FN$,
which is also required to be proportional to $k^2$ or steeper (see discussion in
\Sec{app_largescales} and refs.~\cite{MY75,Caprini:2003vc}).
However, we note that the relation to the velocity spectrum has then
a factor of $1$ instead of $2$ in the denominator, $E_v(k) = k^2 \FN (k) /\pi^2$,
due to the different values of the trace of the tensors $P_{ij}$ and $\hat k_i \hat k_j$.

\subsection{Large-scale limit and causality}
\label{app_largescales}

The asymptotic limit at large scales $k\rightarrow 0$ of the spectral density functions
$\FL (k)$ and $\FN (k)$ is obtained
by Taylor-expanding \Eqq{FNL_BNL} \cite{MY75},
\begin{subequations}
\label{Fs_taylor}
\begin{align}
    F_{\rm A} (k) = &\, f_{\rm A}^{(0)} + k^2 f_{\rm A}^{(2)}  + {\cal O} (k^4)\,,
    \qquad 
    {\rm where \ A} = {\rm L \ and \ N} \,,\label{F_Taylor}
 \end{align}
with
\begin{gather}
    f_{\rm L}^{(0)} = \fN^{(0)} = \frac{4\pi}{3} \int_0^\infty
    \bigl[B_{\rm L} (r) + 2 B_{\rm N} (r) \bigr] \, r^2 \dd r \,,\\
    f_{\rm L}^{(2)} = - \frac{2 \pi}{15} \int_0^\infty
    \bigl[3 \, B_{\rm L} (r) + 2 \, B_{\rm N} (r) \bigr] \, r^4 \dd r  \,,
    \quad 
    f_{\rm N}^{(2)} = - \frac{2 \pi}{15} \int_0^\infty
    \bigl[4 \, \BN (r) + \, \BL (r) \bigr] \, r^4 \dd r \,. \label{fN_0}
\end{gather}
\end{subequations}

Since $\fL^{(0)} = \fN^{(0)}$,
the term $\FL - \FN$ that multiplies $\hat k_{i} \hat k_j$
in \Eq{decomp_Fij} vanishes as $k^2$ or faster in the  $k \to 0$ limit.
This condition requires that the trace
$B_{ii} (r) = \BL (r) + 2 \BN (r)$ decays faster than $r^{-3}$ when $r \to \infty$,
such that the integral for $\fL^{(0)} = \fN^{(0)}$ converges.
More generally, all the coefficients of the Taylor expansions of \Eqq{Fs_taylor}
are bounded if the correlators in coordinate space 
decay faster than
any power law as $r \to \infty$, as expected for any causal field.
This corresponds to the analyticity causal condition described, for example,
in refs.~\cite{MY75,Durrer:2003ja,Durrer:2013pga}.
In particular, as we show in \Sec{potential}, when the field is irrotational or solenoidal, $F_{\rm N}$ or
$F_{\rm L}$ respectively becomes zero, and the resulting non-zero component
then decays as $k^2$ or faster in the $k \to 0$ limit.

\subsection{Irrotational fields}
\label{potential}

An irrotational (or potential) field $\vv$ has zero vorticity
${\pmb \om} = \nab \times \vv = 0$.
Under this condition, the field trivially satisfies $\bra{R_{ij} (\xx) \, v_l (\yy)} = 0$,
where $R_{ij} = \partial_i v_j - \partial_j v_i = \varepsilon_{ijk} \, \omega_k$
is the rotation tensor \cite{2008Kundu}.
From this expression, one can derive an equation for the corresponding two-point
correlation function $B_{ij}(\rr)$ \cite{MY75},
\begin{equation}
    \partial_l B_{ij} (\rr) - \partial_j B_{il} (\rr) = 0 \,.
    \label{generic_potential}
\end{equation}
Using the decomposition of $B_{ij}$ in \Eq{decomp_Fij}
yields a differential equation that relates the longitudinal and normal components of the correlator \cite{MY75},
\begin{equation}
    \BN(r) + r \partial_r \BN(r) = \BL (r)\,,
    \label{diff_eq_pot}
\end{equation}
which must be satisfied when the field is irrotational,
such that
$\BL (r)$ and $\BN (r)$ are both non-zero and related
to each other.

Using the relation in \Eq{diff_eq_pot}, the lateral
component of the spectral density [see \Eq{FN_BL_BN}] becomes zero,
\begin{align}
    \FN (k) = 4 \pi \int_0^\infty
    \biggl[j_0(kr) \BN (r) + \frac{j_1 (kr)}{k} \partial_r \BN(r) \biggr] \,
    r^2 \dd r = 0 \,.
    \label{FN_zero}
\end{align}
This result is found after integrating by
parts \Eq{FN_zero} and using the following property of the spherical Bessel functions,
\begin{align}
    \frac{\dd}{\dd r} \! \left[ \frac{r^{n+1} }{k} j_n(kr) \right] = r^{n+1} j_{n-1}(kr) \,,
    \label{besselth}
\end{align} 
applied to $j_1$.
Therefore, the longitudinal component $\FL(k)$ is the
only non-zero contribution to the spectral density,
and can be obtained from the trace of \Eq{decomp_Fij},
\begin{align}
    \FL (k) =  {\cal F}_{ii} (k) = 4 \pi
    \int_0^\infty j_0 (kr) \bigl[\BL (r) + 2 \BN (r) \bigr] \, r^2 \dd r  \,.
    \label{FL_from_BNBL}
\end{align}
In the same way, the components $\BL(r)$ and $\BN(r)$
can be computed from $\FL (k)$ using \Eq{decomp_Fij},
\begin{equation}
    \BL (r) = \frac{1}{2\pi^2} \int_0^\infty \biggl[j_0 (kr) -
    \frac{2 j_1 (kr)}{kr} \biggr] \, k^2 \FL (k) \dd k \,, \quad
    \BN (r) = \frac{1}{2\pi^2} \int_0^\infty  
    \frac{j_1 (kr)}{kr} \, k^2 \FL (k) \dd k\,,
    \label{BN_BL_irrot}
\end{equation}
such that adding up $\BL (r) + 2 \BN (r)$, we recover the
inverse Fourier transform of ${\cal F}_{ii} (k) = F_{\rm L} (k)$.
In particular, their values at $r = 0$ are
\begin{equation}
    B_{\rm L} (0) = B_{\rm N} (0) = \onethird B_{ii} (0) = \frac{1}{6 \pi^2}
    \int_0^\infty k^2  F_{\rm L} (k) \dd k\,.
    \label{bii0_irrot}
\end{equation}
Then, the spectrum $E_v (k)$, given in \Eq{EK_FK}, is
\begin{equation}
    E_v (k) = \frac{k^2}{2 \pi^2} \FL (k) = \frac{2}{\pi}
    \int_0^\infty \sin (kr) \bigl[2 \BN (r) + \BL (r) \bigr]
    \, kr \dd r \,,
\label{EK_FL_irrot}
\end{equation}
and the two-point correlation function in Fourier space can be expressed as
\begin{align}
    \bra{v_i (\kk) \, v^*_j (\kk')}  =  (2\pi)^3 \,
    \delta^3 (\kk - \kk')\, {\cal F}_{ij} (\kk)
    =  (2 \pi)^6 \, \delta^3 (\kk - \kk') \, \hat k_i \hat k_j \,
    \frac{E_v (k)}{4 \pi k^2} \,.
    \label{two_point_fourier_irrot}
\end{align}

We have shown that for any causal field,
$\FL (k) - \FN (k)$ is proportional to $k^2$
(or steeper) in the $k \to 0$ limit [see text below \Eq{fN_0}].
Then, as $F_{\rm N} = 0$, it immediately follows that the
spectrum of irrotational fields is proportional to $k^4$ (or steeper)
in the large-scale limit \cite{MY75},
\begin{equation}
    E_v (k) = \frac{k^2 \FL (k)}{2 \pi^2} \sim k^{2p}, \quad 
    {\rm with \ } p \geq 2 \,,
    \label{cond_irrot}
\end{equation}
which implies $\PP_v (k) = k \, E_v (k) \sim k^5$ or steeper.
This condition is equivalent to the analyticity condition claimed in refs.~\cite{Durrer:2003ja,Hindmarsh:2019phv}.

A similar result is found for solenoidal fields with $\nab \cdot \vv = 0$
(e.g., magnetic or incompressible velocity fields).
In this case, it can be shown that $F_{\rm L} (k) = 0$ and,
hence, $E_v (k) = k^2 F_{\rm N} (k)/\pi^2 \sim k^{2p}$ with $p \geq 2$
is also required when $k \to 0$ \cite{MY75,Durrer:2003ja}.

\section{Statistical distribution of bubbles in time}
\label{time_dist}

In this section, we review the statistical distribution of
bubble nucleation times and lifetimes.
The probability of nucleating a broken-phase bubble at a time $t$, per unit volume in the symmetric phase, is
computed from the action \cite{Linde:1981zj}
\begin{equation}
    p(t) \sim e^{-S(t)} \,.
\end{equation}
The Euclidean action, $S(t)$, decreases from infinity to a finite value at the critical time $t_c$,
when the two vacua in a first-order
phase transition become degenerate \cite{Enqvist:1991xw}.
The nucleation rate is then $\Gamma(t) = p(t) \, h(t)$, where $h(t)$
is the volume fraction in the symmetric phase \cite{Enqvist:1991xw},
\begin{equation}
    h(t) = \exp \biggl[ -
    \int_{t_c}^t p(t_0)
    \, {\cal V}_w \, (t - t_0)^3  \dd t_0  \,  \Theta(t - t_c) \biggr] \,,
    \label{h_t}
\end{equation}
where we have defined the bubble
self-similar volume ${\cal V}_w \equiv V_w/(t - t_0)^3 = \tfrac{4}{3} \pi \xi_w^3$.

The number density of bubbles in a volume $V$ as a function of time is
\begin{equation}
    n_b(t \geq t_c) =  \int_{t_c}^t \Gamma(t_0)
    \dd t_0 = \int_{t_c}^t p(t_0) \, h(t_0)\dd t_0 \,.
    \label{nb_t}
\end{equation}
Then, at any time,
the mean bubble separation can be computed as $R_\ast (t) = n_b^{-1/3} (t)$.
The asymptotic value of the number density of bubbles $n_b$ is
\begin{equation}
    n_b \equiv \frac{N_b}{V} =
    \lim_{t\to\infty}n_b(t) = \int_{t_c}^\infty p(t_0)\, h(t_0) \dd t_0 \,.
    \label{number_bubbles}
\end{equation}
The asymptotic bubble mean separation, $R_\ast = n_b^{-1/3}$, is the
length scale that is commonly used to
characterize the fluid perturbations and the resulting GW signal
\cite{Caprini:2015zlo,Caprini:2019egz,Hindmarsh:2019phv,Caprini:2024gyk,Caprini:2024hue}. 
On the other hand, refs.~\cite{Athron:2023xlk,Caprini:2024ofd}
suggest the characteristic length scale of the resultant
GW spectrum corresponds to
$R_\ast (t_p)$ evaluated at the percolation time,
$t_p$, which is defined as the time when a connected group of bubbles spans
the entire Universe.
In our work, we investigate the resulting length scales $2\pi \beta/k_1$
and $2\pi \beta/k_2$
that characterize the fluid perturbations (see \Sec{template_FL}),
where $\beta$ is the rate that characterizes the inverse
time scale of the phase transition, defined in \Secs{exp_nucl}{sim_nucl}, respectively,
for exponential and simultaneous nucleations.
We extend the analysis
to the GW spectrum from sound waves in ref.~\cite{part2}.

We define the nucleation time $t_n$ as that at which the average number of bubbles 
in a Hubble volume is one \cite{Enqvist:1991xw},
\begin{equation}
    N_b (t_n) = \frac{4 \pi}{3} H_\ast^{-3} \, n_b (t_n) = 1\,.
    \label{nucleation_time}
\end{equation}
Note that $t_n$ can be significantly different from,
for example, the percolation time $t_p$ \cite{Athron:2023xlk,Caprini:2024ofd},
which provides a better characteristic time for GW production,
which starts only after the broken-phase bubbles start to collide.

\subsection{Exponential nucleation}
\label{exp_nucl}

When the potential barrier is generated by thermal
fluctuations, the nucleation probability, $p(t)$, is often assumed to increase
exponentially in time \cite{Hogan:1986qda,Ignatius:1993qn}.
Indeed, this is found when the action decreases and it is Taylor expanded around
a characteristic time $t_\ast$,
\begin{equation}
    S(t > t_c) \simeq S (t_\ast) - \beta(t - t_\ast) \Rightarrow p(t > t_c) \simeq p_\ast \,
    e^{\beta (t - t_\ast)} = p_\ast\, e^{\tilde t - \tilde t_\ast}\,,
    \label{pt_exp}
\end{equation}
where $\tilde t = \beta t$, $\tilde t_\ast = \beta t_\ast$,
and $p_* \equiv p(t_*) = e^{-S(t_*)}$.
Then, the parameter $\beta = - S' (t_\ast) \simeq \left[ \dd \ln p(t)/\dd t \right]_{t=t_*}$ is the rate
that characterizes the inverse
time scale of the phase transition.
A physical choice of $t_\ast$ is,
for example, the percolation time, $t_p$.
Percolation occurs when the broken-phase bubbles merge to form a connected cluster that spans the entire Universe.
For simultaneous nucleation of uniformly-distributed
bubbles (see \Sec{sim_nucl}),
percolation has been found to occur when $h(t_p) \simeq 0.71$
\cite{Shante01051971,10.1063/1.1338506,LIN2018299,CONSIGLIO200349,LI2020112815}.
The exact value of $t_p$ is not generally known for exponential
nucleation and it can, in principle, differ from the one found for
simultaneous nucleation.
An alternative is to choose $t_\ast = t_f$ with
$h(t_f) = 1/e$ \cite{Enqvist:1991xw,Hindmarsh:2019phv,Hindmarsh:2020hop},
which corresponds to the moment
when the rate of change of the symmetric volume,
$|h'(t)|$, reaches its maximum if $\tilde t_\ast - \tilde t_c
\gg 1$ is satisfied.
For simplicity, we 
restrict our analysis in \Sec{ensemble_times}
to \Sec{FL_template}
to this asymptotic limit.
We show below that our results become independent of the
choice of $t_\ast$ when we are in this limit.
However, for the results presented
in this section, we
consider a generic finite value of
$\tilde t_\ast - \tilde t_c$.

For the exponential nucleation case, we can express $h({t})$
introducing \Eq{pt_exp} into \Eq{h_t},
\begin{equation}
    h(\tilde t) =
    \exp \biggl[-
    \frac{{\cal V}_w}{\beta^4} \,
    p_\ast \,
    e^{\tilde t - \tilde t_\ast}
    \,
    \gamma(4, \tilde t - \tilde t_c) \, 
    \Theta(\tilde t - \tilde t_c) \biggr]\,,
    \label{ht_exp_gen}
\end{equation}
where we have used the change of variable $z = \tilde t - \tilde t'$ to introduce
the lower incomplete Gamma function
$\gamma(4, \tilde t)$,
\begin{equation} \label{Gamma_exp}
    \gamma(4,\tilde t)
    \equiv
    \int_{0}^{ \tilde t} 
    z^3 e^{-z} \dd z = 6 - e^{-\tilde t} \,
    (\tilde t^{\, 3} + 3 \tilde t^{\, 2}
    + 6 \tilde t + 6) \,.
\end{equation}
Note that the
$\gamma(4, \tilde t - \tilde t_c)$
function reduces to the Gamma function $\Gamma(4) = 3! = 6$ when $\tilde t - \tilde t_c \to \infty$,
and it should not be confused
with the normalized nucleation rate
$\gamma(t) = p(t)\, h(t)/(\beta n_b)$ introduced
in \Eq{gamma_def}.

\EEq{ht_exp_gen} can be expressed in
such a way that it manifestly depends only on the choice
of $\ln h_\ast$, with $h_\ast = h(t_\ast)$,
and $\tilde{t}_\ast - \tilde t_c$,
\begin{align}
    h(\tilde t)= &\, \exp \Biggl[
    \ln h_\ast\,
    e^{\tilde{t}-\tilde{t}_\ast} \frac{ 
    \gamma(4,\tilde t - \tilde t_c) 
    }{\gamma(4,\tilde t_* - \tilde t_c) }
    \, \Theta (\tilde t - \tilde t_c)
    \Biggr] \,.    \label{ht_exp_explicit}
\end{align}
The number density of bubbles
is found introducing \Eq{ht_exp_explicit} into \Eq{nb_t},
\begin{align}
    n_b (\tilde t) &= \,
    \frac{p_\ast}{\beta} 
    \int_{\tilde{t}_c}^{\tilde{t}}
    e^{\tilde{t}' -\tilde{t}_\ast} \,
    \exp \Biggl[ \ln h_\ast \, e^{\tilde{t}'-\tilde{t}_\ast}
    \frac{ \gamma(4,\tilde t' - \tilde t_c) 
    }{\gamma(4,\tilde t_* - \tilde t_c) }
    \Biggr] \, \Theta (\tilde t - \tilde t_c) \dd \tilde{t}' \,.
    \label{num_of_bubbles}
\end{align}
Introducing the normalized nucleation rate $\gamma (t) = 
p(t)\, h(t)/(\beta n_b)$,
its integral over $\tilde t$ corresponds to the ratio of $n_b(\tilde t)$ and
its asymptotic value
$n_b = n_b (\tilde t \to \infty)$,
\begin{equation} \label{gamma_def}
    \frac{n_b (\tilde t)}{n_b} = \frac{1}{n_b} \int_{t_c}^t \Gamma(t') \dd t' = \int_{\tilde t_c}^{\tilde t}
    \gamma(\tilde t') \dd \tilde t'\,.
\end{equation}

\begin{figure}
    \centering
     \includegraphics[width=0.48\linewidth]{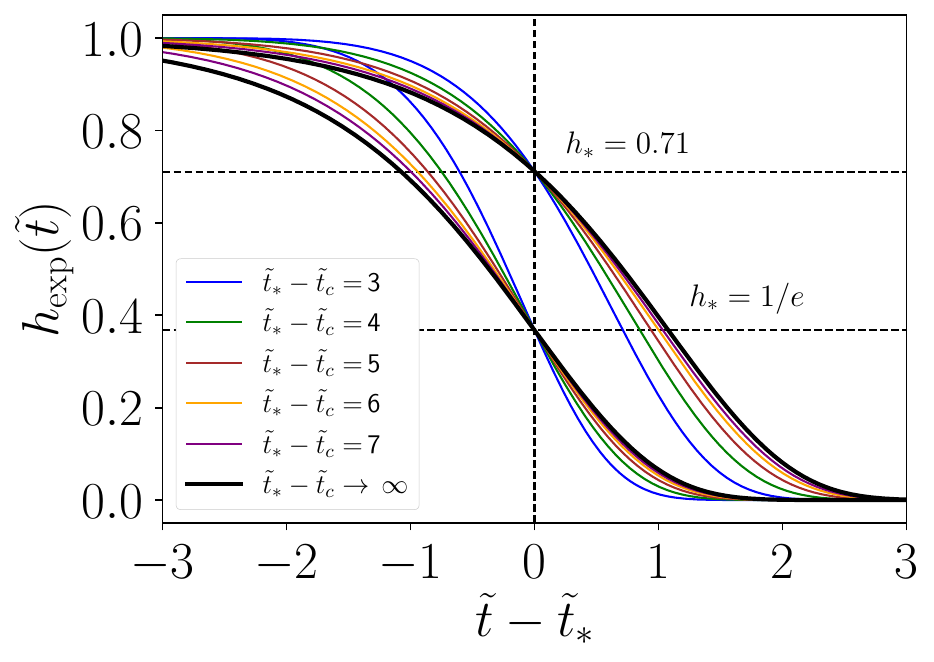}
     \includegraphics[width=0.48\linewidth]{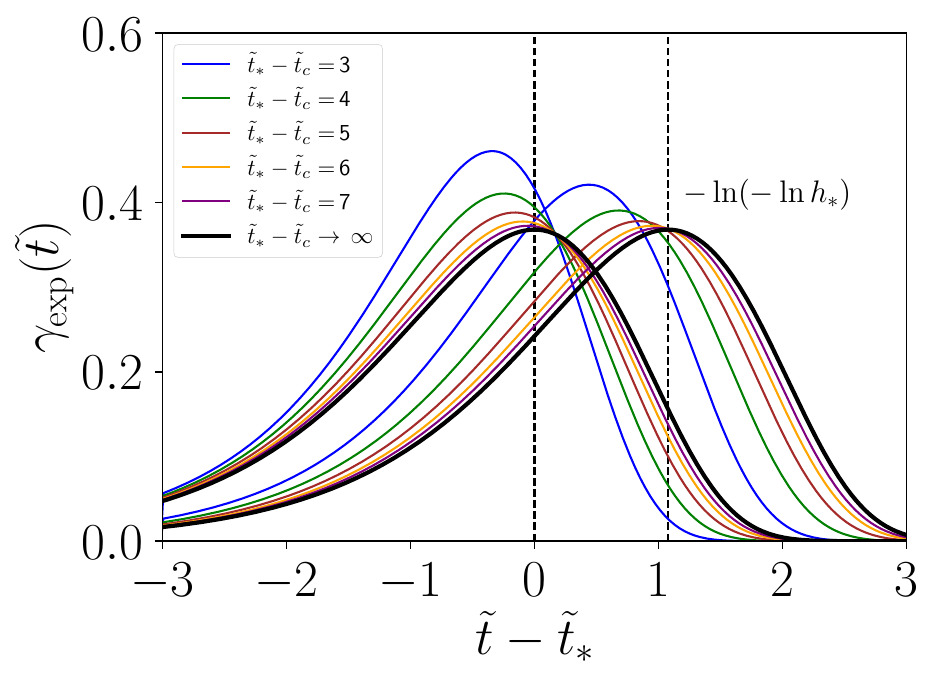}

    \caption{
    {\em Left panel:} Average fraction of the volume in the symmetric phase
    as a function of time $\tilde t - \tilde t_\ast$ for different
    values of $\tilde t_\ast - \tilde t_c$, considering
    an exponentially increasing nucleation distribution function with time,
    for two choices of $t_\ast$ ($t_f$ and $t_p$).
    {\em Right panel:} Normalized nucleation rate $\gamma \equiv \Gamma/(\beta n_b)$
    [see \Eq{gamma_def}].
    We observe that, in the limit $\tilde{t}_\ast - \tilde{t}_c \gg 1$,
    $\gamma$ peaks at $\tilde t_{\rm peak}
    = \tilde t_*+C_{\rm exp}$, where $C_{\rm exp} = - \ln (-\ln h_\ast)$,
    reducing to $C = 0$ and $C \simeq 1.08$
    for $t_\ast = t_f$  and $t_\ast = t_p$, respectively.
    We take $\tilde t_\ast - \tilde t_c \geq 3$
    to ensure $p_c/p_\ast \lesssim 0.05$.
    Black solid lines correspond to the limit $\tilde t_\ast - \tilde t_c \gg 1$.
    We observe that $h_{\rm exp}$ and $\gamma_{\rm exp}$ collapse when shown
    as a function of $\tilde t - \tilde t_{\rm peak}$ only in the $\tilde t_\ast - \tilde t_c \gg 1$ limit,
    as it can be inferred from \Eq{ht_exp_as2},
    with the asymptotic form of $\gamma_{\rm exp}(\tilde{t})$ shown in \Fig{fig:gammaexp_as}.
    }
    \label{hhs}
\end{figure}

\begin{figure}[t]
    \centering
    \includegraphics[width=0.47\linewidth]{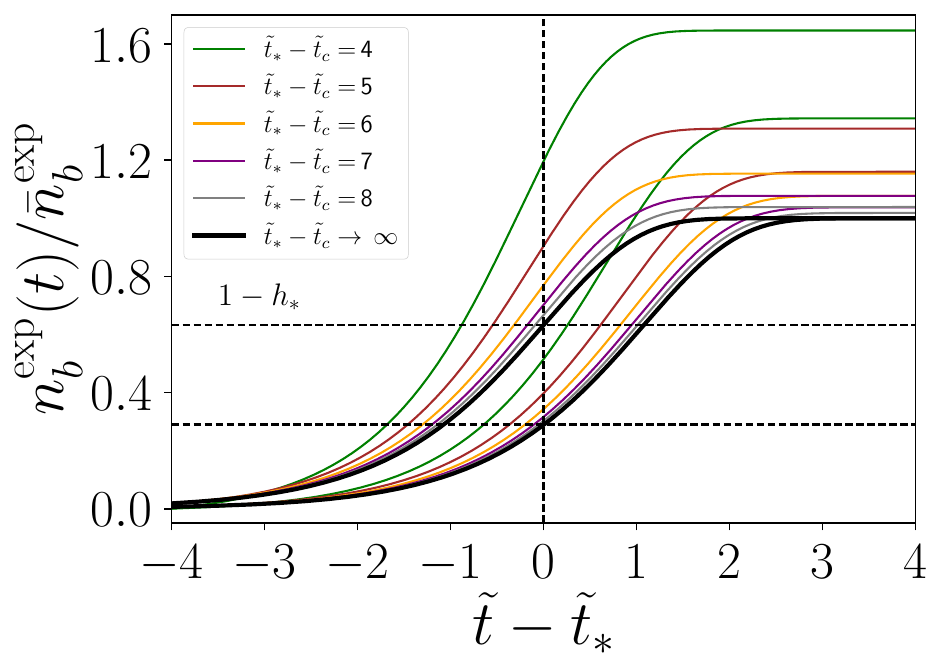}    \includegraphics[width=0.47\linewidth]{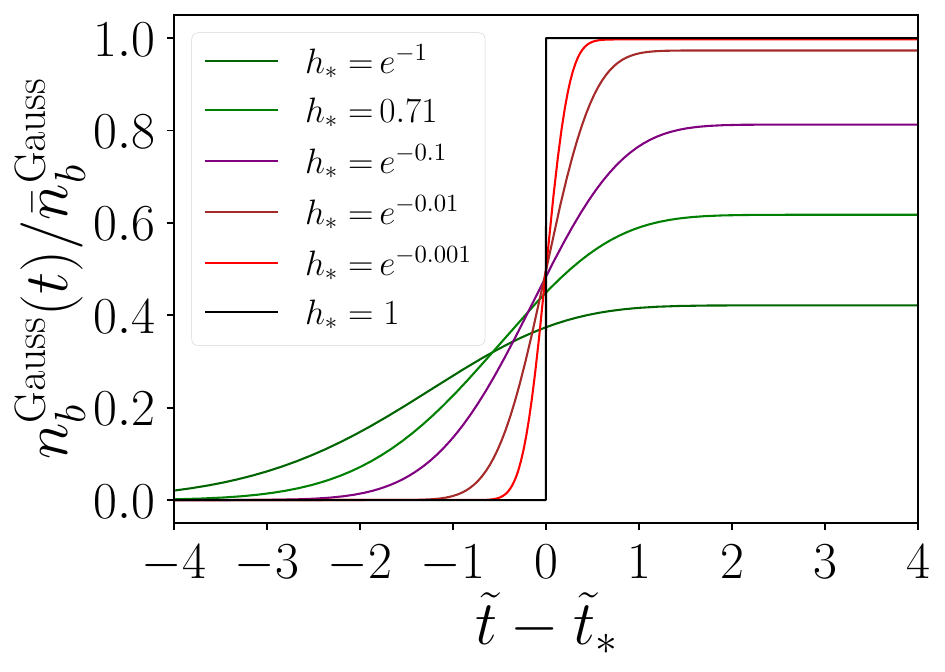}
    \caption{
    Average number of bubbles for exponential (left)
    and Gaussian (right) nucleations 
    as a function of time $\tilde t - \tilde t_\ast$,
    normalized by the asymptotic number of bubbles, $\bar n_b$,
    evaluated at $\tilde{t} \to \infty$ and taken
    either
    in the $\tilde t_\ast - \tilde t_c \gg 1$ limit (left)
    or in the simultaneous limit $h_\ast \to 1$ (right) [see \Eqs{nb_exp2}{nb_sims}].
    In the left panel (exponential), we show $n_b$
    for different values of $\tilde t_\ast - \tilde t_c$,
    and for $t_\ast = t_f$ (upper lines) and $t_p$ (lower lines).
    The black solid line corresponds to
    $n_b^{\rm exp}(\tilde t)/\bar n_b^{\rm exp}$, which becomes $1 - h$
    at $\tilde t_{\rm peak} - \tilde t_c \gg 1$ [see \Eq{nb_exp}].
    In the right panel, we show different choices of $\ln h_\ast$
    for the Gaussian case, in the regime $\beta_2 (t_\ast - t_c) > 3$, where
    we observe no dependence on $\beta_2 (t_\ast - t_c)$.
    $n_b^{\rm Gauss}$ is shown as a function of $\tilde t - \tilde t_\ast$, where times
    are normalized with $\beta_{\rm eff}$ (see details in \Sec{sim_nucl}).
    The black line corresponds to the limit $h_\ast \to 1$, for which
    $n_b$ becomes a step function at $t = t_\ast$ (simultaneous nucleation).
    We observe that the approximations in the corresponding limits
    affect the asymptotic number of bubbles by a factor that can be as large as
    150\% and 80\% for exponential nucleation choosing $t_\ast = t_f$
    and $t_p$ respectively with $\tilde{t}_\ast - \tilde{t}_c = 3$, and as large as 50\% for Gaussian
    nucleation when $h_\ast \geq 1/e$ compared to its asymptotic value
    when $h_\ast \to 1$; see exact values in \Tab{tab:rel_error_nb}.
    }
    \label{nbs_exp_sim}
\end{figure}

We show
the resulting $h(\tilde t)$ and $\gamma(\tilde t)$ in \Fig{hhs},
and $n_b(\tilde t)/n_b$ in the left panel of \Fig{nbs_exp_sim},
for different values of $\tilde t_\ast - \tilde t_c$
and for $t_\ast = t_f$ and $t_p$, i.e., $h_\ast = 1/e$ and $h_\ast \simeq 0.71$.
We take $\tilde t_\ast - \tilde t_c = 3$ as the minimum separation
between $\tilde t_\ast$ and $\tilde t_c$.
For this separation, $p(t_c) \simeq 0.05\,p_\ast$, so it can be
assumed that $p(t_c) \ll p_\ast$.
The dependence on $\beta$ and $\xi_w$ remains then in $n_b$ via
the relation between $p_\ast$ and $\ln h_\ast $,
which can be found from \Eq{ht_exp_gen},
\begin{align}
    p_\ast =-
    \frac{\beta^4}{{\cal V}_w}
    \frac{\ln h_\ast}{
    \gamma(4,\tilde t_* - \tilde t_c)
    } \,.
    \label{pt_exp_explicit}
\end{align}

Both $p(\tilde{t})$ and $h(\tilde{t})$ include the function
$\gamma(4, \tilde t - \tilde t_c)$,
which, for $\tilde t - \tilde t_c \gg 1$, converges to the value
$\Gamma(4)=6$.
We find in \Fig{hhs} that it takes values of around $\tilde t_\ast - \tilde t_c \gtrsim 8$
to reach the $\tilde t_\ast - \tilde t_c \to \infty$
asymptotic forms of $p$ and $h$.
For this time difference, the lower incomplete Gamma function
already takes a value close to 6,
$\gamma(4, 8)
\simeq 5.75$.
The limits $\tilde t - \tilde t_c \gg 1$ and $\tilde t_\ast - \tilde t_c \gg 1$
yield the distribution often considered in the literature
\cite{Ignatius:1993qn,Hindmarsh:2019phv}, with
\begin{align}
    \lim_{\tilde t_\ast - \tilde t_c \gg 1} p(\tilde t)
    = - \frac{\beta^4}{6\, {\cal V}_w} \ln h_\ast \,e^{\tilde t - \tilde t_\ast}
    \,, \qquad
    \lim_{\tilde t - \tilde t_c, \, \tilde t_\ast - \tilde t_c \gg 1}
    h(\tilde t) 
    =
    \exp \bigl(\ln h_\ast \, e^{\tilde t - \tilde t_\ast}
    \bigr)\,,
    \label{ht_exp_as1}
\end{align}
where $6 {\cal V}_w = 8 \pi \xi_w^3$ is the common factor
that determines
$(R_\ast \, \beta)^3$ in GW literature \cite{Caprini:2019egz,Caprini:2024gyk}.
In this limit, the maximum value of the nucleation rate
$\gamma(\tilde t) = p(\tilde t)\, h(\tilde t)/(\beta n_b)$
occurs
at the time $\tilde{t}_{\rm peak} =\tilde{t}_\ast
+ C_{\rm exp}$,
where $C_{\rm exp} = - \ln (- \ln h_\ast)$.
Then, using the time interval $\tilde t - \tilde t_{\rm peak}$, we can
reexpress $p(\tilde t)$ and  $h(\tilde t)$ as
\begin{equation}
    \lim_{\tilde t - \tilde t_c, \tilde t_\ast - \tilde t_c \gg 1}
    p(\tilde t) = \frac{\beta^4}{6 \, {\cal V}_w}
    e^{\tilde t - \tilde t_{\rm peak}} \,, \qquad
    \lim_{\tilde t - \tilde t_c, \tilde t_\ast - \tilde t_c \gg 1}
    h(\tilde t)= \exp
    \bigl( - e^{\tilde t - \tilde t_{\rm peak}} \bigr)\,,
    \label{ht_exp_as2}
\end{equation}
which only depends on $\tilde t - \tilde t_{\rm peak}$,
indicating that the lines in \Figs{hhs}{nbs_exp_sim} for different values of $h_\ast$
in the exponential nucleation case
collapse after a time shift.
Hence, the results presented in our work can be generalized to any
choice of $h_\ast$ when expressed
as a universal function of $\tilde t - \tilde t_{\rm peak}$.
On the other hand, the
integrand $\gamma(\tilde t')$ becomes dominated by
large values of $\tilde t' - \tilde t_c \gg 1$
around its peak value at $\tilde t_{\rm peak}=
\tilde t_\ast + C_{\rm exp}$,
as can be seen in \Fig{hhs},
such that for sufficiently large $\tilde t_{\rm peak} - \tilde t_c$, all relevant non-zero
values of $\gamma$ are going to satisfy the asymptotic limits,
and the bubble number density becomes
\begin{align}
    \bar n_b (\tilde t) \equiv \lim_{\tilde t_{\rm peak} - \tilde t_c \gg 1} n_b (\tilde t)
    = \frac{\beta^3}{6 \, {\cal V}_w}
    \int_{\tilde{t}_c}^{\tilde{t}}
    e^{\tilde t' - \tilde t_{\rm peak}}
    \exp \bigl(- e^{\tilde t' - \tilde t_{\rm peak}}
    \bigr) \dd \tilde{t}' =
      \frac{\beta^3}{6 \, {\cal V}_w}
     \bigl[1 - h(\tilde t) \bigr]\,,
     \label{nb_exp}
\end{align}
where we use a bar to denote $n_b (\tilde t)$ computed in this limit,
with
\begin{equation}
    \bar n_b = \frac{\beta^3}{6 \, {\cal V}_w} = \frac{\beta^3}{8 \pi \xi_w^3}\,.  
    \label{nb_exp2}
\end{equation}
In the limit $\tilde t_\ast - \tilde t_c \gg 1$,
the number of bubbles at $\tilde t \gtrsim \tilde t_{\rm peak}$
can then be approximated as $\bar n_b(\tilde t)$;
see black lines in \Fig{nbs_exp_sim}.
The relative error of the asymptotic $n_b = n_b(\tilde t \to \infty)$
with respect to $\bar n_b = \bar n_b (\tilde t \to \infty)$,
$\varepsilon = 1 - n_b/\bar n_b$,
is given in \Tab{tab:rel_error_nb} for different values of
$\tilde t_\ast - \tilde t_c$
and $\ln h_\ast$.
We can thus explicitly relate the
asymptotic
mean bubble separation
to $\beta$ and $\xi_w$,
\begin{equation}
    R_\ast \equiv \lim_{\tilde t \to \infty} R_\ast (\tilde t) =
    \bar n_b^{-1/3} =
    (8 \pi)^{1/3} \frac{\xi_w}{\beta} \simeq 2.9 \, \frac{\xi_w}{\beta}\,.
    \label{rstar_as}
\end{equation}
Alternatively, if one considers the characteristic length scale
$R_\ast$ as the mean
bubble separation
at the percolation time \cite{Caprini:2024ofd}, we find
\begin{equation}
    R_\ast (\tilde{t}_p) = \bar n_b^{-1/3} (\tilde{t}_p) =
    \biggl[\frac{8 \pi}{1 - h(\tilde{t}_p)} \biggr]^{1/3} \frac{\xi_w}{\beta} \simeq 4.4 \, \frac{\xi_w}{\beta}\,.
\end{equation}

\begin{table}[b]
    \centering
    \begin{tabular}{|c||c|c|c||c|c|c||c|c|c|}
    \hline
    $\tilde t_\ast - \tilde t_c$ & $h_\ast$ &
    $\varepsilon_{\rm exp}$ &
    $\varepsilon_{\rm Gauss}$ & $h_\ast$ &
    $\varepsilon_{\rm exp}$ &
    $\varepsilon_{\rm Gauss}$ & 
    $h_\ast$ &
    $\varepsilon_{\rm exp}$ &
    $\varepsilon_{\rm Gauss}$ \\ \hline
    3 & $ e^{-1} $ & $-1.504$ & 0.577
    & $ 0.71 $ & $-0.812$ & 0.386
    & $ e^{-0.001} $ & 0.640 & 0.004
    \\
    4 & $ e^{-1} $ & $-0.647$ & 0.578
    & $ 0.71 $ & $-0.344$ & 0.384
    & $ e^{-0.001} $ & 0.429 & 0.003
    \\
    5 & $ e^{-1} $ & $-0.308$ & 0.579
    & $ 0.71 $ & $-0.160$ & 0.384
    & $ e^{-0.001} $ & 0.263 & 0.003
    \\
    6 & $ e^{-1} $ & $-0.154$ & 0.579
    & $ 0.71 $ & $-0.077$ & 0.384
    & $ e^{-0.001} $ & 0.150 & 0.003
    \\
    7 & $ e^{-1} $ & $-0.077$ & 0.579
    & $ 0.71 $ & $-0.038$ & 0.384
    & $ e^{-0.001} $ & 0.081 & 0.003
    \\
    8 & $ e^{-1} $ & $-0.039$ & 0.579
    & $ 0.71 $ & $-0.018$ & 0.384
    & $ e^{-0.001} $ & 0.042 & 0.003
    \\  \hline
    \end{tabular}
    \caption{
    Relative difference $\varepsilon \equiv 1 - n_b/\bar n_b$
    of the asymptotic number of bubbles $n_b = n_b(t\to \infty)$
    for finite values of $\tilde t_\ast - \tilde t_c$ (or 
    $\beta_2 (t_\ast - t_c)$ for the Gaussian case)
    and $h_\ast$, compared to their values $\bar n_b$,
    evaluated in
    the following asymptotic limits:
    $\tilde t_{\rm peak} - \tilde t_c \gg 1$
    in the case of exponential nucleation, and $\beta_2 \to \infty$
    (simultaneous) for the case of Gaussian nucleation.
    The asymptotic number of bubbles is
    $\bar n_b = \tfrac{1}{6} \beta^3/{\cal V}_w$ and $\bar n_b = \tfrac{1}{6}
    \beta_{\rm eff}^3/{\cal V}_w$
    for the cases of exponential
    and simultaneous nucleation, respectively.}
    \label{tab:rel_error_nb}
\end{table}

\subsection{Gaussian and simultaneous nucleation}
\label{sim_nucl}

An alternative scenario occurs when all 
the bubbles nucleate almost simultaneously around the characteristic time
$t_\ast$.
This scenario can be used to represent first-order
phase transitions in vacuum, like supercooled phase transitions \cite{Cutting:2018tjt,Hindmarsh:2019phv}.

The action $S(t)$ decreases until it reaches a minimum around $t_\ast$ and then it
increases again to a larger zero-temperature level, yielding
a probability of bubble nucleation that strongly peaks around $t_\ast$.
The action can be Taylor expanded around $t_\ast$ as
\begin{align}
    S(t > t_c) \simeq  S (t_\ast) + \half \beta_2^2 (t - t_\ast)^2 \Rightarrow
    p(t > t_c) \simeq 
    p_\ast
    e^{ - \half \beta_2^2 (t - t_\ast)^2}\,, \label{pt_sim_gen}
\end{align}
with $\beta_2^2 = S''(t_\ast)$.
The volume
fraction in the metastable phase from \Eq{h_t} is
\begin{equation}
    h(t) = \exp\biggl[-
    \frac{{\cal V}_w}{\beta_2^4}
    \, p_\ast
    \, 
    \gamma_{\rm Gauss} (y) \, \Theta(t - t_c)
    \biggr]\,,
    \label{ht_sim_gen2}
\end{equation}
where we have introduced the changes of variables $z = \beta_2 (t' - t_\ast)$ and $y = \beta_2 (t - t_\ast)$,
and defined the following integral,
\begin{align} \label{Gamma_Gauss}
    \gamma_{\rm Gauss} (y) = 
    \int_{y_c}^y
    e^{-\half z^2} (y - z )^3  \dd z \,,
\end{align}
which we name $\gamma_{\rm Gauss}$ in
analogy to the lower incomplete Gamma function found in the exponential case [see \Eq{Gamma_exp}].
This integral can be computed by evaluating the following indefinite integral
\begin{subequations}
\begin{align}
    \int  e^{-\half z^2}   &\,
    (y -  z )^3 \dd z =  e^{-\half z^2}
    (3 y^2 - 3 y z + 2 + z^2)
    + \sqrt{\frac{\pi}{2}} y \,
    (y^2 + 3) \,
    {\rm erf}
    \biggl(\frac{z}{\sqrt{2}}\biggr) \,,
\end{align}
where erf is the error function,
\begin{equation}
    {\rm erf}(z) \equiv \frac{2}{\sqrt{\pi}} \int_0^{z} e^{-x^2} \dd x\,.
\end{equation}
\end{subequations}
Taking the value $\ln h_\ast$ as a parameter,
we can reexpress $h(t)$ as
\begin{equation}
    h(t) = \exp \Biggl[\ln h_\ast \,
    \frac{\gamma_{\rm Gauss}(y)}
    {\gamma_{\rm Gauss} (0)} \, \Theta (t - t_c) \Biggr]\,,  \label{ht_sim_gen}
\end{equation}
where
\begin{equation}
    \gamma_{\rm Gauss} (0) = \int_{0}^{-y_c}
    e^{-\half z^2} z^3 \dd z  = 
    2 - e^{-\half y_c^2}
    (y_c^2 + 2)\,.
\end{equation}
We then also have, as for the exponential nucleation case in \Eq{pt_exp_explicit},
\begin{align}
    p_* = -
    \frac{\beta_2^4}{{\cal V}_w}
    \frac{\ln h_\ast}{ \gamma_{\rm Gauss}}(0)\,.
    \label{pt_sim_explicit}
\end{align}

In the asymptotic limit of large
separation in time, $-y_c \equiv \beta_2 (t_\ast - t_c) \gg 1$,
$\gamma_{\rm Gauss} (0) \to 2$, and the asymptotic limit
of $\gamma_{\rm Gauss}(y) \equiv \gamma_{\rm Gauss} [\beta_2(t - t_\ast)]$
becomes
\begin{align}
    \lim_{y_c \to -\infty}
    \gamma_{\rm Gauss} (y) =  e^{-\half y^2} 
    (y^2 + 2) +
    \sqrt{\frac{\pi}{2}} \, y \, (y^2 + 3)
    \biggl[{\rm erf}
    \biggl(\frac{y}{\sqrt{2}}
    \biggr) + 1 \biggr]\,.
\end{align}
Furthermore, to introduce the simultaneous case,
we also take the limit $y \to \infty$,
\begin{equation}
    \lim_{-y_c, \,y \to \infty}
    \, \gamma_{\rm Gauss} (y) = \sqrt{2 \pi} \, y^3\,,
\end{equation}
as usually considered in the literature, see, e.g.,
ref.~\cite{Hindmarsh:2019phv}.
Both $\beta_2 (t - t_\ast)$ and $\beta_2 (t_\ast - t_c)$
asymptotic limits are
found if one takes $\beta_2 \to \infty$,
such that the probability distribution becomes
\begin{equation}
    \lim_{\beta_2 \to \infty} \, p(t)
    = \frac{\sqrt{2\pi}}{\beta_2} \,
    p_\ast \, \delta (t - t_\ast)\,,
    \label{pt_sim_as}
\end{equation}
and the volume in the symmetric phase is
\begin{equation}
    \lim_{\beta_2 \to \infty} h(\tilde t) = \exp\biggl[-
    \frac{{\cal V}_w}{\beta_2}\, \sqrt{2 \pi} \, p_\ast
    (t - t_\ast)^3 \, \Theta(t-t_\ast) \biggr]\,.
    \label{ht_asympt}
\end{equation}

\begin{figure}
    \centering
    \includegraphics[width=0.45\linewidth]{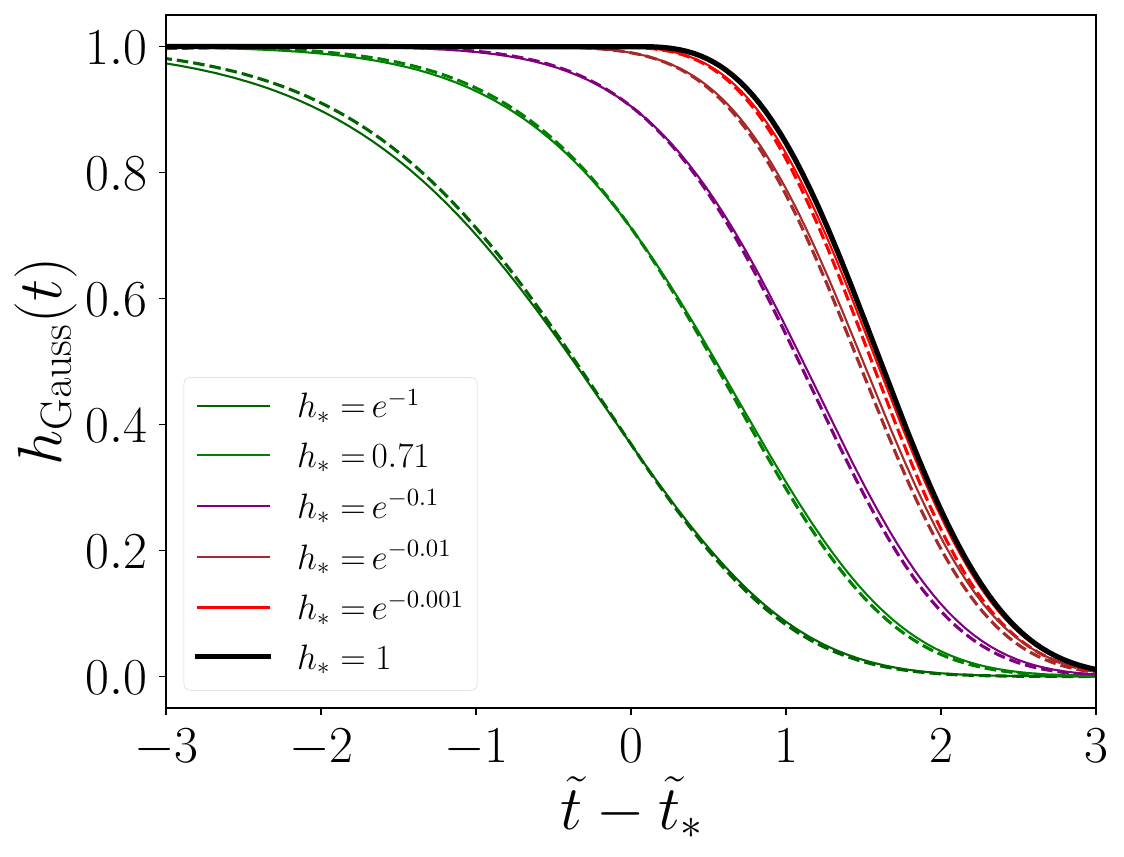}
    \includegraphics[width=0.45\linewidth]{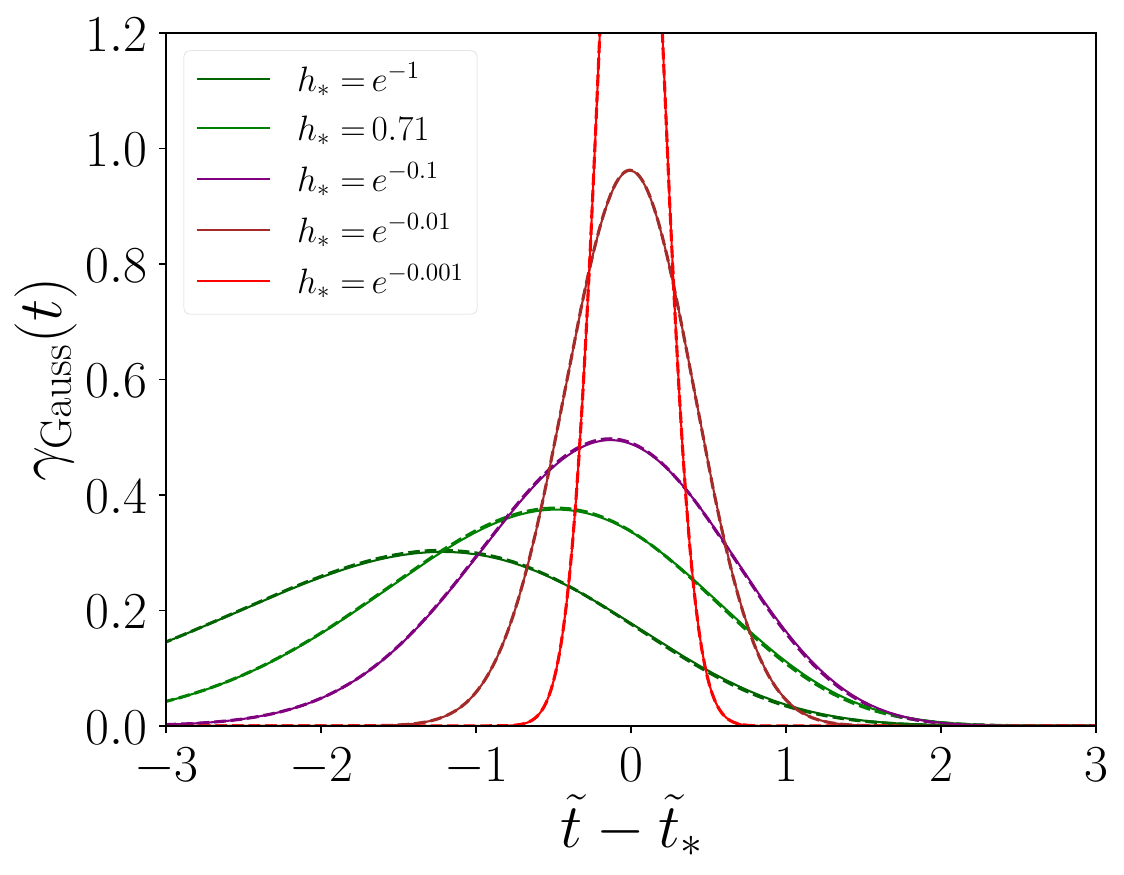}
    \caption{
    {\em Left panel:} Average fraction of the volume in the symmetric phase
    as a function of time $\tilde t - \tilde t_\ast$,
    normalized with $1/\beta_{\rm eff}$ [see \Eq{ratio_betaeff}],
    for $\beta_2 (t_\ast - t_c) = 3$ (dashed lines)
    and 4 (solid lines), considering
    a Gaussian nucleation distribution function with time,
    for different choices of $h_\ast$.
    The functions collapse to a unique dependence on
    $\beta_{\rm eff} (t - t_\ast) = \tilde t - \tilde t_\ast$ as $h_\ast \to 1$:
    $h(\tilde t) = \exp\bigl[-\sixth (\tilde t - \tilde t_\ast)^3
    \, \Theta(\tilde{t}-\tilde{t}_\ast)\bigr]$,
    given in \Eq{ht_asympt_2}.
    {\em Right panel:}
    Same as for left panel for the normalized nucleation rate $\gamma \equiv \Gamma/(\beta_{\rm eff}\, n_b)$.
    As $h_\ast \to 1$,
    $\gamma(\tilde{t}) \to \delta (\tilde{t} - \tilde{t}_\ast)$,
    found using \Eq{pt_sim_as}, is recovered.}
    \label{hs_sims}
\end{figure}

The numerically-computed values
of $h(\tilde t)$ and $\gamma(\tilde t)$ for two choices of $\beta_2 (t_\ast -
t_c)$ = 3 (dashed lines) and 4 (solid lines),
and different values of $h_\ast$,
are shown in \Fig{hs_sims}.
We find that for $\beta_2 (t_\ast - t_c) \gtrsim 4$, the asymptotic
limit $\beta_2 (t_\ast - t_c) \gg 1$
has already been reached.
For the minimum separation considered, $\beta_2 (t_\ast - t_c) = 3$,
the value of $p$ at $t_c$ is negligible,
$p_c/p_\ast \simeq 0.01$.

Using \Eq{nb_t}, the asymptotic number of bubbles in the
simultaneous case becomes
\begin{equation}
    \bar n_b = \lim_{\beta_2 \to \infty} n_b = 
    \lim_{\beta_2 \to \infty}
    \frac{\sqrt{2 \pi}}{\beta_2}  \,
    p_\ast \,.
    \label{nb_sims}
\end{equation}
The numerical errors for different values of $\beta_2 (t_\ast - t_c)$ and $\ln h_\ast$
in the asymptotic number of bubbles,
$\varepsilon = 1 - n_b/\bar n_b$, are given in \Tab{tab:rel_error_nb}.
The normalized distribution function $\gamma(t)$ is shown in
the right panel of \Fig{hs_sims} and the cumulative number of
bubbles, $n_b (t)/\bar n_b$,
in the right panel of \Fig{nbs_exp_sim} for different values of $h_\ast$.

Based on the asymptotic number of bubbles $\bar n_b$ obtained for
exponential nucleation [see \Eq{nb_exp2}],
we can define an effective $\beta_{\rm eff}$, such that $\bar n_b = \tfrac{1}{6}
\,\beta_{\rm eff}^3/{\cal V}_w$.
Then, we
can introduce \Eq{nb_sims} in \Eq{ht_asympt} and find
\begin{equation}
    \lim_{\beta_2 \to \infty} h(\tilde t) = \exp\bigl[- \tfrac{1}{6}
    (\tilde t - \tilde t_\ast)^3 \, \Theta(\tilde{t}-\tilde{t}_\ast)
    \bigr]\,, \qquad \text{where \ } \tilde t = t \, \beta_{\rm eff} \quad
    \text{and \ } \tilde t_\ast = t_\ast \, \beta_{\rm eff}\,.
    \label{ht_asympt_2}
\end{equation}

For any time $\tilde t$, when
the asymptotic limit $\beta_2 (t - t_\ast) \gg 1$ does not
need to hold, we introduce this normalization back in \Eq{ht_sim_gen2} to find
$h(\tilde t)$ in the case of Gaussian nucleation,
\begin{equation}
    h(\tilde t) = \exp\biggl[- \frac{1}{6 \sqrt{2\pi}}
    \frac{\beta_{\rm eff}^3}{\beta_2^3}
    \gamma_{\rm Gauss} (y)
    \biggr]\,.
    \label{ht_sim_general}
\end{equation}
Therefore,
the fraction of the volume in the symmetric phase at the characteristic
time $\tilde t_\ast$ is determined by the ratio $\beta_{\rm eff}/\beta_2$,
\begin{equation}
    \label{ratio_betaeff}
    \frac{\beta_{\rm eff}^3}{\beta_2^3} = -  \ln h_\ast
    \frac{6 \, \sqrt{2 \pi}}{\gamma_{\rm Gauss} (0)} \simeq -
    \ln h_\ast\, 3 \sqrt{2 \pi}\,,
\end{equation}
where the last approximation holds in the limit $\beta_2 (t_\ast - t_c) \gg 1$.
This implies that for a fixed number of nucleated bubbles (hence,
fixed value of $\beta_{\rm eff}$), we require $h_\ast$ to
approximate one in the $\beta_2 \to \infty$ limit, such that most of
the bubbles are almost simultaneously nucleated around $t \to t_\ast^+$.
This limit corresponds to $\beta_{\rm eff}^3/\beta_2^3 = \sqrt{2 \pi} \, 
8\pi \, p_\ast \,
\xi_w^3/\beta_2^4 \ll 1$,
then
the approximation in \Eq{ht_asympt} will hold for a broader range
of the phase transition duration (see \FFig{hs_sims}).

\subsection{Ensemble average over bubble lifetimes}
\label{distribution_lifetimes}

Characterizing the fluid perturbations
across bubble collisions and during the sound-wave regime requires
a precise description of the collisions,
as we discuss in \Secs{EffCollTime}{FL_template}.
According to the sound shell model, the correlation
function of the velocity field is described by averaging over
the lifetimes of the broken-phase
bubbles \cite{Hindmarsh:2019phv}.
Within the sound shell model, the required ingredients for computing 
the lifetime distribution are the bubble nucleation rate per unit 
volume $p(t)$, given in \Eqs{pt_exp}{pt_sim_gen} respectively for exponential
and Gaussian nucleation, and the area per
unit volume of the phase boundary,
\begin{equation}
    A(t) = - \frac{1}{\xi_w} \frac{\dd h}{\dd t}\,,    
\end{equation}
where $h(t)$ is the fraction of the Universe in the metastable 
phase, given in \Eqs{ht_exp_explicit}{ht_sim_gen} for exponential and
Gaussian nucleations, respectively.

One of the
main assumptions of the sound shell model is that the bubbles drive unperturbed fluid
profiles until they disappear due to collisions once that the
bubble wall reaches the nucleation location of another bubble.
Consider all bubbles nucleated between times $[t, t+\dd t]$,
whose walls are separated by a distance $[l, l+\dd l]$.
For these bubbles, the time between nucleation and the first collision is
contained in $[l/(2\xi_{w}),(l+\dd l)/(2\xi_{w})]$,
and their collision time in $[t+l/(2\xi_w), t+\dd t + (l+\dd l)/(2\xi_w)]$,
with their radii in the interval $[l/2,(l+\dd l)/2]$.
The wall of a bubble then reaches the nucleation point of the other bubble in
a time interval between
$[t+r/\xi_{w}, t+\dd t + (r+\dd r)/\xi_w]$,
when the radii of the collided bubbles lie within $[r, r+ \dd r]$.
The approximation proposed in ref.~\cite{Hindmarsh:2019phv},
consists of taking the time $T \equiv r/\xi_{w}$
as the lifetime of the bubbles.
The fraction of volume over which the bubbles disappear at the time $T$ is then given by
$A(t+r/\xi_w)\dd r$.
As a consequence, the differential number of bubbles
nucleated at
$[t, t+\dd t]$ and disappearing with
radii in $[r, r+\dd r]$, is
\begin{align}
    \dd n & = p(t) \dd t \, A\biggl(t + \frac{r}{\xi_w}\biggr) \dd r \,,
\end{align}
and the lifetime distribution is
\begin{align}
    \nu (\tilde{T}\equiv \beta T)  \equiv \frac{\xi_{w}}{\beta n_b} \frac{\dd n}{\dd r} = \frac{\xi_{w}}{\beta n_b} \int_{t_{c}}^{\infty} A
    (t + T) \, p(t) \dd t = - \frac{1}{\beta n_b}
    \int_{t_{c}}^{\infty} \frac{\dd h( t + T)}{\dd  t}
    \, p( t) \dd t \,,
\label{lifetimes_ssm}
\end{align}
which is already normalized by $\beta n_b$,
defined such that $\int_0^\infty \nu(\tilde T)\, \dd \tilde T = 1$
(see \Eq{gamma_def} for an analogous normalization).
Indeed, notice that, using the change of variable $ T' =  t +  T$,
the following relation holds
\begin{align}
    \int_{0}^{\infty} \frac{\dd h({t}+{T})}{\dd {t}} \dd {T}
    = \int_{{t}}^{\infty} \frac{\dd h({T}')}{\dd {T}'} \dd {T}' 
    = h(\infty)-h({t}) = - h({t}) \,,
\end{align}
and we can show that $\nu({\tilde T})$
is properly normalized, 
\begin{align}
    \int_0^{\infty} \nu(\tilde T)  \dd \tilde T
    = - \frac{1}{\beta n_b} \int_{\tilde {t}_c}^{\infty}
    p(\tilde t) \dd \tilde t 
    \int_{0}^{\infty}
    \frac{d h(\tilde{t} 
    +\tilde{T})}{\dd \tilde{t}}  \dd {T}
    = \int_{\tilde t_c}^\infty \gamma(\tilde t) \dd \tilde t 
    = 1\,.
\end{align}
To simplify the notation in this section, we use $\beta$ for both
cases of exponential and simultaneous nucleation, where
for the latter we set $\beta = \beta_{\rm eff}$.

\begin{figure}
    \centering
    \includegraphics[width=0.46\linewidth]{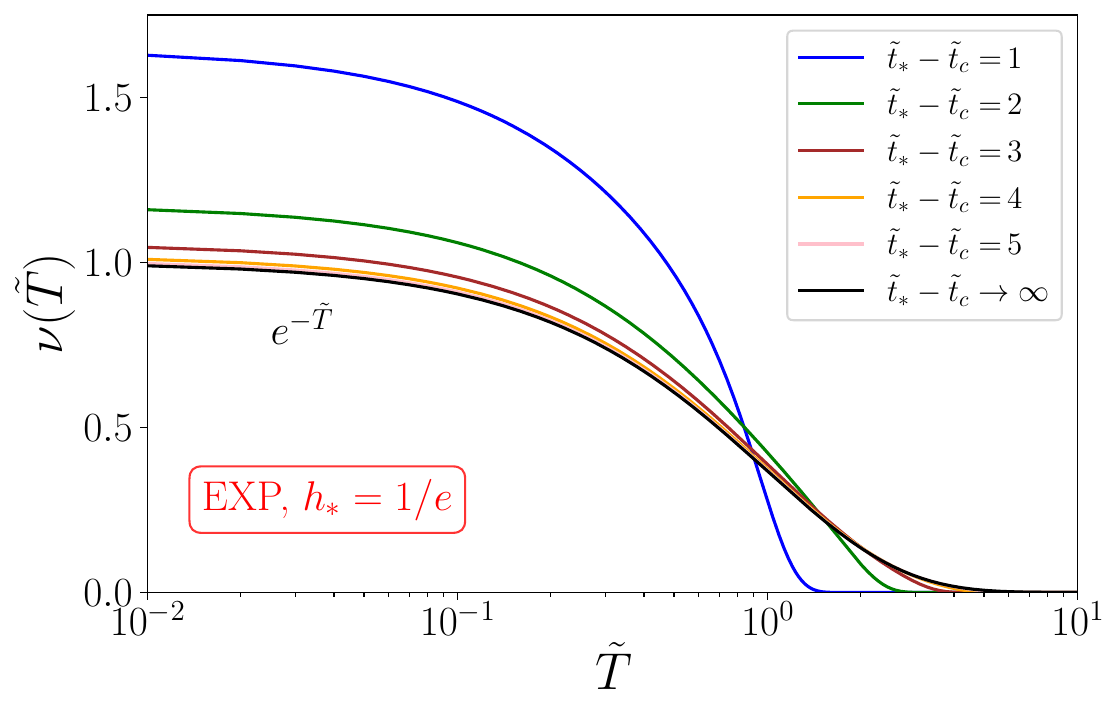}
    \includegraphics[width=0.46\linewidth]{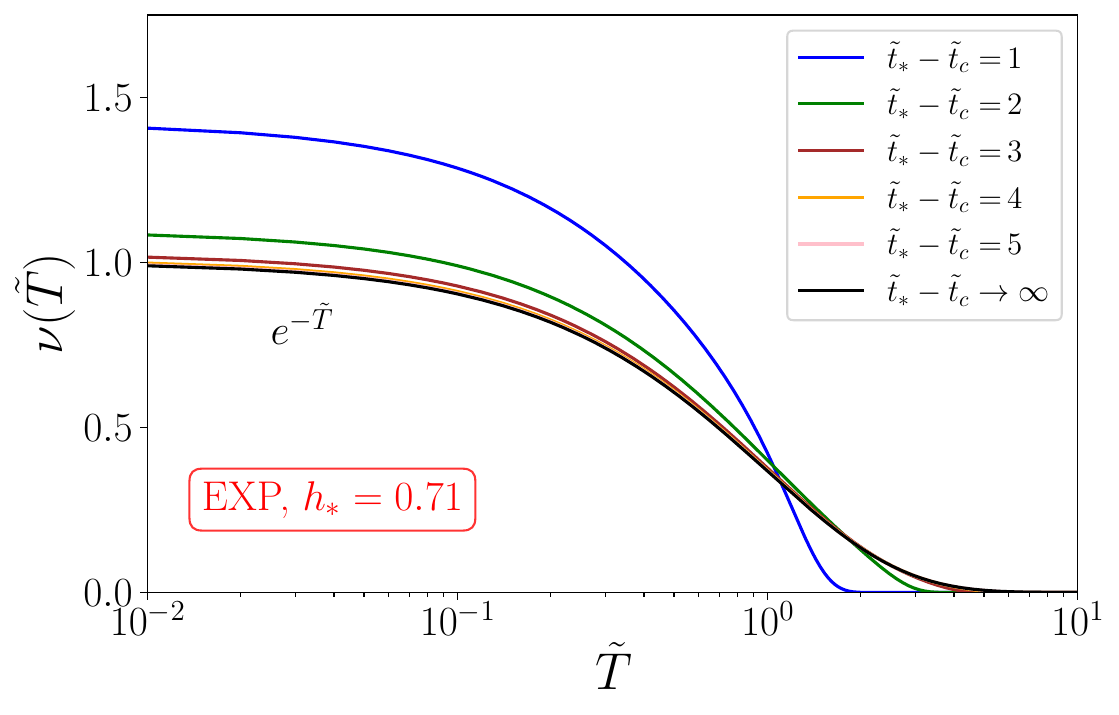}
    \includegraphics[width=0.45\linewidth]{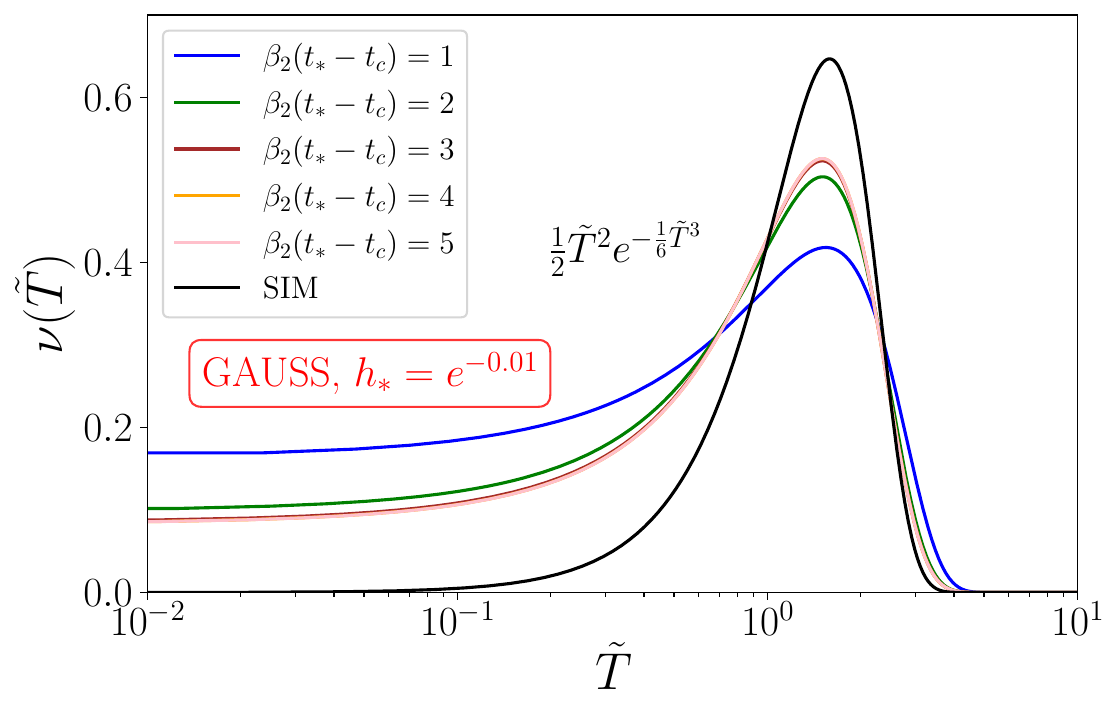}
    \includegraphics[width=0.45\linewidth]{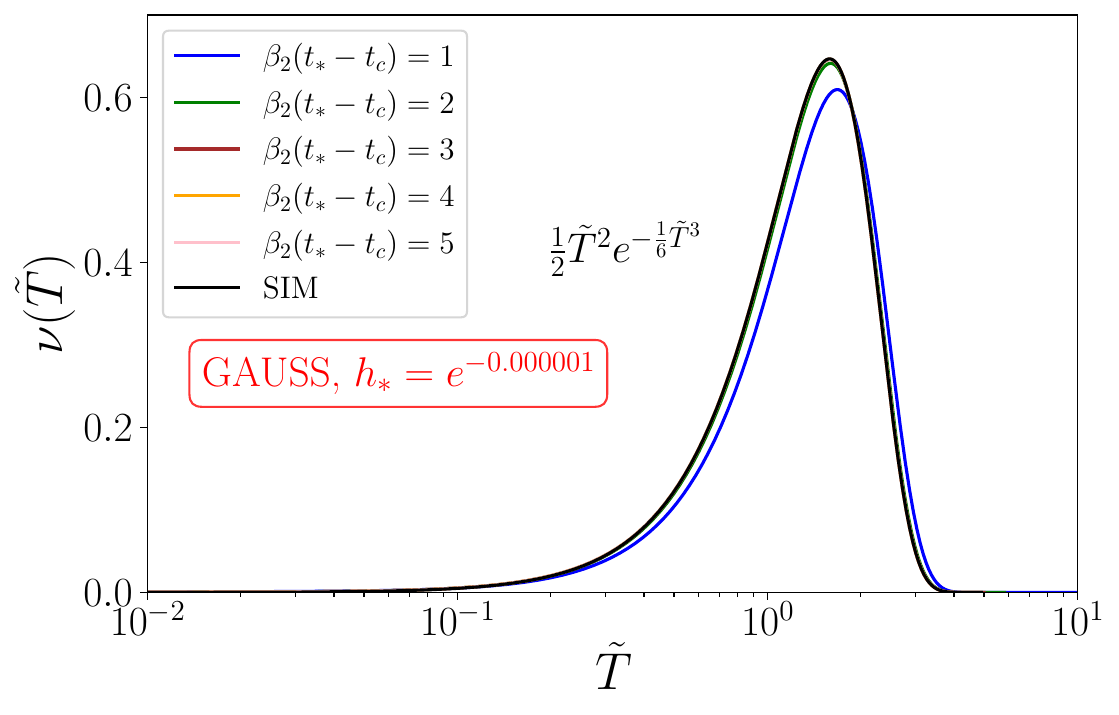}
    \caption{
    Normalized distribution function of the bubble lifetimes for
    an exponentially increasing distribution
    of nucleation times (EXP, upper panels) with the choices $h_\ast = 1/e$ (left) and $h_\ast = 0.71$ (right),
    and for a Gaussian nucleation
    probability (GAUSS, lower panels) with $\ln h_\ast = -0.01$ (left)
    and $\ln h_\ast = - 10^{-6}$ (right), which approaches the simultaneous (SIM) limit
    when $h_\ast \to 1$.
    The distribution of lifetimes is computed for
    different values of
    $\tilde t_\ast - \tilde t_c$
    (or $\beta_2 (t_\ast - t_c)$ for the Gaussian case, in which
    times $\tilde t$ are normalized by $1/\beta_{\rm eff}$).
    Note that for the exponential
    case, the lines tend to the lifetime distribution
    found in the asymptotic limit $\tilde t_\ast - \tilde t_c \gg 1$,
    $\nu_{\rm exp} (\tilde T) = e^{-\tilde T}$ 
    [see \Eq{nu_exp_as}],
    regardless of the choice of $h_\ast$.
    For the case of Gaussian nucleation,
    the asymptotic limit $\beta_2 (t_\ast - t_c) \gg 1$
    tends to the form $\nu_{\rm sim} (\tilde T) = \half \tilde T^2 \exp (- {1 \over 6} \tilde T^3)$ [see \Eq{nu_sim_as}] only
    in the simultaneous case with $h_\ast \to 1$ (bottom right panel).
    Therefore, for other values of $h_\ast$, $\nu_{\rm Gauss} (\tilde T)$ takes a different
    form than $\nu_{\rm sim} (\tilde T)$ in the $\beta_2 (t_\ast - t_c) \gg 1$ limit. 
    }
    \label{nuT_exp}
\end{figure}

Integrating by parts \Eq{lifetimes_ssm}, $\nu$ can be expressed as
\begin{equation} \label{nu_T_aux}
    \nu({\tilde T}) = \frac{1}{\beta \, n_b} \Biggl[
    h({\tilde t_c}  + {\tilde T} ) \, p({\tilde t_c} ) +
    \int_{{\tilde t_c} }^\infty
    h({\tilde t + \tilde T})
    \frac{\dd p ({\tilde t} )}{\dd {\tilde t} }
    \dd {\tilde t}  \Biggr]
    \simeq \frac{1}{\beta \, n_b} \int_{{\tilde t_c} }^\infty
    h({\tilde t + \tilde T} )
    \frac{\dd p ({\tilde t})}{\dd {\tilde t} }
    \dd {\tilde t}  \,,
\end{equation}
where we have assumed that $p(t_c) \ll 1$.
This approximation holds exactly only
in the $\tilde t_\ast - \tilde t_c \gg 1$
limit.
For finite values of $\tilde t_\ast - \tilde t_c$, the boundary term should be included,
but it quickly vanishes as one increases $\tilde t_\ast - \tilde t_c$.
In the case of exponential nucleation, using \Eq{nu_T_aux}, we find
\begin{equation}
    \nu(\tilde T) = \frac{p_\ast}{\beta \, n_b} \int_{\tilde t_c}^\infty
    e^{\tilde t - \tilde t_\ast} h(\tilde t + \tilde T)
    \dd \tilde t\,,
\end{equation}
where $h(\tilde t)$ is given in \Eq{ht_exp_explicit}.
The numerical results, for different choices of $\tilde{t}_\ast - \tilde t_c$ and two values of $h_\ast$,
corresponding to $h_\ast = 1/e$ and $h_\ast = 0.71$,
are shown in the top panels of \Fig{nuT_exp}.

In the asymptotic limit $\tilde{t}-\tilde{t}_c \gg 1$ and $\tilde{t}_\ast-\tilde{t}_c \gg 1$, we can use
the asymptotic expression of $p$ and $h$ in
\Eq{ht_exp_as2}
to compute the bubble lifetime distribution
when the nucleation time distribution
is exponential as
\begin{align}
    \nu_{\rm exp}(\tilde{T}) =
    \frac{1}{6 \, {\cal V}_w} \frac{\beta^3}{\bar n_b}
    \int_{\tilde{t}_c}^{\infty} 
    e^{\tilde{t}-\tilde{t}_{\rm peak}} \, \exp \bigl( -
    e^{\tilde{t}+\tilde{T}-\tilde{t}_{\rm peak}}\bigr)
    \dd \tilde{t} = e^{-\tilde{T}} \int_{x_c}^{\infty} e^{-x} \dd x  \simeq e^{-\tilde{T}} \,,
    \label{nu_exp_as}
\end{align}
where we have defined
$x = \exp(\tilde t + \tilde T - \tilde t_{\rm peak})$
and $x_c = \exp(\tilde t_c + \tilde T -
\tilde t_{\rm peak})$.
In the last approximation, we assume $\tilde{t}_{\rm peak} - \tilde{t}_c \gg 1$,
which implies $x_c \simeq 0$ and yields the normalized bubble lifetime distribution
$\nu_{\rm exp} (\tilde T) \approx e^{-\tilde T}$ \cite{Hindmarsh:2019phv}.
We note that this function is independent of the choice $h_\ast$.
In \Fig{nuT_exp}, we show that choosing $\tilde{t}_\ast - \tilde t_c
\gtrsim 4$ is sufficient to reach the approximation in \Eq{nu_exp_as}.
In this limit, also the choice of $h_\ast$ becomes irrelevant, and for any
choice, one finds $\nu_{\rm exp} (\tilde T) = e^{-\tilde T}$.
We note that in ref.~\cite{Guo:2023gwv}, the lifetime
distribution
$\nu_{\rm exp}$ has been found to slightly
differ with respect to $e^{-\tilde T}$ when computed numerically.
The authors attribute this deviation to the incorrect assumption, in the derivation leading to \Eq{lifetimes_ssm}, that the distance between uncollided bubbles $l$ remains constant, while new bubbles nucleating in between increase the distribution function at shorter lifetimes.

In the case of Gaussian nucleation, $p$ and $h$ are given in \Eqs{pt_sim_gen}{ht_sim_general} respectively.
Hence, \Eq{lifetimes_ssm} yields
\begin{equation}
    \nu_{\rm Gauss} (\tilde T)
    = - \frac{\ln h_\ast}{n_b \, \gamma_{\rm Gauss} (0)}
    \int_{\tilde{t}_c}^{\infty} 
     \, \gamma'_{\rm Gauss}
    (\tilde t + \tilde T -\tilde t_\ast) \, h(\tilde t + \tilde T)\,  p (\tilde t) 
    \dd \tilde t\,,
\end{equation}
where the derivative
$\gamma'_{\rm Gauss}$ [see \Eq{Gamma_Gauss}] is
\begin{align}
    \gamma'_{\rm Gauss} (y) = &\,
    3\, \int_{y_c}^{y}
    e^{-\half z^2} (y - z)^2 \dd z\,,
\end{align}
and can be computed using the following indefinite integral
\begin{equation}
    \int e^{-\half z^2} (y - z)^2 \dd z = 
    \sqrt{\frac{\pi}{2}} {\rm erf} \biggl(\frac{z}{\sqrt{2}}
    \biggr) (y^2 + 1) - e^{-\half z^2}
    (z - 2 y)\,.
\end{equation}
However, in the limit of infinitesimally short phase transitions,
$\beta_2 \rightarrow \infty$, which implies $\beta_2
(t - t_\ast) \gg 1$ and $\beta_2( t_\ast - t_{c}) \gg 1$, we can use the delta approximation for $p(\tilde{t})$, as in \Eq{pt_sim_as},
and $h(\tilde t)$ in \Eq{ht_asympt_2}
to find the
bubble lifetime distribution
when the nucleation time distribution
is simultaneous as
\cite{Hindmarsh:2019phv} 
\begin{align}
    \nu_{\rm sim} (\tilde{T})
    &= - \ln h_\ast \, 3 \pi \frac{p_\ast}{\bar n_b}
    \int_{\tilde{t}_c}^{\infty}
    \delta(t-t_\ast)
    \, \beta_2^2(t+T-t_\ast)^2
    \exp \biggl[ \sqrt{\frac{\pi}{2}} \ln h_\ast \beta_2^3
    (t+T-t_\ast)^3 \biggr] \dd \tilde t
    \nonumber \\
    &= - \ln h_\ast \, 3 \pi \frac{p_\ast}{\bar n_b}
    \, (\beta_2 T)^2 \exp \biggl[
    \sqrt{\frac{\pi}{2}} \ln h_\ast (\beta_2 T)^3 \biggr]
    = \half \tilde T^2 \exp\bigl(- \tfrac{1}{6} \tilde T^3 \bigr)\,.
    \label{nu_sim_as}
\end{align}
The lifetime distribution for Gaussian and simultaneous nucleation
is shown in the bottom panels of \Fig{nuT_exp}, where we see how the simultaneous
expression in \Eq{nu_sim_as} is found only when $\beta_2 (t_\ast - t_c) \gg 1$
and $h_\ast \to 1$ (bottom right panel).

\section{Analytical computation of the integral of the single and double broken power law templates}
\label{analytical_comp}

In this section, we compute analytically the integral over the functions
used to provide templates of the velocity spectrum in \Sec{template_FL}.
In particular, we choose a single (SBPL) and a double (DBPL) broken power law  template
[see \Eqq{SS_funcs}], equivalent to those used in the LISA Cosmology Working group
work on phase transitions \cite{Caprini:2024hue},
\begin{subequations}
\begin{align}
    S_{\rm SBPL} (k) = &\,
    \biggl(\frac{k}{k_1}\biggr)^5
    \Biggl[ 1+ \biggl(\frac{k}{k_1} \biggr)^{b_1} \Biggr]^{-6/b_1}\,,
    \\
    S_{\rm DBPL} (k) = &\, 
    \biggl(\frac{k}{k_2}\biggr)^5
    \Biggl[ 1+ \biggl(\frac{k}{k_1} \biggr)^{b_1} \Biggr]^{\frac{\sigma-2}{b_1}}\, \Biggl[ 1+ \biggl(\frac{k}{k_2} \biggr)^{b_2} \Biggr]^{\frac{-\sigma-4}{b_2}}\,,
    \label{FL_DBPL_app}
\end{align}
\end{subequations}
where $b_1, b_2 \in \mathbb{R}^+$ are smoothness parameters,
$k_1, k_2 \in \mathbb{R}^+$ 
are the locations of the transitions, and
$\sigma \in \mathbb{R}$ is the intermediate slope of the DBPL.

The normalization coefficient ${\cal N}$, included in \Eq{final_temps}, is
defined such that
\begin{equation}
    {\cal N}^{-1} =
    \int_0^\infty S(k) \, \dd \ln k\,.
\end{equation}
Let us start computing this integral for the SBPL case,
\begin{align}
    {\cal N}^{-1}_{\rm SBPL} = &\,
    k_1^{-5}
    \int_0^{\infty} k^4 \bigl[1 + (k/k_1)^{b_1}\bigr]^{-6/b_1}
    \dd k =
    \int_0^{\infty} x^4 \, (1 + x^{b_1})^{-6/b_1} \dd x \nonumber \\
    = &\,
    \frac{1}{b_1}
    \int_0^{\infty} y^{5/b_1 - 1}
    (1 + y)^{-6/b_1} \dd y  =
    \frac{1}{b_1}
    B\bigl(\tfrac{5}{b_1}, \tfrac{1}{b_1}\bigr)
    =
    \frac{1}{b_1}
    \frac{\Gamma\bigl(\tfrac{5}{b_1}\bigr)
    \Gamma\bigl(\tfrac{1}{b_1}\bigr)}{\Gamma\bigl(\tfrac{6}{b_1}\bigr)}\,,
\label{kappav_to_FL_1}
\end{align}
where we have used the changes of variables $x = k/k_1$ and $y = x^{b_1}$,
and we have defined the Euler beta function,
\begin{equation}
    B(a, b) \equiv \int_0^\infty y^{a - 1} (1 + y)^{-(a + b)} = 
    \frac{\Gamma(a)\, \Gamma(b)}{\Gamma(a + b)}\,.
\end{equation}

For the DBPL template,
let us first consider the case in which $b_1 = b_2 = b$ for simplicity.
Then,
the normalization factor is
\begin{align}
    {\cal N}_{\rm DBPL}^{-1} (b_1 = b_2 = b) = &\,
    k_2^{-5}
    \int_0^{\infty} k^4 \bigl[ 1+ (k/k_1)^{b} \bigr]^{(\sigma-2)/b}\, \bigl[ 1+(k/k_2)^{b} \bigr]^{-(\sigma-4)/b}  \dd k \nonumber 
    \\ = &\, 
    \frac{1}{b}
    \int_0^1 t^{5/b - 1} \bigl[
    1 - (1 - k_2^b/k_1^b) \, t \bigr]^{(\sigma - 2)/b}
    (1 - t)^{1/b - 1} \dd t \nonumber \\
    = &\,
    \frac{1}{b}
    B\bigl(\tfrac{5}{b},
    \tfrac{1}{b}\bigr) {}_2F_1\Bigl( \tfrac{2-\sigma}{b}, \tfrac{5}{b} \,; \tfrac{6}{b} \,; 1-\tfrac{k_2^b}{k_1^b} \Bigr)\,,
    \label{kappav_to_FL_2}
\end{align}
where we have used the change of variable $(k/k_2)^{b} = t/(1 - t)$ and
defined the 
integral representation of the Gauss hypergeometric function
\begin{equation}
    \label{hyper}
    {}_2F_1(\alpha, \beta \,; \gamma \,; z)
    =
    \frac{1}{B (\beta,\gamma-\beta)}
    \int_0^1
    t^{\beta-1}\,(1-t)^{\gamma-\beta-1}\,(1- z \,t)^{-\alpha}
    \dd t\,,
\end{equation}
for ${\rm Re}(\gamma)> {\rm Re}(\beta)>0$.
In the limit $k_1 = k_2$, we recover the expression corresponding to the
SBPL in \Eq{kappav_to_FL_1}.
This is due to the fact that the normalized hypergeometric function
is normalized such that ${}_2F_1(\alpha, \beta \,; \gamma \,; 0)
= 1$ for any $\alpha$, $\beta$, $\gamma$, as can be seen in \Eq{hyper}.

Let us now consider the DBPL in
the general case, where we can have $b_1 \neq b_2$.
The integral can be performed by making use of the
Mathai-Saxena-Haubold identity (see Eq.~(1.43) of ref.~\cite{Mathai2009TheHT}),
\begin{equation}
(1+x)^{-\lambda}
=
\frac{1}{\Gamma(\lambda)}
\,H^{1,1}_{1,1}
\!\left[
x
\;\middle|\;
\begin{matrix}
(1-\lambda,\,1)\\
(0,\,1)
\end{matrix}
\right],
\qquad
\text{with \ } {\rm Re}(x)>0,\, {\rm Re}(\lambda) > 0\,,
\end{equation}
by which the factors $\bigl[1+(k/\alpha)^\beta\bigr]^\gamma$ in the integral can be expressed using the Fox $H$-function. This leads to the following integral
\begin{align} \label{Mellin}
    {\cal N}_{\rm DBPL}^{-1} =
    k_2^{-5} &\,
    \int_0^\infty k^4 \bigl[1 + (k/k_1)^{b_1}\bigr]^{(\sigma - 2)/b_1}
    \bigl[1 + (k/k_2)^{b_2}\bigr]^{-(\sigma - 4)/b_2} \dd k
     =
    \frac{1}{\Gamma\bigl(\tfrac{2 - \sigma}
    {b_1}\bigr) \, \Gamma\bigl(\tfrac{\sigma + 4}{b_2}\bigr)}
    \nonumber \\
    \times \, k_2^{-5} &\,
    \int_0^{\infty} k^4 H^{1,1}_{1,1}
    \!\left[
    \biggl(\frac{k}{k_1} \biggr)^{b_1}
    \;\middle|\;
    \begin{matrix}
    \bigl(1-\frac{2-\sigma}{b_1},\,1 \bigr) \\
    (0,\,1)
    \end{matrix}
    \right]
    \,H^{1,1}_{1,1}
    \!\left[
    \biggl(\frac{k}{k_2} \biggr)^{b_2}
    \;\middle|\;
    \begin{matrix}
    \bigl(1-\frac{\sigma+4}{b_2},\,1 \bigr) \\
    (0,\,1)
    \end{matrix}
    \right] \dd k \,.
\end{align}
We then recognize the Mellin transform of the product of two Fox $H$-functions,
for which we have the result presented
in Sec.~2.3 of ref.~\cite{Mathai2009TheHT},
\begin{align} \label{product_Hbox}
    & \int_{0}^{\infty} x^{s-1}
    H_{p,q}^{m,n}\!\left[
    z x^{\sigma}
    \;\middle|\;
    \begin{array}{l}
    (a_p,A_p) \\
    (b_q,B_q)
    \end{array}
    \right]
    H_{p_1,q_1}^{m_1,n_1}\!\left[
    \eta x
    \;\middle|\;
    \begin{array}{l}
    (d_{p_1},D_{p_1}) \\
    (e_{q_1},E_{q_1})
    \end{array}
    \right]
    \dd x  \\
    &= \eta^{-s}
    H_{p+q_1,q+p_1}^{m+n_1,n+m_1}\!\left[
    z \eta^{-\sigma}
    \;\middle|\;
    \begin{array}{l}
    (1-e_{q_1}-sE_{q_1},\,\sigma E_{q_1}),\,(a_p,A_p)
    \\
    (b_m,B_m),\,
    (1-d_{p_1}-sD_{p_1},\,\sigma D_{p_1}),\,
    (b_{m+1},B_{m+1}),\,\ldots,(b_q,B_q)
    \end{array}
    \right]\,. \nonumber 
\end{align}

The integral in \Eq{Mellin} can be expressed using \Eq{product_Hbox} with
the change of variable $x = (k/k_2)^{b_2}$ and for the values
$m=n=p=q=m_1=n_1=p_1=q_1=1$, $z=k_1^{-b_1}$,
$\eta = k_2^{-b_2}$, $x=k^{b_2}$, $s=5/b_2$, $\sigma=b_1/b_2$,
$a_1=1-(2-\sigma)/b_1$, $A_1=1$, $b_1=0$, $B_1=1$,
$d_1=1-(\sigma+4)/b_2$, $D_1=1$, $e_1=0$, and $E_1=1$.
Hence, the 
normalization
factor relating the integral and the amplitude of the DBPL fit becomes
\begin{equation}
    {\cal N}^{-1}_{\rm DBPL} =
    \frac{1}{b_2} \frac{1}{\Gamma\bigl(\tfrac{2 - \sigma}
    {b_1}\bigr) \, \Gamma\bigl(\tfrac{\sigma + 4}{b_2}\bigr)}
    \, H_{2,2}^{2,2}\!\left[
    \left(\frac{k_2}{k_1}\right)^{b_1}
    \;\middle|\;
    \begin{array}{l}
    \bigl(-4,\frac{b_1}{b_2}\bigr),\;
    \bigl(1-\frac{2-\sigma}{b_1},1 \bigr)
    \\
    (0,1),\;
    \bigl(\frac{\sigma+4}{b_2}-5,\frac{b_1}{b_2} \bigr)
    \end{array}
    \right]\,.
    \label{kappav_to_FL_3}
\end{equation}
We note that when $b_1 = b_2 = b$, this expression becomes
\begin{equation}
    {\cal N}^{-1}_{\rm DBPL} (b_1 = b_2 = b) =
    \frac{1}{b} \frac{1}{\Gamma\bigl(\tfrac{2 - \sigma}
    {b}\bigr) \, \Gamma\bigl(\tfrac{\sigma + 4}{b}\bigr)}
    \, H_{2,2}^{2,2}\!\left[
    \left(\frac{k_2}{k_1}\right)^{b}
    \;\middle|\;
    \begin{array}{l}
    \bigl(-4,1 \bigr),\;
    \bigl(1-\frac{2-\sigma}{b},1 \bigr)
    \\
    (0,1),\;
    \bigl(\frac{\sigma+4}{b}-5,1 \bigr)
    \end{array}
    \right]\,.
\end{equation}
Plugging this back into \Eqs{Mellin}{product_Hbox}, we recover the definition
of the hypergeometric function, and, hence, \Eq{kappav_to_FL_2}.

\paragraph{Data availability statement.}

The calculations that are provided in this work have been included in the
public Python package \href{https://github.com/CosmoGW/cosmoGW}{\sc CosmoGW} \cite{CosmoGW_GH,cosmogw_manual,Stomberg:2025kxf}, which allow
to compute the single-bubble velocity profiles,
the velocity spectrum during the bubble expansion phase, and the velocity
spectrum at the initial time of the sound-wave regime.
All the results and plots presented in our work
can be reproduced using {\sc CosmoGW},
where we have also included the libraries
to compute the GW spectrum in the sound shell model, as presented in ref.~\cite{RoperPol:2023dzg}.
Updated documentation and tutorials
can be found at \href{https://cosmogw-manual.readthedocs.io/}{Read the Docs}.

\bibliographystyle{JCAP}
\bibliography{library}

@article{Kirzhnits:1972ut,
    author = "Kirzhnits, D. A. and Linde, Andrei D.",
    title = "{Macroscopic Consequences of the Weinberg Model}",
    doi = "10.1016/0370-2693(72)90109-8",
    journal = "Phys. Lett. B",
    volume = "42",
    pages = "471--474",
    year = "1972"
}

@article{Sazhin:1978myk,
    author = "Sazhin, Mikhail V.",
    title = "{Opportunities for detecting ultralong gravitational waves}",
    journal = "Sov. Astron.",
    volume = "22",
    pages = "36--38",
    year = "1978"
}

@article{Detweiler:1979wn,
    author = "Detweiler, Steven L.",
    title = "{Pulsar timing measurements and the search for gravitational waves}",
    doi = "10.1086/157593",
    journal = "Astrophys. J.",
    volume = "234",
    pages = "1100--1104",
    year = "1979"
}

@article{LISACosmologyWorkingGroup:2022jok,
    author = "Auclair, Pierre and others",
    collaboration = "LISA Cosmology Working Group",
    title = "{Cosmology with the Laser Interferometer Space Antenna}",
    eprint = "2204.05434",
    archivePrefix = "arXiv",
    primaryClass = "astro-ph.CO",
    reportNumber = "LISA CosWG-22-03, FERMILAB-PUB-22-349-SCD",
    doi = "10.1007/s41114-023-00045-2",
    journal = "Living Rev. Rel.",
    volume = "26",
    number = "1",
    pages = "5",
    year = "2023"
}

@article{LISA:2024hlh,
    author = "Colpi, Monica and others",
    collaboration = "LISA",
    title = "{LISA Definition Study Report}",
    eprint = "2402.07571",
    archivePrefix = "arXiv",
    primaryClass = "astro-ph.CO",
    month = "2",
    year = "2024"
}

@article{RoperPol:2025lgc,
    author = "Roper Pol, Alberto and Midiri, Antonino Salvino",
    title = "{Relativistic magnetohydrodynamics in the early Universe}",
    eprint = "2501.05732",
    archivePrefix = "arXiv",
    primaryClass = "gr-qc",
    doi = "10.1088/1475-7516/2026/06/070",
    journal = "J. Cosmol. Astropart. Phys.",
    volume = "06",
    pages = "070",
    year = "2026"
}

@article{Caprini:2024ofd,
    author = "Caprini, Chiara and Pujol{\`a}s, Oriol and Quelquejay-Leclere, Hippolyte and Rompineve, Fabrizio and Steer, Dani{\`e}le A.",
    title = "{Primordial gravitational wave backgrounds from phase transitions with next generation ground based detectors}",
    eprint = "2406.02359",
    archivePrefix = "arXiv",
    primaryClass = "astro-ph.CO",
    reportNumber = "CERN-TH-2024-065",
    doi = "10.1088/1361-6382/ad9a48",
    journal = "Class. Quant. Grav.",
    volume = "42",
    number = "4",
    pages = "045015",
    year = "2025"
}

@article{Hindmarsh:2019phv,
    author = "Hindmarsh, Mark and Hijazi, Mulham",
    title = "{Gravitational waves from first order cosmological phase transitions in the sound shell model}",
    eprint = "1909.10040",
    archivePrefix = "arXiv",
    primaryClass = "astro-ph.CO",
    reportNumber = "NORDITA-2019-083, HIP-2019-29/TH",
    doi = "10.1088/1475-7516/2019/12/062",
    journal = "J. Cosmol. Astropart. Phys.",
    volume = "12",
    pages = "062",
    year = "2019"
}

@article{Hindmarsh:2013xza,
    author = "Hindmarsh, Mark and Huber, Stephan J. and Rummukainen, Kari and Weir, David J.",
    title = "{Gravitational waves from the sound of a first order phase transition}",
    eprint = "1304.2433",
    archivePrefix = "arXiv",
    primaryClass = "hep-ph",
    reportNumber = "HIP-2013-07-TH",
    doi = "10.1103/PhysRevLett.112.041301",
    journal = "Phys. Rev. Lett.",
    volume = "112",
    pages = "041301",
    year = "2014"
}

@article{Hindmarsh:2015qta,
    author = "Hindmarsh, Mark and Huber, Stephan J. and Rummukainen, Kari and Weir, David J.",
    title = "{Numerical simulations of acoustically generated gravitational waves at a first order phase transition}",
    eprint = "1504.03291",
    archivePrefix = "arXiv",
    primaryClass = "astro-ph.CO",
    reportNumber = "HIP-2015-13-TH",
    doi = "10.1103/PhysRevD.92.123009",
    journal = "Phys. Rev. D",
    volume = "92",
    number = "12",
    pages = "123009",
    year = "2015"
}

@article{Hindmarsh:2017gnf,
    author = "Hindmarsh, Mark and Huber, Stephan J. and Rummukainen, Kari and Weir, David J.",
    title = "{Shape of the acoustic gravitational wave power spectrum from a first order phase transition}",
    eprint = "1704.05871",
    archivePrefix = "arXiv",
    primaryClass = "astro-ph.CO",
    reportNumber = "HIP-2017-02-TH, HIP-2017-02/TH",
    doi = "10.1103/PhysRevD.96.103520",
    journal = "Phys. Rev. D",
    volume = "96",
    number = "10",
    pages = "103520",
    year = "2017",
    note = "[Erratum: Phys.~Rev.~D {\bf 101}, 089902 (2020)]"
}

@article{Hindmarsh:2016lnk,
    author = "Hindmarsh, Mark",
    title = "{Sound shell model for acoustic gravitational wave production at a first-order phase transition in the early Universe}",
    eprint = "1608.04735",
    archivePrefix = "arXiv",
    primaryClass = "astro-ph.CO",
    doi = "10.1103/PhysRevLett.120.071301",
    journal = "Phys. Rev. Lett.",
    volume = "120",
    number = "7",
    pages = "071301",
    year = "2018"
}

@article{Cutting:2019zws,
    author = "Cutting, Daniel and Hindmarsh, Mark and Weir, David J.",
    title = "{Vorticity, kinetic energy, and suppressed gravitational wave production in strong first order phase transitions}",
    eprint = "1906.00480",
    archivePrefix = "arXiv",
    primaryClass = "hep-ph",
    reportNumber = "HIP-2019-15/TH",
    doi = "10.1103/PhysRevLett.125.021302",
    journal = "Phys. Rev. Lett.",
    volume = "125",
    number = "2",
    pages = "021302",
    year = "2020"
}

@article{Jinno:2017fby,
    author = "Jinno, Ryusuke and Takimoto, Masahiro",
    title = "{Gravitational waves from bubble dynamics: Beyond the envelope}",
    eprint = "1707.03111",
    archivePrefix = "arXiv",
    primaryClass = "hep-ph",
    reportNumber = "CTPU-17-26, KEK-TH-1986",
    doi = "10.1088/1475-7516/2019/01/060",
    journal = "J. Cosmol. Astropart. Phys.",
    volume = "01",
    pages = "060",
    year = "2019"
}

@article{Jinno:2016vai,
    author = "Jinno, Ryusuke and Takimoto, Masahiro",
    title = "{Gravitational waves from bubble collisions: An analytic derivation}",
    eprint = "1605.01403",
    archivePrefix = "arXiv",
    primaryClass = "astro-ph.CO",
    reportNumber = "KEK-TH-1900",
    doi = "10.1103/PhysRevD.95.024009",
    journal = "Phys. Rev. D",
    volume = "95",
    number = "2",
    pages = "024009",
    year = "2017"
}

@article{RoperPol:2023bqa,
    author = "Roper Pol, A. and Neronov, A. and Caprini, C. and Boyer, T. and Semikoz, D.",
    title = "{LISA and {\ensuremath{\gamma}}-ray telescopes as multi-messenger probes of a first-order cosmological phase transition}",
    eprint = "2307.10744",
    archivePrefix = "arXiv",
    primaryClass = "astro-ph.CO",
    doi = "10.1051/0004-6361/202658936",
    journal = "Astron. Astrophys.",
    volume = "708",
    pages = "A337",
    year = "2026"
}

@article{Isserlis:1916,
    author = {Isserlis, L.},
    title = "{On certain probable errors and correlation of multiple frequency distributions with skew regression}",
    journal = {Biometrika},
    volume = {11},
    number = {3},
    pages = {185},
    year = {1916},
    month = {05},
    issn = {0006-3444},
    doi = {10.1093/biomet/11.3.185},
}

@article{CONSIGLIO200349,
title = {Continuum percolation thresholds for mixtures of spheres of different sizes},
journal = {Physica A: Statistical Mechanics and its Applications},
volume = {319},
pages = {49-55},
year = {2003},
issn = {0378-4371},
doi = {https://doi.org/10.1016/S0378-4371(02)01501-7},
url = {https://www.sciencedirect.com/science/article/pii/S0378437102015017},
author = {R. Consiglio and D.R. Baker and G. Paul and H.E. Stanley},
}

@article{Shante01051971,
author = {Vinod K.S. Shante and Scott Kirkpatrick},
title = {An introduction to percolation theory},
journal = {Advances in Physics},
volume = {20},
number = {85},
pages = {325--357},
year = {1971},
publisher = {Taylor \& Francis},
doi = {10.1080/00018737100101261},
URL = { 
        https://doi.org/10.1080/00018737100101261
},
eprint = { 
        https://doi.org/10.1080/00018737100101261
}
}

@article{LI2020112815,
title = {Numerical study for the percolation threshold and transport properties of porous composites comprising non-centrosymmetrical superovoidal pores},
journal = {Computer Methods in Applied Mechanics and Engineering},
volume = {361},
pages = {112815},
year = {2020},
issn = {0045-7825},
doi = {https://doi.org/10.1016/j.cma.2019.112815},
url = {https://www.sciencedirect.com/science/article/pii/S0045782519307078},
author = {Mingqi Li and Huisu Chen and Jianjun Lin},
}

@article{LIN2018299,
title = {Continuum percolation of porous media via random packing of overlapping cube-like particles},
journal = {Theoretical and Applied Mechanics Letters},
volume = {8},
number = {5},
pages = {299-303},
year = {2018},
issn = {2095-0349},
doi = {https://doi.org/10.1016/j.taml.2018.05.007},
url = {https://www.sciencedirect.com/science/article/pii/S209503491830196X},
author = {Jianjun Lin and Huisu Chen},
}

@article{10.1063/1.1338506,
    author = {Lorenz, Christian D. and Ziff, Robert M.},
    title = {Precise determination of the critical percolation threshold for the three-dimensional ``Swiss cheese'' model using a growth algorithm},
    journal = {The Journal of Chemical Physics},
    volume = {114},
    number = {8},
    pages = {3659-3661},
    year = {2001},
    month = {02},
    issn = {0021-9606},
    doi = {10.1063/1.1338506},
    url = {https://doi.org/10.1063/1.1338506},
}

@article{Aoki:2006we,
    author = "Aoki, Y. and Endrodi, G. and Fodor, Z. and Katz, S. D. and Szabo, K. K.",
    title = "{The Order of the quantum chromodynamics transition predicted by the standard model of particle physics}",
    eprint = "hep-lat/0611014",
    archivePrefix = "arXiv",
    doi = "10.1038/nature05120",
    journal = "Nature",
    volume = "443",
    pages = "675--678",
    year = "2006"
}

@misc{vorticity,
    author = "Garg, Deepen and Roper Pol, Alberto and Caprini, Chiara",
    title = "{Vorticity generation in relativistic perfect fluids with application to
first-order phase transitions}",
    note = "{in preparation}",
    year = "2026"
}

@article{Barni:2024lkj,
    author = "Barni, Giulio and Blasi, Simone and Vanvlasselaer, Miguel",
    title = "{The hydrodynamics of inverse phase transitions}",
    eprint = "2406.01596",
    archivePrefix = "arXiv",
    primaryClass = "hep-ph",
    doi = "10.1088/1475-7516/2024/10/042",
    journal = "J. Cosmol. Astropart. Phys.",
    volume = "10",
    pages = "042",
    year = "2024"
}

@misc{part2,
    author = "Roper Pol, Alberto and Midiri, Antonino Salvino and Caprini, Chiara",
    title = "{Theoretical interpretation of the acoustic gravitational wave spectrum from first-order phase transitions}",
    howpublished = "{{\em In preparation}}",
    year = "2026"
}

@article{Caprini:2024hue,
    author = "Caprini, Chiara and Jinno, Ryusuke and Lewicki, Marek and Madge, Eric and Merchand, Marco and Nardini, Germano and Pieroni, Mauro and Roper Pol, Alberto and Vaskonen, Ville",
    collaboration = "LISA Cosmology Working Group",
    title = "{Gravitational waves from first-order phase transitions in LISA: reconstruction pipeline and physics interpretation}",
    eprint = "2403.03723",
    archivePrefix = "arXiv",
    primaryClass = "astro-ph.CO",
    reportNumber = "LISA-COSWG-24-01, CERN-TH-2024-029",
    doi = "10.1088/1475-7516/2024/10/020",
    journal = "J. Cosmol. Astropart. Phys.",
    volume = "10",
    pages = "020",
    year = "2024"
}

@article{Gowling:2022pzb,
    author = "Gowling, Chloe and Hindmarsh, Mark and Hooper, Deanna C. and Torrado, Jes\'us",
    title = "{Reconstructing physical parameters from template gravitational wave spectra at LISA: First order phase transitions}",
    eprint = "2209.13551",
    archivePrefix = "arXiv",
    primaryClass = "astro-ph.CO",
    doi = "10.1088/1475-7516/2023/04/061",
    journal = "J. Cosmol. Astropart. Phys.",
    volume = "04",
    pages = "061",
    year = "2023"
}

@article{Jinno:2022mie,
    author = "Jinno, Ryusuke and Konstandin, Thomas and Rubira, Henrique and Stomberg, Isak",
    title = "{Higgsless simulations of cosmological phase transitions and gravitational waves}",
    eprint = "2209.04369",
    archivePrefix = "arXiv",
    primaryClass = "astro-ph.CO",
    reportNumber = "DESY 22-148, IFT-UAM/CSIC-22-100, TUM-HEP-1416/22",
    doi = "10.1088/1475-7516/2023/02/011",
    journal = "J. Cosmol. Astropart. Phys.",
    volume = "02",
    pages = "011",
    year = "2023"
}

@article{Coleman:1977py,
    author = "Coleman, Sidney R.",
    title = "{The fate of the false vacuum. 1. Semiclassical theory}",
    reportNumber = "HUTP-77-A004",
    doi = "10.1103/PhysRevD.16.1248",
    journal = "Phys. Rev. D",
    volume = "15",
    pages = "2929",
    year = "1977",
    note = "[Erratum: Phys.~Rev.~D {\bf 16}, 1248 (1977)]"
}

@article{Kosowsky:1992rz,
    author = "Kosowsky, Arthur and Turner, Michael S. and Watkins, Richard",
    title = "{Gravitational waves from first order cosmological phase transitions}",
    reportNumber = "FERMILAB-PUB-91-333-A-REV, FERMILAB-PUB-91-333-A",
    doi = "10.1103/PhysRevLett.69.2026",
    journal = "Phys. Rev. Lett.",
    volume = "69",
    pages = "2026--2029",
    year = "1992"
}

@article{Sharma:2023mao,
    author = "Sharma, Ramkishor and Dahl, Jani and Brandenburg, Axel and Hindmarsh, Mark",
    title = "{Shallow relic gravitational wave spectrum with acoustic peak}",
    eprint = "2308.12916",
    archivePrefix = "arXiv",
    primaryClass = "gr-qc",
    reportNumber = "NORDITA-2023-051, HIP-2023-13/TH",
    doi = "10.1088/1475-7516/2023/12/042",
    journal = "J. Cosmol. Astropart. Phys.",
    volume = "12",
    pages = "042",
    year = "2023"
}

@article{Konstandin:2017sat,
    author = "Konstandin, Thomas",
    title = "{Gravitational radiation from a bulk flow model}",
    eprint = "1712.06869",
    archivePrefix = "arXiv",
    primaryClass = "astro-ph.CO",
    reportNumber = "DESY-17-227",
    doi = "10.1088/1475-7516/2018/03/047",
    journal = "J. Cosmol. Astropart. Phys.",
    volume = "03",
    pages = "047",
    year = "2018"
}

@article{Steinhardt:1981ct,
    author = "Steinhardt, Paul Joseph",
    title = "{Relativistic detonation waves and bubble growth in false vacuum decay}",
    reportNumber = "UPR-0181T",
    doi = "10.1103/PhysRevD.25.2074",
    journal = "Phys. Rev. D",
    volume = "25",
    pages = "2074",
    year = "1982"
}

@article{Bodeker:2009qy,
    author = "Bodeker, Dietrich and Moore, Guy D.",
    title = "{Can electroweak bubble walls run away?}",
    eprint = "0903.4099",
    archivePrefix = "arXiv",
    primaryClass = "hep-ph",
    doi = "10.1088/1475-7516/2009/05/009",
    journal = "J. Cosmol. Astropart. Phys.",
    volume = "05",
    pages = "009",
    year = "2009"
}

@article{Bodeker:2017cim,
    author = "Bodeker, Dietrich and Moore, Guy D.",
    title = "{Electroweak bubble wall speed limit}",
    eprint = "1703.08215",
    archivePrefix = "arXiv",
    primaryClass = "hep-ph",
    doi = "10.1088/1475-7516/2017/05/025",
    journal = "J. Cosmol. Astropart. Phys.",
    volume = "05",
    pages = "025",
    year = "2017"
}

@article{Huber:2008hg,
    author = "Huber, Stephan J. and Konstandin, Thomas",
    title = "{Gravitational wave production by collisions: More bubbles}",
    eprint = "0806.1828",
    archivePrefix = "arXiv",
    primaryClass = "hep-ph",
    doi = "10.1088/1475-7516/2008/09/022",
    journal = "J. Cosmol. Astropart. Phys.",
    volume = "09",
    pages = "022",
    year = "2008"
}

@article{Cutting:2018tjt,
    author = "Cutting, Daniel and Hindmarsh, Mark and Weir, David J.",
    title = "{Gravitational waves from vacuum first-order phase transitions: From the envelope to the lattice}",
    eprint = "1802.05712",
    archivePrefix = "arXiv",
    primaryClass = "astro-ph.CO",
    reportNumber = "HIP-2018-4-TH",
    doi = "10.1103/PhysRevD.97.123513",
    journal = "Phys. Rev. D",
    volume = "97",
    number = "12",
    pages = "123513",
    year = "2018"
}

@article{Enqvist:1991xw,
    author = "Enqvist, K. and Ignatius, J. and Kajantie, K. and Rummukainen, K.",
    title = "{Nucleation and bubble growth in a first order cosmological electroweak phase transition}",
    reportNumber = "HU-TFT-91-35",
    doi = "10.1103/PhysRevD.45.3415",
    journal = "Phys. Rev. D",
    volume = "45",
    pages = "3415",
    year = "1992"
}

@article{RoperPol:2022iel,
    author = "Roper Pol, Alberto and Caprini, Chiara and Neronov, Andrii and Semikoz, Dmitri",
    title = "{Gravitational wave signal from primordial magnetic fields in the Pulsar Timing Array frequency band}",
    eprint = "2201.05630",
    archivePrefix = "arXiv",
    primaryClass = "astro-ph.CO",
    doi = "10.1103/PhysRevD.105.123502",
    journal = "Phys. Rev. D",
    volume = "105",
    number = "12",
    pages = "123502",
    year = "2022"
}

@book{Rezzolla:2013dea,
    author = "Rezzolla, Luciano and Zanotti, Olindo",
    title = "{Relativistic Hydrodynamics}",
    doi = "10.1093/acprof:oso/9780198528906.001.0001",
    isbn = "978-0-19-174650-5, 978-0-19-852890-6",
    publisher = "Oxford University Press",
    month = "9",
    year = "2013"
}

@inproceedings{Stomberg:2025kxf,
    author = "Stomberg, Isak and Roper Pol, Alberto",
    title = "{Gravitational wave spectra for cosmological phase transitions with non-linear decay of the fluid motion}",
    booktitle = "{59th Rencontres de Moriond on Gravitation}: {Moriond 2025 Gravitation}",
    eprint = "2508.04263",
    archivePrefix = "arXiv",
    primaryClass = "gr-qc",
    month = "8",
    year = "2025"
}

@article{Caprini:2024gyk,
    author = "Caprini, Chiara and Jinno, Ryusuke and Konstandin, Thomas and Roper Pol, Alberto and Rubira, Henrique and Stomberg, Isak",
    title = "{Gravitational waves from first-order phase transitions: from weak to strong}",
    eprint = "2409.03651",
    archivePrefix = "arXiv",
    primaryClass = "gr-qc",
    doi = "10.1007/JHEP07(2025)217",
    journal = "J. High Energy Phys.",
    volume = "07",
    pages = "217",
    year = "2025"
}

@article{Brandenburg:2017neh,
    author = "Brandenburg, Axel and Kahniashvili, Tina and Mandal, Sayan and Roper Pol, Alberto and Tevzadze, Alexander G. and Vachaspati, Tanmay",
    title = "{Evolution of hydromagnetic turbulence from the electroweak phase transition}",
    eprint = "1711.03804",
    archivePrefix = "arXiv",
    primaryClass = "astro-ph.CO",
    reportNumber = "NORDITA-2017-116",
    doi = "10.1103/PhysRevD.96.123528",
    journal = "Phys. Rev. D",
    volume = "96",
    number = "12",
    pages = "123528",
    year = "2017"
}

@article{NANOGrav:2021flc,
    author = "Arzoumanian, Zaven and others",
    collaboration = "NANOGrav",
    title = "{Searching for Gravitational Waves from Cosmological Phase Transitions with the NANOGrav 12.5-Year Dataset}",
    eprint = "2104.13930",
    archivePrefix = "arXiv",
    primaryClass = "astro-ph.CO",
    doi = "10.1103/PhysRevLett.127.251302",
    journal = "Phys. Rev. Lett.",
    volume = "127",
    number = "25",
    pages = "251302",
    year = "2021"
}

@article{Brandenburg:2018ptt,
    author = "Brandenburg, Axel and Durrer, Ruth and Kahniashvili, Tina and Mandal, Sayan and Yin, Weichen Winston",
    title = "{Statistical Properties of Scale-Invariant Helical Magnetic Fields and Applications to Cosmology}",
    eprint = "1804.01177",
    archivePrefix = "arXiv",
    primaryClass = "astro-ph.CO",
    reportNumber = "NORDITA-2018-024",
    doi = "10.1088/1475-7516/2018/08/034",
    journal = "J. Cosmol. Astropart. Phys.",
    volume = "08",
    pages = "034",
    year = "2018"
}

@article{Jinno:2020eqg,
    author = "Jinno, Ryusuke and Konstandin, Thomas and Rubira, Henrique",
    title = "{A hybrid simulation of gravitational wave production in first-order phase transitions}",
    eprint = "2010.00971",
    archivePrefix = "arXiv",
    primaryClass = "astro-ph.CO",
    reportNumber = "DESY-20-170, DESY 20-170",
    doi = "10.1088/1475-7516/2021/04/014",
    journal = "J. Cosmol. Astropart. Phys.",
    volume = "04",
    pages = "014",
    year = "2021"
}

@article{Caprini:2015zlo,
    author = "Caprini, Chiara and others",
    title = "{Science with the space-based interferometer eLISA. II: Gravitational waves from cosmological phase transitions}",
    eprint = "1512.06239",
    archivePrefix = "arXiv",
    primaryClass = "astro-ph.CO",
    reportNumber = "DESY-15-246",
    doi = "10.1088/1475-7516/2016/04/001",
    journal = "J. Cosmol. Astropart. Phys.",
    volume = "04",
    pages = "001",
    year = "2016"
}

@article{Espinosa:2010hh,
    author = "Espinosa, Jose R. and Konstandin, Thomas and No, Jose M. and Servant, Geraldine",
    title = "{Energy budget of cosmological first-order phase transitions}",
    eprint = "1004.4187",
    archivePrefix = "arXiv",
    primaryClass = "hep-ph",
    reportNumber = "CERN-PH-TH-2010-027",
    doi = "10.1088/1475-7516/2010/06/028",
    journal = "J. Cosmol. Astropart. Phys.",
    volume = "06",
    pages = "028",
    year = "2010"
}

@article{Kahniashvili:2005qi,
    author = "Kahniashvili, Tina and Gogoberidze, Grigol and Ratra, Bharat",
    title = "{Polarized cosmological gravitational waves from primordial helical turbulence}",
    eprint = "astro-ph/0505628",
    archivePrefix = "arXiv",
    doi = "10.1103/PhysRevLett.95.151301",
    journal = "Phys. Rev. Lett.",
    volume = "95",
    pages = "151301",
    year = "2005"
}

@article{Caprini:2019egz,
    author = "Caprini, Chiara and others",
    title = "{Detecting gravitational waves from cosmological phase transitions with LISA: An update}",
    eprint = "1910.13125",
    archivePrefix = "arXiv",
    primaryClass = "astro-ph.CO",
    reportNumber = "DESY-19-159, IPPP/19/27, HIP-2019-14/TH, MITP/19-066, IFT-UAM/CSIC-19-139",
    doi = "10.1088/1475-7516/2020/03/024",
    journal = "J. Cosmol. Astropart. Phys.",
    volume = "03",
    pages = "024",
    year = "2020"
}

@article{Niksa:2018ofa,
    author = {Niksa, Peter and Schlederer, Martin and Sigl, G\"unter},
    title = "{Gravitational waves produced by compressible MHD turbulence from cosmological phase transitions}",
    eprint = "1803.02271",
    archivePrefix = "arXiv",
    primaryClass = "astro-ph.CO",
    doi = "10.1088/1361-6382/aac89c",
    journal = "Classical Quantum Gravity",
    volume = "35",
    number = "14",
    pages = "144001",
    year = "2018"
}

@article{Brandenburg:2021bvg,
    author = "Brandenburg, Axel and Gogoberidze, Grigol and Kahniashvili, Tina and Mandal, Sayan and Roper Pol, Alberto and Shenoy, Nakul",
    title = "{The scalar, vector, and tensor modes in gravitational wave turbulence simulations}",
    eprint = "2103.01140",
    archivePrefix = "arXiv",
    primaryClass = "gr-qc",
    reportNumber = "NORDITA-2021-019",
    doi = "10.1088/1361-6382/ac011c",
    journal = "Classical Quantum Gravity",
    volume = "38",
    number = "14",
    pages = "145002",
    year = "2021"
}

@article{Kahniashvili:2020jgm,
    author = "Kahniashvili, Tina and Brandenburg, Axel and Gogoberidze, Grigol and Mandal, Sayan and Roper Pol, Alberto",
    title = "{Circular polarization of gravitational waves from early-Universe helical turbulence}",
    eprint = "2011.05556",
    archivePrefix = "arXiv",
    primaryClass = "astro-ph.CO",
    reportNumber = "NORDITA-2020-102",
    doi = "10.1103/PhysRevResearch.3.013193",
    journal = "Phys. Rev. Res.",
    volume = "3",
    number = "1",
    pages = "013193",
    year = "2021"
}

@article{RoperPol:2019wvy,
    author = "Roper Pol, Alberto and Mandal, Sayan and Brandenburg, Axel and Kahniashvili, Tina and Kosowsky, Arthur",
    title = "{Numerical simulations of gravitational waves from early-universe turbulence}",
    eprint = "1903.08585",
    archivePrefix = "arXiv",
    primaryClass = "astro-ph.CO",
    reportNumber = "NORDITA-2019-024",
    doi = "10.1103/PhysRevD.102.083512",
    journal = "Phys. Rev. D",
    volume = "102",
    number = "8",
    pages = "083512",
    year = "2020"
}

@article{RoperPol:2018sap,
    author = "Roper Pol, Alberto and Brandenburg, Axel and Kahniashvili, Tina and Kosowsky, Arthur and Mandal, Sayan",
    title = "{The timestep constraint in solving the gravitational wave equations sourced by hydromagnetic turbulence}",
    eprint = "1807.05479",
    archivePrefix = "arXiv",
    primaryClass = "physics.flu-dyn",
    reportNumber = "NORDITA-2018-054",
    doi = "10.1080/03091929.2019.1653460",
    journal = "Geophys. Astrophys. Fluid Dyn.",
    volume = "114",
    number = "1-2",
    pages = "130",
    year = "2020"
}

@article{Durrer:2013pga,
    author = "Durrer, Ruth and Neronov, Andrii",
    title = "{Cosmological Magnetic Fields: Their Generation, Evolution and Observation}",
    eprint = "1303.7121",
    archivePrefix = "arXiv",
    primaryClass = "astro-ph.CO",
    doi = "10.1007/s00159-013-0062-7",
    journal = "Astron. Astrophys. Rev.",
    volume = "21",
    pages = "62",
    year = "2013"
}

@article{EPTA:2023fyk,
    author = "Antoniadis, J. and others",
    collaboration = "EPTA-InPTA Collaborations",
    title = "{The second data release from the European Pulsar Timing Array III. Search for gravitational wave signals}",
    eprint = "2306.16214",
    archivePrefix = "arXiv",
    primaryClass = "astro-ph.HE",
    doi = "10.1051/0004-6361/202346844",
    journal = "Astron. Astrophys.",
    volume = "678",
    pages = "A50",
    year = "2023"
}

@article{Reardon:2023gzh,
    author = "Reardon, Daniel J. and others",
    title = "{Search for an isotropic gravitational-wave background with the Parkes Pulsar Timing Array}",
    eprint = "2306.16215",
    archivePrefix = "arXiv",
    primaryClass = "astro-ph.HE",
    doi = "10.3847/2041-8213/acdd02",
    journal = "Astrophys. J. Lett.",
    volume = "951",
    number = "1",
    pages = "L6",
    year = "2023"
}

@article{EPTA:2023xxk,
    author = "Antoniadis, J. and others",
    collaboration = "EPTA, InPTA",
    title = "{The second data release from the European Pulsar Timing Array - IV. Implications for massive black holes, dark matter, and the early Universe}",
    eprint = "2306.16227",
    archivePrefix = "arXiv",
    primaryClass = "astro-ph.CO",
    doi = "10.1051/0004-6361/202347433",
    journal = "Astron. Astrophys.",
    volume = "685",
    pages = "A94",
    year = "2024"
}

@article{NANOGrav:2023hvm,
    author = "Afzal, Adeela and others",
    collaboration = "NANOGrav Collaboration",
    title = "{The NANOGrav 15 yr data set: Search for signals from new physics}",
    eprint = "2306.16219",
    archivePrefix = "arXiv",
    primaryClass = "astro-ph.HE",
    doi = "10.3847/2041-8213/acdc91",
    journal = "Astrophys. J. Lett.",
    volume = "951",
    number = "1",
    pages = "L11",
    year = "2023"
}

@article{NANOGrav:2023gor,
    author = "Agazie, Gabriella and others",
    collaboration = "NANOGrav Collaboration",
    title = "{The NANOGrav 15 yr data set: Evidence for a gravitational-wave background}",
    eprint = "2306.16213",
    archivePrefix = "arXiv",
    primaryClass = "astro-ph.HE",
    doi = "10.3847/2041-8213/acdac6",
    journal = "Astrophys. J. Lett.",
    volume = "951",
    number = "1",
    pages = "L8",
    year = "2023"
}

@article{Wygas:2018otj,
    author = {Wygas, Mandy M. and Oldengott, Isabel M. and B\"odeker, Dietrich and Schwarz, Dominik J.},
    title = "{Cosmic QCD epoch at nonvanishing lepton asymmetry}",
    eprint = "1807.10815",
    archivePrefix = "arXiv",
    primaryClass = "hep-ph",
    doi = "10.1103/PhysRevLett.121.201302",
    journal = "Phys. Rev. Lett.",
    volume = "121",
    number = "20",
    pages = "201302",
    year = "2018"
}

@article{Schwarz:2009ii,
    author = "Schwarz, Dominik J. and Stuke, Maik",
    title = "{Lepton asymmetry and the cosmic QCD transition}",
    eprint = "0906.3434",
    archivePrefix = "arXiv",
    primaryClass = "hep-ph",
    reportNumber = "BI-TP-2009-14",
    doi = "10.1088/1475-7516/2009/11/025",
    journal = "J. Cosmol. Astropart. Phys.",
    volume = "11",
    pages = "025",
    year = "2009",
    note = "[Erratum: J. Cosmol. Astropart. Phys. {\bf 10}, E01 (2010)]"
}

@article{Stephanov:2006dn,
    author = "Stephanov, M. A.",
    title = "{QCD critical point and complex chemical potential singularities}",
    eprint = "hep-lat/0603014",
    archivePrefix = "arXiv",
    doi = "10.1103/PhysRevD.73.094508",
    journal = "Phys. Rev. D",
    volume = "73",
    pages = "094508",
    year = "2006"
}

@article{Kajantie:1996qd,
    author = "Kajantie, K. and Laine, M. and Rummukainen, K. and Shaposhnikov, Mikhail E.",
    title = "{A nonperturbative analysis of the finite T phase transition in SU(2) x U(1) electroweak theory}",
    eprint = "hep-lat/9612006",
    archivePrefix = "arXiv",
    reportNumber = "BI-TP-96-54, CERN-TH-96-334A, HD-THEP-96-48",
    doi = "10.1016/S0550-3213(97)00164-8",
    journal = "Nucl. Phys. B",
    volume = "493",
    pages = "413",
    year = "1997"
}

@article{Dahl:2021wyk,
    author = "Dahl, Jani and Hindmarsh, Mark and Rummukainen, Kari and Weir, David J.",
    title = "{Decay of acoustic turbulence in two dimensions and implications for cosmological gravitational waves}",
    eprint = "2112.12013",
    archivePrefix = "arXiv",
    primaryClass = "gr-qc",
    reportNumber = "HIP-2021-29/TH",
    doi = "10.1103/PhysRevD.106.063511",
    journal = "Phys. Rev. D",
    volume = "106",
    number = "6",
    pages = "063511",
    year = "2022"
}

@article{Caprini:2009fx,
    author = "Caprini, Chiara and Durrer, Ruth and Konstandin, Thomas and Servant, Geraldine",
    title = "{General properties of the gravitational wave spectrum from phase transitions}",
    eprint = "0901.1661",
    archivePrefix = "arXiv",
    primaryClass = "astro-ph.CO",
    doi = "10.1103/PhysRevD.79.083519",
    journal = "Phys. Rev. D",
    volume = "79",
    pages = "083519",
    year = "2009"
}

@article{Caprini:2018mtu,
    author = "Caprini, Chiara and Figueroa, Daniel G.",
    title = "{Cosmological backgrounds of gravitational waves}",
    eprint = "1801.04268",
    archivePrefix = "arXiv",
    primaryClass = "astro-ph.CO",
    doi = "10.1088/1361-6382/aac608",
    journal = "Classical Quantum Gravity",
    volume = "35",
    number = "16",
    pages = "163001",
    year = "2018"
}

@article{Hogan:1986qda,
    author = "Hogan, C. J.",
    title = "{Gravitational radiation from cosmological phase transitions}",
    journal = "Mon. Not. R. Astron. Soc.",
    volume = "218",
    pages = "629",
    year = "1986"
}

@article{Witten:1984rs,
    author = "Witten, Edward",
    title = "{Cosmic separation of phases}",
    reportNumber = "PRINT-84-0400 (IAS,PRINCETON)",
    doi = "10.1103/PhysRevD.30.272",
    journal = "Phys. Rev. D",
    volume = "30",
    pages = "272",
    year = "1984"
}

@article{Auclair:2022jod,
    author = "Auclair, Pierre and Caprini, Chiara and Cutting, Daniel and Hindmarsh, Mark and Rummukainen, Kari and Steer, Dani\`ele A. and Weir, David J.",
    title = "{Generation of gravitational waves from freely decaying turbulence}",
    eprint = "2205.02588",
    archivePrefix = "arXiv",
    primaryClass = "astro-ph.CO",
    reportNumber = "HIP-2021-35/TH",
    doi = "10.1088/1475-7516/2022/09/029",
    journal = "J. Cosmol. Astropart. Phys.",
    volume = "09",
    pages = "029",
    year = "2022"
}

@book{MY75,
    author = "Monin, A. S. and Yaglom, A. M.",
    title = "{Statistical fluid mechanics:
    Mechanics of turbulence}",
    volume = "2",
    address = "MIT press, Cambridge, MA, USA.",
    year = "1975"
}

@article{Hindmarsh:2020hop,
    author = {Hindmarsh, Mark B. and L\"uben, Marvin and Lumma, Johannes and Pauly, Martin},
    title = "{Phase transitions in the early universe}",
    eprint = "2008.09136",
    archivePrefix = "arXiv",
    primaryClass = "astro-ph.CO",
    reportNumber = "MPP-2020-163, HIP-2020-27/TH",
    doi = "10.21468/SciPostPhysLectNotes.24",
    journal = "SciPost Phys. Lect. Notes",
    volume = "24",
    pages = "1",
    year = "2021"
}

@article{Caprini:2009yp,
    author = "Caprini, Chiara and Durrer, Ruth and Servant, Geraldine",
    title = "{The stochastic gravitational wave background from turbulence and magnetic fields generated by a first-order phase transition}",
    eprint = "0909.0622",
    archivePrefix = "arXiv",
    primaryClass = "astro-ph.CO",
    doi = "10.1088/1475-7516/2009/12/024",
    journal = "J. Cosmol. Astropart. Phys.",
    volume = "12",
    pages = "024",
    year = "2009"
}

@article{RoperPol:2023dzg,
    author = "Roper Pol, Alberto and Procacci, Simona and Caprini, Chiara",
    title = "{Characterization of the gravitational wave spectrum from sound waves within the sound shell model}",
    eprint = "2308.12943",
    archivePrefix = "arXiv",
    primaryClass = "gr-qc",
    doi = "10.1103/PhysRevD.109.063531",
    journal = "Phys. Rev. D",
    volume = "109",
    number = "6",
    pages = "063531",
    year = "2024"
}

@article{Chodos:1974je,
    author = "Chodos, A. and Jaffe, R. L. and Johnson, K. and Thorn, Charles B. and Weisskopf, V. F.",
    title = "{A New Extended Model of Hadrons}",
    reportNumber = "MIT-CTP-387-REV, MIT-CTP-387",
    doi = "10.1103/PhysRevD.9.3471",
    journal = "Phys. Rev. D",
    volume = "9",
    pages = "3471--3495",
    year = "1974"
}

@article{Giese:2020rtr,
    author = "Giese, Felix and Konstandin, Thomas and van de Vis, Jorinde",
    title = "{Model-independent energy budget of cosmological first-order phase transitions\textemdash{}A sound argument to go beyond the bag model}",
    eprint = "2004.06995",
    archivePrefix = "arXiv",
    primaryClass = "astro-ph.CO",
    reportNumber = "DESY-20-064",
    doi = "10.1088/1475-7516/2020/07/057",
    journal = "J. Cosmol. Astropart. Phys.",
    volume = "07",
    number = "07",
    pages = "057",
    year = "2020"
}

@article{Giese:2020znk,
    author = "Giese, Felix and Konstandin, Thomas and Schmitz, Kai and van de Vis, Jorinde",
    title = "{Model-independent energy budget for LISA}",
    eprint = "2010.09744",
    archivePrefix = "arXiv",
    primaryClass = "astro-ph.CO",
    reportNumber = "DESY-20-173, DESY 20-173, CERN-TH-2020-170",
    doi = "10.1088/1475-7516/2021/01/072",
    journal = "J. Cosmol. Astropart. Phys.",
    volume = "01",
    pages = "072",
    year = "2021"
}

@article{Tian:2024ysd,
    author = "Tian, Chi and Wang, Xiao and Bal{\'a}zs, Csaba",
    title = "{Gravitational waves from cosmological first-order phase transitions with precise hydrodynamics}",
    eprint = "2409.14505",
    archivePrefix = "arXiv",
    primaryClass = "hep-ph",
    doi = "10.1140/epjc/s10052-025-14826-2",
    journal = "Eur. Phys. J. C",
    volume = "85",
    number = "10",
    pages = "1091",
    year = "2025"
}

@article{Tian:2025zlo,
    author = "Tian, Chi and Wang, Xiao and Bal{\'a}zs, Csaba",
    title = "{DeepSSM: an emulator of gravitational wave spectra from sound waves during cosmological first-order phase transitions}",
    eprint = "2501.10244",
    archivePrefix = "arXiv",
    primaryClass = "astro-ph.CO",
    doi = "10.1088/1475-7516/2025/08/060",
    journal = "J. Cosmol. Astropart. Phys.",
    volume = "08",
    pages = "060",
    year = "2025"
}

@article{LIGOScientific:2025kry,
    author = "Abac, A. G. and others",
    collaboration = "LIGO Scientific, VIRGO, KAGRA",
    title = "{Cosmological and High Energy Physics implications from gravitational-wave background searches in LIGO-Virgo-KAGRA's O1-O4a runs}",
    eprint = "2510.26848",
    archivePrefix = "arXiv",
    primaryClass = "gr-qc",
    reportNumber = "LIGO-PP2500150",
    month = "10",
    year = "2025"
}

@article{Cai:2023guc,
    author = "Cai, Rong-Gen and Wang, Shao-Jiang and Yuwen, Zi-Yan",
    title = "{Hydrodynamic sound shell model}",
    eprint = "2305.00074",
    archivePrefix = "arXiv",
    primaryClass = "gr-qc",
    doi = "10.1103/PhysRevD.108.L021502",
    journal = "Phys. Rev. D",
    volume = "108",
    number = "2",
    pages = "L021502",
    year = "2023"
}

@article{Ignatius:1993qn,
    author = "Ignatius, J. and Kajantie, K. and Kurki-Suonio, H. and Laine, M.",
    title = "{The growth of bubbles in cosmological phase transitions}",
    eprint = "astro-ph/9309059",
    archivePrefix = "arXiv",
    reportNumber = "HU-TFT-93-43",
    doi = "10.1103/PhysRevD.49.3854",
    journal = "Phys. Rev. D",
    volume = "49",
    pages = "3854",
    year = "1994"
}

@article{Giese:2021dnw,
    author = "Giese, Felix and Konstandin, Thomas and van de Vis, Jorinde",
    title = "{Finding sound shells in LISA mock data using likelihood sampling}",
    eprint = "2107.06275",
    archivePrefix = "arXiv",
    primaryClass = "astro-ph.CO",
    reportNumber = "DESY-21-109, DESY 2021-02966",
    doi = "10.1088/1475-7516/2021/11/002",
    journal = "J. Cosmol. Astropart. Phys.",
    volume = "11",
    pages = "002",
    year = "2021"
}

@article{Xu:2023wog,
    author = "Xu, Heng and others",
    title = "{Searching for the nano-Hertz stochastic gravitational wave background with the Chinese Pulsar Timing Array data release I}",
    eprint = "2306.16216",
    archivePrefix = "arXiv",
    primaryClass = "astro-ph.HE",
    doi = "10.1088/1674-4527/acdfa5",
    journal = "Res. Astron. Astrophys.",
    volume = "23",
    number = "7",
    pages = "075024",
    year = "2023"
}

@article{Madge:2023dxc,
    author = "Madge, Eric and Morgante, Enrico and Puchades-Ib\'a\~nez, Cristina and Ramberg, Nicklas and Ratzinger, Wolfram and Schenk, Sebastian and Schwaller, Pedro",
    title = "{Primordial gravitational waves in the nano-Hertz regime and PTA data \textemdash{} towards solving the GW inverse problem}",
    eprint = "2306.14856",
    archivePrefix = "arXiv",
    primaryClass = "hep-ph",
    reportNumber = "MITP-23-029",
    doi = "10.1007/JHEP10(2023)171",
    journal = "J. High Energy Phys.",
    volume = "10",
    pages = "171",
    year = "2023"
}

@article{Figueroa:2023zhu,
    author = "Figueroa, Daniel G. and Pieroni, Mauro and Ricciardone, Angelo and Simakachorn, Peera",
    title = "{Cosmological Background Interpretation of Pulsar Timing Array Data}",
    eprint = "2307.02399",
    archivePrefix = "arXiv",
    primaryClass = "astro-ph.CO",
    reportNumber = "CERN-TH-2023-132",
    doi = "10.1103/PhysRevLett.132.171002",
    journal = "Phys. Rev. Lett.",
    volume = "132",
    number = "17",
    pages = "171002",
    year = "2024"
}

@article{Pen:2015qta,
    author = "Pen, Ue-Li and Turok, Neil",
    title = "{Shocks in the early universe}",
    eprint = "1510.02985",
    archivePrefix = "arXiv",
    primaryClass = "astro-ph.CO",
    doi = "10.1103/PhysRevLett.117.131301",
    journal = "Phys. Rev. Lett.",
    volume = "117",
    number = "13",
    pages = "131301",
    year = "2016"
}

@article{Kahniashvili:2008er,
    author = "Kahniashvili, Tina and Gogoberidze, Grigol and Ratra, Bharat",
    title = "{Gravitational Radiation from Primordial Helical MHD Turbulence}",
    eprint = "0802.3524",
    archivePrefix = "arXiv",
    primaryClass = "astro-ph",
    reportNumber = "KSUPT\_08-1",
    doi = "10.1103/PhysRevLett.100.231301",
    journal = "Phys. Rev. Lett.",
    volume = "100",
    pages = "231301",
    year = "2008"
}

@article{Caprini:2007xq,
    author = "Caprini, Chiara and Durrer, Ruth and Servant, Geraldine",
    title = "{Gravitational wave generation from bubble collisions in first-order phase transitions: An analytic approach}",
    eprint = "0711.2593",
    archivePrefix = "arXiv",
    primaryClass = "astro-ph",
    reportNumber = "CERN-PH-TH-2007-206, SACLAY-T07-142",
    doi = "10.1103/PhysRevD.77.124015",
    journal = "Phys. Rev. D",
    volume = "77",
    pages = "124015",
    year = "2008"
}

@article{Caprini:2006jb,
    author = "Caprini, Chiara and Durrer, Ruth",
    title = "{Gravitational waves from stochastic relativistic sources: Primordial turbulence and magnetic fields}",
    eprint = "astro-ph/0603476",
    archivePrefix = "arXiv",
    doi = "10.1103/PhysRevD.74.063521",
    journal = "Phys. Rev. D",
    volume = "74",
    pages = "063521",
    year = "2006"
}

@article{Nicolis:2003tg,
    author = "Nicolis, Alberto",
    title = "{Relic gravitational waves from colliding bubbles and cosmic turbulence}",
    eprint = "gr-qc/0303084",
    archivePrefix = "arXiv",
    reportNumber = "IEM-FT-230-03",
    doi = "10.1088/0264-9381/21/4/L05",
    journal = "Classical Quantum Gravity",
    volume = "21",
    pages = "L27",
    year = "2004"
}

@article{Dolgov:2002ra,
    author = "Dolgov, Alexander D. and Grasso, Dario and Nicolis, Alberto",
    title = "{Relic backgrounds of gravitational waves from cosmic turbulence}",
    eprint = "astro-ph/0206461",
    archivePrefix = "arXiv",
    doi = "10.1103/PhysRevD.66.103505",
    journal = "Phys. Rev. D",
    volume = "66",
    pages = "103505",
    year = "2002"
}

@article{Kosowsky:2001xp,
    author = "Kosowsky, Arthur and Mack, Andrew and Kahniashvili, Tinatin",
    title = "{Gravitational radiation from cosmological turbulence}",
    eprint = "astro-ph/0111483",
    archivePrefix = "arXiv",
    reportNumber = "RAP-334",
    doi = "10.1103/PhysRevD.66.024030",
    journal = "Phys. Rev. D",
    volume = "66",
    pages = "024030",
    year = "2002"
}

@article{Kamionkowski:1993fg,
    author = "Kamionkowski, Marc and Kosowsky, Arthur and Turner, Michael S.",
    title = "{Gravitational radiation from first order phase transitions}",
    eprint = "astro-ph/9310044",
    archivePrefix = "arXiv",
    reportNumber = "IASSNS-HEP-93-44, FERMILAB-PUB-93-235-A",
    doi = "10.1103/PhysRevD.49.2837",
    journal = "Phys. Rev. D",
    volume = "49",
    pages = "2837--2851",
    year = "1994"
}

@article{Gowling:2021gcy,
    author = "Gowling, Chloe and Hindmarsh, Mark",
    title = "{Observational prospects for phase transitions at LISA: Fisher matrix analysis}",
    eprint = "2106.05984",
    archivePrefix = "arXiv",
    primaryClass = "astro-ph.CO",
    doi = "10.1088/1475-7516/2021/10/039",
    journal = "J. Cosmol. Astropart. Phys.",
    volume = "10",
    pages = "039",
    year = "2021"
}

@article{CosmoGW_GH,
    author = "Roper Pol, Alberto",
    publisher = {GitHub},
    year = {2024},
    journal = {GitHub repository},
    doi = "10.5281/zenodo.6045844",
    howpublished = {\url{https://github.com/CosmoGW/CosmoGW}},
    title = "{{\sc CosmoGW}}",
}

@book{Mathai2009TheHT,
  author    = {A. M. Mathai and Ram Kishore Saxena and Hans J. Haubold},
  title     = {The H-Function: Theory and Applications},
  publisher = {Springer Science+Business Media},
  address   = {New York, NY},
  year      = {2009},
  isbn      = {978-1-4419-0916-9},
  doi       = {10.1007/978-1-4419-0916-9}
}

@misc{cosmogw_manual,
    author = "Roper Pol, Alberto and Gurgenidze, Murman and Midiri, Antonino Salvino
    and Puro, Touko and Salom\'e, Madeline and Stomberg, Isak",
    title = "{{\sc CosmoGW v1.0}: {A} public library for cosmological gravitational wave backgrounds. {Part I --- Sound waves and MHD turbulence}}",
    note = "{in preparation}",
    year = "2026"
}

@article{Athron:2023xlk,
    author = "Athron, Peter and Bal{\'a}zs, Csaba and Fowlie, Andrew and Morris, Lachlan and Wu, Lei",
    title = "{Cosmological phase transitions: From perturbative particle physics to gravitational waves}",
    eprint = "2305.02357",
    archivePrefix = "arXiv",
    primaryClass = "hep-ph",
    doi = "10.1016/j.ppnp.2023.104094",
    journal = "Prog. Part. Nucl. Phys.",
    volume = "135",
    pages = "104094",
    year = "2024"
}

@article{Deryagin:1986qq,
    author = "Deryagin, D. V. and Grigoriev, Dmitri Yu. and Rubakov, V. A. and Sazhin, M. V.",
    title = "{Possible Anisotropic Phases in the Early Universe and Gravitational Wave Background}",
    doi = "10.1142/S0217732386000750",
    journal = "Mod. Phys. Lett. A",
    volume = "1",
    pages = "593--600",
    year = "1986"
}

@article{Kosowsky:1991ua,
    author = "Kosowsky, Arthur and Turner, Michael S. and Watkins, Richard",
    title = "{Gravitational radiation from colliding vacuum bubbles}",
    reportNumber = "FERMILAB-PUB-91-323-A",
    doi = "10.1103/PhysRevD.45.4514",
    journal = "Phys. Rev. D",
    volume = "45",
    pages = "4514",
    year = "1992"
}

@article{Kosowsky:1992vn,
    author = "Kosowsky, Arthur and Turner, Michael S.",
    title = "{Gravitational radiation from colliding vacuum bubbles: Envelope approximation to many bubble collisions}",
    eprint = "astro-ph/9211004",
    archivePrefix = "arXiv",
    reportNumber = "FERMILAB-PUB-92-295-A",
    doi = "10.1103/PhysRevD.47.4372",
    journal = "Phys. Rev. D",
    volume = "47",
    pages = "4372",
    year = "1993"
}

@article{Neronov:2020qrl,
    author = "Neronov, Andrii and Roper Pol, Alberto and Caprini, Chiara and Semikoz, Dmitri",
    title = "{NANOGrav signal from magnetohydrodynamic turbulence at the QCD phase transition in the early Universe}",
    eprint = "2009.14174",
    archivePrefix = "arXiv",
    primaryClass = "astro-ph.CO",
    doi = "10.1103/PhysRevD.103.L041302",
    journal = "Phys. Rev. D",
    volume = "103",
    number = "4",
    pages = "041302",
    year = "2021"
}

@article{LISA:2017pwj,
    author = "Amaro-Seoane, Pau and others",
    collaboration = "LISA Collaboration",
    title = "{Laser Interferometer Space Antenna}",
    eprint = "1702.00786",
    archivePrefix = "arXiv",
    primaryClass = "astro-ph.IM",
    month = "2",
}

@article{RoperPol:2021xnd,
    author = "Roper Pol, Alberto and Mandal, Sayan and Brandenburg, Axel and Kahniashvili, Tina",
    title = "{Polarization of gravitational waves from helical MHD turbulent sources}",
    eprint = "2107.05356",
    archivePrefix = "arXiv",
    primaryClass = "gr-qc",
    reportNumber = "NORDITA-2021-062",
    doi = "10.1088/1475-7516/2022/04/019",
    journal = "J. Cosmol. Astropart. Phys.",
    volume = "04",
    number = "04",
    pages = "019",
    year = "2022"
}

@article{Caprini:2003vc,
    author = "Caprini, Chiara and Durrer, Ruth and Kahniashvili, Tina",
    title = "{The cosmic microwave background and helical magnetic fields: The tensor mode}",
    eprint = "astro-ph/0304556",
    archivePrefix = "arXiv",
    doi = "10.1103/PhysRevD.69.063006",
    journal = "Phys. Rev. D",
    volume = "69",
    pages = "063006",
    year = "2004"
}

@BOOK{2008Kundu,
       author = {{Kundu}, Pijush K. and {Cohen}, Ira M.},
        title = "{Fluid Mechanics: Fourth Edition}",
         year = 2008,
       adsurl = {https://ui.adsabs.harvard.edu/abs/2008flme.book.....K}
}

@article{Brandenburg:2016odr,
    author = "Brandenburg, Axel and Kahniashvili, Tina",
    title = "{Classes of hydrodynamic and magnetohydrodynamic turbulent decay}",
    eprint = "1607.01360",
    archivePrefix = "arXiv",
    primaryClass = "physics.flu-dyn",
    reportNumber = "NORDITA-2016-82",
    doi = "10.1103/PhysRevLett.118.055102",
    journal = "Phys. Rev. Lett.",
    volume = "118",
    number = "5",
    pages = "055102",
    year = "2017"
}

@article{Durrer:2003ja,
    author = "Durrer, Ruth and Caprini, Chiara",
    title = "{Primordial magnetic fields and causality}",
    eprint = "astro-ph/0305059",
    archivePrefix = "arXiv",
    doi = "10.1088/1475-7516/2003/11/010",
    journal = "J. Cosmol. Astropart. Phys.",
    volume = "11",
    pages = "010",
    year = "2003"
}

@article{Linde:1981zj,
    author = "Linde, Andrei D.",
    title = "{Decay of the false vacuum at finite temperature}",
    reportNumber = "LEBEDEV-81-265",
    doi = "10.1016/0550-3213(83)90072-X",
    journal = "Nucl. Phys. B",
    volume = "216",
    pages = "421",
    year = "1983",
    note = "[Erratum: Nucl.~Phys.~B {\bf 223}, 544 (1983)]"
}

@article{Ai:2023see,
    author = "Ai, Wen-Yuan and Laurent, Benoit and van de Vis, Jorinde",
    title = "{Model-independent bubble wall velocities in local thermal equilibrium}",
    eprint = "2303.10171",
    archivePrefix = "arXiv",
    primaryClass = "astro-ph.CO",
    reportNumber = "KCL-PH-TH/2023-19",
    doi = "10.1088/1475-7516/2023/07/002",
    journal = "J. Cosmol. Astropart. Phys.",
    volume = "07",
    pages = "002",
    year = "2023"
}

@article{vandeVis:2025plm,
    author = "van de Vis, Jorinde and Schicho, Philipp and Niemi, Lauri and Laurent, Benoit and Hirvonen, Joonas and Gould, Oliver",
    title = "{WallGo investigates: Theoretical uncertainties in the bubble wall velocity}",
    eprint = "2510.27691",
    archivePrefix = "arXiv",
    primaryClass = "hep-ph",
    reportNumber = "CERN-TH-2025-221",
    doi = "10.1007/JHEP04(2026)041",
    journal = "J. High Energy Phys.",
    volume = "04",
    pages = "041",
    year = "2026"
}

@article{Ekstedt:2024fyq,
    author = "Ekstedt, Andreas and Gould, Oliver and Hirvonen, Joonas and Laurent, Benoit and Niemi, Lauri and Schicho, Philipp and van de Vis, Jorinde",
    title = "{How fast does the WallGo? A package for computing wall velocities in first-order phase transitions}",
    eprint = "2411.04970",
    archivePrefix = "arXiv",
    primaryClass = "hep-ph",
    reportNumber = "CERN-TH-2024-174, DESY-24-162, HIP-2024-21/TH",
    doi = "10.1007/JHEP04(2025)101",
    journal = "J. High Energy Phys.",
    volume = "04",
    pages = "101",
    year = "2025"
}

@article{Sesana:2025udx,
    author = "Sesana, Alberto and Figueroa, Daniel G.",
    title = "{Nanohertz Gravitational Waves}",
    eprint = "2512.18822",
    archivePrefix = "arXiv",
    primaryClass = "astro-ph.CO",
    doi = "10.1146/annurev-astro-122325-025132",
    journal = "Ann. Rev. Astron. Astrophys.",
    volume = "64",
    number = "1",
    pages = "537--586",
    year = "2026"
}

@article{Ajith:2024mie,
    author = "Ajith, Parameswaran and others",
    title = "{The Lunar Gravitational-wave Antenna: mission studies and science case}",
    eprint = "2404.09181",
    archivePrefix = "arXiv",
    primaryClass = "gr-qc",
    doi = "10.1088/1475-7516/2025/01/108",
    journal = "J. Cosmol. Astropart. Phys.",
    volume = "01",
    pages = "108",
    year = "2025"
}

@mastersthesis{Maki:2025ykv,
    author = {M{\"a}ki, Mika},
    title = "{The effect of sound speed on the gravitational wave spectrum of first order phase transitions in the early universe}",
    eprint = "2511.20436",
    archivePrefix = "arXiv",
    primaryClass = "astro-ph.CO",
    month = "November",
    year = "2025"
}

@article{Guo:2024gmu,
    author = "Guo, Huai-ke and Hajkarim, Fazlollah and Sinha, Kuver and White, Graham and Xiao, Yang",
    title = "{A precise fitting formula for gravitational wave spectra from the sound shell model}",
    eprint = "2407.02580",
    archivePrefix = "arXiv",
    primaryClass = "hep-ph",
    doi = "10.1088/1475-7516/2025/02/056",
    journal = "J. Cosmol. Astropart. Phys.",
    volume = "02",
    pages = "056",
    year = "2025"
}

@article{Barni:2025gnm,
    author = "Barni, Giulio and Blasi, Simone and Madge, Eric and Vanvlasselaer, Miguel",
    title = "{Gravitational waves from the sound shell model: direct and inverse phase transitions in the early Universe}",
    eprint = "2510.21439",
    archivePrefix = "arXiv",
    primaryClass = "hep-ph",
    reportNumber = "DESY-25-144, IFT-UAM/CSIC-25-118",
    doi = "10.1088/1475-7516/2026/03/066",
    journal = "J. Cosmol. Astropart. Phys.",
    volume = "03",
    pages = "066",
    year = "2026"
}

@article{Laurent:2020gpg,
    author = "Laurent, Benoit and Cline, James M.",
    title = "{Fluid equations for fast-moving electroweak bubble walls}",
    eprint = "2007.10935",
    archivePrefix = "arXiv",
    primaryClass = "hep-ph",
    doi = "10.1103/PhysRevD.102.063516",
    journal = "Phys. Rev. D",
    volume = "102",
    number = "6",
    pages = "063516",
    year = "2020"
}

@article{Laurent:2022jrs,
    author = "Laurent, Benoit and Cline, James M.",
    title = "{First principles determination of bubble wall velocity}",
    eprint = "2204.13120",
    archivePrefix = "arXiv",
    primaryClass = "hep-ph",
    doi = "10.1103/PhysRevD.106.023501",
    journal = "Phys. Rev. D",
    volume = "106",
    number = "2",
    pages = "023501",
    year = "2022"
}

@article{vandeVis:2025efm,
    author = "van de Vis, Jorinde and de Vries, Jordy and Postma, Marieke",
    title = "{Bubble trouble: A review on electroweak baryogenesis}",
    eprint = "2508.09989",
    archivePrefix = "arXiv",
    primaryClass = "hep-ph",
    reportNumber = "CERN-TH-2025-161, Nikhef 2025-012",
    doi = "10.1016/j.ppnp.2026.104244",
    journal = "Prog. Part. Nucl. Phys.",
    volume = "150",
    pages = "104244",
    year = "2026"
}

@article{Correia:2025qif,
    author = "Correia, Jos{\'e} and Hindmarsh, Mark and Rummukainen, Kari and Weir, David J.",
    title = "{Gravitational waves from strong first-order phase transitions}",
    eprint = "2505.17824",
    archivePrefix = "arXiv",
    primaryClass = "astro-ph.CO",
    doi = "10.1103/8wmq-f635",
    journal = "Phys. Rev. D",
    volume = "112",
    number = "12",
    pages = "123546",
    year = "2025"
}

@article{Giombi:2024kju,
    author = "Giombi, Lorenzo and Dahl, Jani and Hindmarsh, Mark",
    title = "{Signatures of the speed of sound on the gravitational wave power spectrum from sound waves}",
    eprint = "2409.01426",
    archivePrefix = "arXiv",
    primaryClass = "gr-qc",
    reportNumber = "HIP-2024-19/TH",
    doi = "10.1088/1475-7516/2025/01/100",
    journal = "J. Cosmol. Astropart. Phys.",
    volume = "01",
    pages = "100",
    year = "2025"
}

@article{Giombi:2025tkv,
    author = "Giombi, Lorenzo and Dahl, Jani and Hindmarsh, Mark",
    title = "{Acoustic gravitational waves beyond leading-order in bubble over Hubble radius}",
    eprint = "2504.08037",
    archivePrefix = "arXiv",
    primaryClass = "gr-qc",
    reportNumber = "HIP-2025-13/TH",
    doi = "10.1088/1475-7516/2026/03/024",
    journal = "J. Cosmol. Astropart. Phys.",
    volume = "03",
    pages = "024",
    year = "2026"
}

@inproceedings{RoperPol:2022hxn,
    author = "Roper Pol, Alberto",
    title = "{Gravitational waves from MHD turbulence at the QCD phase transition as a source for Pulsar Timing Arrays}",
    booktitle = "{56th Rencontres de Moriond on Gravitation}",
    eprint = "2205.09261",
    archivePrefix = "arXiv",
    primaryClass = "gr-qc",
    month = "5",
    year = "2022"
}

@inproceedings{RoperPol:2021gjc,
    author = "Roper Pol, Alberto",
    title = "{Gravitational radiation from MHD turbulence in the early universe}",
    booktitle = "{55th Rencontres de Moriond on Gravitation}",
    eprint = "2105.08287",
    archivePrefix = "arXiv",
    primaryClass = "gr-qc",
    month = "5",
    year = "2021"
}

@article{Dahl:2024eup,
    author = "Dahl, Jani and Hindmarsh, Mark and Rummukainen, Kari and Weir, David J.",
    title = "{Primordial acoustic turbulence: Three-dimensional simulations and gravitational wave predictions}",
    eprint = "2407.05826",
    archivePrefix = "arXiv",
    primaryClass = "gr-qc",
    reportNumber = "HIP-2024-16/TH",
    doi = "10.1103/PhysRevD.110.103512",
    journal = "Phys. Rev. D",
    volume = "110",
    number = "10",
    pages = "103512",
    year = "2024"
}

@article{Guo:2023gwv,
    author = "Guo, Huai-Ke and Li, Song and Xiao, Yang and Yang, Jin Min and Zhang, Yang",
    title = "{Estimating the uncertainty of cosmological first order phase transitions with numerical simulations of bubble nucleation}",
    eprint = "2310.04654",
    archivePrefix = "arXiv",
    primaryClass = "hep-ph",
    doi = "10.1103/PhysRevD.110.063541",
    journal = "Phys. Rev. D",
    volume = "110",
    number = "6",
    pages = "063541",
    year = "2024"
}

\label{RealLastPage}

\end{document}